  \documentclass{aa}
   \usepackage{graphicx}
  \begin{document}


  \title{Possible evidence for double precessing nozzle structure in QSO 3C345}
     
  \author{S.J.~Qian\inst{1}}

 \institute{National Astronomical Observatories, Chinese Academy of Sciences,
     Beijing, 100012, China}
 \date{Compiled by using A\&A latex}
 \abstract{The precessing jet-nozzle scenario previously proposed has been
   tentatively applied to interpret the VLBI-kinematics of twenty-seven 
    superluminal components in blazar 3C345 measured during a $\sim$38-year
     period.} {The superluminal components could be divided into two
   groups ascribed to jet-A (13 knots) and jet-B (14 knots), having different
   kinematic behaviors. They could be ejected from a double jet-nozzle system
   forming a double-jet structure.}{Through model-simulation of kinematic
   behavior of the knots, it was tentatively found that both nozzle
    could precess with same period of $\sim$7.30\,yr (4.58\,yr
   in the source frame) and in the same direction. The precession of jet-A
   was simulated over four periods, while that of jet-B was simulated over
   two periods.} {For both jets a steady  precessing common trajectory could
    exist along which different superluminal knots moved according to their 
    precession phases and their kinematics could be well interpreted.
    Most superluminal knots  were model simulated to 
    be accelerated with their bulk Lorentz factor in the range of
    $\sim$4 to $\sim$30. The radio light curves of knot C9 was found to be
    extraordinarily well coincident with its
    Doppler boosting profile, implying that its Lorentz factor and Doppler
    factor profiles were correctly derived and that superluminal 
    components could be recognized
    as relativistic shocks moving along helical trajectories toward us.}
    {The kinematic features observed in 3C345 and interpreted in terms of our 
    precessing-nozzle scenario can be understood in the framework of MHD theory 
   for formation of relativistic jets, although formation of double-jet 
   structure in black hole binaries seems to be a new theoretical field to be 
   investigated. The double precessing nozzle scenario has now been applied to 
   interpret the VLBI-kinematics of 
   superluminal knots for four blazars (3C279, OJ287, 3C454.3 
   and 3C345). The characteristic parameters of the four putative supermassive
    binary black hole systems (including hole masses, mass ratio,orbital
   separation, post-Newtonian parameter, gravitational radiation lifetime, 
   etc.) were tentatively derived and compared, showing that they are in
   physically reasonable ranges, and well consistent with some theoretical
    arguments for close black hole binaries. These results seem providing 
    some enlightening clues that keplerian motion of  supermassive  black 
    hole binaries in blazars  could be discovered through VLBI-observations
    over sufficient long periods.}
  \maketitle
  \section{Introduction}
   3C345 (z=0.595) is an archetypal quasar and one of the best-studied 
  blazars (e.g., Biretta et al. \cite{Bi86}, Hardee et al. \cite{Ha87},
  Homan et al.  \cite{Ho14},  Jorstad et al. \cite{Jo05}, \cite{Jo13},
    \cite{Jo17}, Klare \cite{Kl03}, Klare et al. \cite{Kl05}, Lobanov \& 
    Roland \cite{Lo05}, Qian et al. \cite{Qi91a}, \cite{Qi96}, \cite{Qi09},
    Schinzel et al. \cite{Sc10a}, Schinzel \cite{Sc11a}, Schinzel et al.
    \cite{Sc11b}, \cite{Sc11c}, Steffen et al.
    \cite{St96}, Unwin et al. \cite{Un97}, Zensus et al. \cite{Ze97}). 
   Its emission  spreads over the entire electromagnetic spectrum, from radio,
   IR, optical, UV, X-ray  to high-energy $\gamma$-rays. Prominent and complex
   variabilities in all these wavebands and the spectral energy distributions
   have been extensively monitored and studied, leading to many important
    results on the properties of the
   emitting sources.  Studies of correlation among the variabilities at
    multi-frequencies (from radio to $\gamma$-rays) play an important role. 
    The connection between radio flares and high-energy $\gamma$-rays was
    observed (Schinzel et al. \cite{Sc10b}).\\
   It is a remarkable compact flat-spectrum radio source
    with a relativistic jet, from which superluminal components are emanated.
    VLBI-observations reveal the parsec structures of its jet and track the
    motion of its superluminal components ejected from its radio core.
    It has shown that flaring activities in multi-frequencies (from radio to
    $\gamma$-rays) are closely connected with the jet-activity and ejection of
    superluminal knots. In addition, VLBI-monitoring observations have shown
    that its relativistic jet may be precessing with  quasi-periods.
    This phenomenon may be very important for understanding the properties
    of the central energy engine in its nucleus. The VLBI-kinematics
    of superluminal components in 3C345 has been analyzed by Schinzel
     (\cite{Sc11a}) for a $\sim$30 years period, showing no trend for 
    jet-precession. In this paper we investigate the VLBI-kinematics of 
    3C345 extending to $\sim$38\,yr time-interval, yielding some new 
    significant results.\\
      Since 1991 (Qian et al. \cite{Qi91a}, \cite{Qi09}), we 
    have tried to explain the 
    VLBI-kinematics of superluminal components in 3C345 in terms of a
     precessing nozzle scenario. Our scenario not only considered the 
    precession of the jet-nozzle to explain the position angle swing of its
    superluminal components, but also considered the possible existence of a 
    common (helical) trajectory pattern, which could produce the trajectories
    of the  knots ejected at different times through its precession. 
    It was found that model simulations of the observed trajectories of its
     superluminal knots by using the precession of the common trajectory 
   pattern could quite effectively find the period of jet-nozzle precession.\\
    Helical motion has been widely invoked to interpret the 
    VLBI-kinematics of superluminal components in radio quasars, especially in
    blazars (referring to Perucho et al. (\cite{Peru12a}, \cite{Peru12b}
    : S5 0836+710),  Lister et al. (\cite{Lis13a}, \cite{Lis13b}: 
    BL Lacertae) and Qian et al. (\cite{Qi21}, \cite{Qi09}). 
    Cohen et al (\cite{Co14},
    \cite{Co15}) introduced the concept of relativistic Alfv\'en waves to
    explain jet structure and structural evolution (BL Lacertae). 
   Some authors suggested that jet instabilities (e.g. Kelvin-Helmholtz
     instability) could play significant role in forming helical trajectories 
   in outer jet regions (e.g., Perucho \cite{Peru12c}, Schinzel et al. 
   \cite{Sc10a}, \cite{Sc11b}).\\
    Our precessing nozzle scenario has been previously applied to 
    analyze the VLBI-kinematics of superluminal knots in 
    several QSOs, e.g., 3C279, B1308+328, PG1302+202, NRAO150, 3C454.3 and 
    OJ287 (Qian et al. \cite{Qi14}, \cite{Qi17}, \cite{Qi18a}, \cite{Qi19a}, 
    Qian \cite{Qi13}, \cite{Qi16},
     \cite{Qi18b}). These studies revealed that jet-nozzle 
    precession  may exist in these sources. Through model-fitting of
    the VLBI-kinematics in terms of the precessing nozzle scenario, possible
    periods of precession  and other kinematic parameters for the superlumial
    knots (bulk Lorentz factor, viewing angle, apparent velocity and  
    Doppler factor vs time) were derived. In particular, in two cases 
    (3C279 and OJ287) possible evidence
    has been obtained that double-jet systems might exist in their nuclei,
    which could be mostly produced  by binary black hole/accretion-disk systems.
    In the case of OJ287, we might speculate that its quasi-periodic optical
    variability is connected with its double-jet activity (Villata et al.
   \cite{Vi98}, Qian \cite{Qi18b}, \cite{Qi19b}, \cite{Qi19c}, \cite{Qi20};
   also referring to Qian et al. \cite{Qi07}).\\
     Search for periodicities in optical and radio light-curves
   (e.g. Sillanp\"a\"a et al. \cite{Si88}, Babadzhanyants et al. \cite{Ba95},
     Kudryavtseva et al.\cite{Ku06}, Qian et al. \cite{Qi07})  are important
    and could provide key information on the nature of the central engine
     in blazars.\\
     The position angle swings of superluminal components on parsec-scales
    observed by VLBI-monitoring observations could be used to search for
    periodicities in ejection of superluminal knots (e.g., Britzen et al.
    \cite{Br01}, Tateyama \& Kingham \cite{Ta04}, Klare \cite{Kl05},
    Schinzel et al. \cite{Sc12b}, Qian et al. \cite{Qi09}).\\
     In the case of QSO 3C345 and  based on the position angle swing of 
    its superluminal components, some authors argued for the existence of
     a jet precessing period: e.g., $\sim$8-10\,yr (Lobanov \& Zensus
     \cite{Lo99}, Klare et al. \cite{Kl05},
     Klare \cite{Kl03}), $\sim$9.5\,yr (Lobanov \& Roland \cite{Lo05}).\\ 
      In earlier studies we already found that the observed tracks of
    knots C4 and C5 could be reproduced by the rotation of a common helical
    trajectory (Qian et al. \cite{Qi91a}, \cite{Qi91b}, Qian \& Zhang
     \cite{Qi99}).
    Qian et al. (\cite{Qi09}) analyzed the  distribution of the position 
    angles for seven superluminal components (C4 to C10) at different core 
    separations of 0.15\,mas, 0.20\,mas and 0.25\,mas, and found that their 
     inner trajectories (within core separation 
    $r_n{\stackrel{<}{_\sim}}$0.4\,mas) could be explained in terms of the
    precession of a common trajectory and  a precession period of its
    jet-nozzle of $\sim$7.36\,yr was derived. \footnote{It is noted that
    this is an averaged value: according to equations (15), (16) and (17)
     of that paper,  a precession period of 7.44\,yr, 7.34\,yr and 7.31\,yr
    were obtained for core distance 0.10\,mas, 0.15\,mas and 0.20\,mas,
     respectively .}\\
    In this paper we  further analyze the kinematics of 27 superluminal
     knots, spreading over a time-range of $\sim$38 years (1980--2018) and 
    show that the kinematics of these superluminal knots could be consistently
    explained in terms of our precessing-nozzle scenario, if   a double-jet
     system is assumed to be existing in its nucleus. Obviously, if this
     result is verified QSO 3C345 should host a binary black hole system 
    in its nucleus.\\
    We point out that our precessing jet-nozzle scenario is well 
    consistent with the magnetohydrodynamic theories for the formation and 
    collimation of relativistic jets in AGN (e.g., Blandford \& Payne 
    \cite{Bl82}, Blandford \& Znajek \cite{Bl77}, Camenzind \cite{Ca86},
   \cite{Ca87}, \cite{Ca90}, Li et al. \cite{Li92},
    Lovelace et al. \cite{Lo86},Meier 
    \& Nakamura \cite{Me06}, Nakamura \& Asada \cite{Na13},
     Valhakis \& K\"onigl \cite{Vl03}, \cite{Vl04}).\\
   \section{Observational data}
     In this paper we made use of the data collected from the literature:
    (1) Data presented in Klare (\cite{Kl03}; for period 1980.5-2001.9);
    (2) Data presented in Schinzel (\cite{Sc11a}; for period 1980.5-2010.8).
     Part of the data were re-calculated to make the compact core at 
    the unified  origin of coordinates;
    (3) Data kindly provided by Jorstad (private communication); for period
      2011.1-2018.8). This dataset extended the time-interval for 
      our model fitting of the kinematics in 3C345 to $\sim$38 years 
       (about five  precession periods).\\
     Considering the core-shift effects we only used 43GHz and 22GHz 
    observational data (except for knot C5,
   for which only 15GHz data are available). Generally, we would not mark 
   the observational errors for the positions of individual knots in figures
     obtained from model-fittings for clarity, but one should keep
    in mind that errors in measurements of knot's position are in the range of 
   $\sim$0.05--0.1\,mas.\\
    We will apply the concordant cosmological model (Spergel et al.\cite{Sp03},
    Hogg \cite{Hog99})
   with ${\Omega}_{\lambda}$=0.73 and ${\Omega}_m$=0.27, and 
    $H_0$=71\,km${s^{-1}}{{Mpc}^{-1}}$. Thus 
   the luminosity distance of 3C345 is $D_L$=3.49Gpc, angular-diameter distance
    $D_a$=1.37\,Gpc, 1\,mas=6.65\,pc, 1\,mas/yr=34.6c. 1\,c is equivalent to
    an apparent angular velocity 0.046 mas/yr.\\
    \begin{figure*}
    \centering
    \includegraphics[width=10cm,angle=90]{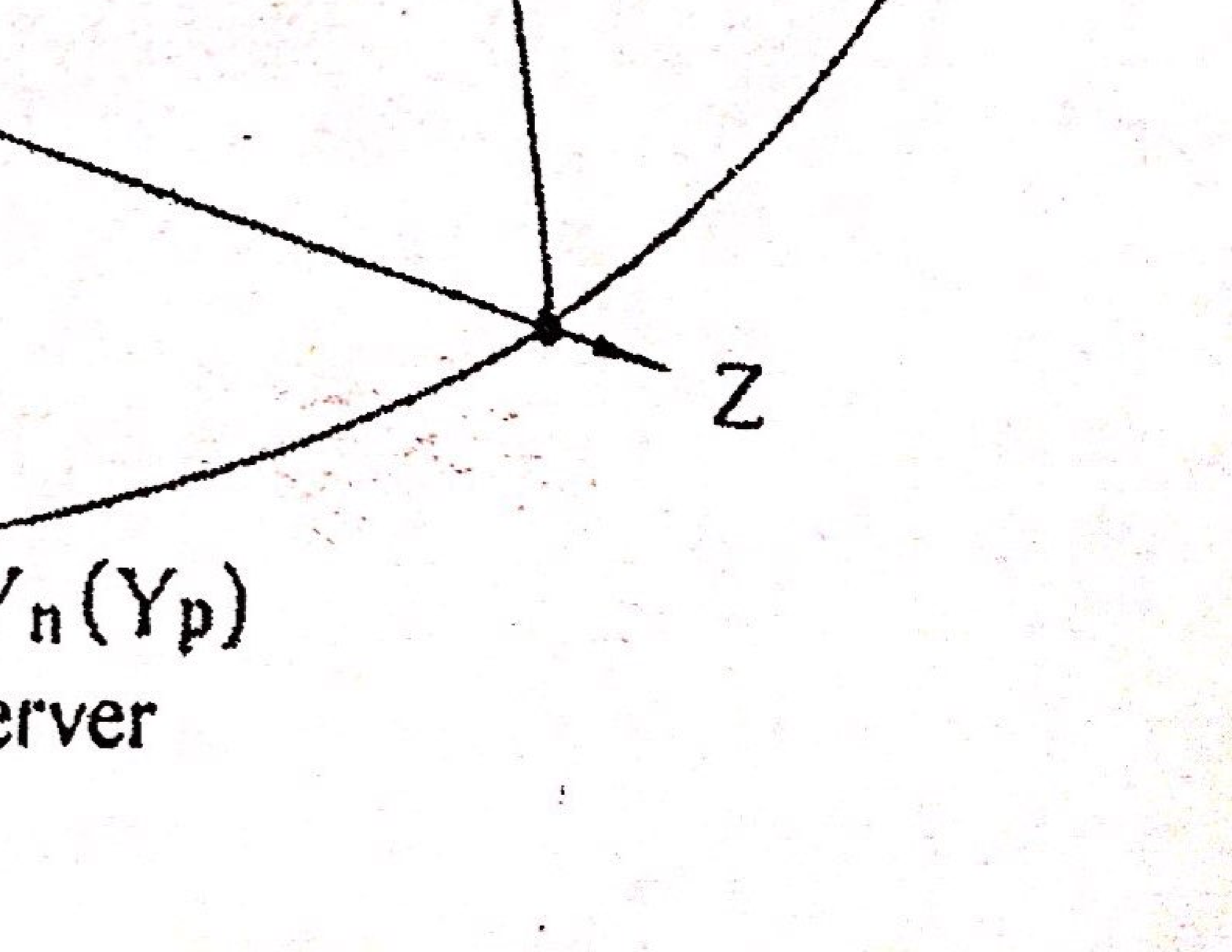}
    \caption{Geometry of the precessing jet-nozzle scenario for 3C345. 
    The jet-axis is defined in the $(X,Z)$-plane by parameters
    ($\epsilon$, $\psi$) and function $\it{x_0(z_0)}$. The common helical 
    trajectory pattern is defined by functions A(Z) and $\phi$(Z) given 
    in Section 3.}
    \end{figure*}
    \begin{figure*}
    \centering
    \includegraphics[width=5.5cm,angle=-90]{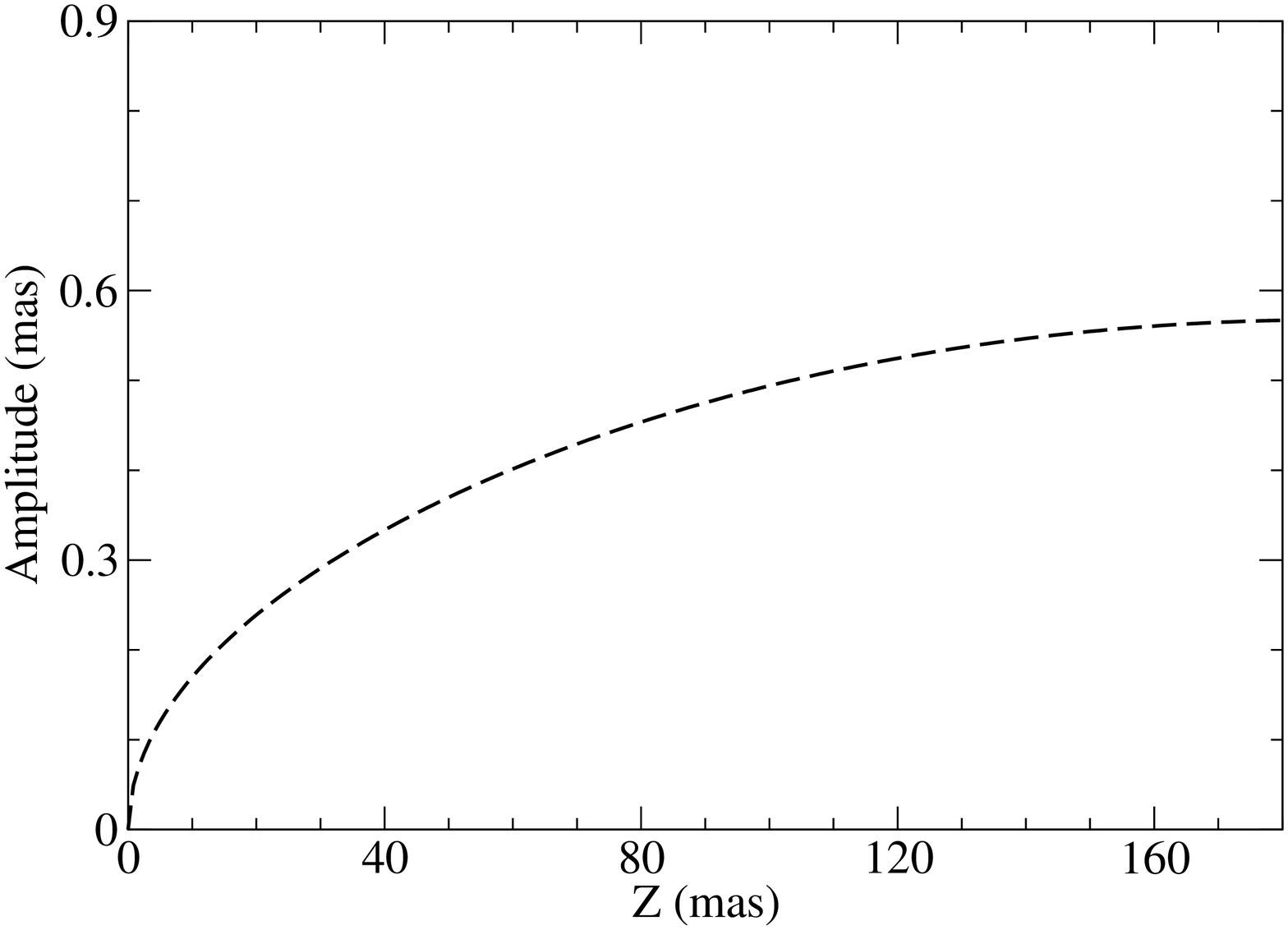}
    \includegraphics[width=5.5cm,angle=-90]{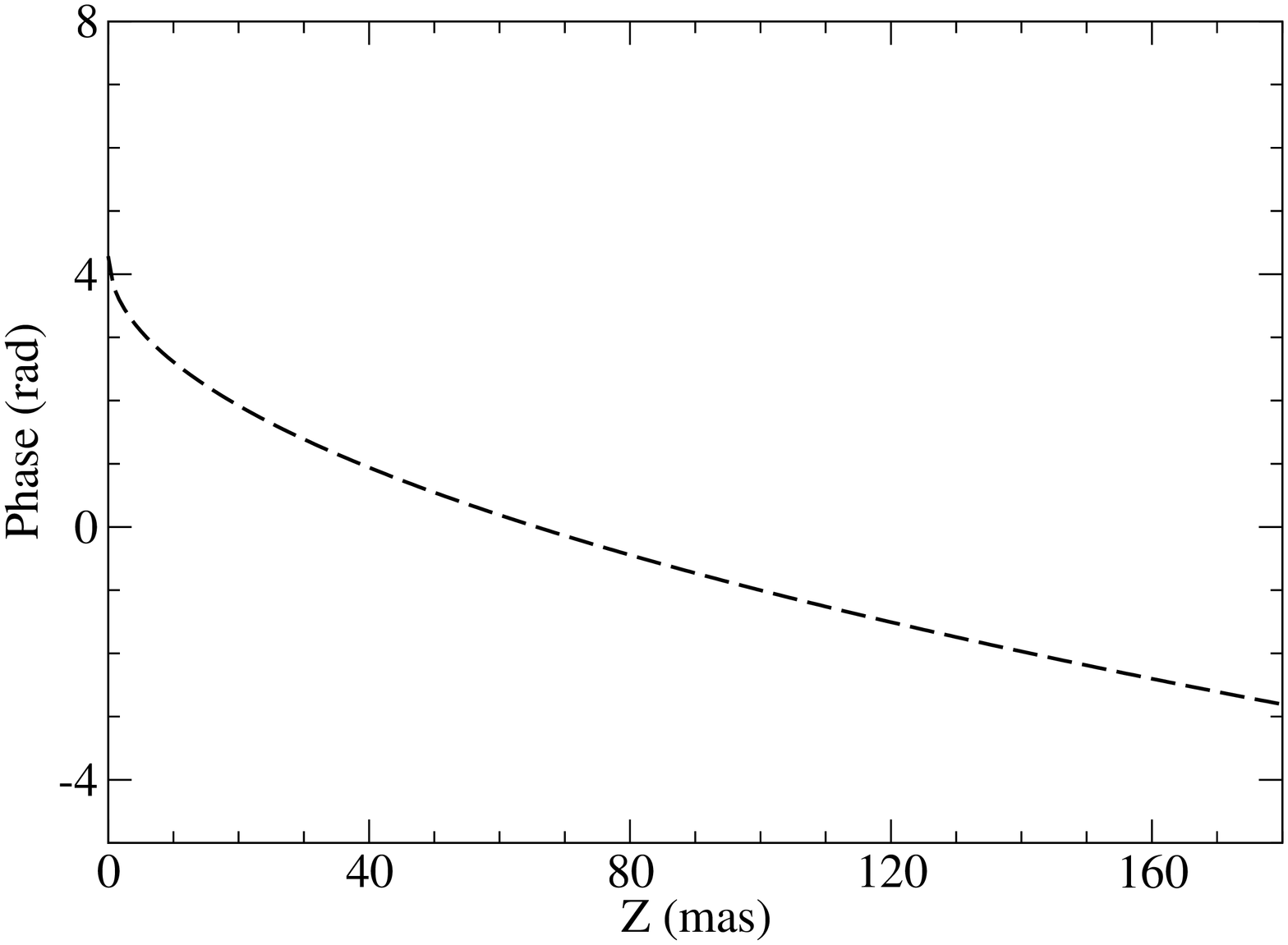}
    \caption{Knot C4. Model functions describing its helical trajectory:
    amplitude A(Z) and phase $\phi$(Z).}
    \end{figure*}
    \begin{figure*}
    \centering
    \includegraphics[width=5.5cm,angle=-90]{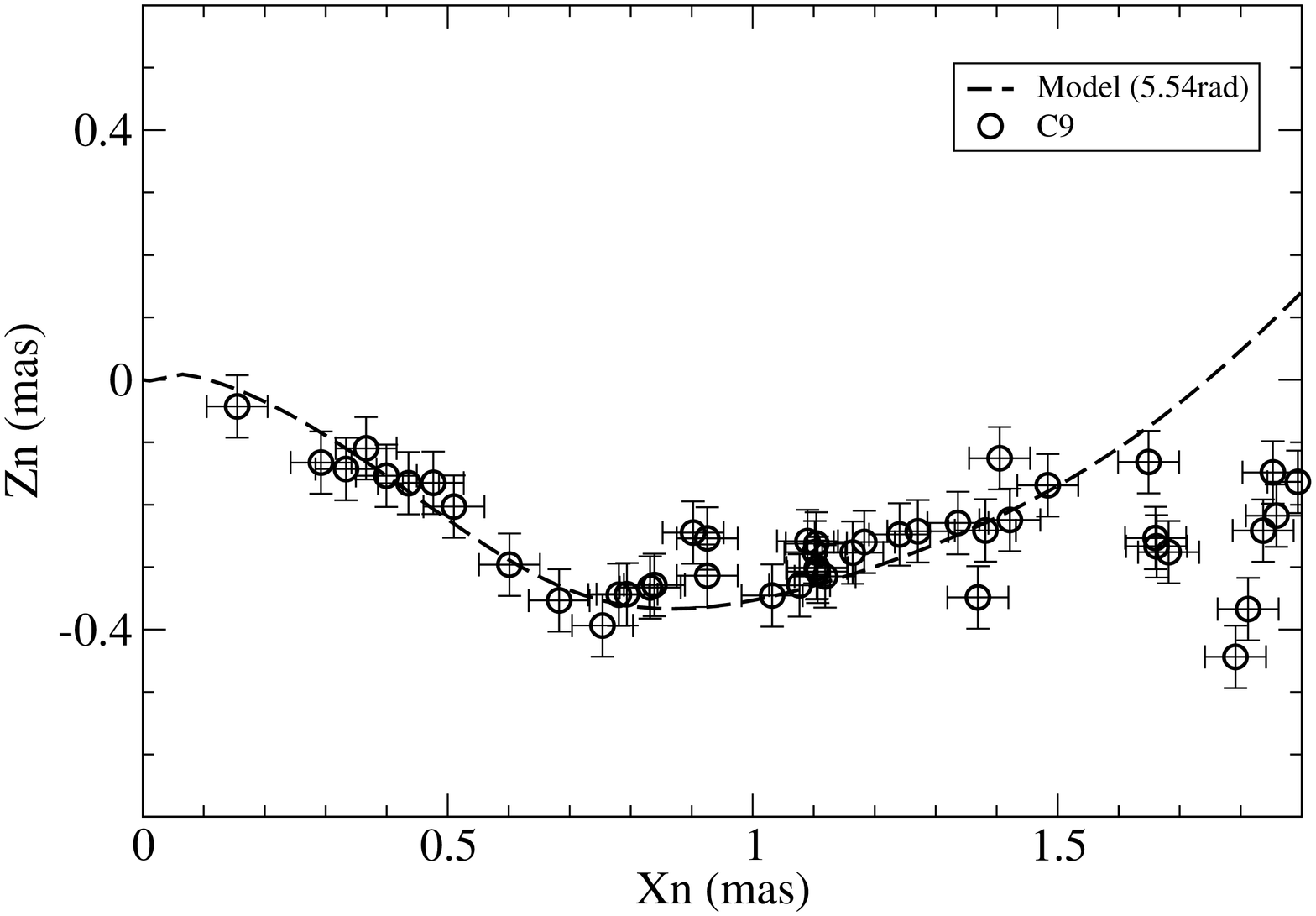}
    \includegraphics[width=5.5cm,angle=-90]{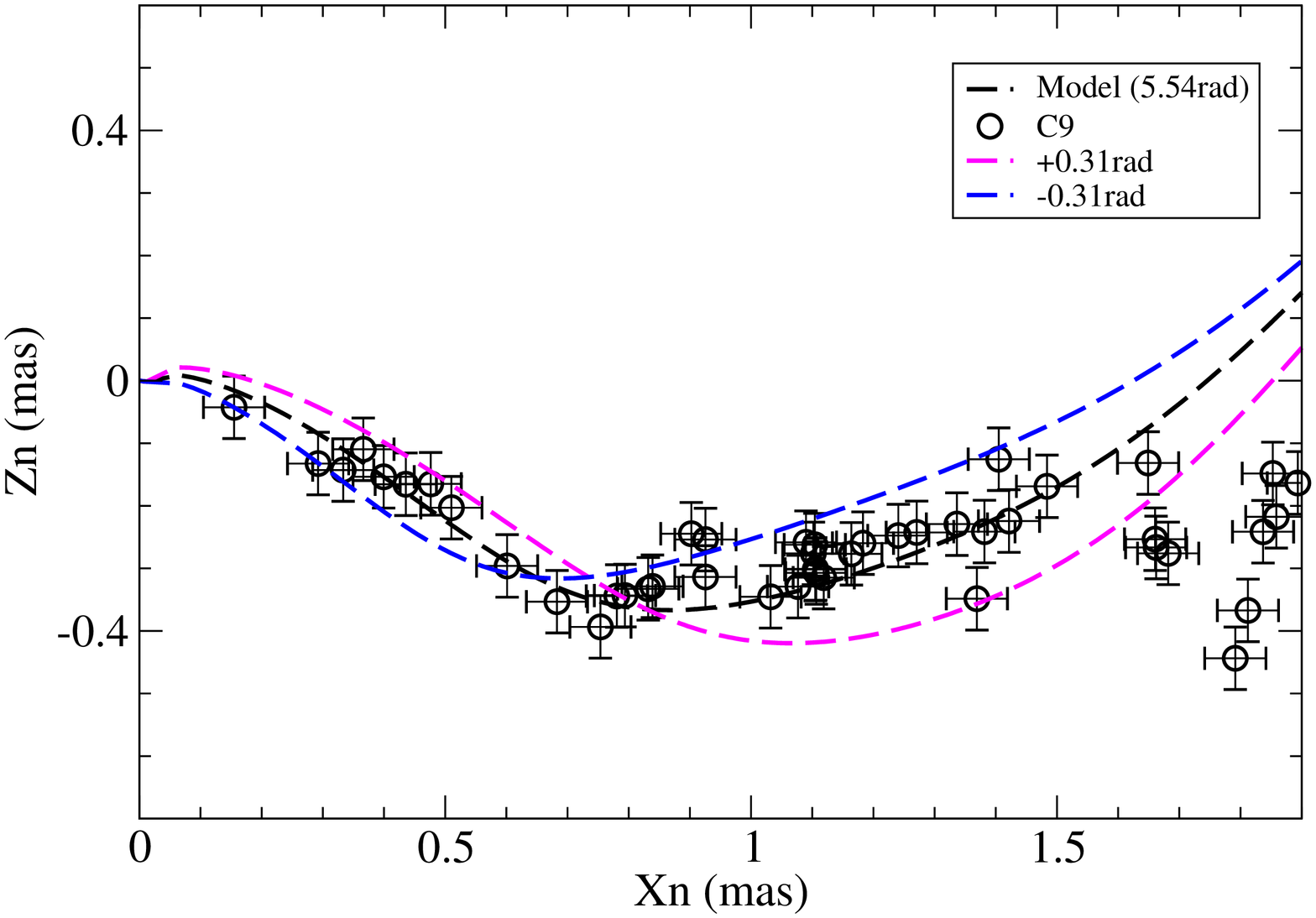}
    \caption{Model fit of the trajectory of knot C9 with the precession
    phase $\phi_0$=5.54\,rad+4$\pi$, equivalent to ejection time 
     $t_0$=1995.06. Left panel shows its curved trajectory being well 
    model-fitted till coordinate $X_n$$\sim$1.6\,mas. Right panel shows that
    the precessing common trajectories (in magenta and blue) defined 
    by $\pm$5\% of its precession
     phase could well demonstrate the accuracy for the trajectory model-fits.
     This is a new criterion introduced in this paper for 
     judging the trajectory model-fitting accuracy,
     which well express the quality of model-simulation for the trajectories
      of the superluminal knots in blazar 3C345 (see text). For  a simple 
     and clear presentation 
     the errors in data-points were taken to be $\pm$0.05\,mas. Obviously,
     errors in individual data-points did not affect the estimation of the 
     quality of the trajectory model-fits.}
    \end{figure*}
   \section{Geometry of the model}
    We will apply the precessing jet-nozzle model previously proposed by 
    Qian et al. (\cite{Qi19a}, also \cite{Qi09}, \cite{Qi91a}) to investigate
    the kinematics and distribution of trajectory of superluminal components
    on parsec scales in the QSO 3C345.\\
     We will use a special geometry consisting of four coordinate systems as 
    shown in Figure 1. We assume that the superluminal components move along
    helical trajectories around the curved jet axis (i.e. axis of the helix).\\
    We use coordinate system ($X_n,Y_n,Z_n$) to define the plane of the sky 
    ($X_n,Z_n$) and the direction of observer ($Y_n$), with $X_n$-axis pointing
    toward the negative right ascension and $Z_n$-axis toward the north pole.\\
    We use coordinate system ($X,Y,Z$) to locate the curved jet-axis in the 
    plane ($X,Z$), where $\epsilon$ represents the angle between $Z$-axis and
    $Y_n$-axis and $\psi$ the angle between $X$-axis and $X_n$-axis. 
    Thus parameters $\epsilon$ and $\psi$ are used to define the plane where 
    the jet-axis locates relative to the coordinate system ($X_n,Y_n,Z_n$).\\
    We use coordinate system (${\it{x}}'$,${\it{y}}'$,${\it{z}}'$) along the
     jet-axis to define the helical trajectory pattern for a knot, introducing
    parameters $A(s_0)$ (amplitude) and $\phi(s_0)$ (phase), where $s_0$
    represents the arc-length along the axis of helix (or curved jet-axis).
    ${\it{z}}'$-axis is along the tangent of the axis of helix.
    ${\it{y}}'$-axis is parallel to the $Y$-axis and $\eta$ is the angle
    between ${\it{x}}'$-axis and $X$-axis (see Figure 1).\\
      In general, we assume that the jet-axis can be defined by a function
     $x_0(z_0)$ in the $(X,Z)$-plane as follows.
    \begin{equation}
    {x_0}=p({z_0}){{z_0}^{\zeta}}
    \end{equation}
     where
    \begin{equation}
    p({z_0})={p_1}+{p_2}[1+\exp(\frac{{z_t}-{z_0}}{{z_m}})]^{-1}
    \end{equation}
     $\zeta$, $p_1$, $p_2$, $z_t$ and $z_m$ are constants.
    \begin{equation}
    {s_0}=\int_{0}^{z_0}{\sqrt{1+(\frac{d{x_0}}{d{z_0}})^2}}{d{z_0}}
    \end{equation}
    Therefore, the helical trajectory of a knot can be described in the (X,Y,Z)
    system as follows.
    \begin{equation}
    X({s_0})=A({s_0}){\cos{\phi({s_0})}}{\cos{\eta({s_0})}}+{x_0}
    \end{equation}
    \begin{equation}
    Y({s_0})=A({s_0}){\sin{\phi({s_0})}}
    \end{equation}
    \begin{equation}
    Z({s_0})=-A({s_0}){\cos{\phi({s_0})}}{\sin{\eta({s_0})}}+{z_0}
    \end{equation}
    where $\tan{\eta({s_0})}$=$\frac{d{x_0}}{d{z_0}}$. The projection of
   the helical trajectory on the sky-plane (or the apparent trajectory)
     is represented by
   \begin{equation}
    {X_n}={X_p}{\cos{\psi}}-{Z_p}{\sin{\psi}}
    \end{equation}
    \begin{equation}
    {Z_n}={X_p}{\sin{\psi}}+{Z_p}{\cos{\psi}}
    \end{equation}
    where
    \begin{equation}
       {X_p}=X({s_0})
    \end{equation}
    \begin{equation}
     {Z_p}={Z({s_0})}{\sin{\epsilon}}-{Y({s_0})}{\cos{\epsilon}}
    \end{equation}
    (All coordinates and amplitude (A) are measured in units of mas).
    Introducing the functions
    \begin{equation}
    {\Delta}=\arctan[(\frac{dX}{dZ})^2+(\frac{dY}{dZ})^2]^{-\frac{1}{2}}
    \end{equation}
   \begin{equation}
    {{\Delta}_p}=\arctan(\frac{dY}{dZ})
    \end{equation}
    \begin{equation}
    {{\Delta}_s}=\arccos[(\frac{dX}{d{s_0}})^2+(\frac{dY}{d{s_0}})^2+
                        (\frac{dZ}{d{s_0}})^2]^{-\frac{1}{2}}
    \end{equation}
    we can then calculate the viewing angle $\theta$, apparent transverse
    velocity ${\beta}_a$, Doppler factor $\delta$ and the elapsed time T,
    at which the knot reaches distance $z_0$ as follows: 
    \begin{equation}
     {\theta}=\arccos[{\cos{\epsilon}}(\cos{\Delta}+
               \sin{\epsilon}\tan{{\Delta}_p})]
    \end{equation}
    \begin{equation}
     {\Gamma}=(1-{\beta}^2)^{-\frac{1}{2}}
    \end{equation}
    \begin{equation}
    {\delta}=[{\Gamma}(1-{\beta}{\cos{\theta}})]^{-1}
    \end{equation}
    \begin{equation}
     {{\beta}_a}={{\beta}{\sin{\theta}}/(1-{\beta}{\cos{\theta}})}
    \end{equation}
    \begin{equation}
    T=\int^{{s_0}}_{0}{\frac{(1+z)}{{\Gamma\delta}{v}{\cos{{\Delta}_s}}}}
                       {d{s_0}}
    \end{equation}
    The amplitude and phase of the helical trajectory for superluminal  knots 
   are defined as follows (Figure 2).
    \begin{equation}
    {A({Z})} = {{A_0}[\sin({\pi}{Z}/{Z_1})]^{1/2}}{\exp(-{Z}/{Z_2})} 
    \end{equation}
    \begin{equation}
    {\phi}({Z})={{\phi}_0}-{({Z}/{Z_3})^{1/2}}
     \end{equation}
      $A_0$ represents the amplitude coefficient of the common helical
     trajectory pattern and  ${\phi}_0$ is the precession phase of an 
     individual knot, which is related to its ejection time (see below).\\
     The aim of our model fitting of the kinematics of the superluminal
     components observed in 3C345 is to show that most components have
    their observed trajectories following the precessing common trajectory 
    and their kinematics can be interpreted in terms of our precessing
    jet nozzle scenario, indicating the possible presence of a supermassive
    black hole binary in its nucleus.
    \section{Model-fitting results for jet-A}
    It was found that 3C345 might comprise a double precessing jet structure:
    jet-A and jet-B. The former consists of knots C4--C14, C22 and C23 and
    the latter comprises knots C15, C15a, C16--C21, B5--B8, B11 and B12.
    Both jets precess
    with the same period of 7.30\,yr and in the same direction: anti-clockwise 
    seen along the line of sight.\\
    \begin{figure*}
   \centering
   \includegraphics[width=5.5cm,angle=-90]{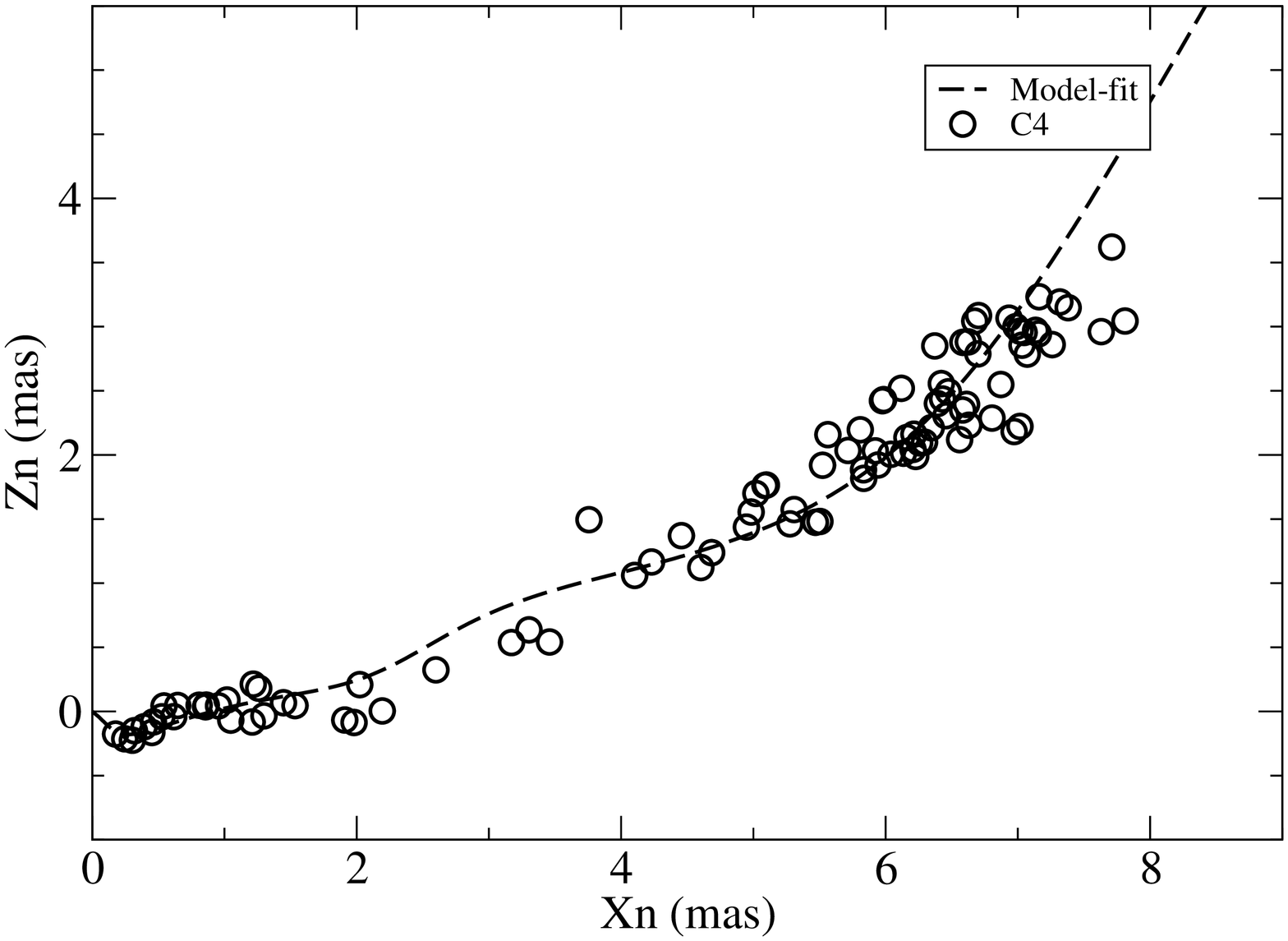}
   \includegraphics[width=5.5cm,angle=-90]{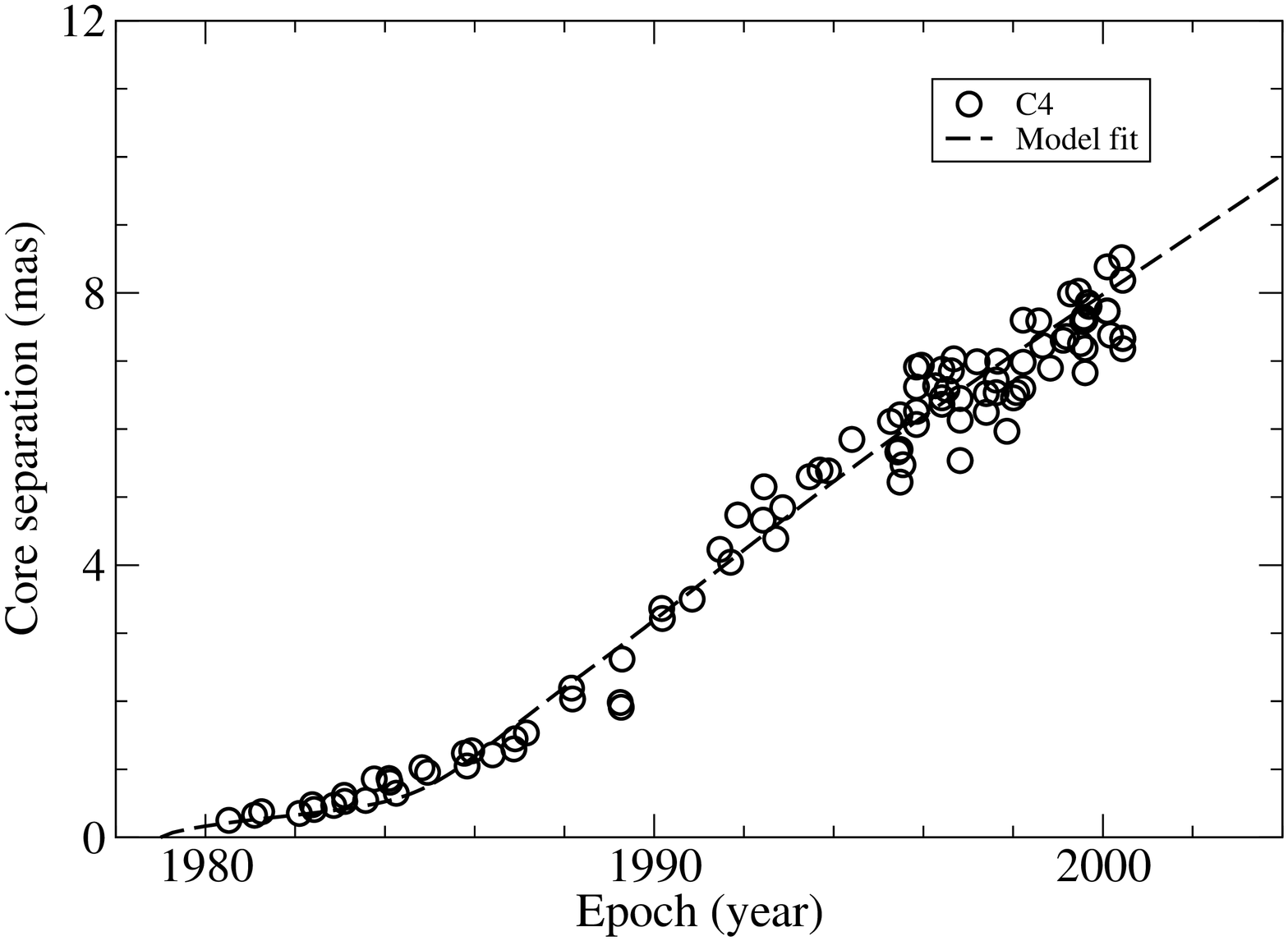}
   \includegraphics[width=5.5cm,angle=-90]{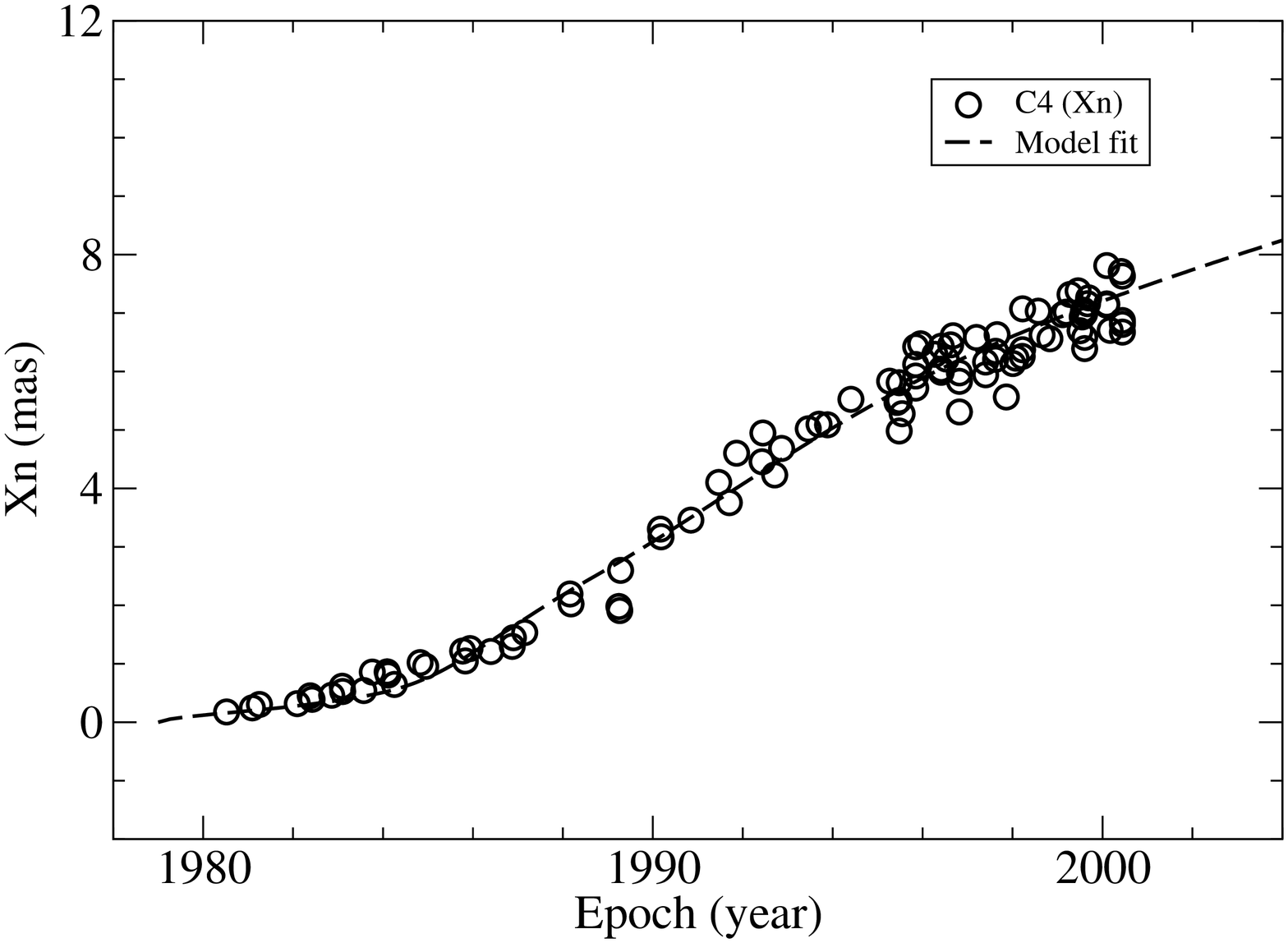}
   \includegraphics[width=5.5cm,angle=-90]{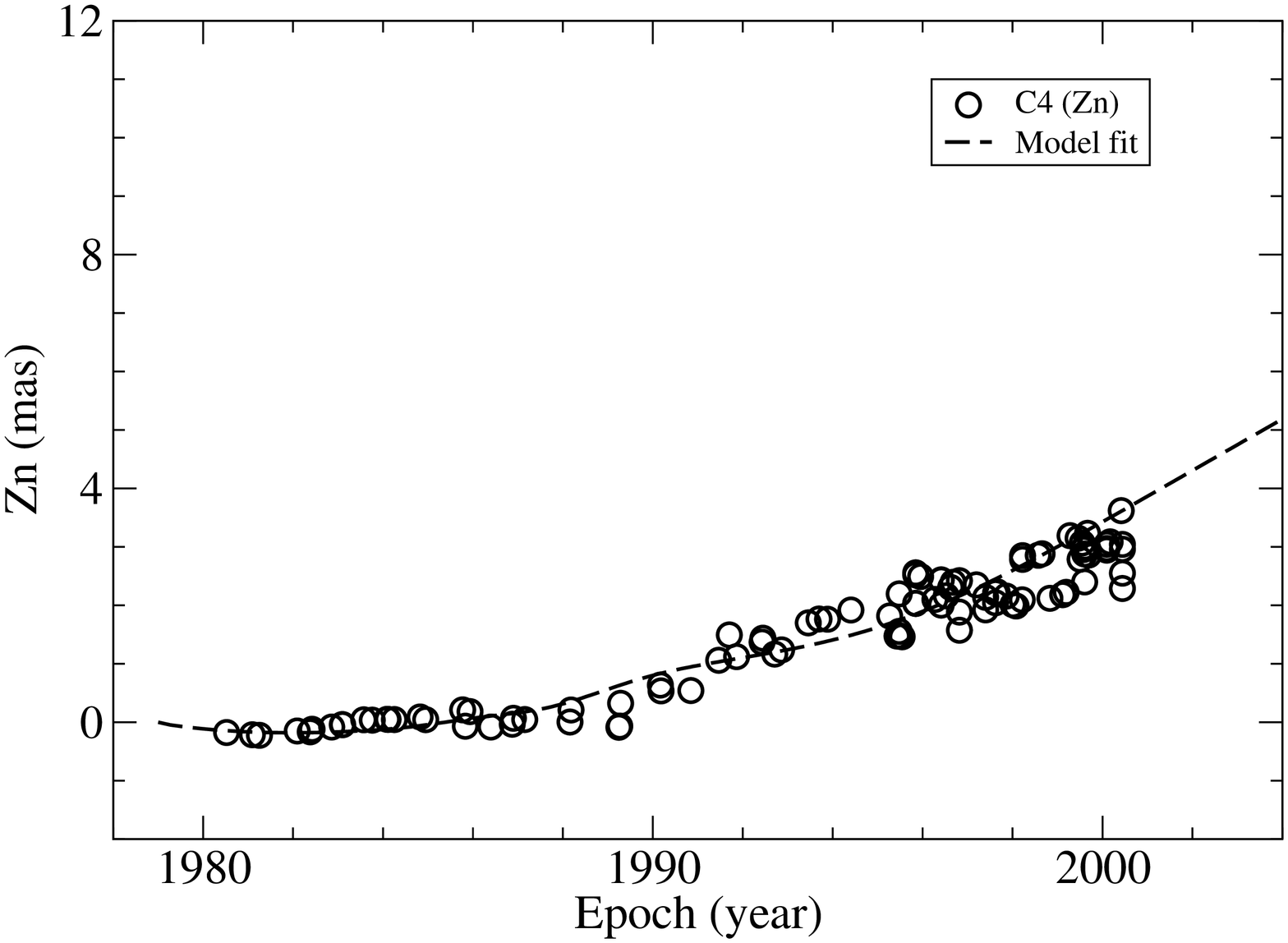}
   \includegraphics[width=5.5cm,angle=-90]{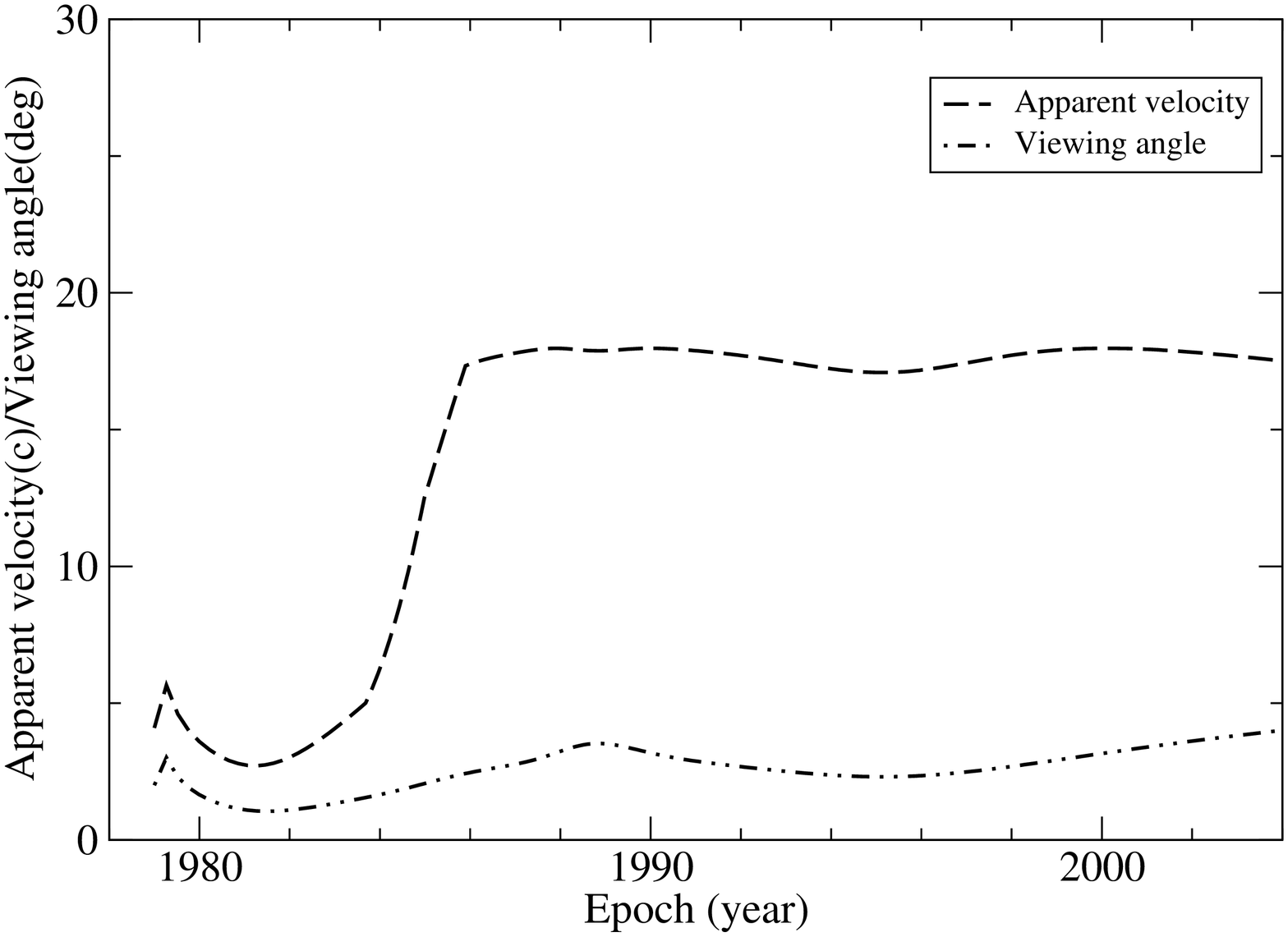}
   \includegraphics[width=5.5cm,angle=-90]{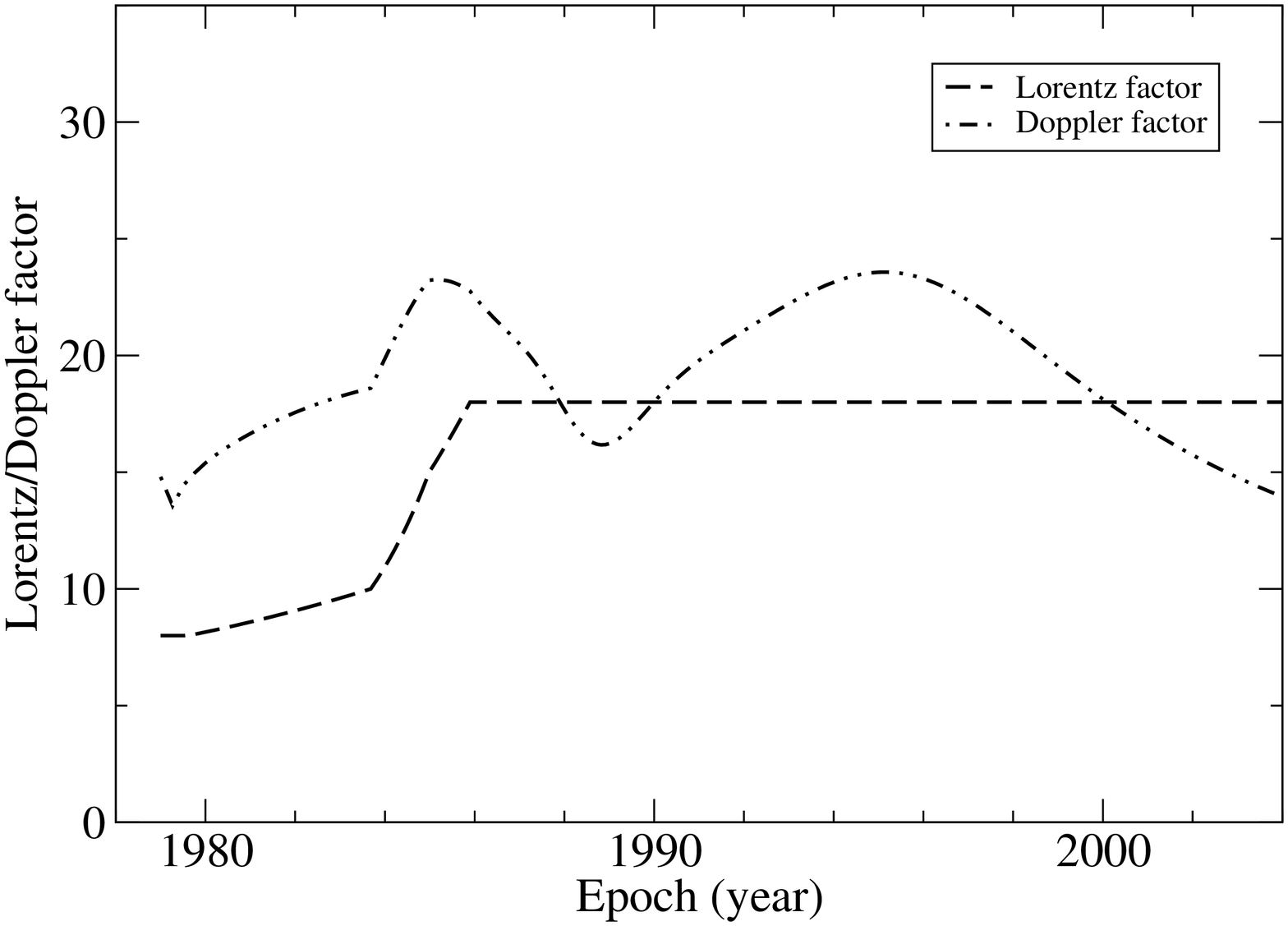}
   \caption{Model-fitting of the entire kinematic behavior within core 
   separation of $\sim$8\,mas for knot C4: trajectory $Z_n$($X_n$), 
    coordinates $X_n$(t) and $Z_n$(t), core separation $r_n$(t), and 
    modeled apparent velocity $\beta_a$(t), viewing angle $\theta$(t),
    bulk Lorentz factor $\Gamma$(t) and Doppler factor $\delta(t)$.
    Due to using appropriate mathematical formulas, functions and
    model-parameters for describing the curved jet-axis and its helical 
    trajectory, th entire trajectory of knot C4 is well model-fitted till
    $\sim$8\,mas from the core.}
   \end{figure*}
       As shown in Section 3, in our precessing jet-nozzle scenario 
    the jet-axis is defined in the ($X,Z$)-plane 
    by parameters ($\epsilon$, $\psi$) and formulas (1) and (2). For jet-A 
    we assume the following  parameters:\\
      $\epsilon$=0.0349\,rad=$2^{\circ}$; $\psi$=0.125\,rad=$7.16^{\circ}$; 
      $\zeta$=2.0, $p_1$=0; $p_2$=1.34$\times{10^{-4}}$; $z_t$=66\,mas; 
     $z_m$=6\,mas.\\
     The amplitude of the helical trajectory is defined by formula (19)
       in Section 3. For jet-A we assume the following parameters:
      $A_0$=0.605\,mas, $Z_1$=396\,mas; $Z_2$=3000\,mas. \\
     The phase of the helical trajectory is defined in Section 3 by
     formula (20). For jet-A we assume $Z_3$=3.58\,mas and the precession
     phase $\phi_0$ is related to the ejection time of a knot as follows:
      \begin{equation}
       {\phi_0}=4.28+{\frac{2\pi}{T_0}}({t_0}-1979.00)
      \end{equation}
      where $T_0$=7.30\,yr--precession period of the jet-A nozzle.
     \subsection{A note on the model-parameters}
      In our previous studies of the VLBI-kinematics of superluminal
      components  in blazar 3C345 we found three distinct features:
     (1) the motion of some knots could be model-simulated as moving along
     helical trajectories (Qian et al. \cite{Qi91a}); (2) there could exist
     a common trajectory which precesses with a period of $\sim$7.4 \,yr, 
     producing the trajectories for different knots at corresponding 
     precession phases (Qian et al. \cite{Qi09}); (3) there were some 
     clues showing that the knots could possibly be divided into two groups
     having different kinematic behaviors. Thus we needed some new methods 
     to further investigate the VLBI-kinematics of superluminal knots in 
     3C345.\\ 
     In order to interpret the kinematics of superluminal components in 3C345
    in terms of our precessing nozzle scenario, we used model-simulation
      methods to model-fit the observed trajectories of its 27 superluminal
     components. Thus a large amount of model parameters were involved:
     e.g., parameters for describing the jet-direction in space, jet-cone size
     and curved jet-axis; jet precession period, ejection time and trajectory
     pattern of the 27 superluminal components, etc.   \footnote{Specifically,
     the model-parameters include $\epsilon, \psi$, $p_1, p_2, \zeta$, $t_0$
     (ejection time), A and $\phi$ (defining helical trajectory), and so on.}
     Based on the formulation 
      (Section 3) of the precessing jet-nozzle model (more details 
     referring to the recent paper (for 3C454.3) published in A\&A:
      Qian et al. \cite{Qi21}) and through trial and error\footnote{
      "Trial and error" method may be the most feasible and effective one 
     to disentangle the two precessing jets and to find appropriate 
     model-parameters to describe their geometric features 
     (jet direction, jet-cone size, precession period, phase distribution
     and direction of nozzle-precessing)
     and kinematic properties (helical trajectory pattern (mathematical
     function) and  bulk Lorentz factor, etc.). It seems very difficult to
     use usual statistical methods to fulfill this complex task.} 
      we found two specific sets of model parameters
     (one for jet-A and other for jet-B) and
     associated functions to model-fit (or model-simulate) the kinematics of
     its superluminal knots in terms of the precessing nozzle scenario 
     (Qian et al. \cite{Qi19a}, \cite{Qi21}). \\       
     In this work we made two assumptions: (1) jets in blazars have 
     the distinct feature: precession; (2) superluminal components move
     along certain precessing common trajectory. These assumptions  
     greatly decreased the number of model-parameters describing the 
    trajectory patterns and ejection times of the 27 superluminal knots.\\     
     We would like to indicate that the values selected for the model 
     parameters and associated functions for both jet-A and jet-B are not 
     statistical samples and not unique either. They are specific and 
     physically appropriate and applicable sets of working ingredients. 
     However, as shown in the main text below  the model-simulation
     methods with these specific model-parameter values could be well applied
    to analyze the distribution of the observed trajectories and kinematics of 
     superluminal components in blazar 3C345 on VLBI-scales, especially 
     discovering the possible separation of its superluminal knots into 
     two groups with different kinematic properties. By using the 
     model-simulation  methods we reached our aims: (1) seeking
      for possible jet-precession;
     (2) searching for double-jet structure; (3) disentangling the 
     observed superluminal components into two groups ascribed to respective
     jets;  (4) determining precession periods;
     and (5) studying the properties of the putative supermassive binary 
     black holes in its nucleus. \\
     Similar methods
     have also been applied to blazar 3C279 (Qian et al. \cite{Qi18a}), OJ287
     (Qian \cite{Qi18b}) and 3C454.3 (Qian et al. \cite{Qi21}). Interestingly,
     we found that the four blazars could all have double-jet structure with 
     their jets precessing with the same period in the same direction.\\ 
     Since we dealt with model-fittings (or model simulations) of the 
     kinematics of superluminal components involving multiple parameters and 
     functions,
      mainly involving the model fits of the observed 
     trajectories of the 27 knots, 
      we introduced a new criterion to judge the validity of the
     model-fitting results. That is, a reasonable and effective model-fit
     was required to satisfy the 
     condition: its observed trajectory (or the related data-points) 
     had to be fitted to follow the precessing common trajectory
     predicted by the scenario within $\pm$5$\%$ of the precession period.
     Taking the model-fit of the trajectory of knot C9 as an example,
      in Figure 3 two plots are shown: the left panel represents that 
     the observed trajectory marked by observational errors was well fitted 
      by the model trajectory. The observed data-points were well
     concentrated around the model trajectory. The right panel represents 
     that the observed trajectory was well within the region limited
      by the model trajectories defined by the $\pm$5\% of the precession
      period. It can be seen that since the model
     fitted the entire trajectory of C9, fitting quality of the entire 
     observed trajectory  was determined by the systematic deviations of 
     the data-points
     from the model trajectory (or the concentration of 
     the data-points relative to the model trajectory), not much
     depending on the observational errors of individual data-points.
     \footnote{So the figures showing the model-fitting of the observed 
     trajectories are not marked with observational errors of the data-points.
     Error estimate of $\pm$5\% of precession period is an effective
     criterion for judging the accuracy of trajectory model-fits. 
     } Thus the new  criterion is a simple and quite effective one 
     for assessing the validity of the  model-simulation results as a whole.
      The model-fit of the trajectory for knot C9 
     is a good example, where almost all the observational data-points were
      well within the region defined by the common trajectories 
     at $\pm$5\% precession period. Obviously, adopting $\pm$3\% of period 
     would let quite a number of data-points becoming outliers. Adopting 
      $\pm$10\% of period would let the derived precession-period having
      larger error. In the case of adopting +/-5\% of period, more than
      85\% of the knots had their trajectories well 
      model-simulated. (See the [status column] in Table 1 and Table 2 
     below). This is a very high probability of success.\\ 
     Thus, based on our precessing nozzle scenario, possible evidence for 
     double  precessing relativistic jets and a putative 
     black-hole binary in 3C345
     could be tentatively investigated. The kinematics of the superluminal 
     components
     of both jets have been well model-fitted and  a precession period of 
     7.30$\pm$0.36\,yr for both the jet-nozzles has been derived.  We 
     emphasize that these results should be tested by  VLBI-observations
     in the future years.\\
     \subsection{Entire kinematic behavior's model-simulation of knot C4}
      We first discuss the model-fitting of the entire kinematics of knot 
     C4 within the core separation $\sim$8\,mas, including its trajectory, and
     core separation, coordinates, apparent velocity, viewing angle, 
     bulk Lorentz factor and Doppler factor versus time.\\
     We assume that its ejection time $t_0$=1979.00 and  the corresponding
     precession phase $\phi_0$=4.28\,rad. \\
     It is worth-while to note that, due to adopting appropriate mathematical 
     formulas, functions and model-parameters to describe the curved jet-axis 
     and its helical trajectory pattern (see Sec.3 and Fig.2), the entire
     trajectory of knot C4 extending to core separation of $\sim$8\,mas was
     well fitted  (Figure 4). For the model-fitting of the entire
    kinematics of knot C4,  the exponential factor in the expression
    of p($z_0$) (see equations (1) and (2) 
    of Section 3) plays a significant role,
     resulting in the modeled trajectory steadily curving northward. 
    Its helical trajectory is described by amplitude A($z_0$) and 
    phase $\phi(z_0)$, which are presented in Figure 3.\\
     The mode-fitting results are shown in Figure 4. It can be seen that
     its trajectory, coordinates, core separation are well fitted. And 
     its apparent velocity, viewing angle, Lorentz factor and Doppler factor
     are all derived as functions of time along the whole trajectory. \\
     Bulk acceleration is required and its bulk Lorentz factor ($\Gamma$)
    changes as: for Z$\leq$2.0\,mas: $\Gamma$=8; for Z=2--20\,mas,
    $\Gamma$=8+2(Z-2)/(20-2); for Z=20--30\,mas
    $\Gamma$=10+5(Z-20)/(30-20); for Z=30-40\,mas
    $\Gamma$=15+3(Z-30)/(40-30); for Z$>$40\,mas $\Gamma$=18.\\
    The fitting results in Figure 4 clearly
     show that the entire kinematics of knot C4 can be 
    well explained in terms of our precessing nozzle model. This may be 
    the first time for a superluminal knot being well fitted by a 3-dimensional
    helical motion to core-separation of $\sim$8.0\,mas, corresponding to
    a spatial distance $\sim$1.1\,kpc from the galaxy center.
    Our model-fitting of the
    kinematics for knot C4 is physical and thus its viewing angle and 
    bulk Lorentz factor/Doppler factor vs time can be derived (or simulated).
    When  the Doppler factor vs time for knot C4 is derived, 
     the characteristic features of its radiation and evolution can then 
    be fully investigated and  the physical parameters of its emitting regions
    can be determined (e.g., Qian et al. \cite{Qi91a}, \cite{Qi96}).\\
    Although we have quite successfully performed the model-fitting of the 
    entire kinematics of knot C4, we mostly concern about the model-fit of its 
    innermost trajectory: whether it could be model-fitted by the precessing
    common trajectory.    
    \begin{figure*}
     \centering
     \includegraphics[width=5.5cm,angle=-90]{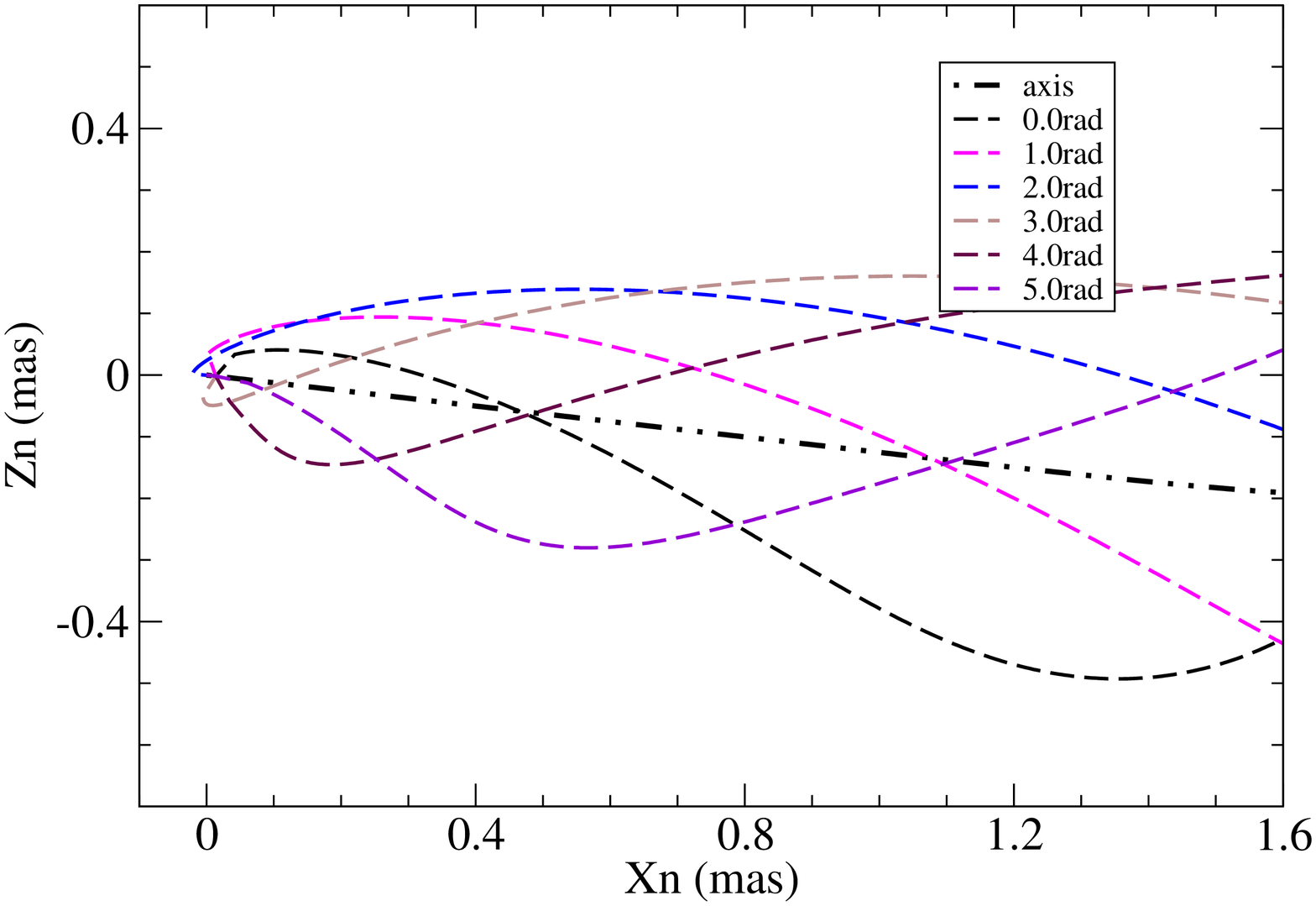}
     \includegraphics[width=5.5cm,angle=-90]{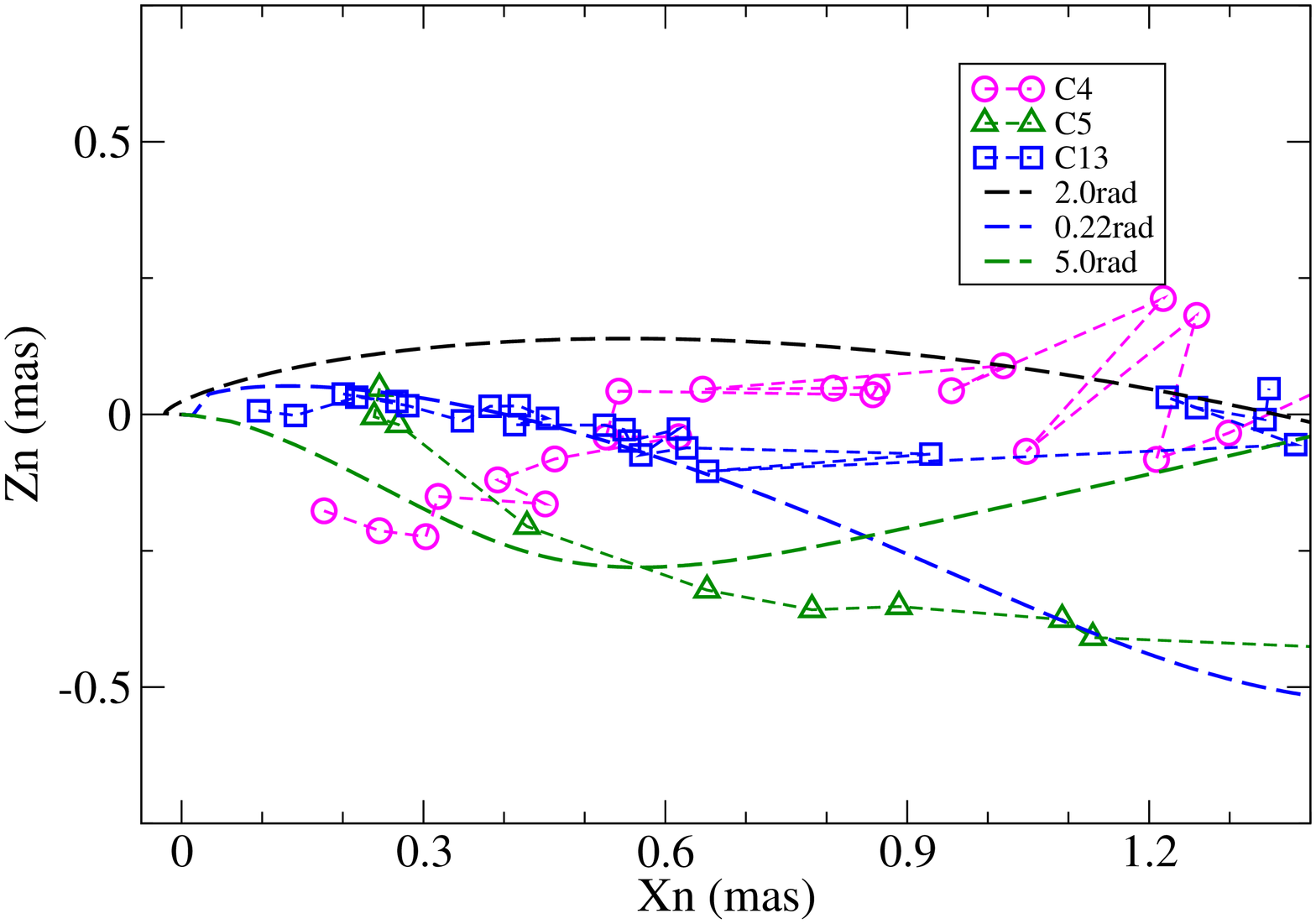}
     \caption{Left: Distribution of the precessing common trajectory for jet-A.
    The jet axis is at position angle $\sim{-97.2^{\circ}}$ with its cone 
    aperture $\sim{42.5^{\circ}}$ (at core separation 0.5\,mas). The opening 
    angle of the jet is $\sim{1.12^{\circ}}$ in space.  Right: the observed
    trajectories of knots C4, C5 and C13 in the jet are shown for comparison.}
     \end{figure*}
    \begin{figure*}
    \centering
    \includegraphics[width=5.5cm,angle=-90]{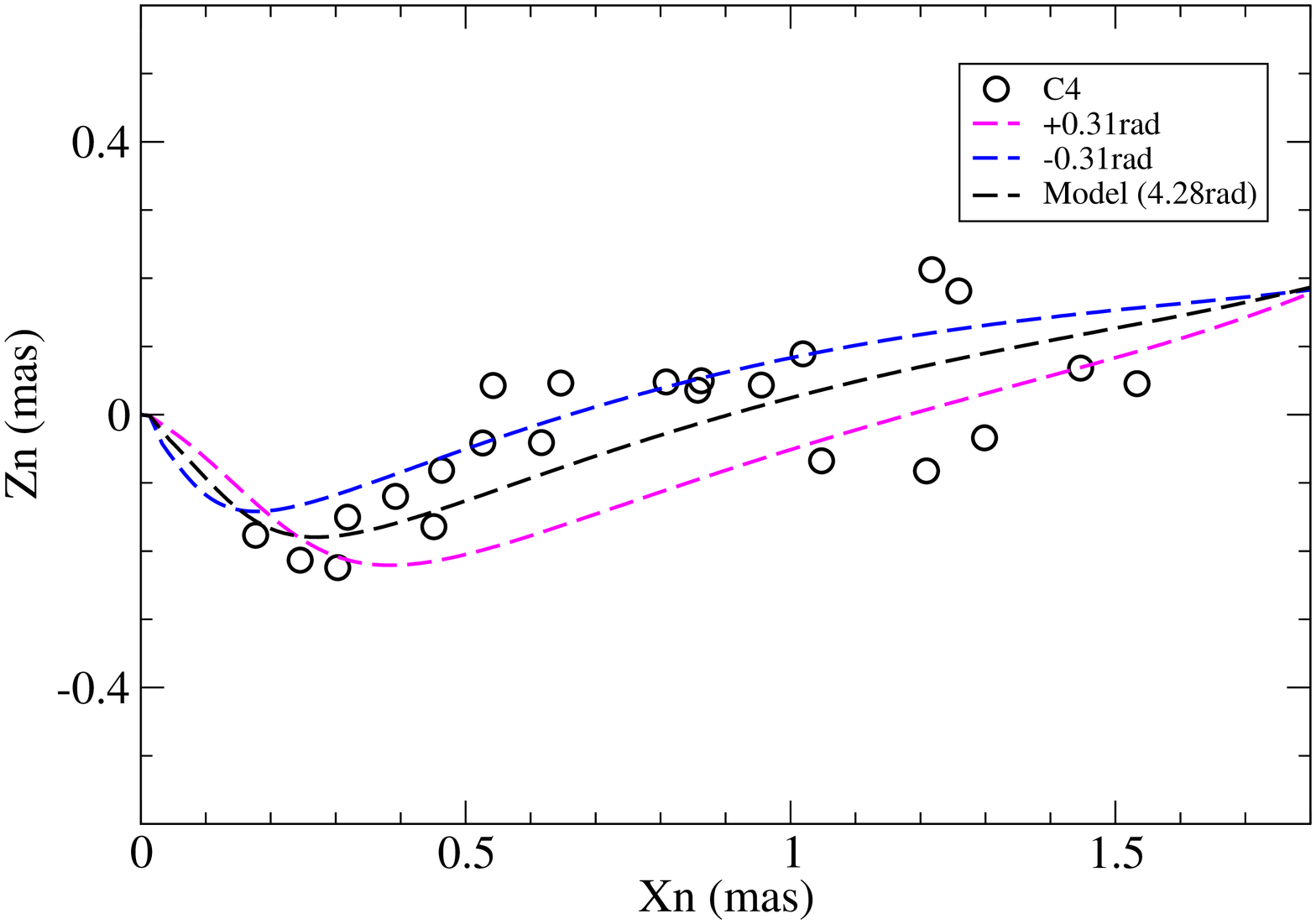}
    \includegraphics[width=5.5cm,angle=-90]{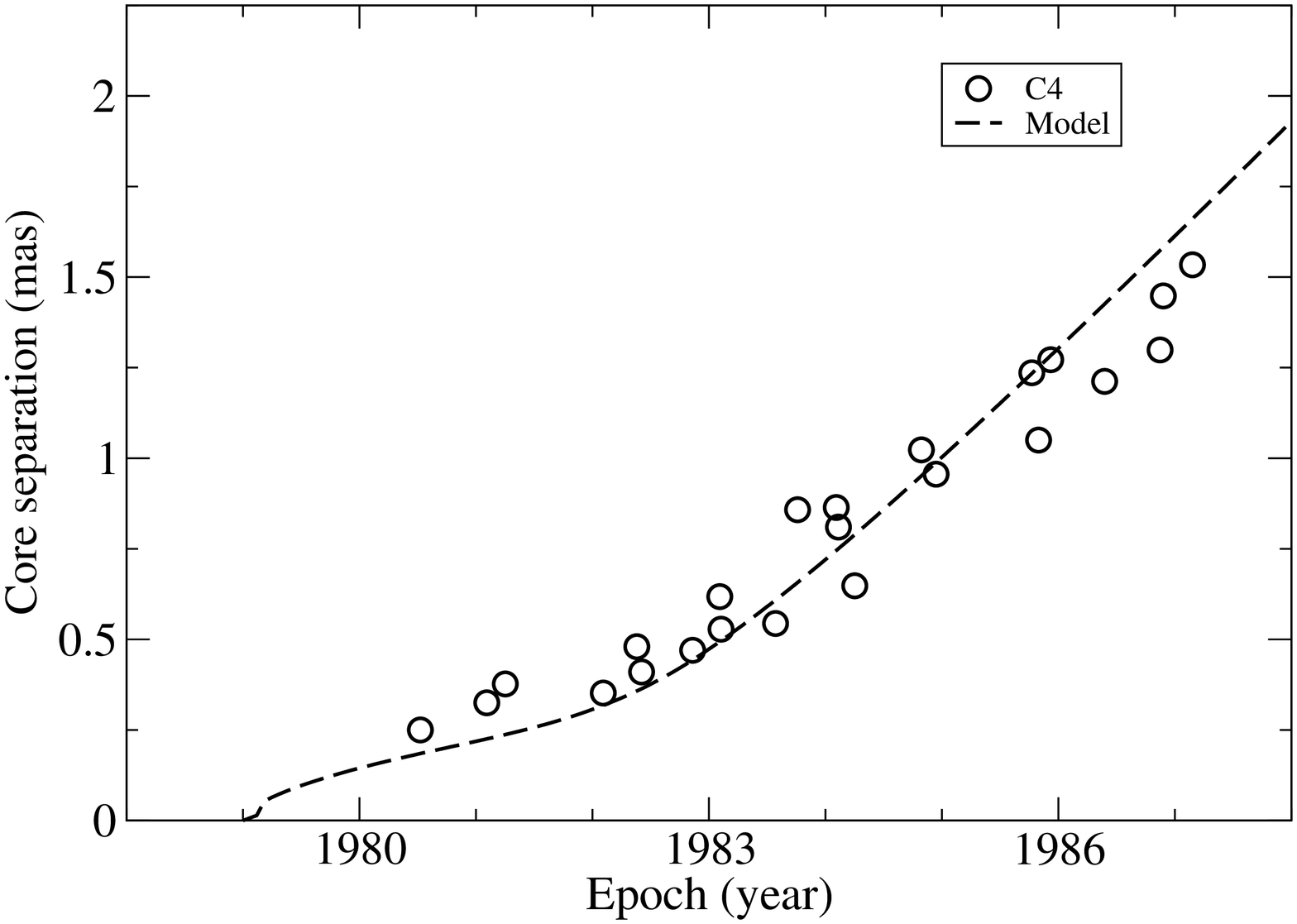}
    \includegraphics[width=5.5cm,angle=-90]{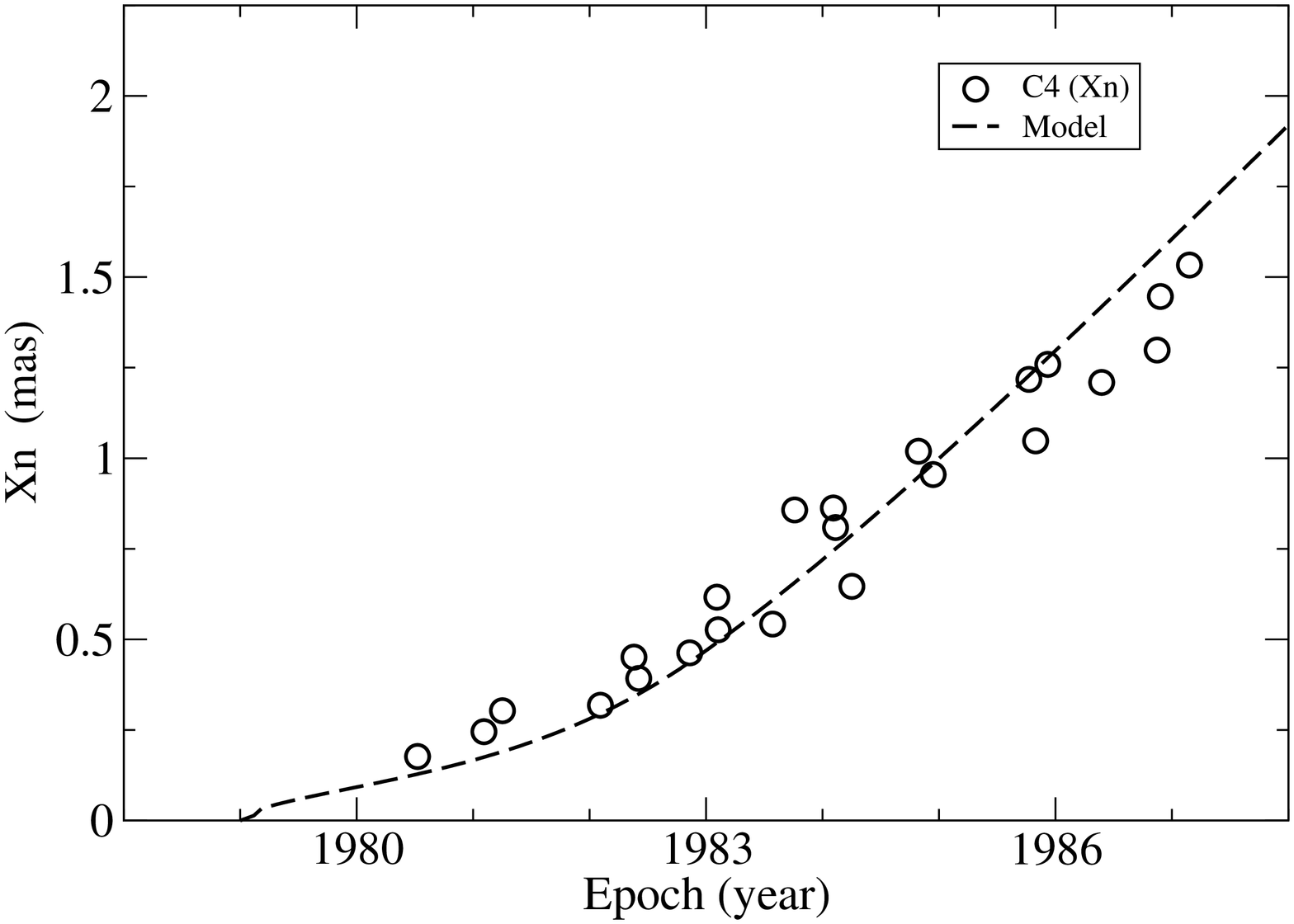}
    \includegraphics[width=5.5cm,angle=-90]{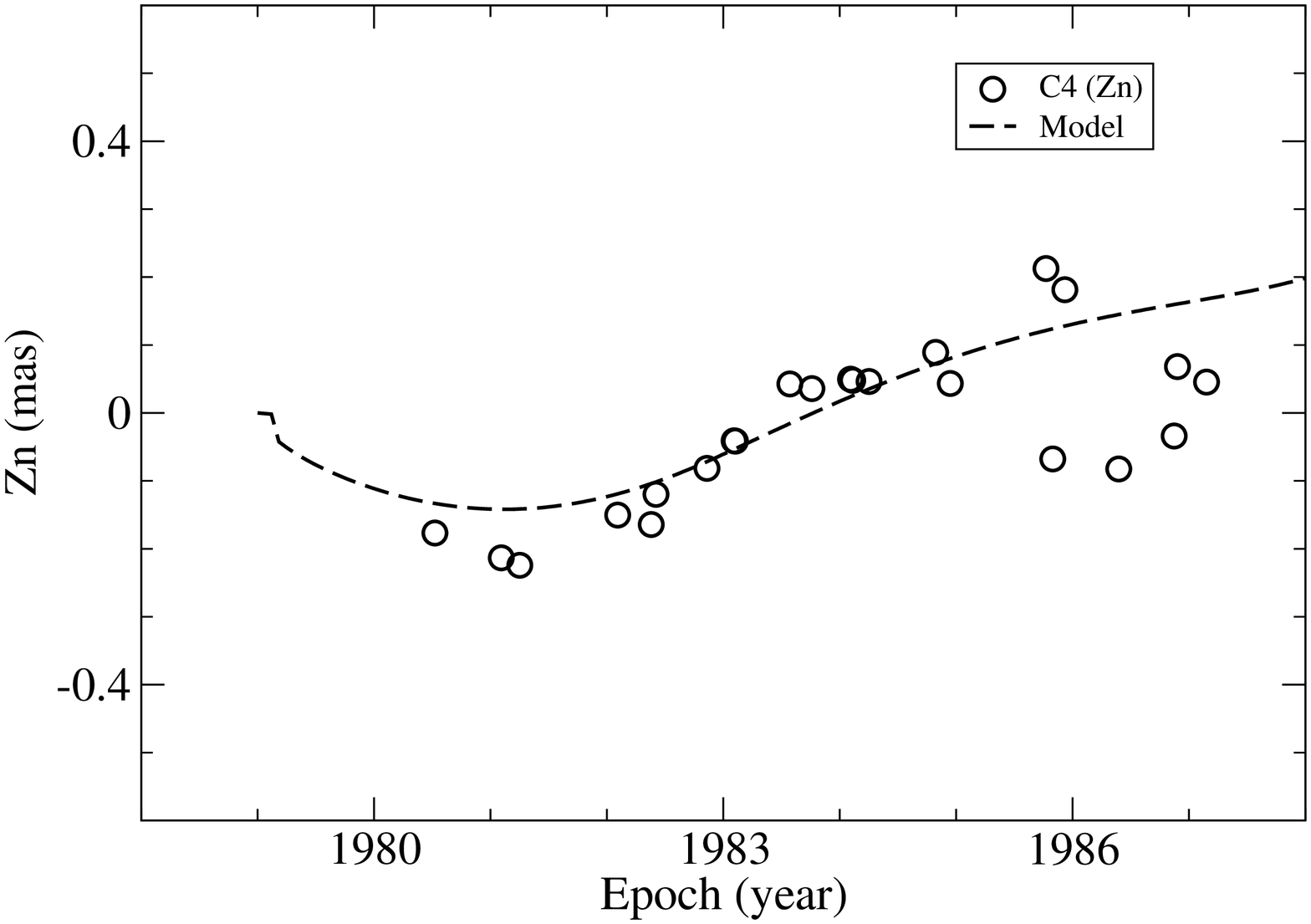}
    \includegraphics[width=5.5cm,angle=-90]{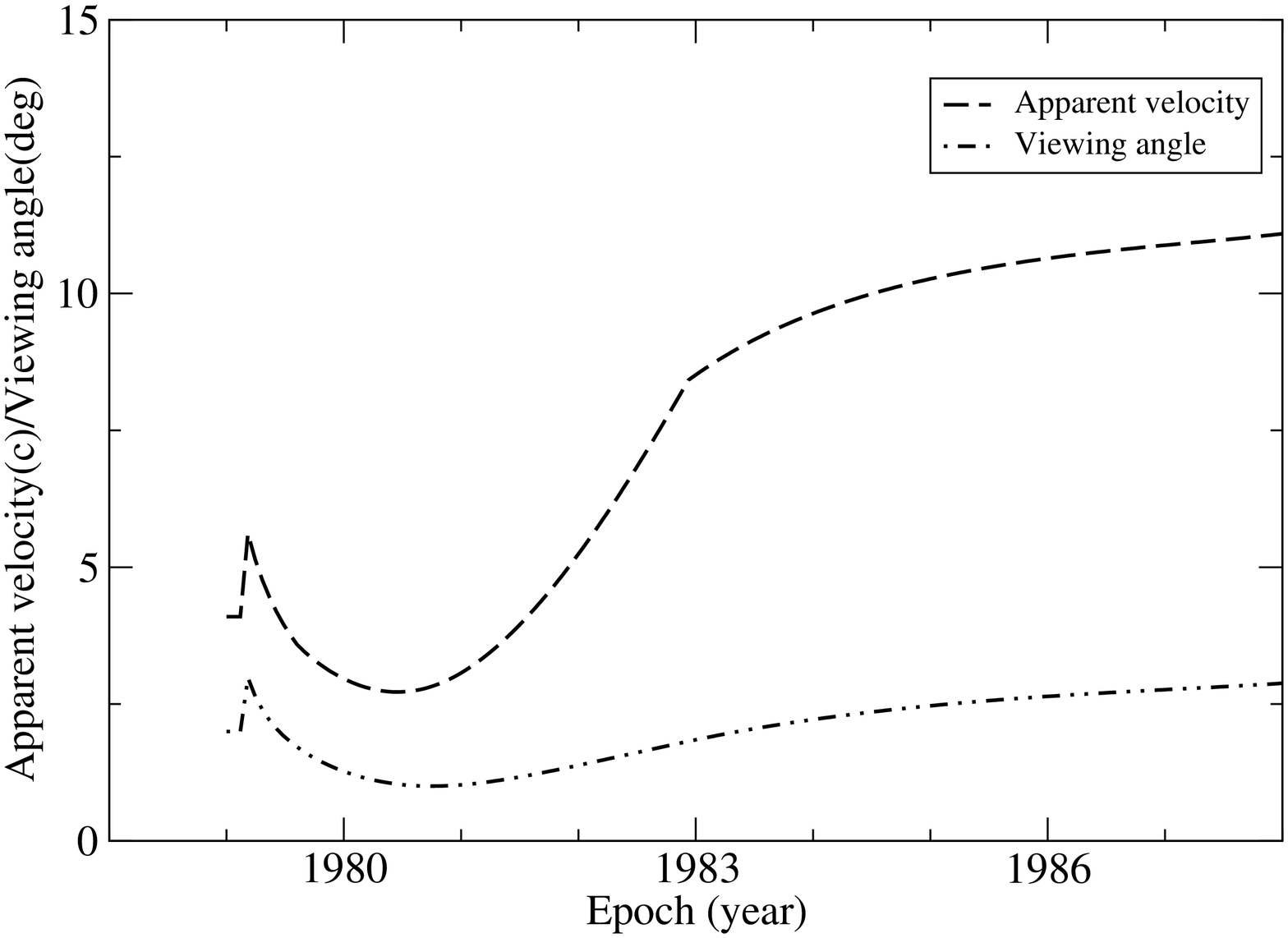}
    \includegraphics[width=5.5cm,angle=-90]{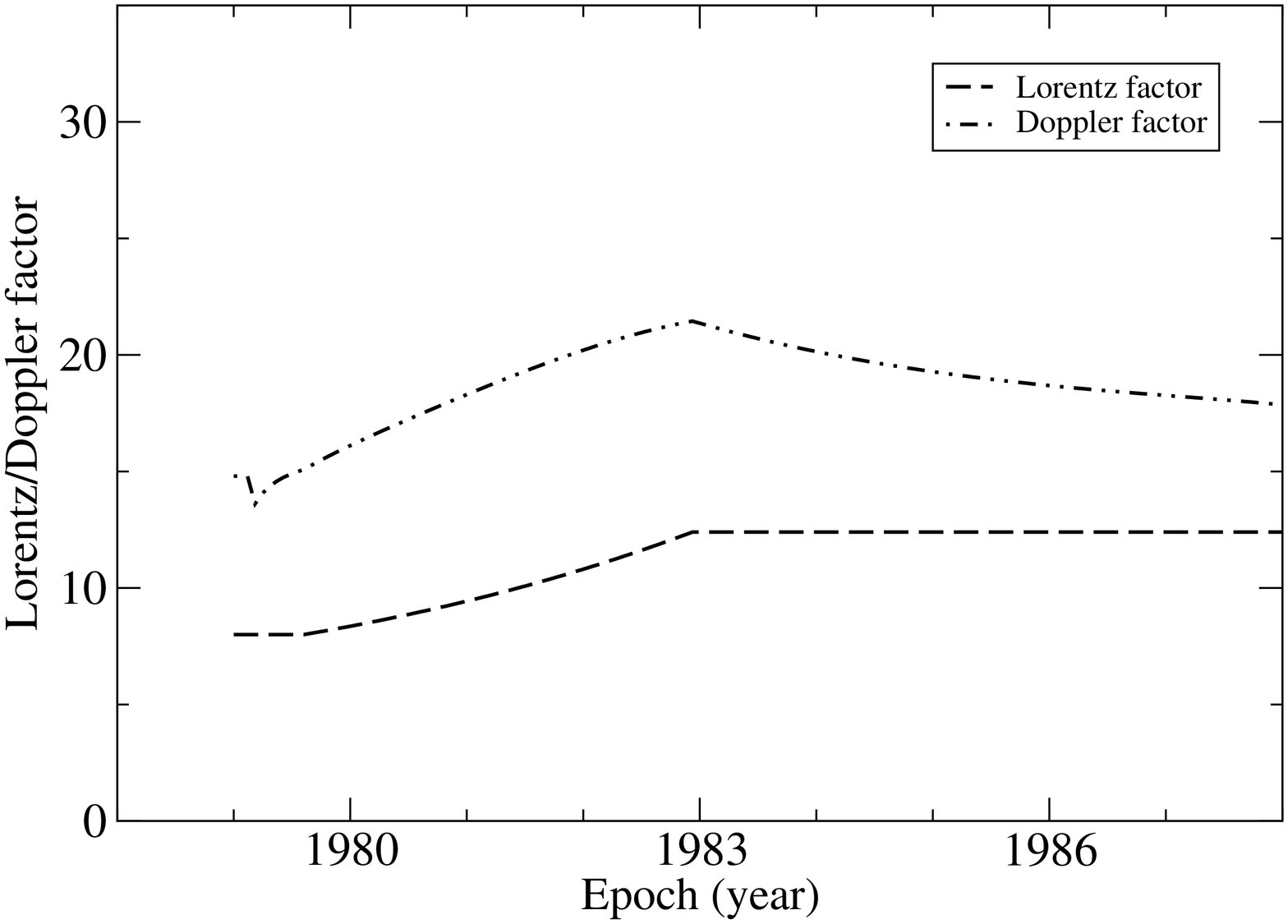}
    \caption{Model fitting of the kinematics of knot C4 in the inner jet 
    (within $r_n{\simeq}$1.8\,mas). The precessing common 
    trajectory is shown by the black dashed-line in upper/left panel 
    (with precession phase $\phi_0$=4.28\,rad, corresponding to
    ejection time $t_0$=1979.00). The 
     dashed-lines in magenta and blue show the precessing common trajectories
     at precession phases $\phi_0$(rad)=4.28+0.31 and 4.28-0.31, respectively
     (0.31\,rad=5\% of the precession period). Most data-points are distributed
     near the precessing common trajectory. Note: The Doppler factor curve
     $\delta$(t), derived in the model-simulation, had a bump-structure 
    during the period 1980--1985 with a 
     maximum at $\sim$1983 and produced the Doppler boosting effect which can
    explain its radio flux evolution (Qian, in preparation).}
    \end{figure*}
    \subsection{Precessing common trajectory for knot C4} 
    Although the entire kinematics of knot C4  has been model-fitted as
     described above,  we do not know how far its precessing common trajectory
     extending from the core. Obviously only its inner trajectory within 
    a certain core-separation could join in with the other superluminal 
    components commonly having the precessing trajectory. In comparison with 
    the observed trajectory of knot C9 (having a very long precessing 
    common trajectory) we assumed that for knot C4 its trajectory
    section within core separation ${r_n}\stackrel{<}{_\sim}$1.8\,mas 
   (equivalent to spatial  angular distance $\sim$52\,mas or 346\,pc 
   from the core; Fig.6) could be ascribed to 
    the precessing common trajectory and model-fitted consistently with other
    superluminal knots of jet-A in terms of our precessing nozzle scenario.\\
    The distribution of precessing common trajectory for jet-A is shown
     in Figure 5 (left panel) and the observed trajectories of knots C4, 
    C5 and C13 are shown in right panel for comparison.\\
     In Figure 6 the model-fitting results of the kinematics for knot C4 
    within core separation ${r_n}\sim$1.8\,mas is presented by using its
    precession phase $\phi_0$=4.28\,rad and ejection epoch $t_0$=1979.0. 
    In order to model-fit its core separation more appropriately, we need
    to slightly adjust its bulk Lorentz factor: for Z$\leq$2\,mas 
    $\Gamma$=8.0; for 2--20\,mas $\Gamma$=8+4.4(Z-2)/(20-2); for Z$>$20\,mas
    $\Gamma$=12.4. It can be seen that its kinematics within core separation
    ${r_n}\stackrel{<}{_\sim}$1.8\,mas can be well explained in terms of 
    our precessing nozzle scenario. Knot C4 has the second 
    longest precessing common trajectory among the knots of group-A with a
     spatial extension of ${Z_{c,m}}\sim$52.0\,mas, corresponding to a spatial distance  
    $Z_{c,p}$$\sim$346\,pc (see Table 1 below). During the period 
    1980.1-1987.5 its bulk Lorentz factor $\Gamma$, Doppler factor $\delta$,
    apparent velocity $\beta_a$ and viewing angle $\theta$ vary over the
    respective ranges: [8.3,12.4], [16.1-(21.7)-17.8], [3.0,11.1] and
     [1.28,2.88]. (Note: as shown in Fig.6 (bottom/left panel) the Doppler 
     factor curve $\delta$(t) has a bump structure during the period
     1980--1985 with a maximum at $\sim$1983. It produced the Doppler
     boosting effect which could well interpret its radio flux evolution
     (Qian, in preparation).\\
    
   \subsection{Model-fitting of kinematics for knot C5-C14, C22 and C23}
    \subsubsection{A brief introduction}
     We would like to note that the observed inner 
    trajectories of knots C5, C7, C10 and C11--C13 are similar, 
    revealing  the  precession of jet-A nozzle. Moreover,  
    the observed inner trajectories of knots C6, C9, C22 and C23 are 
   similar, also showing the precession of jet-A nozzle. Interestingly,
   both knot-sets demonstrate the same nozzle-precession period $\sim$7.30\,yr,
   which was derived about ten years ago in Qian et al. 
   (\cite{Qi09}). We had already found that the kinematics of 
   superluminal knots in QSO 3C345 could be understood in terms of our 
   precessing jet nozzle scenario.  Now we have found the  second jet (jet-B)
   and its precessing nozzle, ejecting knots C15--C21 and B5--B8, B11 and B12, 
   the kinematics of which could also be explained in terms of the 
   precessing nozzle scenario for jet-B. Therefore, the kinematic behavior
    of superluminal knots in QSO 3C345 may likely imply that QSO 3C345 has
    a double relativistic jet system and host a binary black hole
    in its nucleus.\\
   In the following we will first present the model-fitting results for
   the  superluminal knots of jet-A in detail.\\
   \subsubsection{Model-fitting results for knot C5 }
   The model-fitting results of the kinematics of knot C5 are shown in 
   Figure A.1. Its  ejection epoch 
   $t_0$=1980.80 and  corresponding precession phase $\phi_0$=5.83\,rad.\\
    Bulk acceleration is required and its Lorentz factor is modeled as:
    for Z$\leq$3.0\,mas $\Gamma$=5.3; for Z=3--20\,mas 
    $\Gamma$=5.3+(Z-3)(15-5.3)/(20-3);for Z$>$20\,mas $\Gamma$=15.\\
     It can be seen in Figure A.1 that the entire kinematic behavior within 
    core separation ${r_n}\stackrel{<}{_\sim}$1.2\,mas can be  well 
    fitted, implying that its observed
     precessing common trajectory may extend to a spatial distance of 
   $Z_{c,m}$=39.0\,mas (or $Z_{c,p}$$\sim$259.3\,pc) from the core 
    (see Table 1).\\
    During the period 1983.0--1990.0 its Lorentz factor $\Gamma$, Doppler 
    factor $\delta$, apparent velocity $\beta_a$ and viewing angle $\theta$
    vary over the respective ranges: [5.5,15.0], [10.1,(24.8),23.6], 
    [2.9,(10.0),(8.3),12.2] and [3.03,1.98](deg).
   \subsubsection{Model-fitting results for knot C6}
   According to the precessing-nozzle scenario for jet-A, the kinematics 
   of knot C6 is model-fitted with precession phase $\phi_0$=5.74\,rad+2$\pi$
   and ejection epoch $t_0$=1987.99.\\
   In this case the observed  precessing common trajectory may only extend to 
    $r_n{\sim}$0.30\,mas. The kinematics in its outer trajectory has to be
   explained by introducing changes in parameter $\psi$ (or rotation of 
   its trajectory): for Z$\leq$6.0\,mas $\psi$=0.125\,rad (just the same as
   for the precessing common trajectory); 
   for Z=6-20\,mas $\psi$(rad)=0.125-0.225(Z-6)/(20-6);
    for Z$>$20\,mas $\psi$=-0.1.\\
    Acceleration in its motion is required and its bulk Lorentz factor is
    modeled as: for Z$\leq$6\,mas $\Gamma$=7.9; for Z=6-20\,mas
    $\Gamma$=9.7+6.3(Z-6)/(20-6); for Z$>$20\,mas $\Gamma$=16.
    The model-fitting results of kinematic  behavior of knot C6 are shown
    in Figure A.2. Due to lack of observational data-points within 
    $X_n{<}$0.3\,mas, the model fitting of its kinematic behavior along the 
    precessing common trajectory is only marginal.\\
    Its observed precessing common trajectory might be assumed to extend to
    $r_n{\sim}$0.40\,mas, corresponding to a spatial distance 
    $Z_{c,m}$=7.47\,mas (or $Z_{c,p}$=49.7\,pc) from the core.\\
    During the period 1988.2--1990.0 its Lorentz factor $\Gamma$, Doppler 
    factor $\delta$, apparent velocity $\beta_a$ and viewing angle $\theta$
    vary over the following ranges respectively: [9.7,11.1], [13.5,18.7], 
    [8.9,8.0] and [3.90,2.21](deg).
   \subsubsection{Model-fitting results for knot C7}
   According to the precessing jet nozzle scenario for jet-A, the kinematic
    behavior of knot C7  could be fitted with parameters:
   precession phase $\phi_0$(rad)=6.14+2$\pi$ and 
   ejection epoch $t_0$=1988.46.\\
   The model-fitting results of its kinematic behavior along the precessing
   common trajectory are shown in Figure A.3. It can be seen that the observed
   precessing common trajectory  may extend to $r_n{\sim}$0.7\,mas, 
   corresponding to a spatial distance of $Z_{c,p}\sim$99.3\,pc 
   (or corresponding $Z_{c,m}{\sim}$14.9\,mas) from the core.\\
   Bulk acceleration is required. Its Lorentz factor is modeled as: For
     Z$\leq$1.6\,mas $\Gamma$=3.2; for Z=1.6--5\,mas
     $\Gamma$=3.2+(Z-1.6)(13-3.2)/(5-1.6); for Z$>$5\,mas $\Gamma$=13.0.\\
   During the period 1989.0--1994.5 its Lorentz factor $\Gamma$, Doppler factor
   $\delta$, apparent velocity $\beta_a$ and viewing angle $\theta$ vary over
   the respective ranges: [3.2,13.0], [6.2,21.8], [0.65,(11.9),9.5] and
    [2.00,1.92](deg).
   \begin{figure*}
   \centering
   \includegraphics[width=5.5cm,angle=-90]{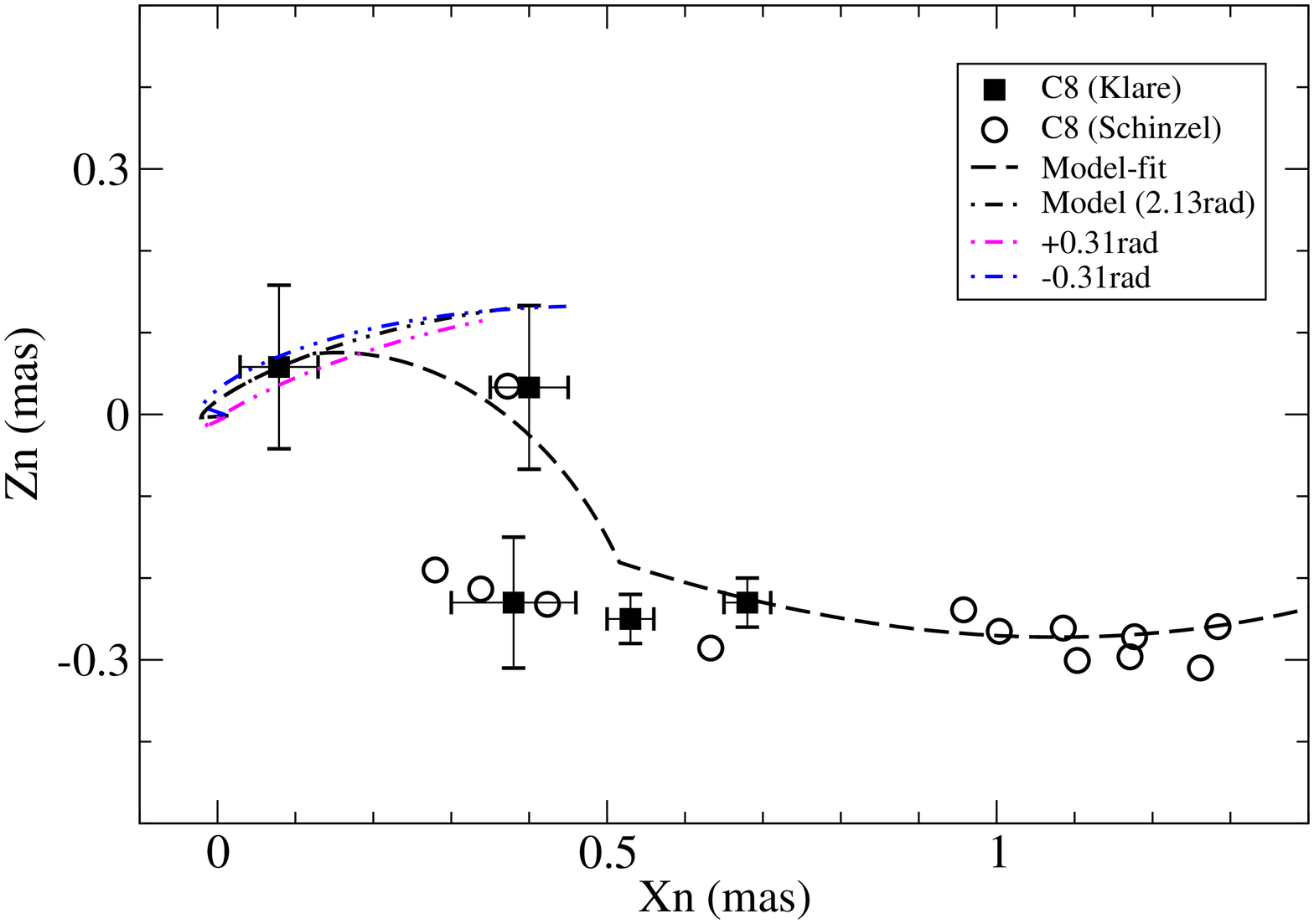}
   \includegraphics[width=5.5cm,angle=-90]{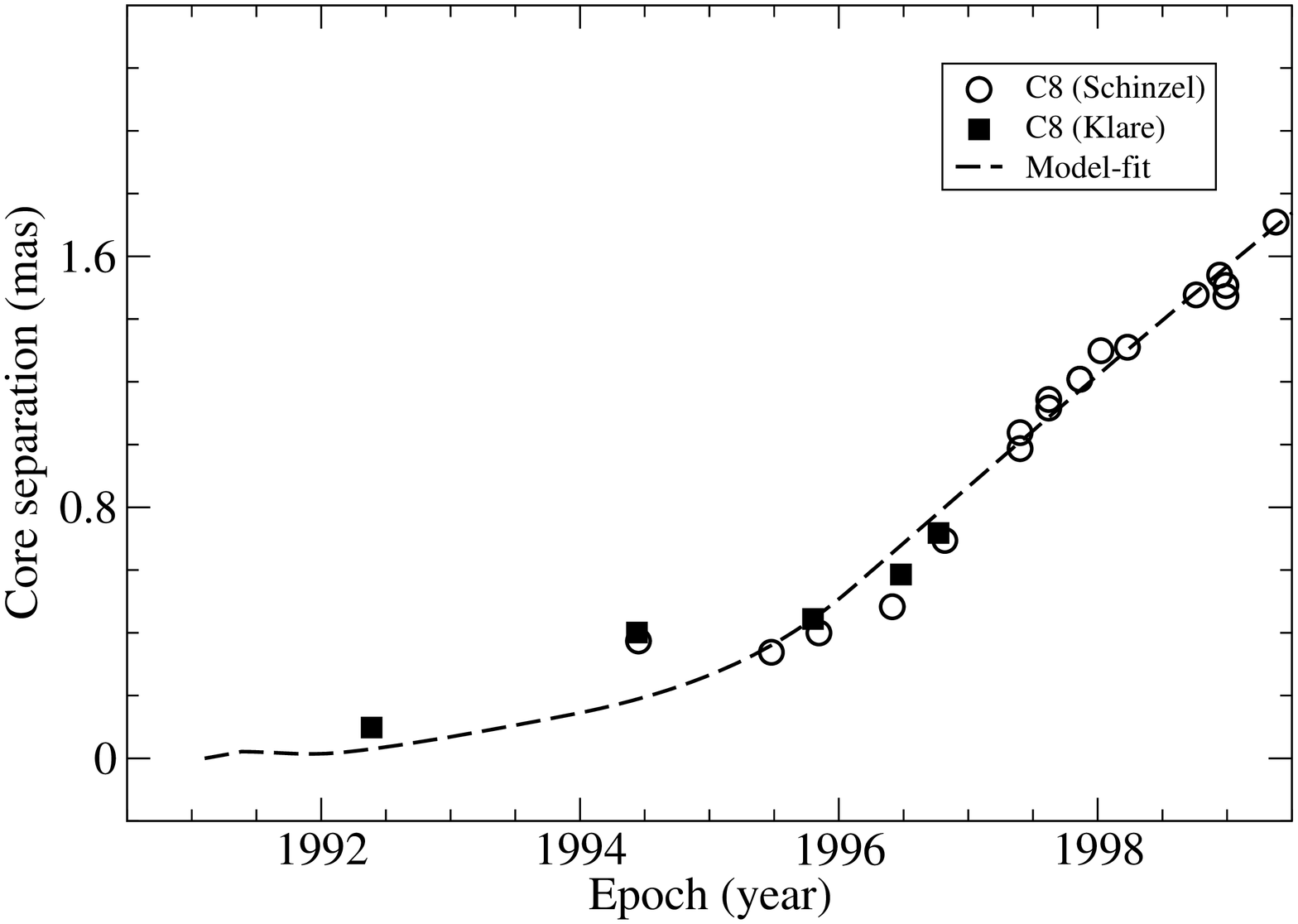}
   \includegraphics[width=5.5cm,angle=-90]{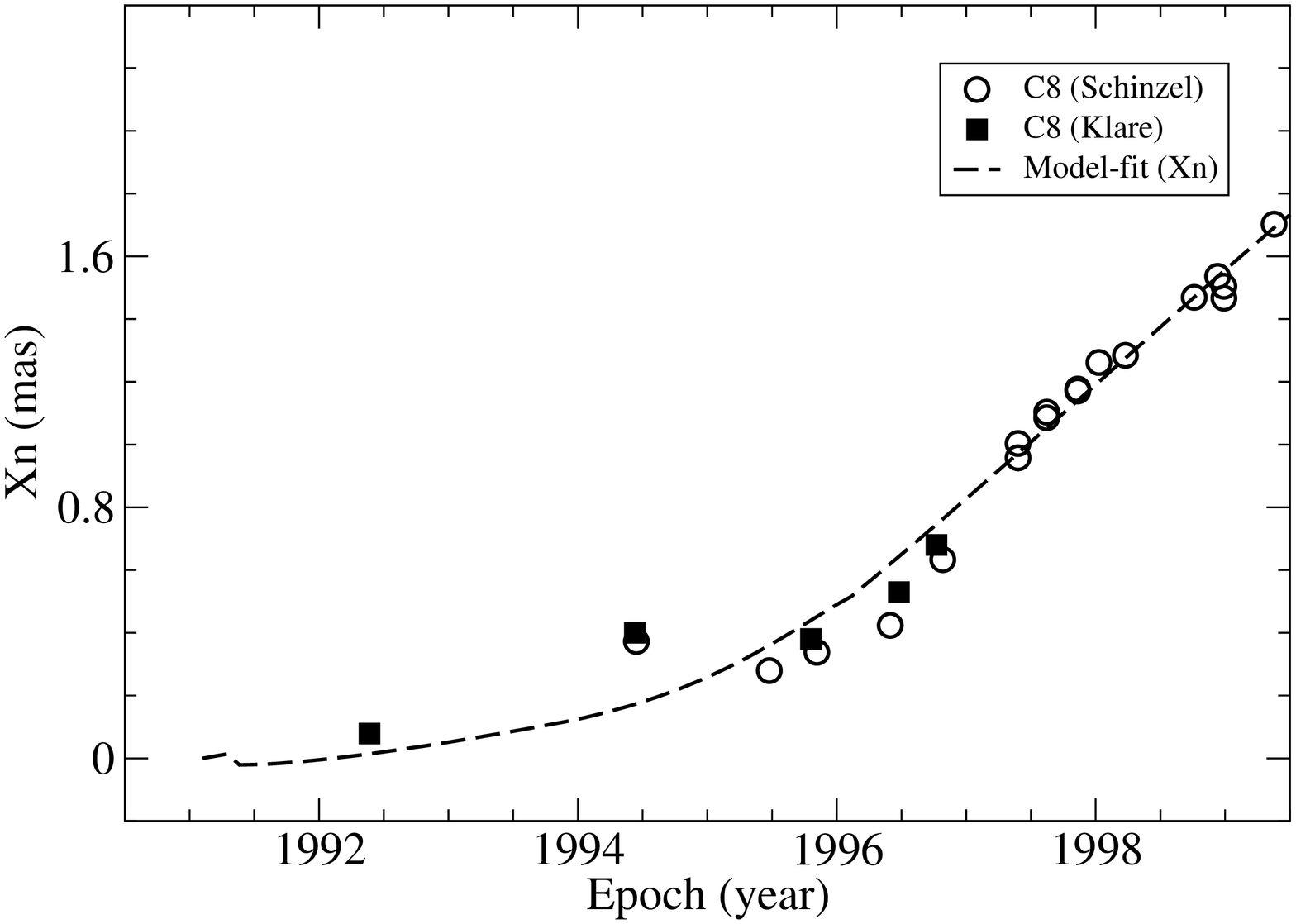}
   \includegraphics[width=5.5cm,angle=-90]{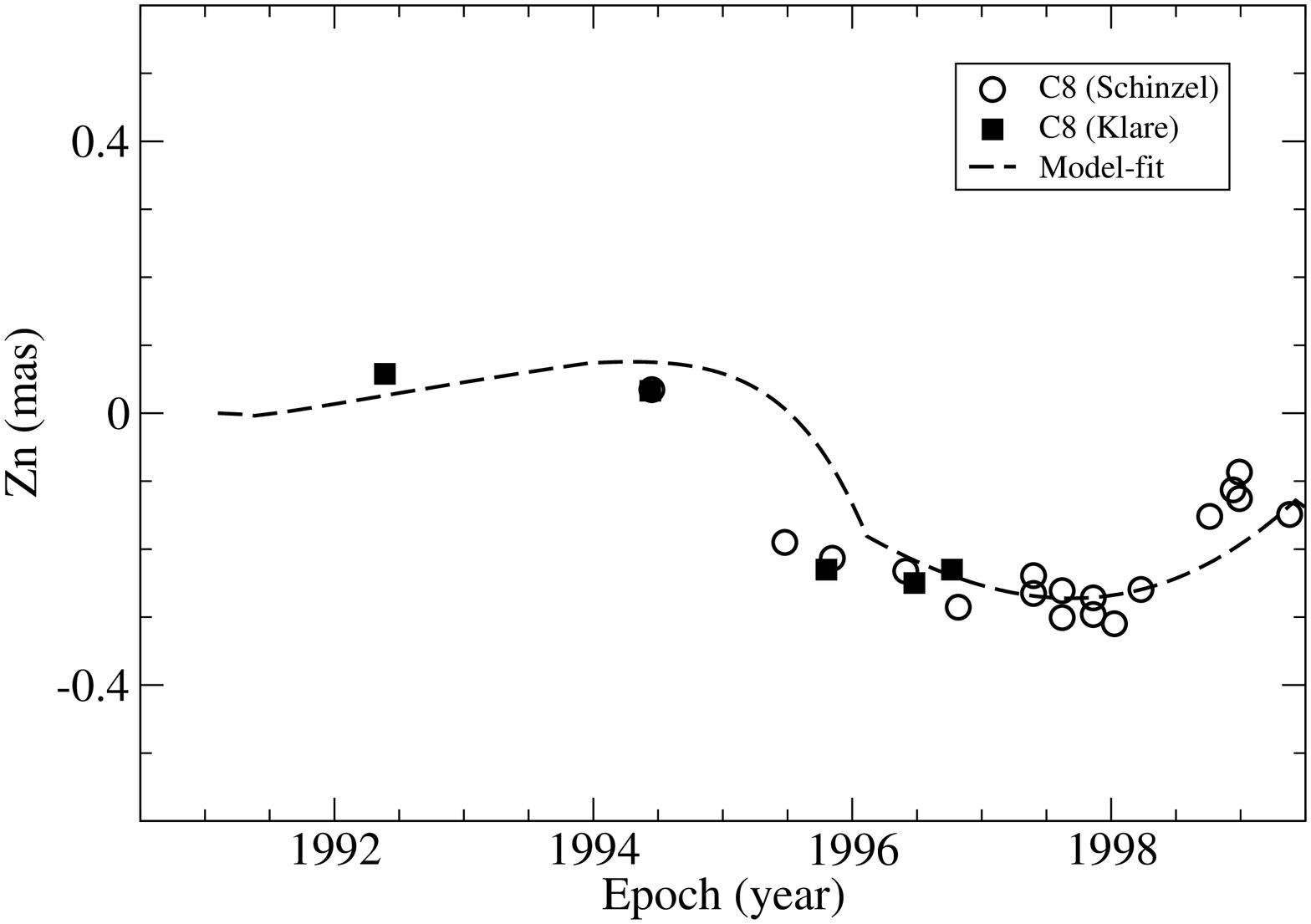}
   \includegraphics[width=5.5cm,angle=-90]{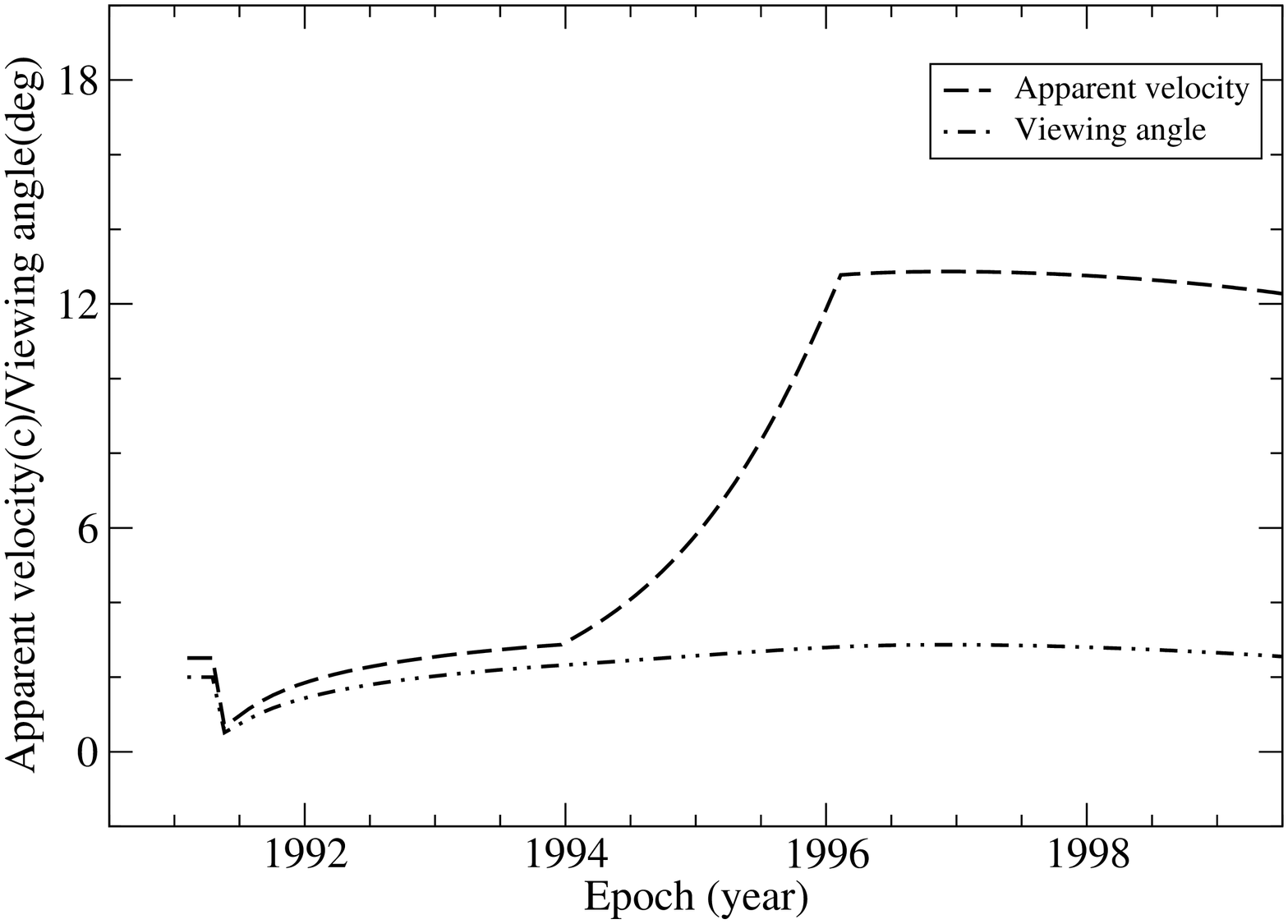}
   \includegraphics[width=5.5cm,angle=-90]{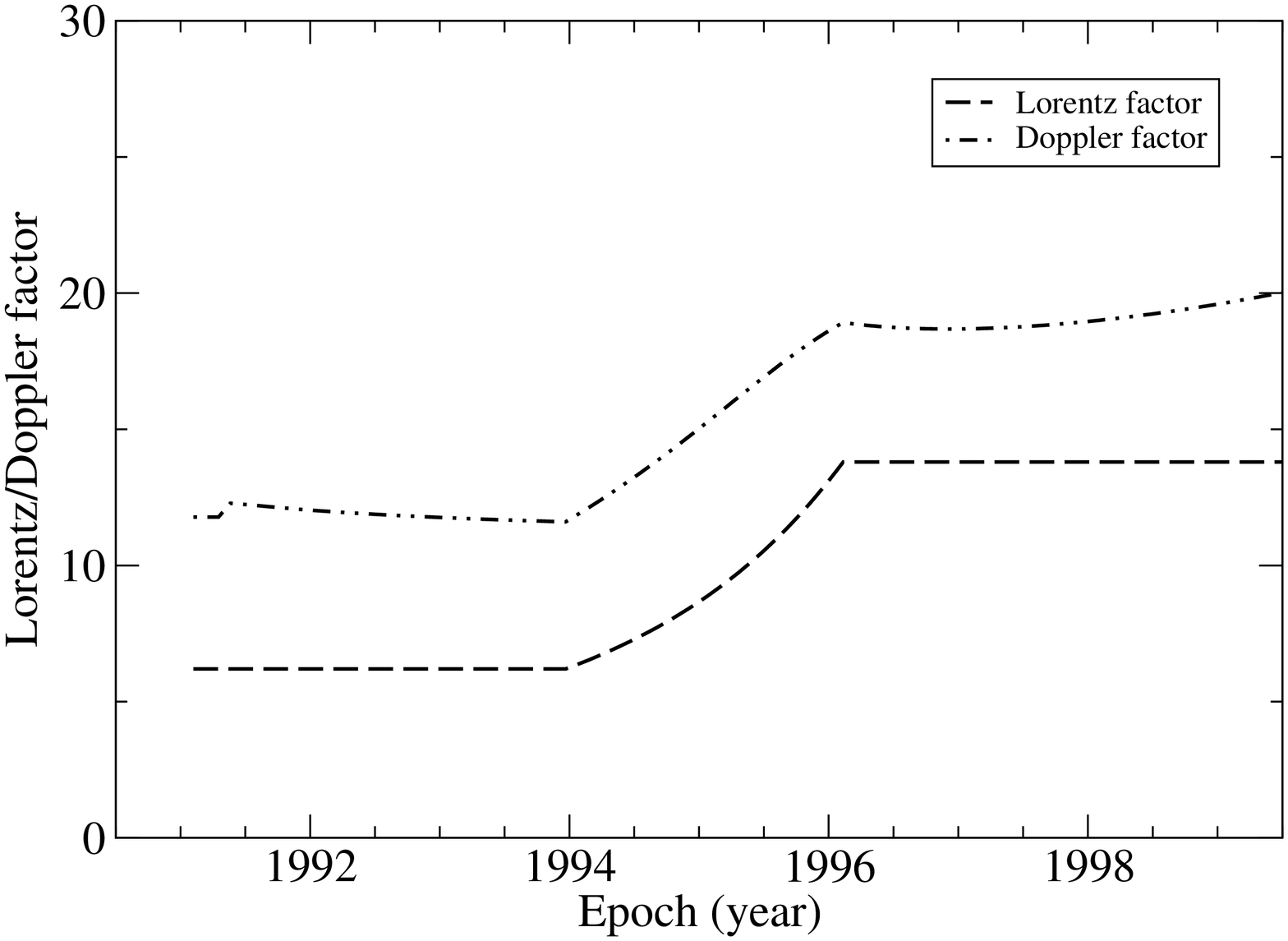}
   \caption{Model fitting results for knot C8: precession phase 
    $\phi_0$=2.13\,rad+4$\pi$ and ejection time $t_0$=1991.10. Trajectory 
    $Z_n$($X_n$), coordinates $X_n$(t) and $Z_n$(t), core separation $r_n$(t), 
    apparent velocity $\beta_a$(t), viewing angle $\theta$(t),
   bulk Lorentz factor $\Gamma$(t) and Doppler factor $\delta$(t).
    The upper/left plot also shows the precessing common trajectories
    for precession phases $\phi_0$=2.13\,rad$\pm$0.31\,rad.}
   \end{figure*}
   \subsubsection{Model fitting results for knot C8 : A particular case}
    The observed kinematic behavior of knot C8 is exceptional, but instructive
    revealing the complex structure of its innermost trajectory. 
    Only one data-point obtained at 22/43\,GHz (1992.4) by Klare (\cite{Kl03}),
    showing its initial trajectory following  the precessing common
    trajectory (precession phase $\phi_0$=2.13\,rad+4$\pi$ and 
    corresponding ejection
    epoch $t_0$=1991.10; see Figure 7). Other observational data-points
    indicate that its trajectory  changed rapidly at core separation 
    of $\sim$0.12\,mas. This observational fact clearly demonstrates that knot
    C8 might have been regarded as not following the precessing common 
    trajectory if no available observations at core separations 
    $r_n{<}$0.1\,mas (like that given by Klare). We will further indicate the 
    importance of VLBI-observations at core separations 
   $r_n{\leq}$0.05-0.1\,mas for our precessing nozzle scenario.\\
    We model-fit its outer trajectory by introducing changes in 
    parameter $\psi$:
   For Z$\leq$6\,mas $\psi$=0.125\,rad which are the same for knot C4; 
   for Z=6-15\,mas $\psi$(rad)=0.125+0.6(Z-6)/(15-6); for Z=15-40\,mas 
   $\psi$(rad)=0.725-0.580(Z-15)/(40-15); for Z$>$40\,mas $\psi$=0.145\,rad.\\
      Bulk acceleration is  required.  For Z$\leq$6\,mas 
   $\Gamma$=6.2; for Z=6--15\,mas $\Gamma$=6.2+7.6(Z-6)/(15-6);
   for Z$>$15\,mas $\Gamma$=13.8.\\
   Its observed precessing common trajectory might be assumed to extend to
   $\sim$0.15\,mas, equivalent to a spatial distance $Z_{c,m}$=6.2\,mas (or
   $Z_{c,p}$=41.2\,pc) from the core.\\
   During the period 1992.0--1998.0 its Lorentz factor $\Gamma$, Doppler factor
   $\delta$, apparent velocity $\beta_a$ and viewing angle vary over the 
   following ranges: [6.2,13.8], [12.0,19.0], [1.9,(12.9),12.8] and 
   [1.48,(2.87),2.80](deg). \\
   Here We would like to emphasize that  rapid changes
   of trajectory as in knot C8 is a significant ingredient which also 
   occurred in other blazars like 3C279 and 3C454.3 (Qian et al. 2013, 2014).
   This ingredient makes the fitting of the kinematics on pc-parsec scales 
   more subtle and difficult.  The cause of these sudden curvatures of 
   trajectory is unclear.
   \begin{figure*}
   \centering
   \includegraphics[width=5.5cm,angle=-90]{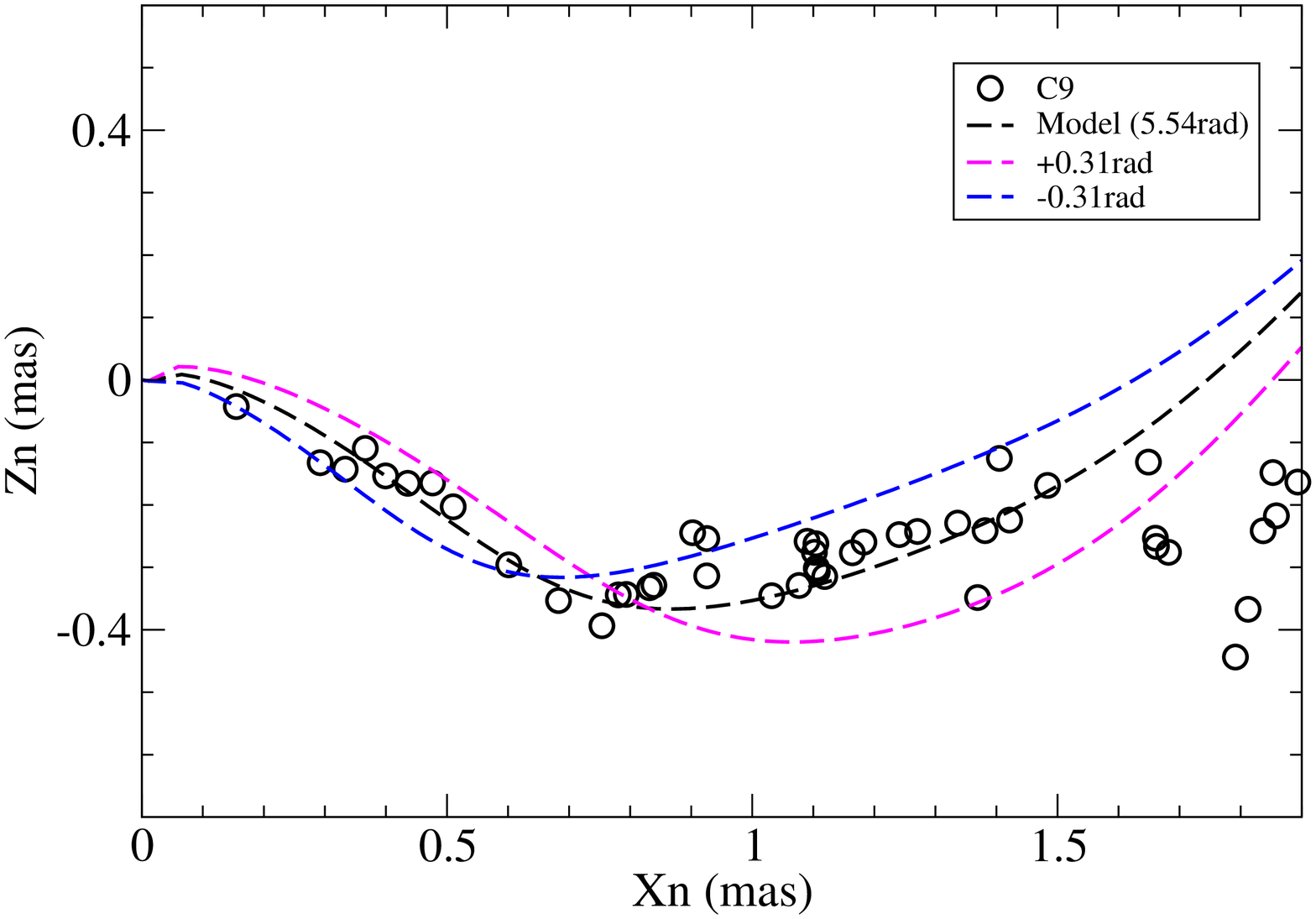}
   \includegraphics[width=5.5cm,angle=-90]{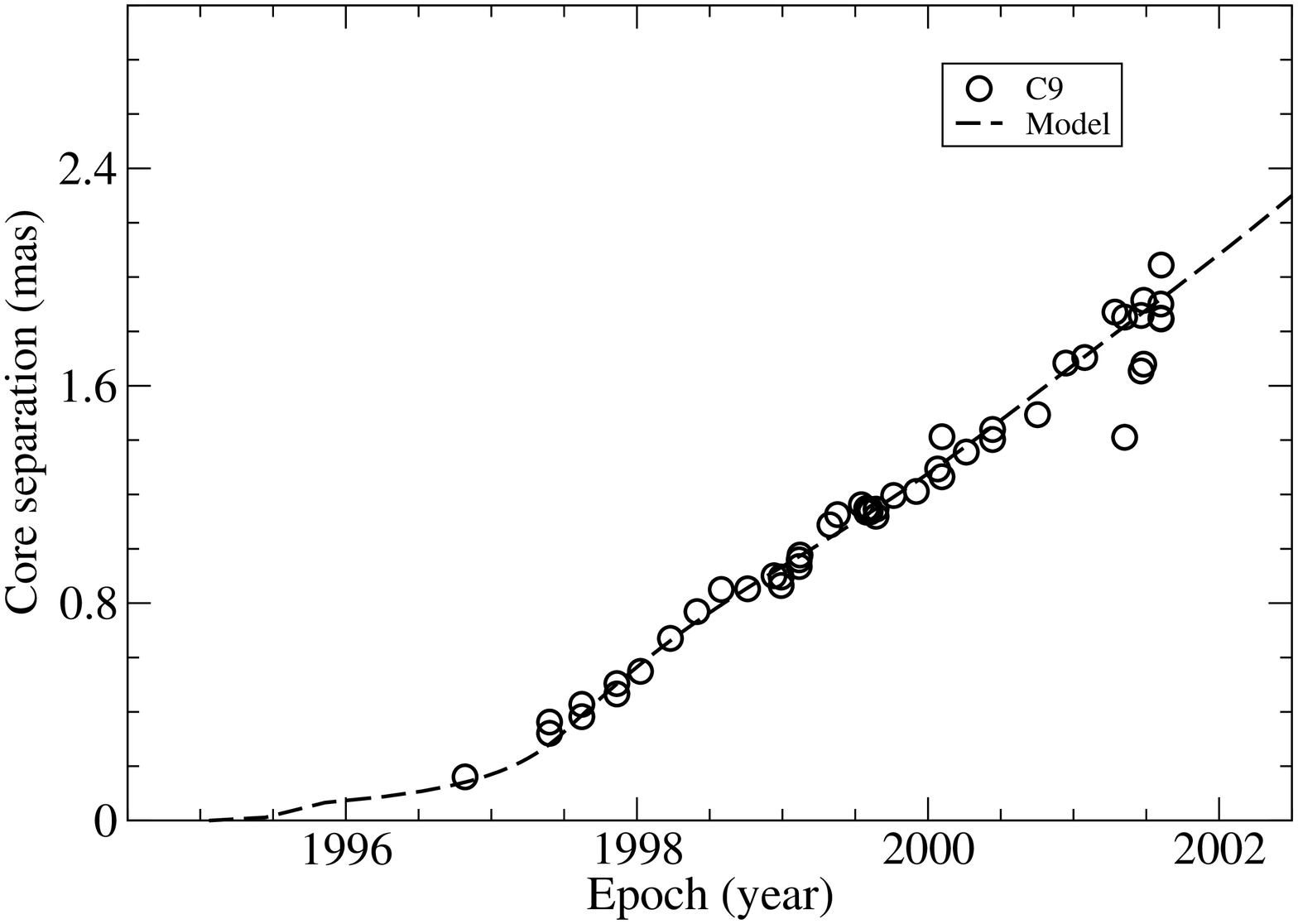}
   \includegraphics[width=5.5cm,angle=-90]{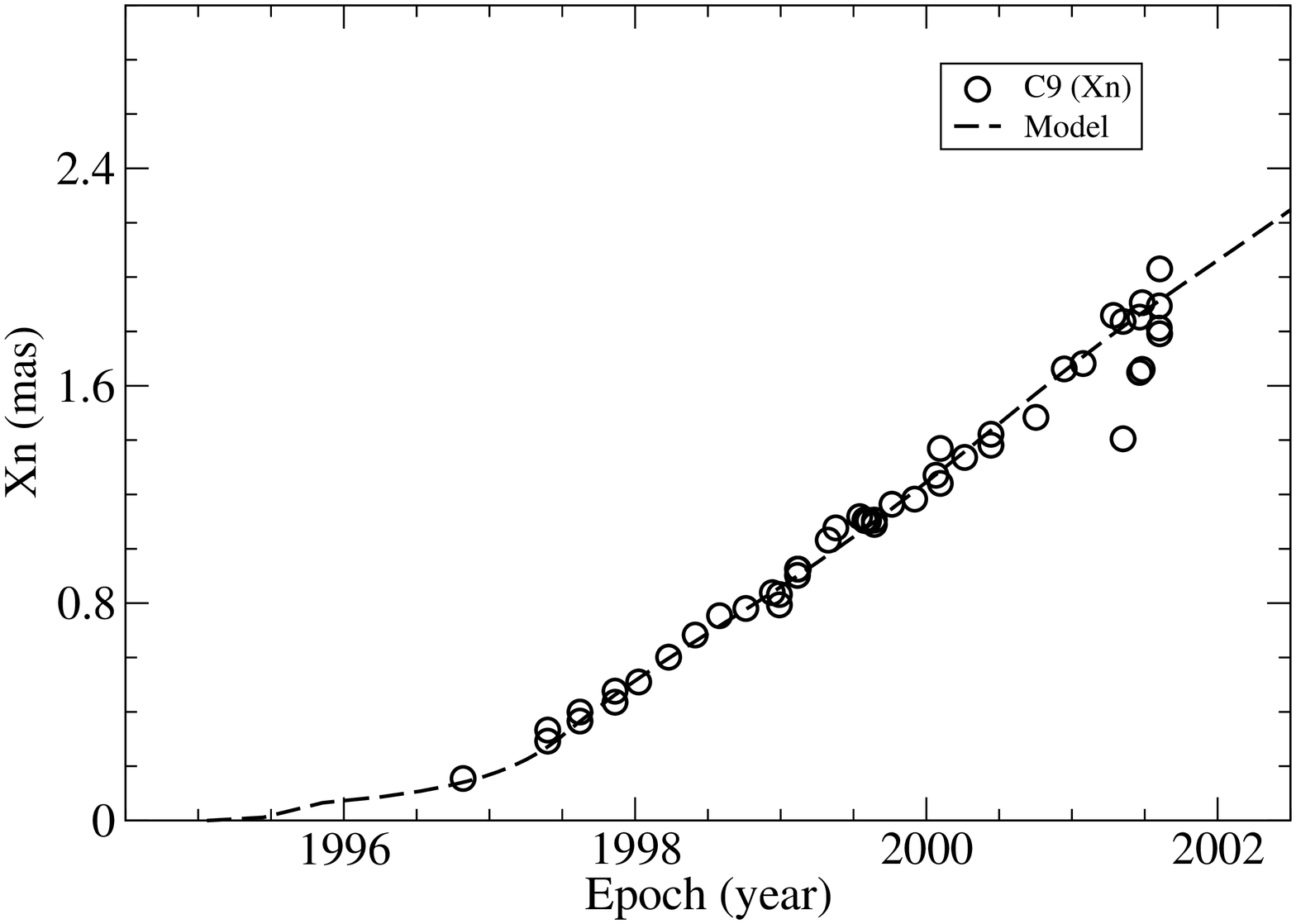}
   \includegraphics[width=5.5cm,angle=-90]{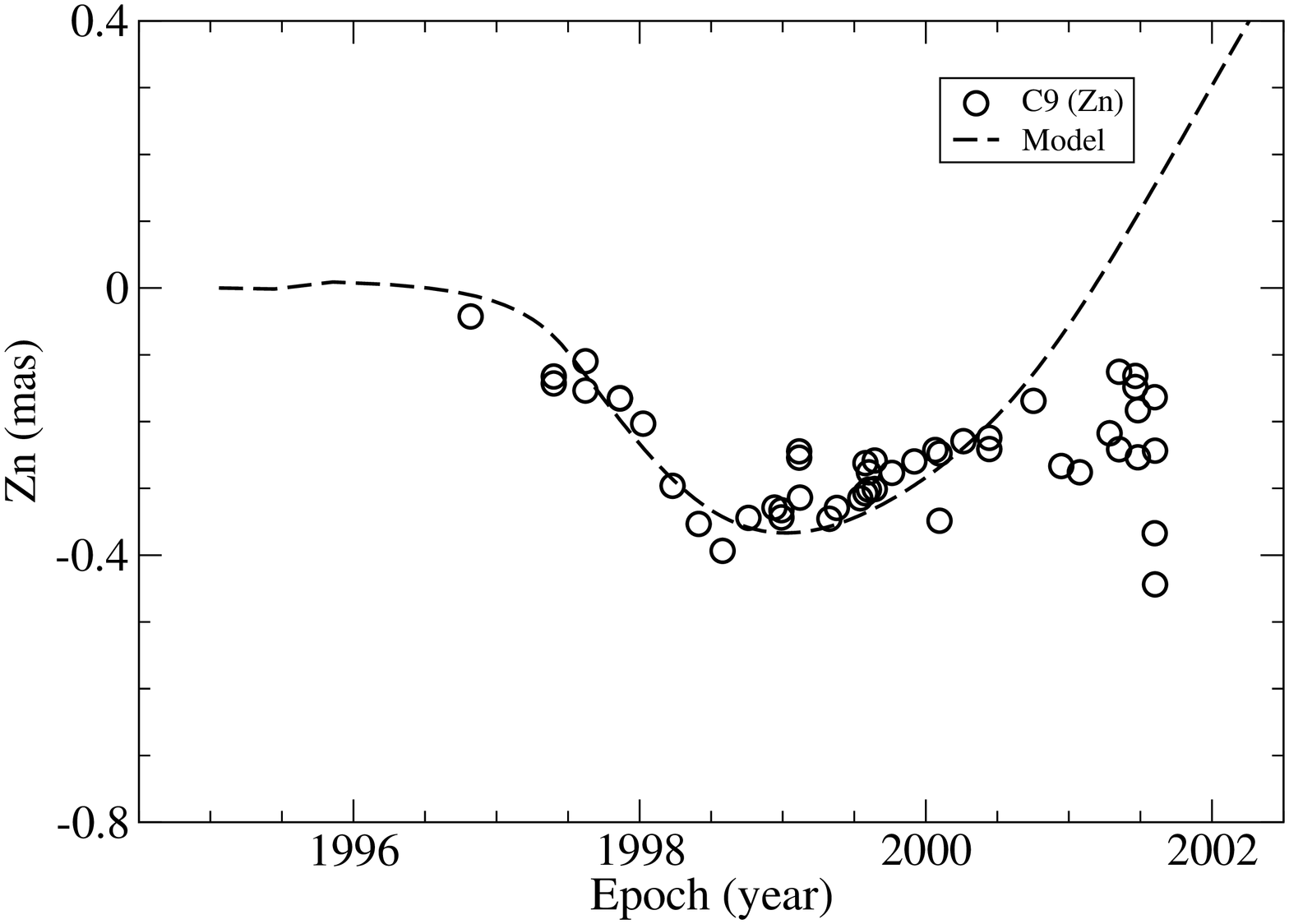}
   \includegraphics[width=5.5cm,angle=-90]{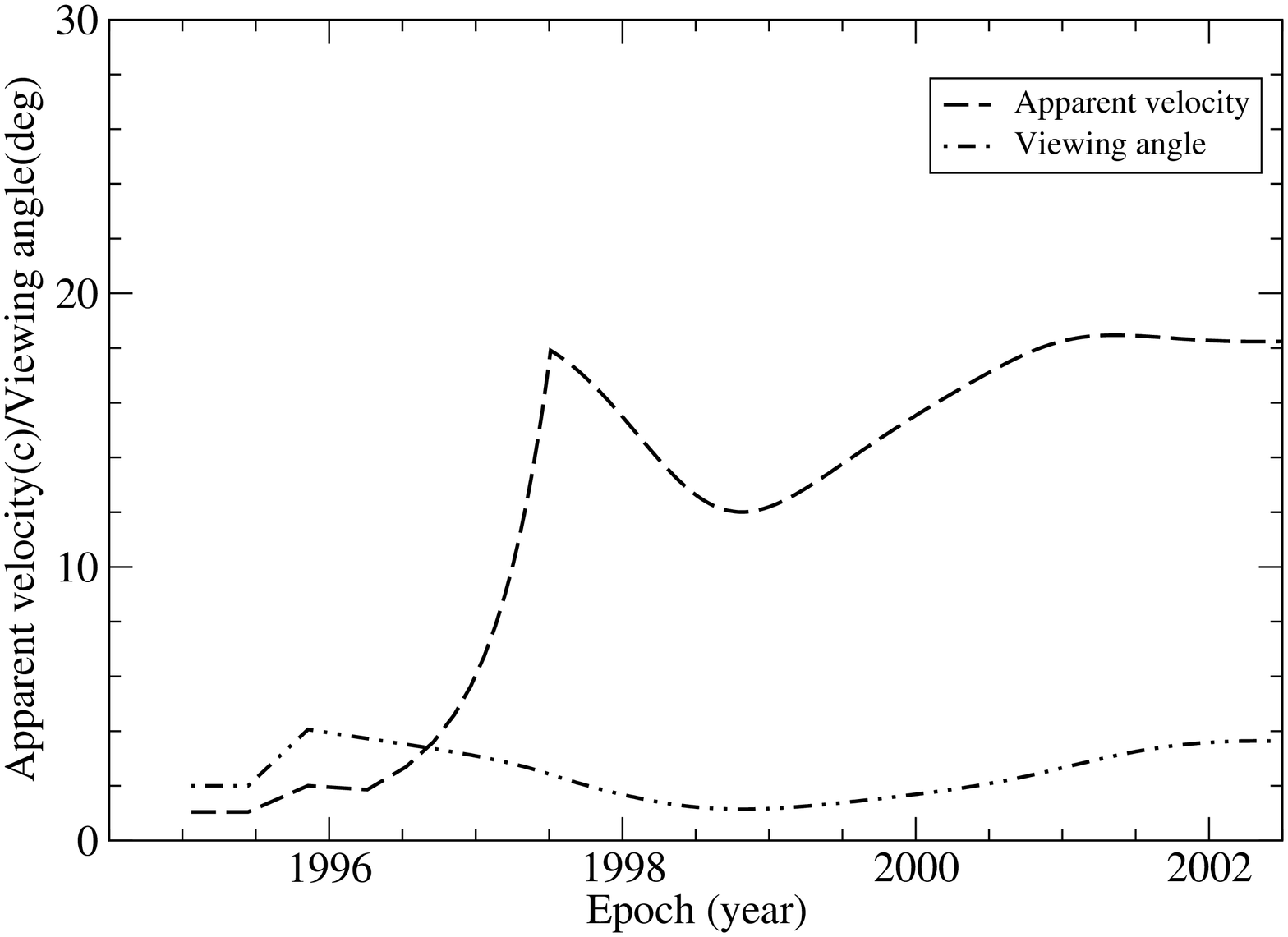}
   \includegraphics[width=5.5cm,angle=-90]{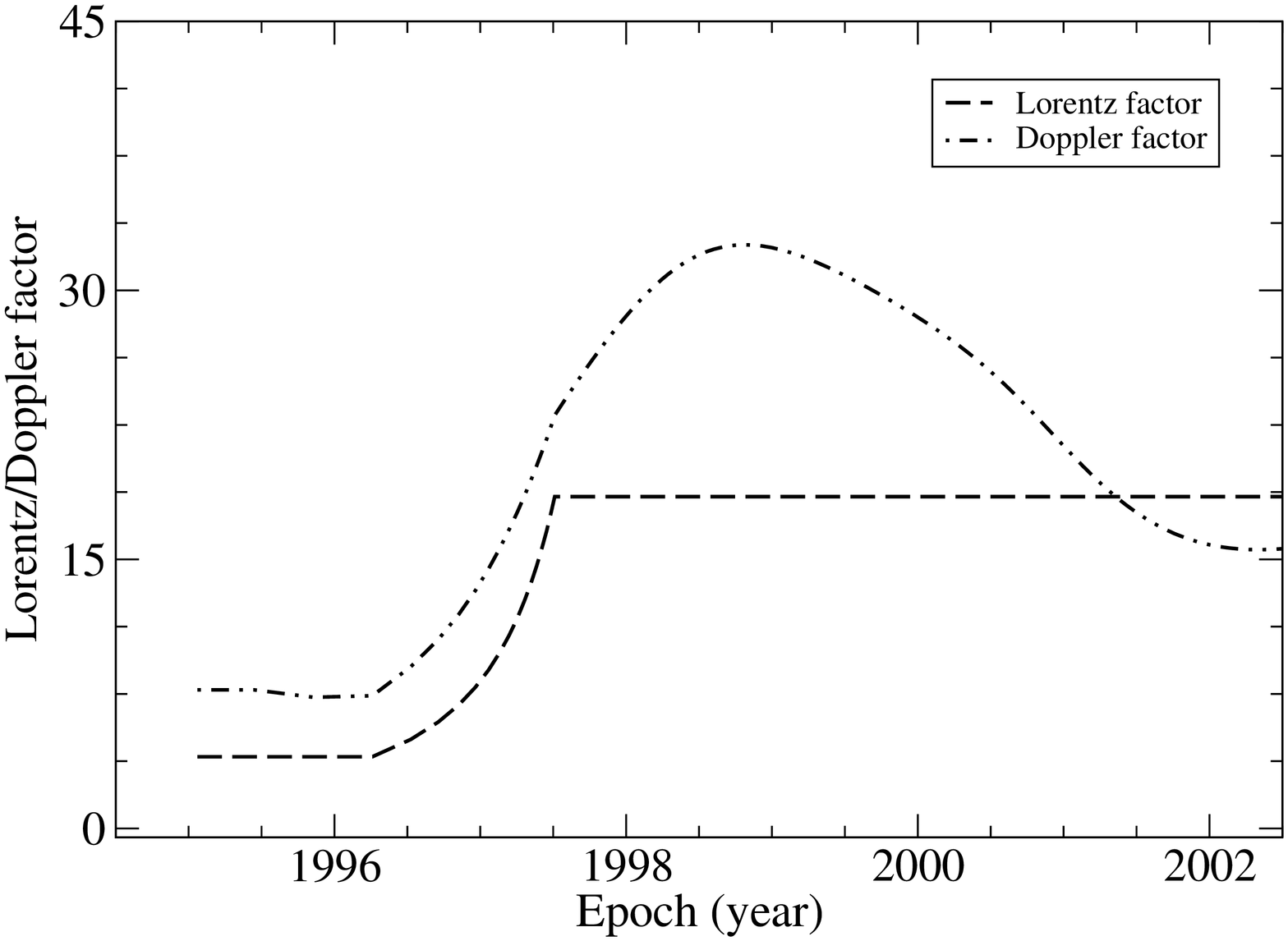}
   \caption{Knot C9: Fitting results are given for trajectory
     $Z_n$($X_n$), coordinates $X_n$(t) and $Z_n$(t) and   
      core separation $r_n$(t). The trajectory can  be well fitted
     by the model within $X_n$$\sim$1.7\,mas. 
    The core separation and coordinates 
   ($X_n$(t), $Z_n$(t)) can be well fitted till $\sim$2001. The upper/left 
   panel shows the precessing common trajectories for precessing phases
   5.54\,rad and 5.54$\pm$0.31\,rad (lines in magenta and blue), indicating
   most data-points are within the area defined by the two lines
    within ${X_n}\sim$1.7\,mas. Precession phase $\phi_0$=5.54\,rad+4$\pi$ and
    $t_0$=1995.06 approximately  equal to 1987.99 ($t_0$ for knot C6)
     plus one period (7.3yr). Note that the Doppler factor curve $\delta$(t)
    ,which was produced by the accelerated and helical motion of knot C9, 
    has a smooth bump structure during period 1997--2001. Such a pattern of
    Doppler factor curve can well interpret the radio flux evolution of knot
    C9 through Doppler boosting effect. See text.}
   \end{figure*}
   \subsubsection{Model-fitting of kinematics for knot C9}
    The model-fitting of the kinematics of knot C9 is very important and
    encouraging for our precessing nozzle scenario for blazar 3C345.\\
     Its kinematic behavior can be very well 
   explained in terms of our scenario as shown in Figure 8.\\
    Its ejection epoch is modeled to be $t_0$=1995.06, corresponding to  
    precession phase $\phi_0$(rad)=5.54+4$\pi$.\\
    It can be seen from Fig.8 that its observed  precessing common trajectory
   may extend to $r_n\sim$1.8\,mas, corresponding to a spatial distance of
    $Z_{c,m}$=62.3\,mas (or $Z_{c,p}$$\sim$414.5\,pc) from the core.\\ 
     Bulk acceleration is  required and its Lorentz factor is modeled as:
    For Z$\leq$1.0\,mas $\Gamma$=4; for Z=1--6\,mas
    $\Gamma$=4+14.5(Z-1)/(6-1);for Z$>$6\,mas $\Gamma$=18.5.\\
     During the period 1996.5--2001.25 its Lorentz factor $\Gamma$, Doppler 
    factor $\delta$, apparent velocity $\beta_a$ and viewing angle $\theta$
    vary over the following ranges: [5.0,18.5], [9.0,(32.5),21.3], 
    [2.7,(17.9),(12.0),18.3] and [3.52,(1.14),2.67](deg.). As shown in
    Figure 8 (bottom left panel) during the period 1997.5-2001.0 the derived
    apparent velocity is $\sim$14$\pm$2\,c, which is very well consistent with 
    the observed apparent velocity 13.2$\pm$1.2c (Schinzel \cite{Sc11a}).
    The model-fitting results for knot C9 show favor toward our precessing
    nozzle scenario:
    \begin{itemize}
    \item The modeled ejection epoch 1995.06 is very close to that (1994.75)
    derived by Klare (\cite{Kl03}) using polynomial extrapolation.\\
    \item Its entire trajectory observed during $\sim$1996.5--2001.5 (in
    a time range of 5 years) can 
    extraordinary well be fitted by the precessing nozzle model with its
   precession phase $\phi_0$=5.54+4$\pi$.
    \item The observational data-points are all concentrated  around the
    precessing common trajectory predicted by the scenario, showing its
     clear regularity, especially considering its curved structure and
    quite long extension of $\sim$1.80\,mas from the core (equivalent to a 
    spatial distance 414.5\,pc).
    \item Around the precessing common trajectory almost all the observational
    data-points locates within the region defined by the magenta and blue 
    lines (Figure 8) which represent the criterion of validity with the
     model-fitting accuracy of $\pm{5\%}$ of the precession period for 
    its precession phase (or $\pm$0.36\,yr for its ejection time). This 
    result for knot C9 is very helpful for verifying the assumption of 
    precessing common trajectory for blazar 3C345.
    \item More important, its precessing common trajectory is very similar
    to that of knots C5 and C22/C23 (also C7 and C12, see below), showing
    some recurrent occurrence of periodic ejection of knots, not only in 
    ejection periodicity, but also in similar curved trajectory structure.
    Such a kind of phenomenon may not be interpreted in terms of instabilities
    of jets during their propagation through surrounding medium.
    \end{itemize}
    All these results were consistently obtained by using physically 
    reasonable assumptions and methods, demonstrating that 
       the model-fitting of the
     kinematics of knot C9 has provided very encouraging and confident 
     results for justifying our precessing nozzle scenario for blazar 3C345.\\
     However, there still remains a question: whether the Lorentz factor 
     and viewing angle ($\Gamma$(t) and $\theta$(t) as functions of time)
     derived in the model-simulation are correct \footnote{$\Gamma$(t) 
     represents the acceleration of knot C9 and $\theta$(t) represents the 
     variation in viewing angle due to its helical motion.}. This would be 
     a (or the last) determinative test for validating our precessing 
     nozzle scenario. Thus we investigated the radio (at 43GHz and 15GHz) flux 
     evolution of knot C9 and its relation to the Doppler boosting
     effect.\footnote{This is the only way to check the correctness of the 
     derived Lorentz factor and viewing angle curves.}
     As its Doppler factor  $\delta(t)$ has been derived from the 
     model-fitting of its kinematics, showing a bump-structure during
     the period 1997--2001 (Fig.8, bottom/right panel), it would be a vital
    test to see whether its radio light-curves could
    be interpreted in terms of Doppler boosting effect.
     In order to investigate the relation between the radio flux evolution 
    and Doppler boosting we used both 15\,GHz and 43\,GHz light-curves and took
    spectral effect into consideration. The light curves observed at 15\,GHz
    and 43\,GHz are shown in Figure 9 (left panel). They have similar
    profiles with a similar peaking time at $\sim$1999.5 and similar rising 
    and decreasing phases. But the 15\,GHz light-curve 
    had enough data-points to 
    determine the shape of its rising phase, while the 43\,GHz light-curve had
    enough data-points to determine the shape of its decaying phase. The 
    spectral index between 15GHz  and 43GHz is $\sim$0.80 
    ($S\propto{{\nu}^{-\alpha}}$). Thus we reformed the rising shape of the
    43\,GHz light-curve by using the 15\,GHz flux during the period 
    1997.62-1999.28 and reformed the decreasing phase of the 15\,GHz 
   light-curve by using the 43GHz flux during the period 1999.64-2000.26.
    Both reformed light-curves are shown in Figure 9 (right panel), which
    have quite regular patterns. We have also taken the spectral effect
    into account to derive the Doppler boosting profile: 
    $S(t)\propto{[\delta(t)]^{3+\alpha}}$. In Figure 10 are shown the 
    comparison between the reformed 15GHz and 43GHz light-curves and
    the Doppler boosting profile.\footnote{The radio light-curves and Doppler
     boosting profile were normalized to their respective peaking values.}
    It can be seen that both the light-curves
    are extremely well coincident with the Doppler boosting profile during 
    the period of $\sim$1997.50--2000.25 and thus be fully  interpreted
    in terms of the Doppler boosting effect derived by our precessing nozzle
    model.  This may imply that the relativistic shock producing the radio 
    emission of knot C9 during this period is extremely stable without
    distinct variations in its intrinsic radiation.  Because the Doppler factor
    curve $\delta(t)$ has already been determined by the model-fitting of the
     kinematics of knot C9 (Figure 8, bottom/right panel),
     thus the reformed 15GHz and 43GHz light-curves for knot C9 being
    well coincident with the Doppler boosting profile is very important.
    This coincidence may imply that we have made a vital test justifying our 
    precessing nozzle scenario, demonstrating that it not only can be used to
    interpret the kinematics of superluminal knots (including trajectory, 
    coordinates, core separation and apparent velocity), but also the 
    Lorentz-factor and Doppler-factor profiles of superluminal components
    can be correctly derived, thus providing effective and applicable ways 
    to investigate their radio flux evolution and intrinsic variations.\\ 
    The almost perfect interpretations of the VLBI-kinematics, 
     flux evolution and nature of the superluminal component C9 in 3C345 
    \footnote{Similar results for more superluminal components in 3C345
    for validating our precessing nozzle scenario
    will be presented elsewhere (Qian, in preparation). Correspondingly,
    the model parameters selected for describing the jet direction, jet cone
     size and pattern of the precessing common helical trajectory, etc. must 
    be correct and appropriate.}
    certainly imply that our precessing jet-nozzle scenario has stood all 
    the observational tests for blazar 3C345 and could also be 
    effectively applicable to study other blazars.\\
     Moreover, these results certainly justify the traditional scenario
    (or common viewpoint) for blazars: superluminal knots move relativistically
    along helical trajectories toward us with acceleration. Obviously, our
     results may not be favorable to
    \begin{figure*}
    \centering
    \includegraphics[width=5.5cm,angle=-90]{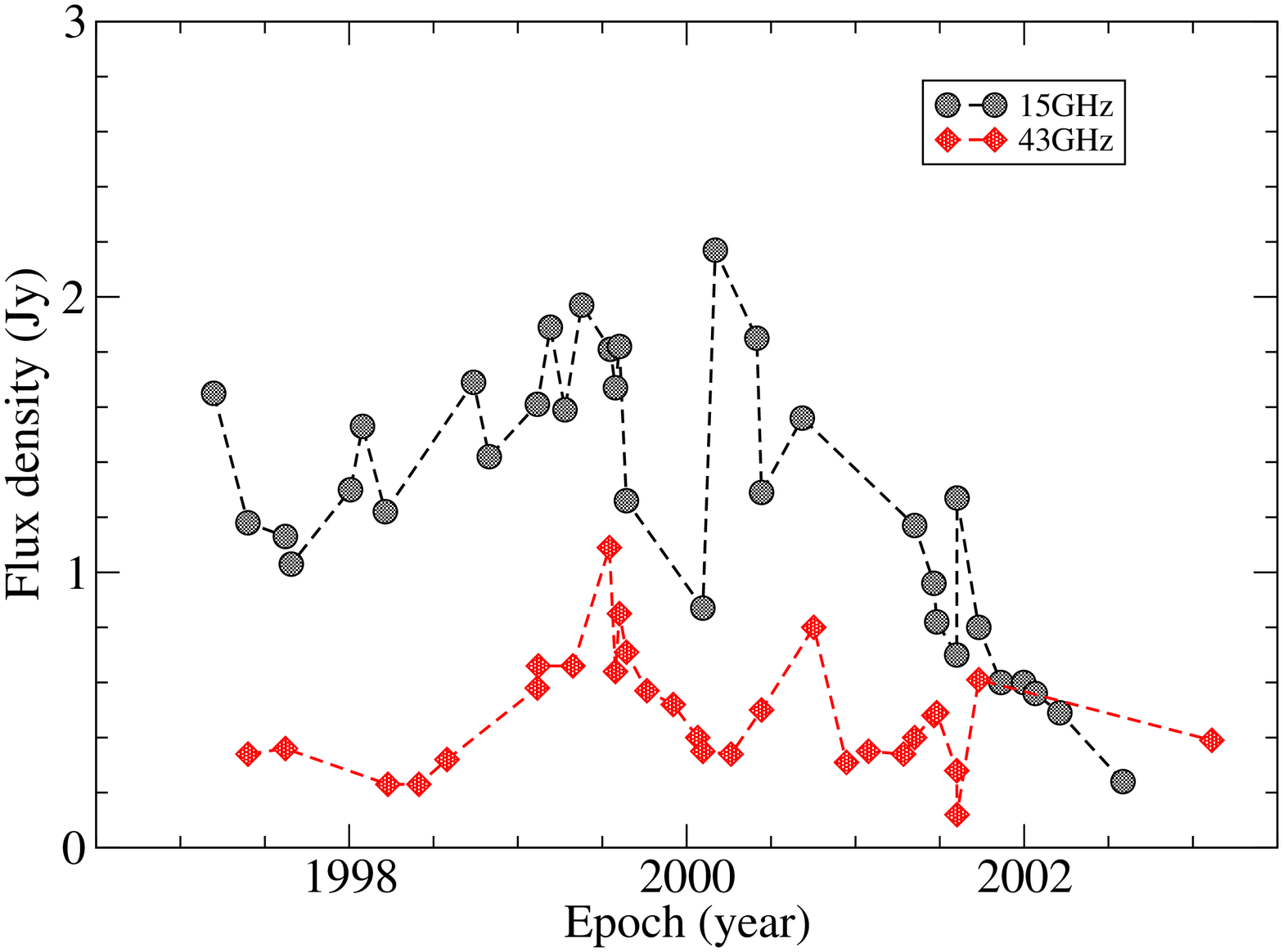}
    \includegraphics[width=5.5cm,angle=-90]{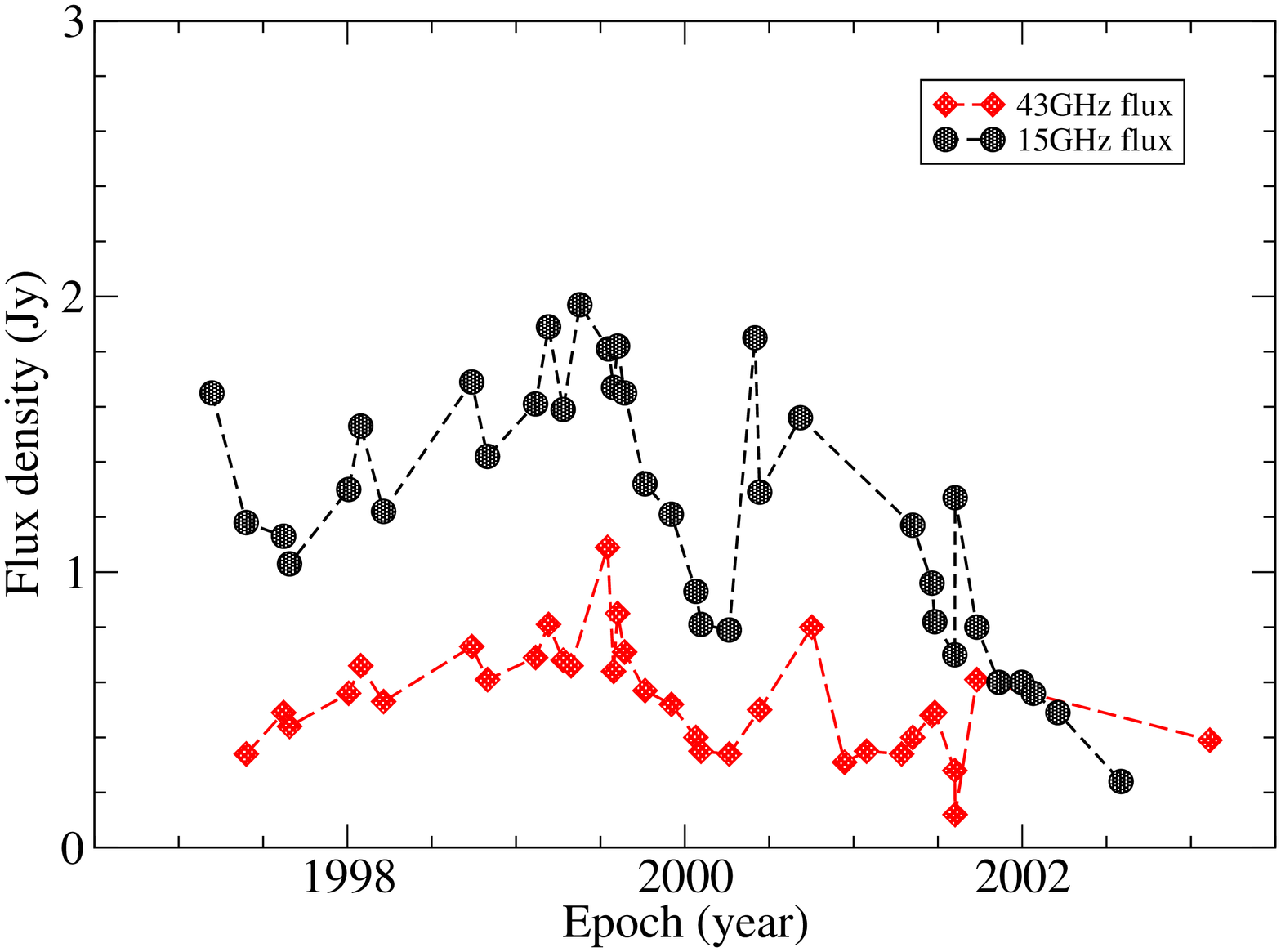} 
    \caption{Knot C9. Left: Correlation between the observed 15GHz and 43GHz 
    light-curves, having similar rising and decreasing phases. 
    Both indicate an epoch for the peak at $\sim$1999.5. Right: the reformed
    15GHz and 43GHz light-curves with enough data-points in the rising and 
    decaying phases to show their regular patterns.}
    \end{figure*}
   \begin{figure*}
   \centering
   \includegraphics[width=5.5cm,angle=-90]{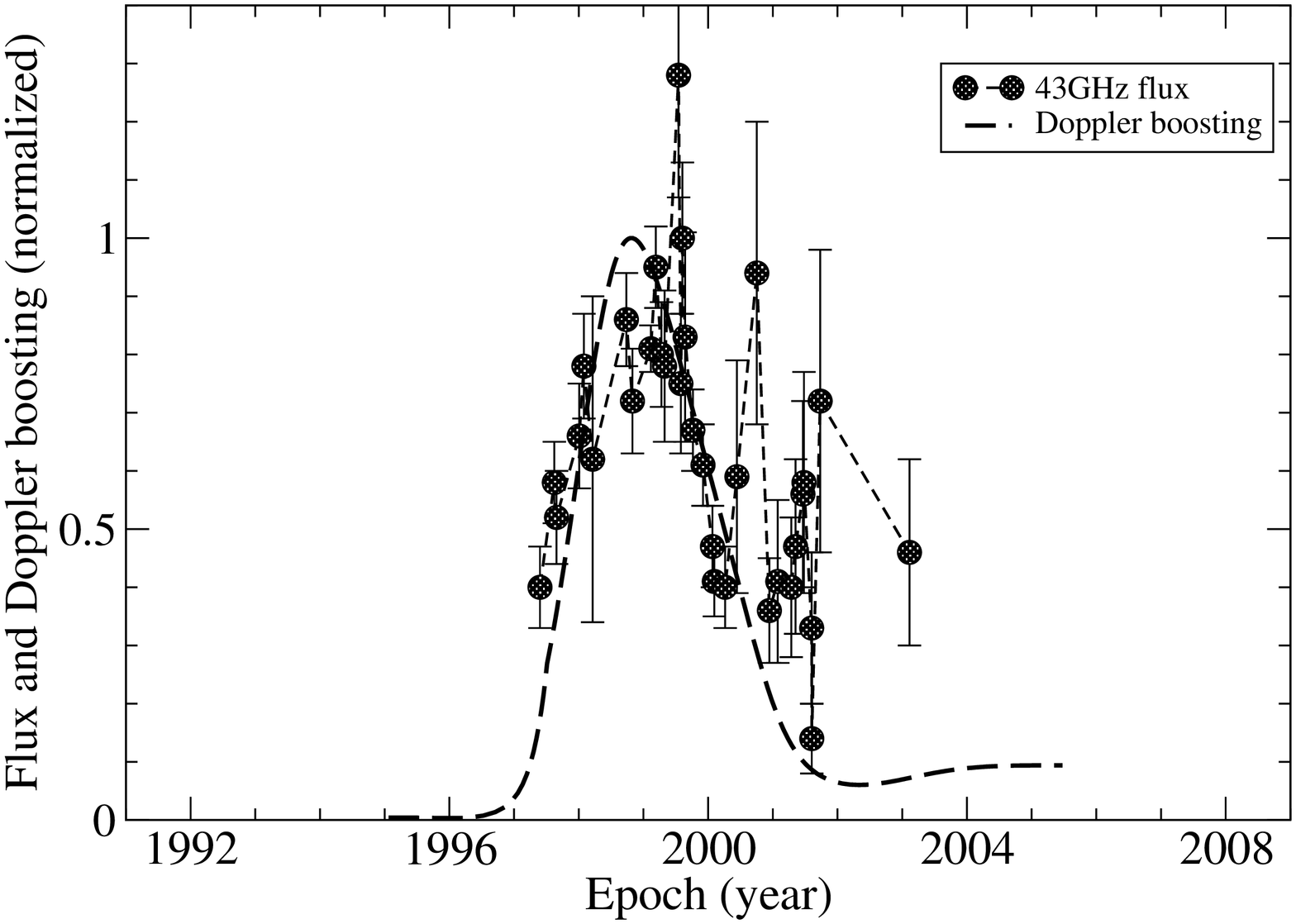}
   \includegraphics[width=5.5cm,angle=-90]{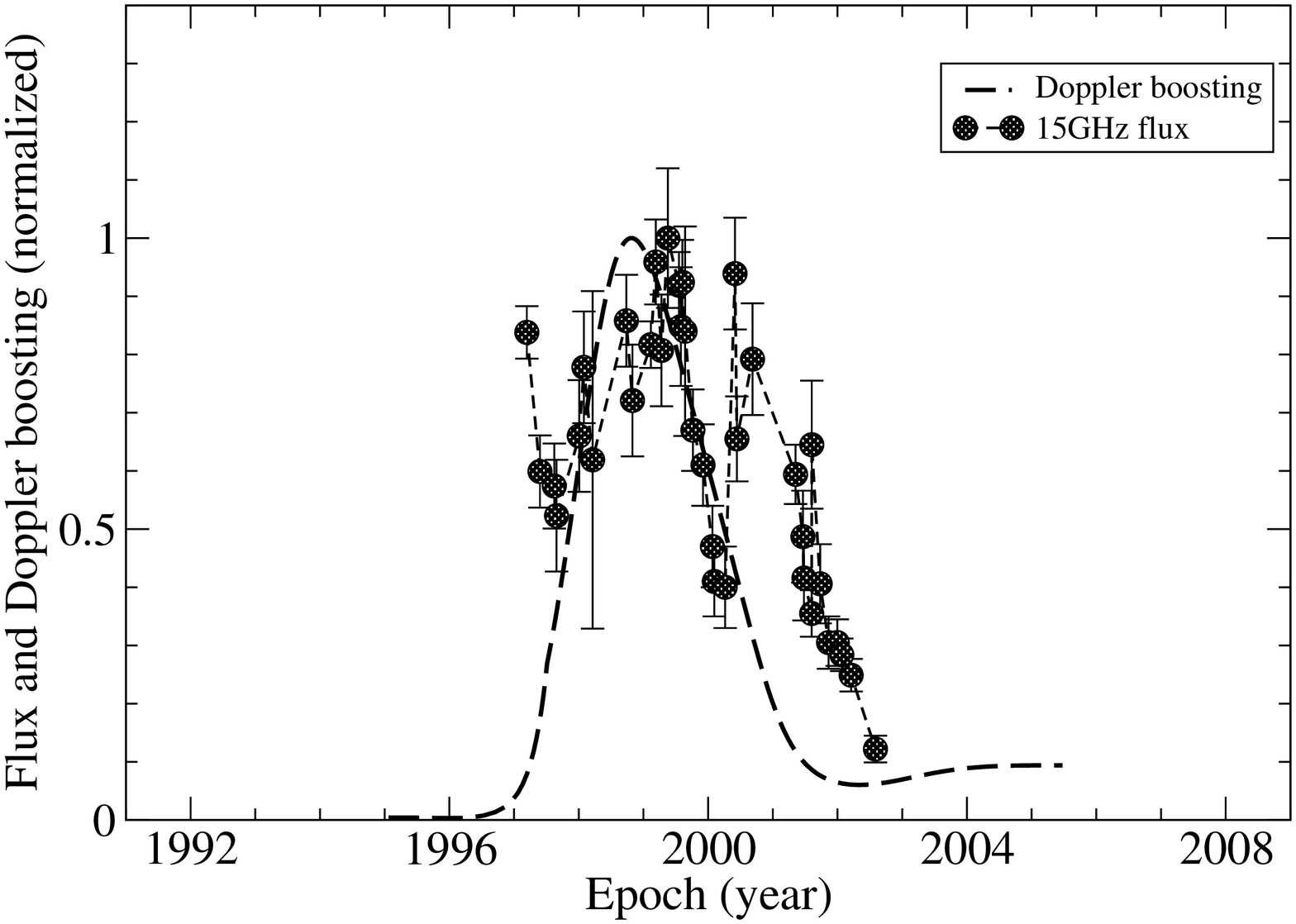}
   \caption{Knot C9. Left: The coincidence of the reformed 43GHz light-curve 
   with the Doppler boosting profile; Right: the coincidence of the reformed
   15GHz light-curve with the Doppler boosting profile. Some sub-flares on 
   smaller time-scales could be the intrinsic variations produced by its
   passages through local standing re-collimation shocks.}
   \end{figure*}
    some scenarios which do not take relativistic motion of superluminal 
    components (as entities) toward us into consideration, e.g., the lit-up
    underlying-pattern scenario, which suggested that the apparent 
    trajectories of superluminal components could result from the jet internal
    structure lit-up by plasma condensations ejected during nuclear flares
    (Schinzel et al. \cite{Sc12a}, \cite{Sc12b}).\\
    However, except the main flare during $\sim$1997.50--2002.25 produced
    by Doppler boosting effect, there remain
     some flux variations (or sub-flares) on smaller time-scales to be
    explained: for example, the variations at 1997.2, 1999.17 and 2000.1 
   and during period 2000.5-2002.6.  These flux variations might be 
   intrinsic due to its  passages through local standing re-collimation shocks.
   \subsubsection{Model-fitting of kinematics for knot C10}
   According to the precessing jet-nozzle scenario of jet-A, the kinematic
   behavior of knot C10 could be explained by assuming its precession phase
    $\phi_0$(rad)=6.14+4$\pi$
   and ejection epoch $t_0$=1995.76. The model-fitting results are shown 
   in Figure A.4. Its kinematics within $r_n{\sim}$0.8\,mas can be interpreted
   in terms of the precessing nozzle model. This corresponds to a spatial
   distance of $Z_{c,m}{\sim}$18.0\,mas, equivalent to $Z_{c,p}$$\sim$119.7\,pc
     from the core.\\
     Knot C10 is found to be accelerated. Its bulk Lorentz factor is
    modeled as: for Z$\leq$2.0\,mas $\Gamma$=4.5; for Z=2--15\,mas 
    $\Gamma$=4.5+24.5(Z-2)/(15-2); for Z$>$15\,mas $\Gamma$=29.0.\\
    During the period 1997.0--2000.0 its Lorentz factor $\Gamma$, Doppler 
    factor $\delta$, apparent velocity $\beta_a$ and viewing angle $\theta$
    vary over the respective ranges: [4.5,29.0], [8.3,34.6], [2.2,(29.0),28.4] 
    and [3.53,1.62](deg.).
   \subsubsection{Model fitting of kinematics for knot C11}
   Model fitting results of the kinematic behavior of knot C11 
    are shown in Figure A.5. Its precession phase and ejection epoch are 
    assumed to be $\phi_0$(rad)=5.88+4$\pi$ and $t_0$=1995.46.\\
   It can be seen from Figure A.5 that its observed precessing common 
  trajectory  may extend to $r_n\sim$0.75\,mas, which corresponds to 
   a spatial distance of $Z_{c,m}{\sim}$=15.7\,mas, equivalent to $Z_{c,p}$= 
   $\sim$104.2\,pc from the core.\\
     Knot C11 is found to be accelerated and its bulk Lorentz factor is 
   modeled as: for Z$\leq$1.0\,mas $\Gamma$=2.5; for Z=1--3\,mas
   $\Gamma$=2.5+13.5(Z-1.0)/(3-1); for Z$>$3.0\,mas $\Gamma$=16.0.\\
   During the period 1998.0--2000.5 its Lorentz factor $\Gamma$, Doppler factor
   $\delta$, apparent velocity $\beta_a$ and viewing angle $\theta$ vary over
   the respective ranges: [2.7,15.0], [5.1,22.3], [0.8,(14.7),13.1] and
    [3.70,2.24](deg.).
   \subsubsection{Model fitting of kinematics for knot C12}
   According to the precessing jet-nozzle scenario for jet-A, the kinematic
   behavior of knot C12 can be model-fitted by assuming its precession phase
   $\phi_0$(rad)=6.30+4$\pi$ and ejection epoch $t_0$=1995.95. The 
   model-fitting results are shown in Figure A.6.\\
    It can be seen from Figure A.6 that its observed precessing common 
   trajectory may extend to $r_n{\sim}$0.50\,mas from the core,
    corresponding to a spatial distance $Z_{c,m}{\sim}$9.67\,mas,
   equivalent to $Z_{c,p}$$\sim$64.3\,pc from the core.\\
    The acceleration of its apparent motion can be modeled as: for 
    Z$\leq$1.2\,mas $\Gamma$=2.7; for Z=1.2--3\,mas 
    $\Gamma$=2.7+10.3(Z-1.2)/(3-1.2); for Z$>$3\,mas $\Gamma$=13.0.\\
    During the period 1999.0--2001.0 its Lorentz factor $\Gamma$, Doppler 
    factor $\delta$, apparent velocity $\beta_a$ and viewing angle $\theta$
    vary over the following ranges respectively: [3.5,13.0], [6.5,19.9], 
    [1.3,10.9] and [3.48,2.43](deg.).
   \subsubsection{Model-fitting of kinematics for knot C13}
   The model-fitting results of the kinematic behavior for knot C13 are shown
    in Figure A.7. Its observed  precessing common trajectory may be assumed to
    extend to $r_n{\sim}$0.70\,mas, corresponding to a spatial distance 
    $Z_{c,m}$$\sim$14.3\,mas or $Z_{c,p}{\sim}$95.3\.pc from the core.\\
   Its precession phase and ejection epoch are modeled as: $\phi_0$=6.50+4$\pi$
   and $t_0$=1996.18. Due to plenty of data-points Figure A.7 indicates that the
   model fitting of its kinematic behavior is very good: most of data-points
   are located within the region limited by the lines in magenta and blue
   which are the modeled trajectories for precession phases 
   $\phi_0$=6.50+0.31\,rad and 6.50-0.31\,rad. This seems strongly justifying 
   the periodicity (7.30\,yr) of  knot-ejection for jet-A.\\
    The acceleration of knot C13 is modeled as: for Z$\leq$1.6\,mas 
   $\Gamma$=3.5; for Z=1.6--13\,mas $\Gamma$=3.5+17.5(Z-1.6)/(13-1.6);
   for Z$>$13\,mas $\Gamma$=21.0.\\
   During the period 1999.0--2002.0 its Lorentz factor $\Gamma$, Doppler factor
   $\delta$, apparent velocity $\beta_a$ and viewing angle $\theta$ vary over
   the respective ranges: [4.1,21.0], [7.7,25.7], [1.8,(20.9),20.5] and
    [3.30,2.19](deg.).
   \subsubsection{Model fitting of kinematics for knot C14}
   The kinematic behavior of knot C14 seems exceptional and instructive.
   Its precession phase and ejection epoch are modeled as: 
   $\phi_0$(rad)=3.16+6$\pi$ and $t_0$=1999.61. The model-fitting results
    are shown in Figure A.8. It can be seen that the modeled
     initial trajectory has a complex pattern of curvature:
    firstly curved southward and then northward. Interestingly, the
   data-points within $X_n{\sim}$0.3\,mas are well located in the region
   limited by the lines in magenta and blue defined by precession phases
   $\phi_0$(rad)=3.16$\pm$0.33,respectively.\\ 
   Its observed  precessing common trajectory  might be assumed to extend to
   $r_n{\sim}$0.5\,mas, corresponding to spatial distance $Z_{c,m}{\sim}$
   18.8\,mas, equivalent to $Z_{c,p}{\sim}$ 125.0\,pc.\\
    Its  motion has been modeled as accelerated: for Z$\leq$0.3\,mas $\Gamma$=
   9.0; for Z=0.3-10\,mas $\Gamma$=9+1.8(Z-0.3)/(10-0.3); For Z$>$10\,mas
   $\Gamma$=10.8.\\
   During the period 2000.0--2003.5 its Lorentz factor $\Gamma$, Doppler factor
   $\delta$, apparent velocity $\beta_a$ and viewing angle $\theta$ vary over
   the following ranges respectively: [9.3,10.8], [18.2,17.6], [2.2,8.1] and
     [0.74,2.40](deg.).
   \begin{figure*}
   \centering
   \includegraphics[width=5.5cm,angle=-90]{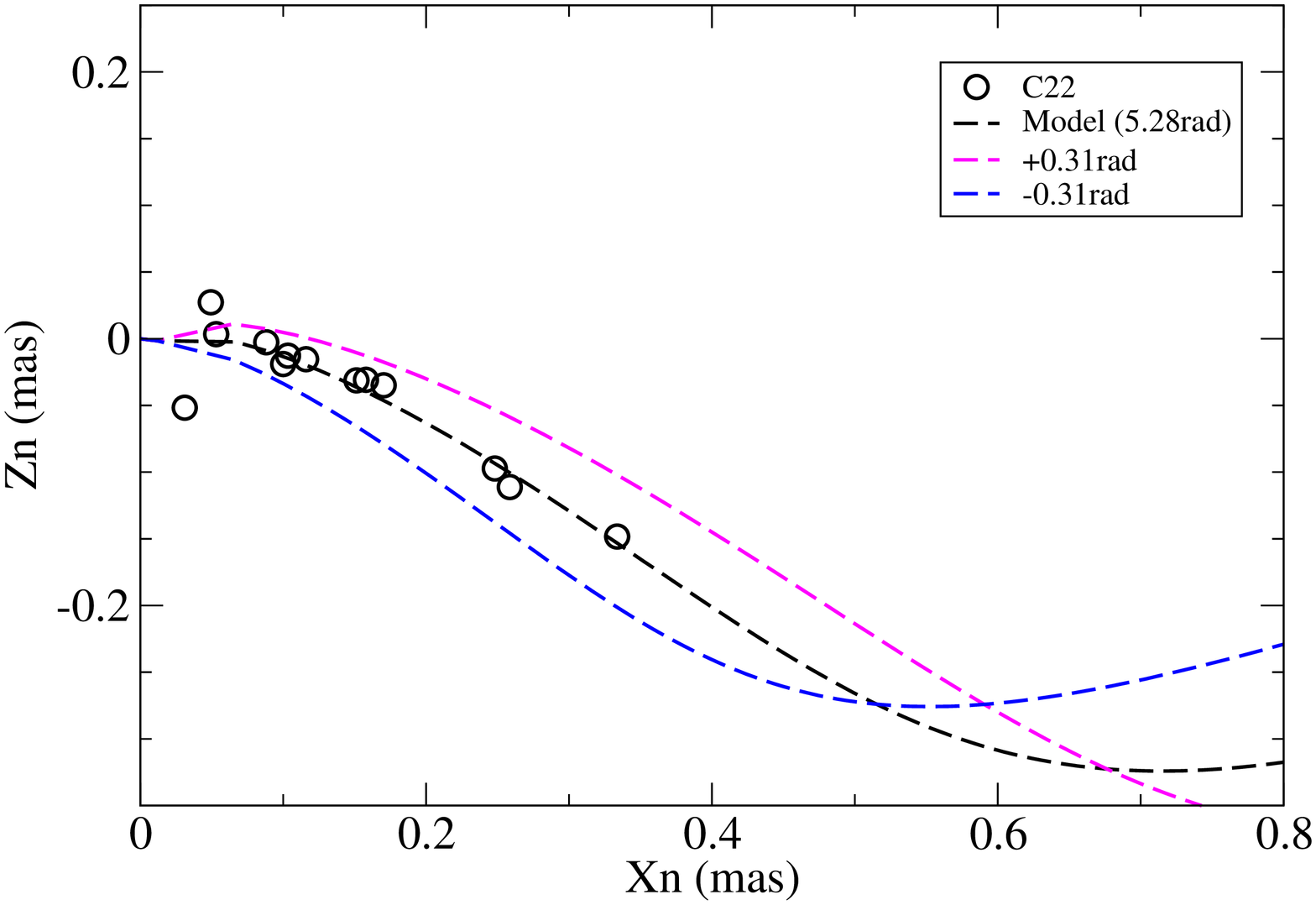}
   \includegraphics[width=5.5cm,angle=-90]{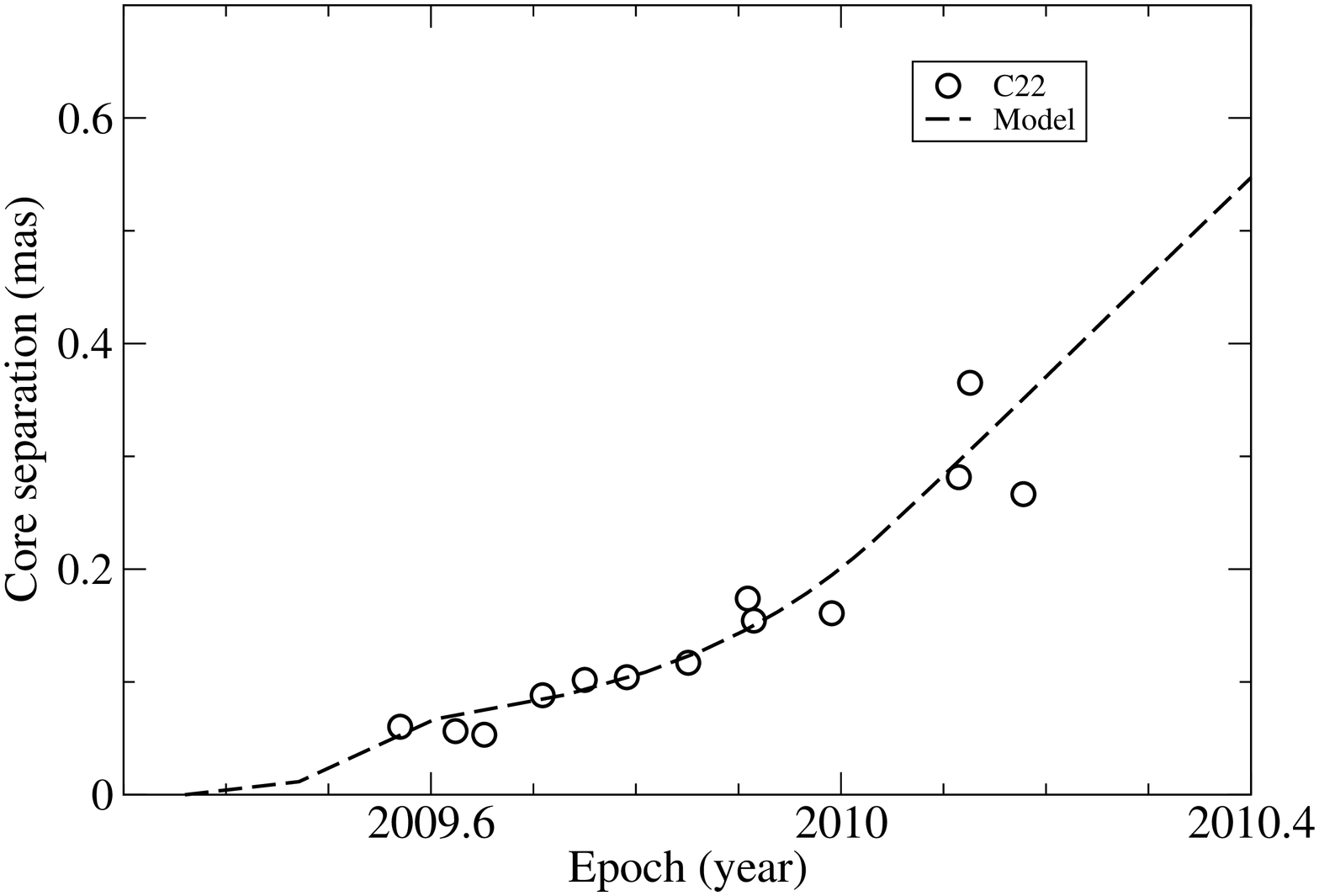}
   \includegraphics[width=5.5cm,angle=-90]{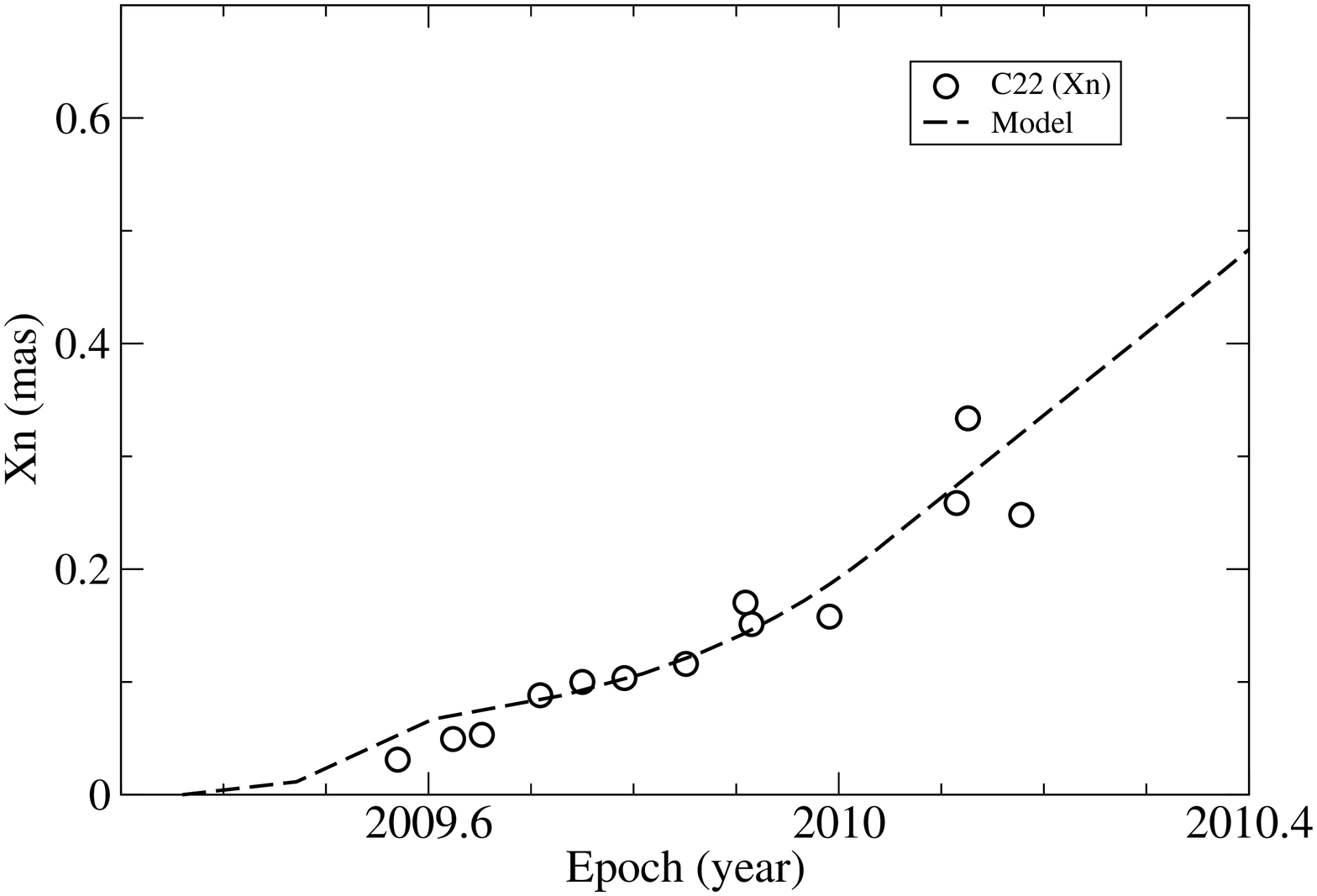}
   \includegraphics[width=5.5cm,angle=-90]{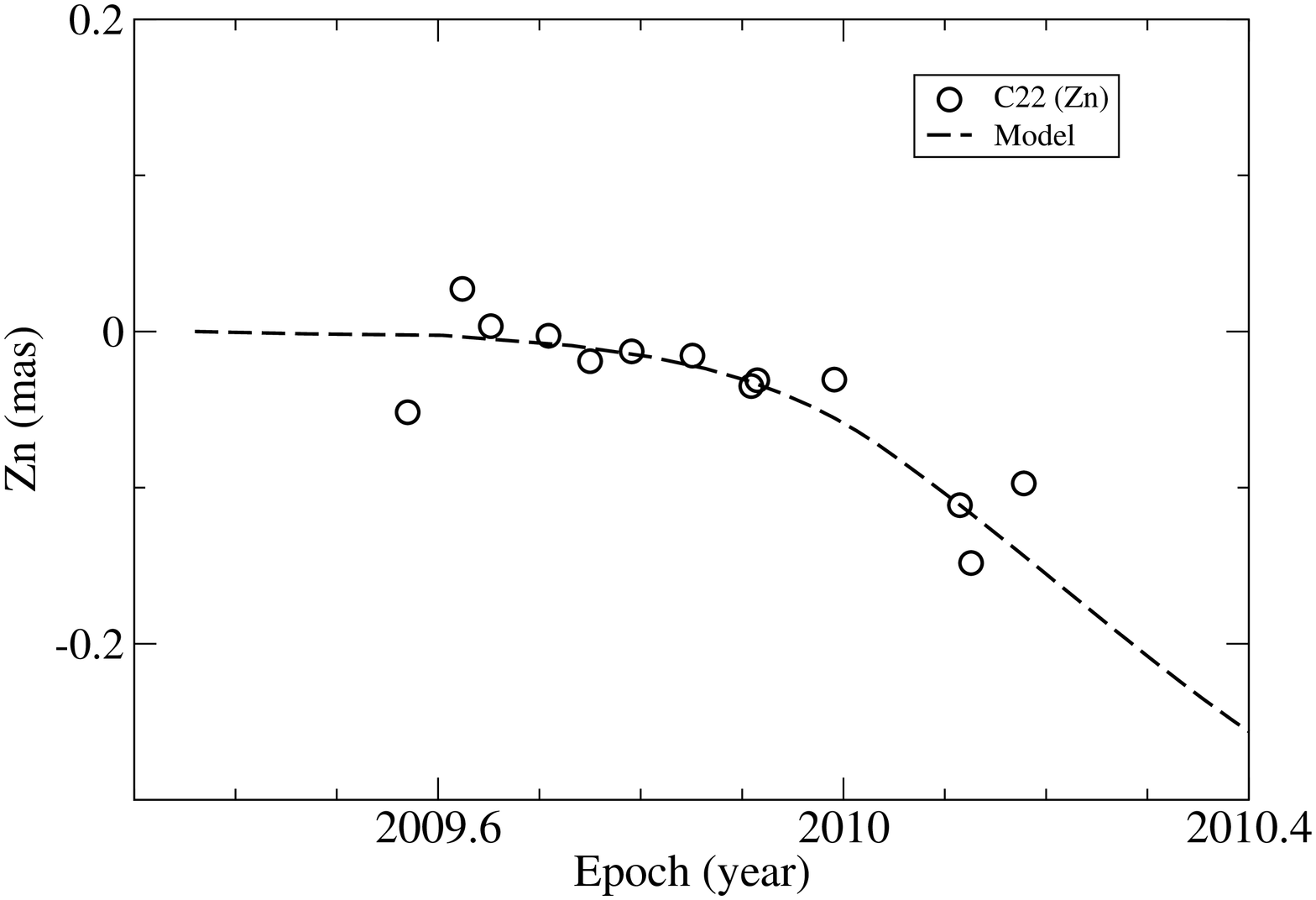}
   \includegraphics[width=5.5cm,angle=-90]{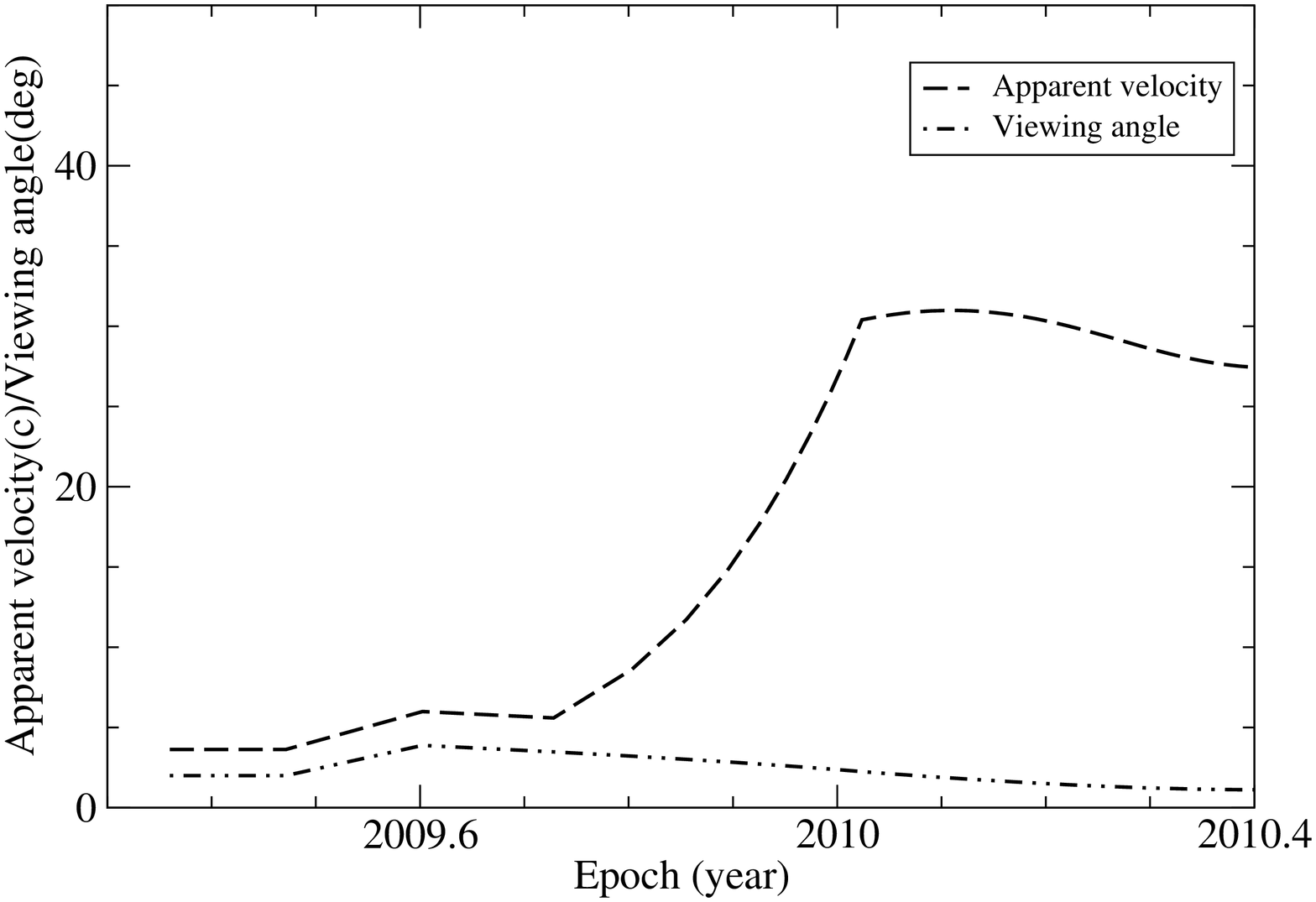}
   \includegraphics[width=5.5cm,angle=-90]{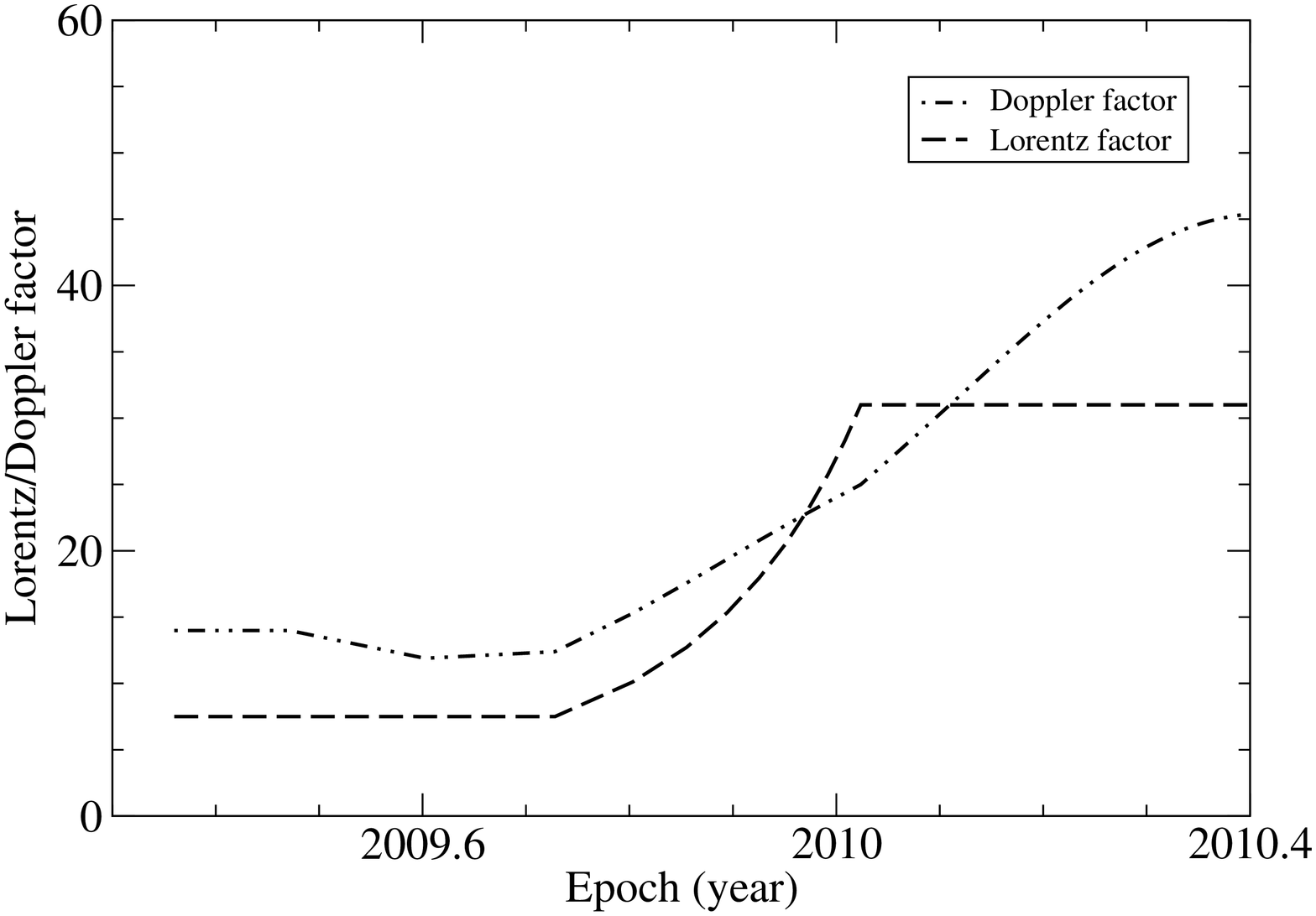}
   \caption{Model fitting of kinematics for knot C22: precession phase 
   $\phi_0$=5.28\,rad+8$\pi$, $t_0$=2009.36.}
   \end{figure*}
   \subsubsection{Model fitting of kinematics for knot C22}
   The model-fitting of the kinematics for knots C22 and C23 is very important
   for justifying the precessing nozzle scenario of jet-A, because the 
   VLBI-observations of knot C22 and C23 have
    extended the periodic behavior of jet-A to 4 precession periods relative 
   to ejection of knot C4 (from $\sim$1979 to 2009, about 30 years).
    In the following we present the fitting results of
   the kinematic behavior for knots C22 and C23.\\
     According to the precessing nozzle scenario the kinematics of C22 may be 
   modeled by using precession phase $\phi_0$(rad)=5.28+8$\pi$ and
    ejection epoch $t_0$=2009.36. The model-fitting  results are shown in 
   Figure 11.\\
    Its observed  precessing common trajectory may be assumed to extend 
   to core separation $r_n{\sim}$0.4\,mas, corresponding to spatial distance 
   $Z_{c,m}{\sim}$9.67\,mas, equivalent to $Z_{c,p}$$\sim$64.3\,pc
    from the core.\\ 
    The  model-fitting of its trajectory is very good, because most of 
    the data-points within $X_n{\sim}$0.4\,mas are closely concentrated
    around the precessing common trajectory ($\phi_0$=5.28\,rad) predicted 
    by the precessing nozzle scenario.\\
    The motion of knot C22 was accelerated and its Lorentz factor was modeled
    as: for Z$\leq$1.0\,mas $\Gamma$=7.5; 
   for Z=1--4\,mas $\Gamma$=7.5+23.5(Z-1)/(4-1); for Z$>$4\,mas
   $\Gamma$=31.0.\\
    During the period 2009.5--2010.2  its Lorentz factor $\Gamma$, Doppler
   factor $\delta$, apparent velocity $\beta_a$ and viewing angle $\theta$
   vary over the following ranges respectively: [7.5,31.0], [14.0,37.4], 
   [3.6,30.3] and [2.00,1.95](deg.).
  \begin{figure*}
   \centering
   \includegraphics[width=5.5cm,angle=-90]{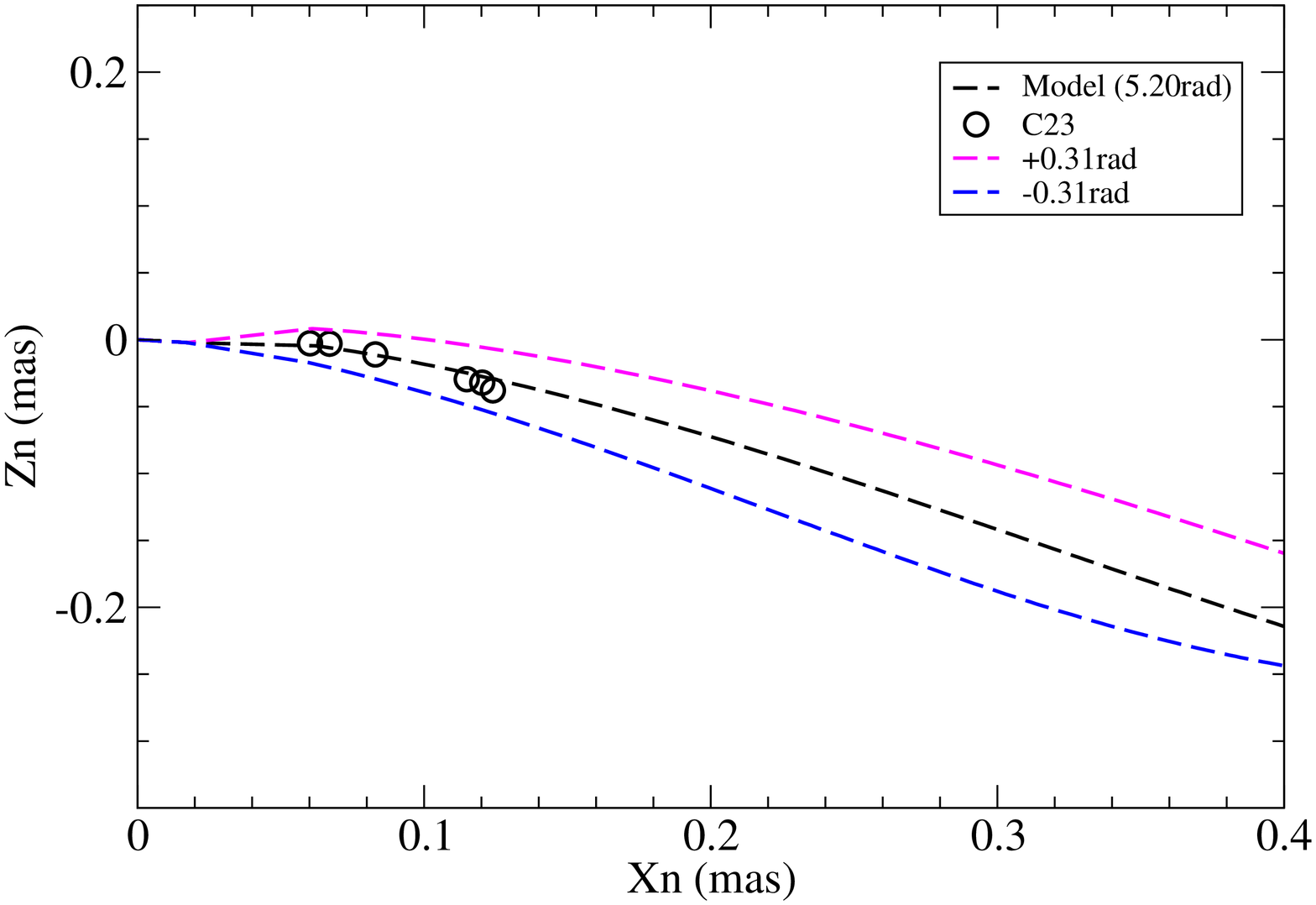}
   \includegraphics[width=5.5cm,angle=-90]{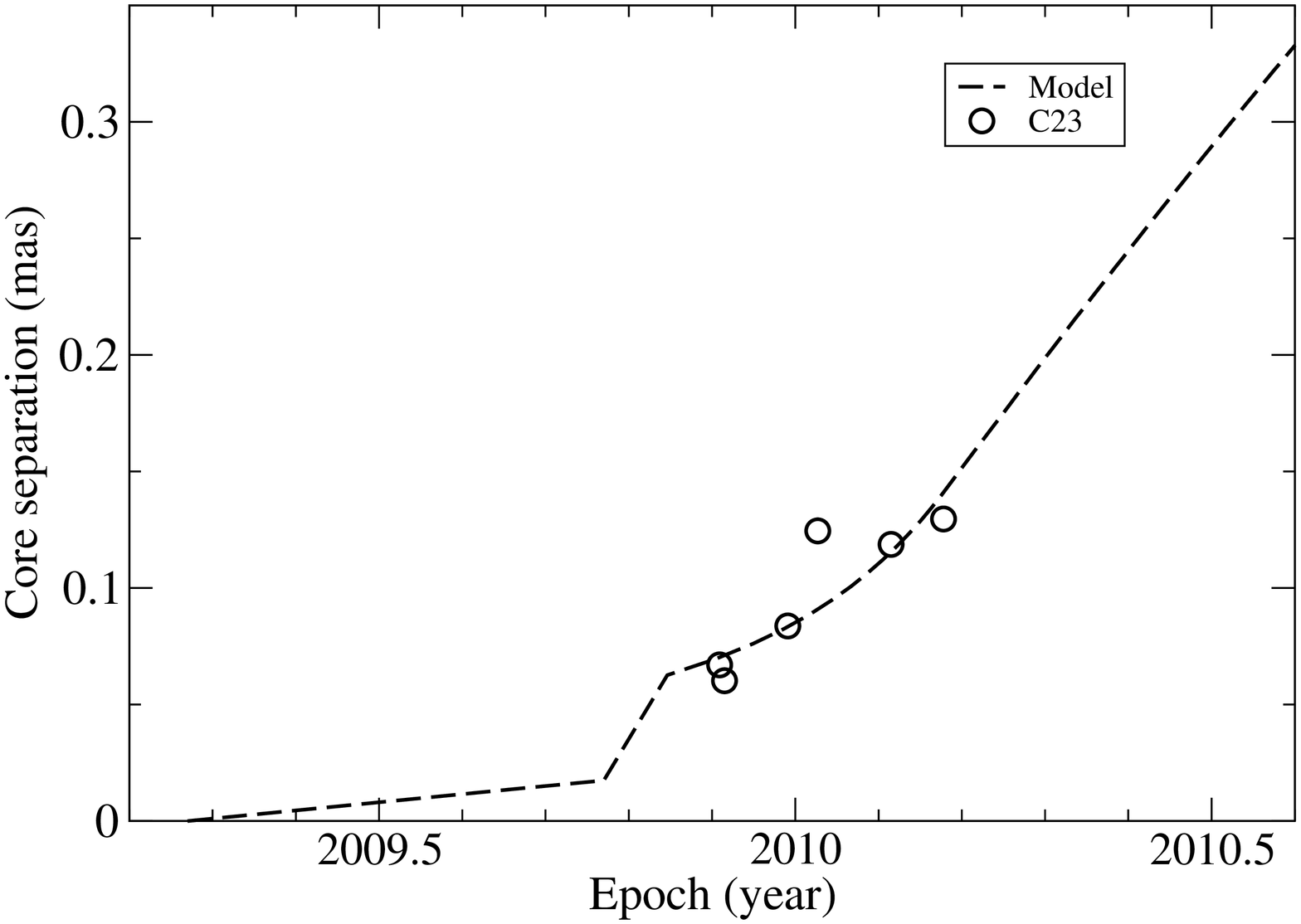}
   \includegraphics[width=5.5cm,angle=-90]{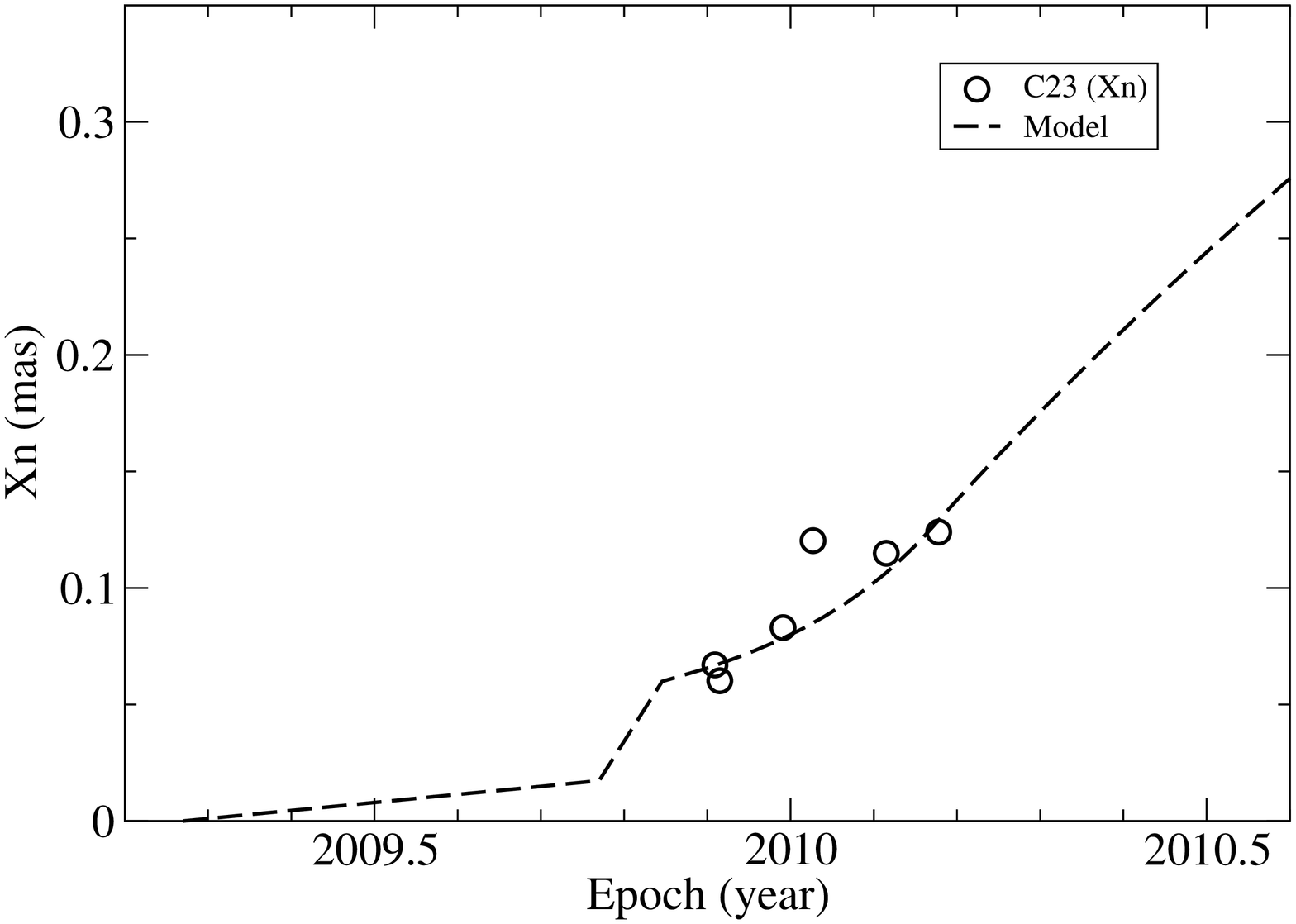}
   \includegraphics[width=5.5cm,angle=-90]{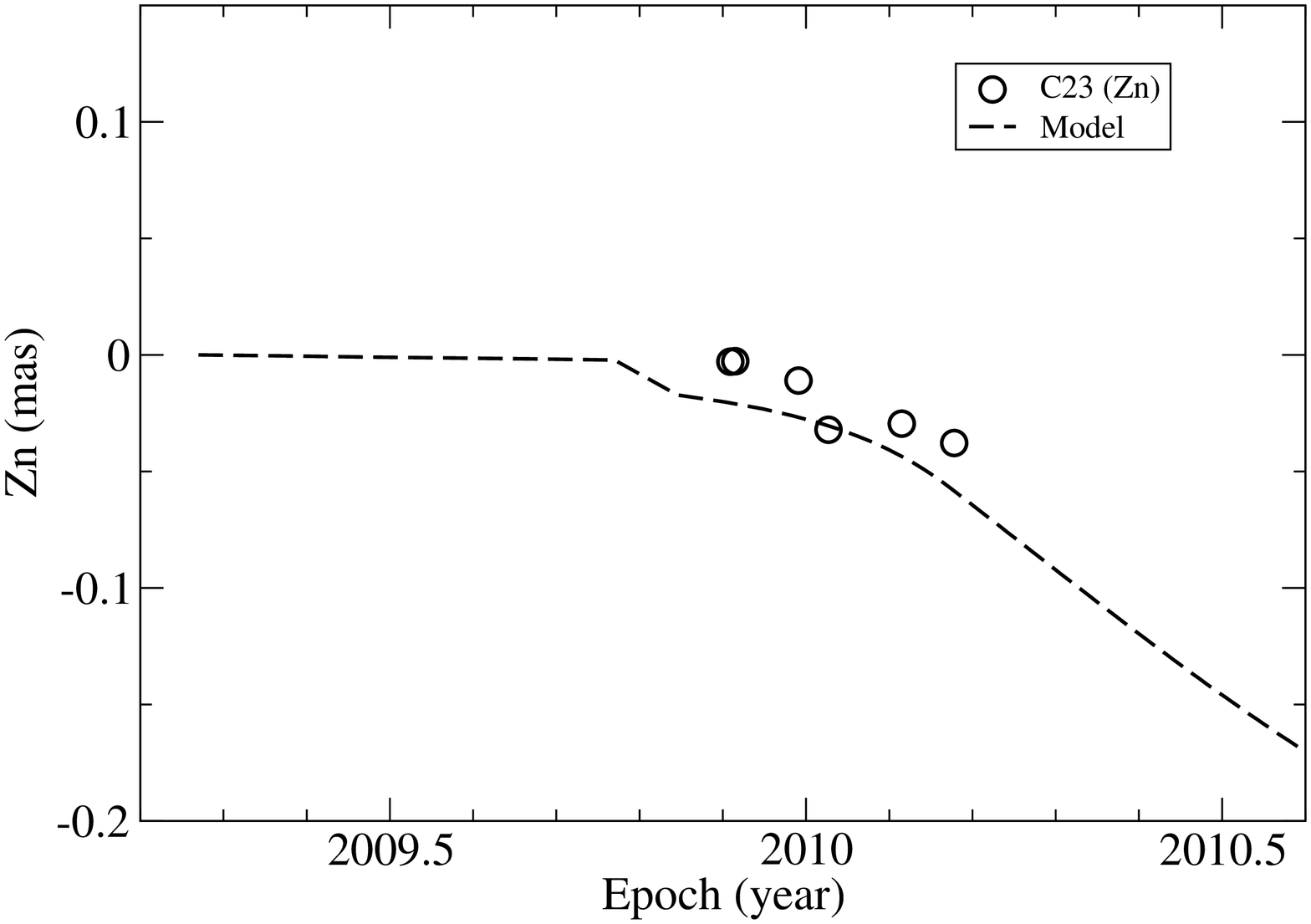}
   \includegraphics[width=5.5cm,angle=-90]{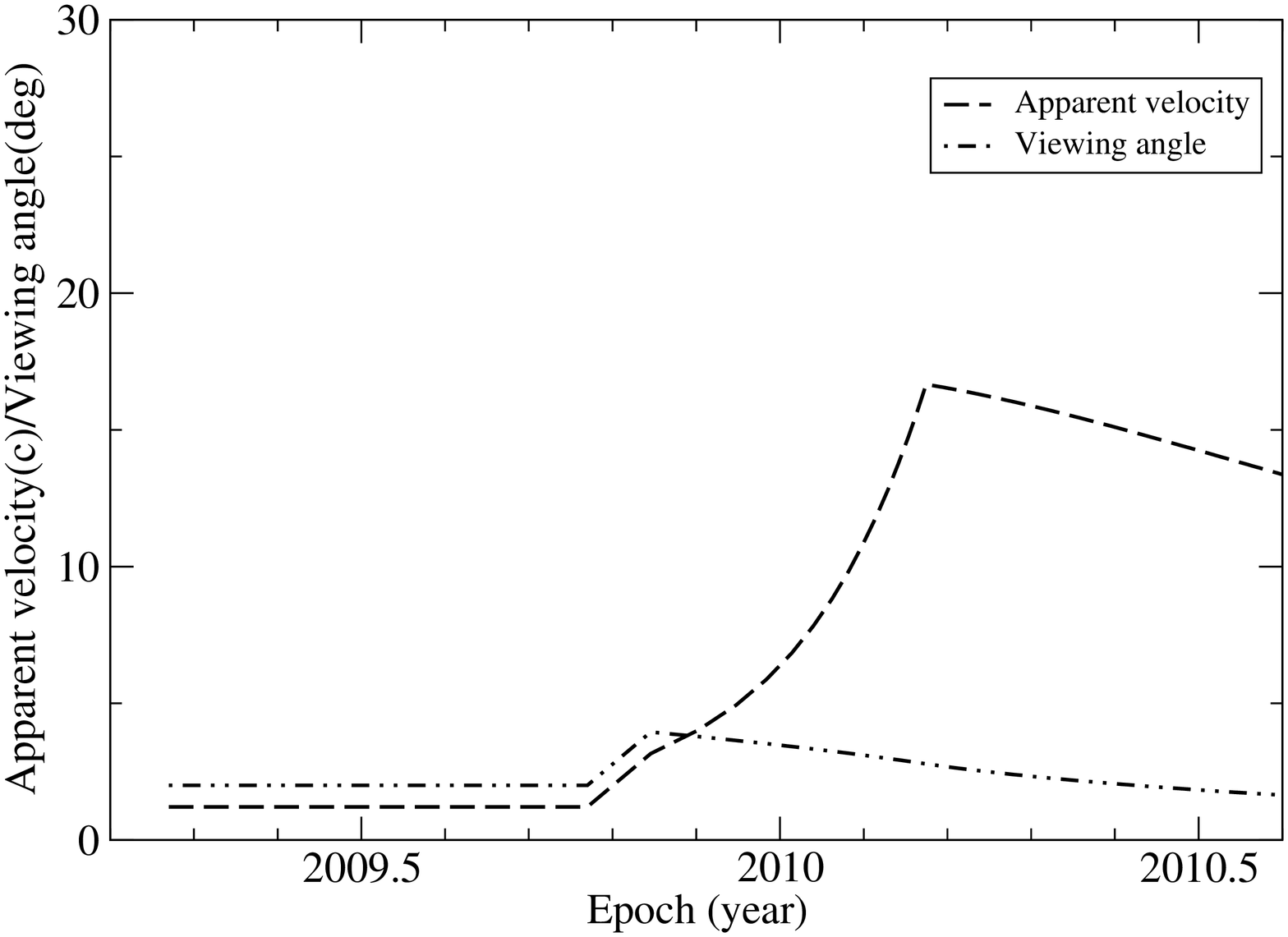}
   \includegraphics[width=5.5cm,angle=-90]{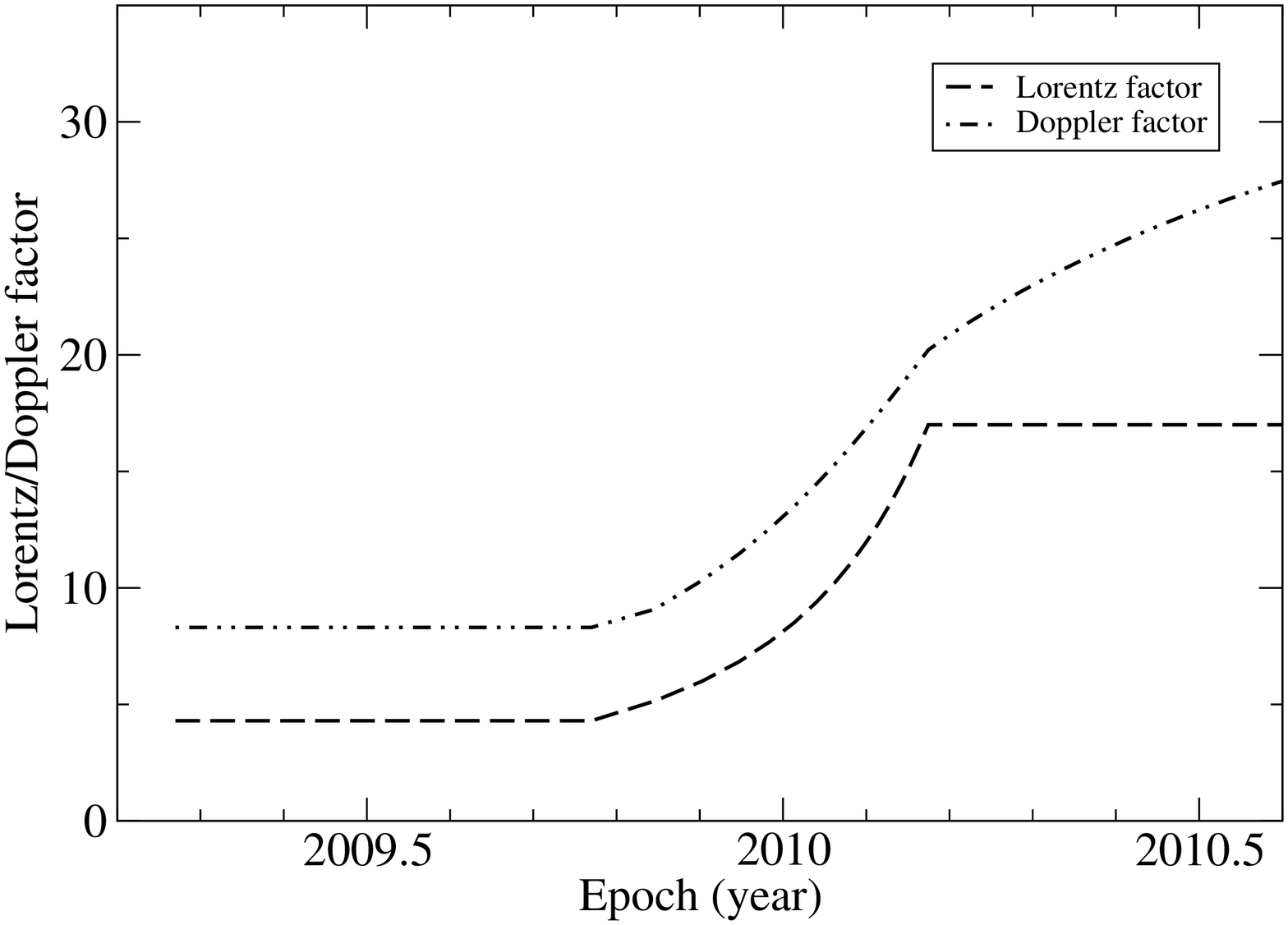}
   \caption{Model fitting of kinematics for knot C23: precession phase 
   $\phi_0$=5.20\,rad+8$\pi$,$t_0$=2009.27.}
   \end{figure*}
   \subsubsection{Model fitting of kinematics for knot C23}
   The kinematic behavior of knot C23 has also been model-fitted 
   in terms of our precessing nozzle model for jet-A. The model-fitting 
   results are shown in Figure 12. Although the data-points  only showed 
   its initial trajectory within $X_n{\sim}$0.15\,mas, they are all closely
   concentrated around the modeled precessing common trajectory, strongly
    justifying our precessing
    nozzle scenario for jet-A with a precession period of 7.30\,yr.\\
    Its precession phase $\phi_0$(rad)=5.20+8$\pi$ and ejection
    epoch $t_0$=2009.27. This ejection time was a bit earlier than that 
    of knot C22 by $\sim$0.09\,yr (about one month), but it is a more accurate
    fit to its observed trajectory.\\
    It can be seen from Figure 12 that its observed  precessing common
    trajectory may be assumed to extend to core separation 
    ${r_n}\sim$0.20\,mas, corresponding to spatial distance 
    $Z_{c,m}$$\sim$3.4\,mas or $Z_{c,p}$$\sim$22.6\,pc from the core.\\
      The motion of knot C23 was also modeled as accelerated.
     Its bulk Lorentz factor
    was modeled as: for Z$\leq$0.5\,mas $\Gamma$=4.3; for Z=0.5--2\,mas
    $\Gamma$=4.3+12.7(Z-0.5)/(2-0.5); for Z$>$2\,mas $\Gamma$=17.0.\\
    During the period 2009.8--2010.2 its Lorentz factor $\Gamma$, Doppler
    factor $\delta$, apparent velocity $\beta_a$ and viewing angle $\theta$
    vary over the respective ranges: [4.3,17.0], [8.3,20.9], [1.2,16.5] 
    and [2.00,2.66].\\
    We would like to point out the characteristic features of kinematic
    behavior of knots C22 and C23 as follows.
    \begin{itemize}
    \item Their ejection times (2009.36 and 2009.27) are close to those
    determined by using linear extrapolation methods in Schinzel 
   (\cite{Sc11a}): 2009.544$\pm$0.018 and 2009.651$\pm$0.042.\\
    \item The trajectories observed for knots C22 and C23 are extraordinarily
    well fitted by the predicted precessing common trajectories, strongly
    favoring our precessing nozzle scenario.
    \item Most importantly, their curved trajectories are very similar to those
    observed for knots C5 and C9, not only implying the existence of a 
    precession period of 7.30\,yr, but also exhibiting the recurrence of 
    similar curved trajectory structures.
    \end{itemize}
    All these observational facts are very helpful to justify our precessing
    nozzle scenario for jet-A. 
   \begin{table*}
   \centering
   \caption{Parameters of the precessing common trajectory for 13 superluminal
    components (C4-C14, C22 and C23: ejection epoch $t_0$, 
     precession phase ${\phi}_0$(rad), 
     , extension of precessing common trajectory from the core 
    $r_{n,c}$(mas), corresponding spatial distance $Z_{c,m}$(mas) and
     $Z_{c,p}$(pc), bulk Lorentz factor $\Gamma$, Doppler factor $\delta$,
    apparent velocity $\beta_a$, viewing angle $\theta$(deg) and status.
    Values in parentheses represent intermediate (or local) maxima or minima.
    Column 'status' denotes the quality of trajectory model-fits:
     + for well fitted cases, -- for not fitted cases.}
    \begin{flushleft}
    \centering
   \begin{tabular}{lllllllllll}
   \hline
   Knot & $t_0$ & ${\phi}_0$ & $r_{n,c}$ & $Z_{c,m}$ & $Z_{c,p}$ & $\Gamma$ &
          $\delta$ & $\beta_a$ & $\theta$ & status\\
   \hline
   C4 & 1979.00 & 4.28 & 1.80  & 52.0 & 345.8  &  8.3-12.4 &
         16.1-(21.7)-17.8 & 3.0-11.1 & 1.28-2.88 & +\\
   C5 & 1980.80 & 5.83 & 1.20  & 39.0  & 259.3 &  5.5-15.0 &
          10.1-(24.8)-23.6 & 2.9-(10.0)-(8.3)-12.2  & 3.03-1.98 & +\\
   C6 & 1987.99 & 5.74+2$\pi$ & 0.40  & 7.5 & 49.7 &  9.7-11.1 &
          13.5-18.7 & 8.9-8.0  & 3.90-2.21 & +\\
   C7 & 1988.46 & 6.14+2$\pi$ & 0.70  & 14.9  & 99.3 & 3.2-13.0 &
          6.2-21.8 & 0.65-(11.9)-9.5  & 2.00-1.92 & +\\
   C8 & 1991.10 & 2.13+4$\pi$ & 0.15  & 6.2 & 41.2  & 6.2-13.8 &
          12.0-19.0 & 1.9-(12.9)-12.8   & 1.48-(2.87)-2.80 & +\\
   C9 & 1995.06 & 5.54+4$\pi$ & 1.80  & 62.3 & 414.5  & 5.0-18.5 &
          9.0-(32.5)-21.3 & 2.7-(17.9)-(12.0)-18.3  & 3.52-(1.14)-2.67 & +\\
   C10 & 1995.76 & 6.14+4$\pi$ & 0.80 & 18.0 & 119.7 & 4.5-29 & 
         8.3-34.6 & 2.2-(29.0)-28.4 & 3.53-1.62 &  +\\
   C11 & 1995.46 & 5.88+4$\pi$ & 0.75 & 15.7 & 104.2  & 2.7-15.0 &
          5.1-22.3 & 0.82-(14.7)-13.1  & 3.70-2.24 & +\\
   C12 & 1995.95 & 6.30+4$\pi$ & 0.50 & 9.7 & 64.3  & 3.5-13.0 &
         6.5-19.9 & 1.3-10.9 & 3.48-2.43 & +\\
   C13 & 1996.18 & 6.50+4$\pi$ & 0.70 & 14.3 & 95.3  & 4.1-21.0 &
          7.7-25.7  & 1.8-(20.9)-20.5  & 3.30-2.19 &  +\\
   C14 & 1999.61 & 3.16+6$\pi$ & 0.50 & 18.8 & 125.0  & 9.3-10.8 &
         18.2-17.6  & 2.2-8.1 & 0.74-2.40 & +\\
   C22 & 2009.36 & 5.28+8$\pi$ & 0.40 & 9.7  & 64.3  & 7.5-31.0 &
         14.0-37.4  & 3.6-30.3 & 2.00-1.95 & +\\
   C23 & 2009.27 & 5.20+8$\pi$ & 0.20 & 3.4  &  22.6  & 4.3-17.0 &
         8.3-20.9 & 1.2-16.5 & 2.00-2.66 & +\\
   \hline
   \end{tabular}
   \end{flushleft}
   \end{table*}
   \section{A brief summary for jet-A}
   The model-fitting results of the kinematic behavior of the thirteen knots 
   (C4--C14, C22 and C23) of jet-A are summarized in Table 1. The main features
   are:
    \begin{itemize}
    \item The kinematics of the 13 knots can be consistently explained in terms
    of our precessing nozzle scenario for jet-A.
    \item the precession period of the jet-nozzle may be 7.30\,yr (or 
    4.60\,yr in the rest frame), which has been observed
      during a time-interval of $\sim$30 years (or in four periods). 
    \item There exists a common trajectory pattern which precesses to 
        produce the observed initial trajectories of  individual knots
         defined by their precession phases. 
     \item This common trajectory pattern may
        be very steady and  controlled by the strong magnetic fields in
        the collimation-acceleration zone of jet-A which is formed in
        the magnetosphere of the central black hole. 
     \item However, for different knots their initial common precessing 
     trajectory sections were observed to
       extend to different distances from the core as given 
     in Table 1. This phenomenon might imply that their outer trajectories
     have been curved due to changes in their status of helical motion.
      \item All the superluminal knots were observed to be accelerated and 
     the modeled maximal bulk Lorentz factors were in the range of 20--30.
     Accelerated motions can be understood in the MHD theories for
     jet-formation in the magnetosphere of strongly magnetized 
     black-hole/accretion-disk systems in the nuclei of blazars (e.g.,
     Vlahakis \& K\"onigl \cite{Vl03}, \cite{Vl04},
      Blandford \& Znajek \cite{Bl77}, Blandford \& Payne \cite{Bl82},
    Camenzind \cite{Ca90}, Meier \& Nakamura \cite{Me06}).
    \end{itemize}
    \begin{figure*}
    \centering
    \includegraphics[width=5.5cm,angle=-90]{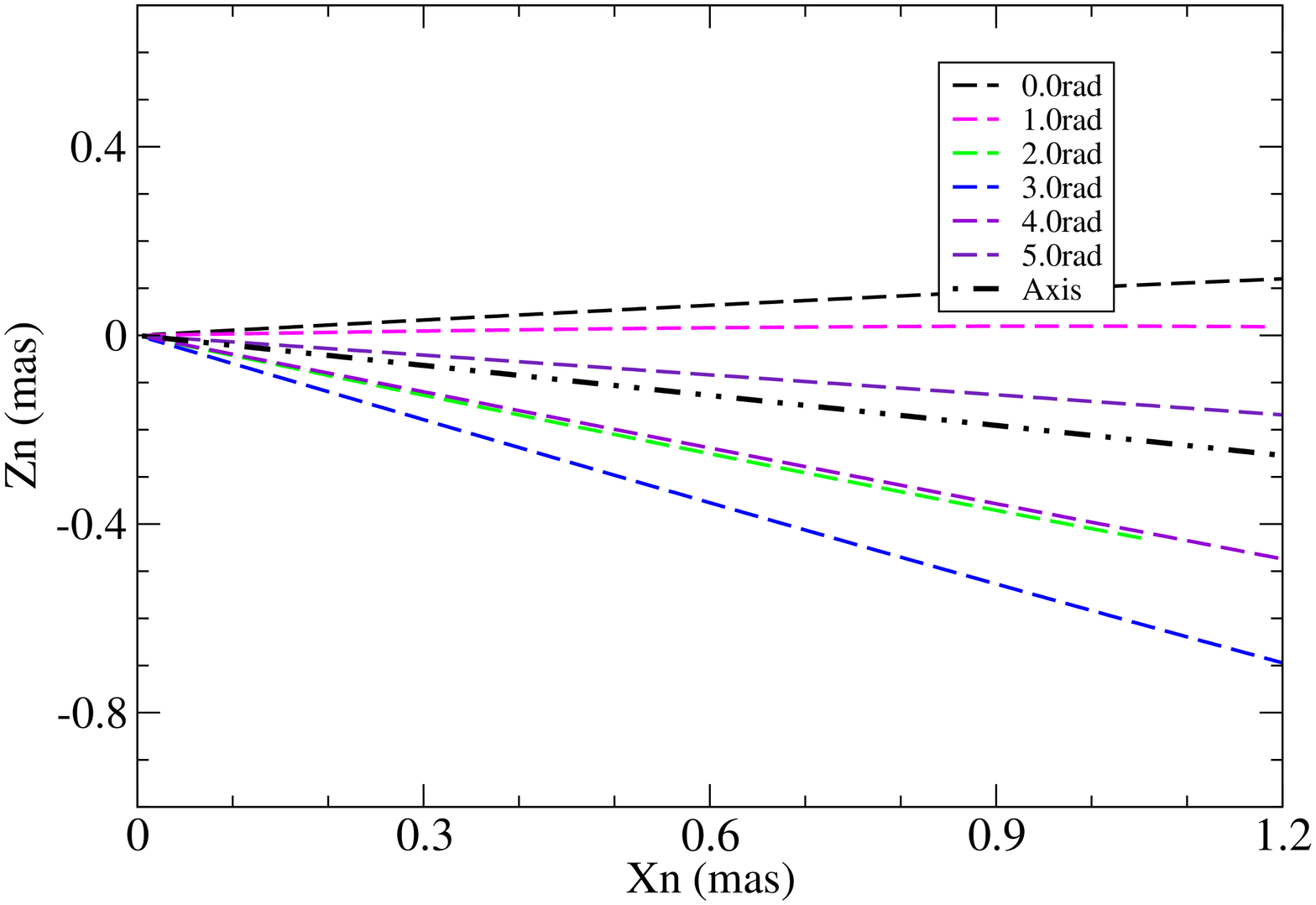}
    \includegraphics[width=5.5cm,angle=-90]{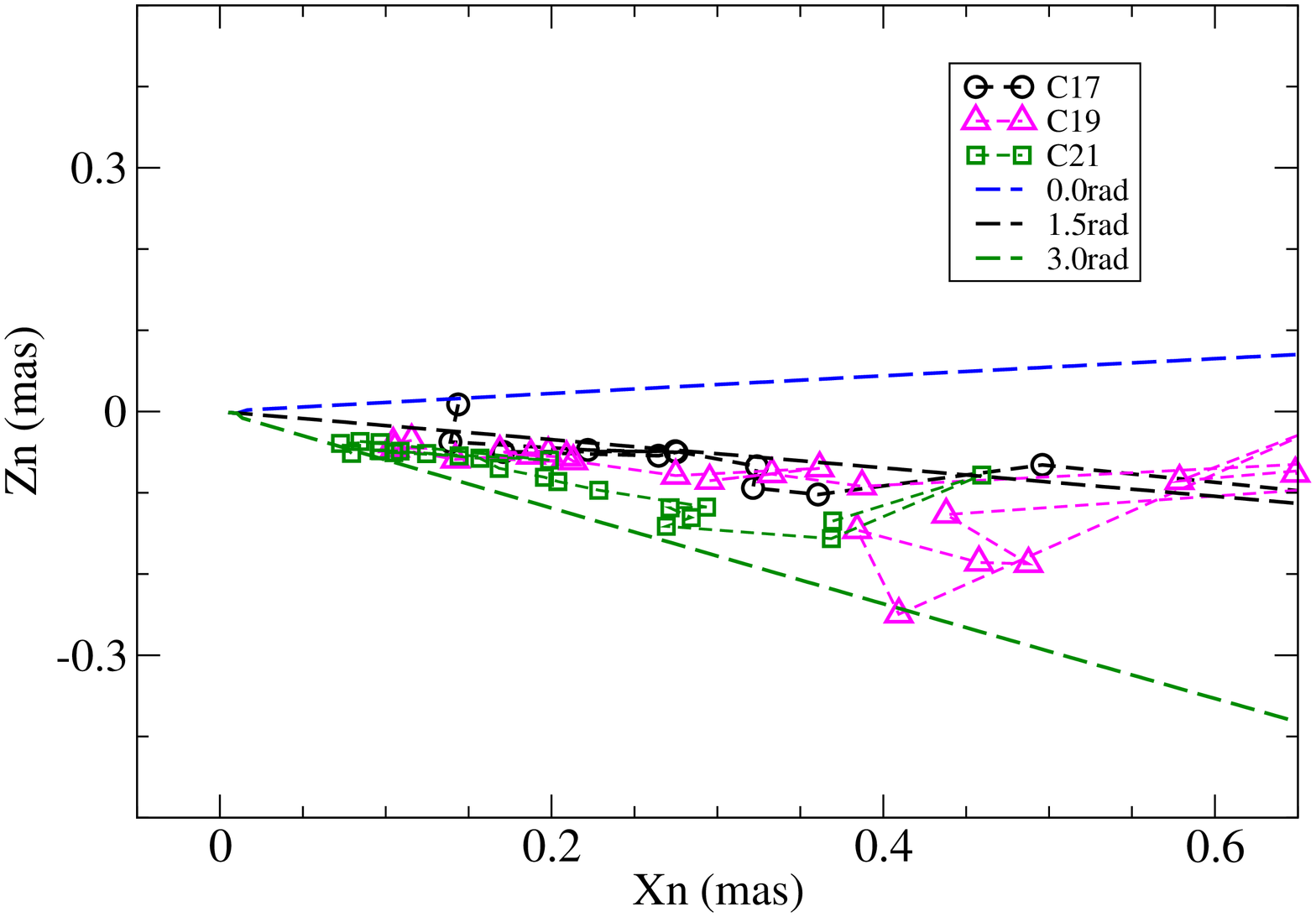}
    \caption{Left: distribution of the precessing common trajectory for jet B.
    The jet axis is at position angle ${\sim}-102.0^{\circ}$ with its 
    cone aperture $\sim{36.8^{\circ}}$ (at core separation $\sim$0.5\,mas).
    The opening angle of the jet is $\sim{0.98^{\circ}}$ in space.
    Right: the observed trajectories of knots C17, C19 and C21 in the jet
    are shown for comparison.}
    \end{figure*}
    \section{Model fitting results for jet-B}
   Our double jet scenario assumes that fourteen superluminal knots (C15--C21,
   B5-B8, B11 and B12) were ejected from jet-B. This is because we found that
   the ejection epochs of these knots can not be explained in terms of 
   the precessing nozzle model for jet-A. Moreover, most of the  trajectories  
   of the knots seem ballistic by visual inspection, obviously different
    from the curved trajectories
   of superluminal knots of group-A. Thus we had to take a different set 
   of parameters to describe the properties of jet-B.\\
    We have taken the model parameters as follows.\\
    $\epsilon$=0.0262\,rad=$1.5^{\circ}$; $\psi$=0.209\,rad=$12^{\circ}$\\
    $p_1$=0; $p_2$=1.341$\times{10^{-4}}$${\rm{mas}}^{-1}$;$Z_1$=396\,mas;
    $Z_2$=3000\,mas; $Z_3$=3.581\,mas; $z_t$=66\,mas; $z_m$=6$\times$$10^5$.\\
     $\it{x_0}$ and p(i) are similarly defined by equations (1) and (2), but
    taking parameter $\zeta$=1.0.\\
    The amplitude and phase of the helical trajectory are defined as:\\
      \begin{equation}
       A(Z)={A_0}{\sin[{\pi}{z_0}/{Z_1}]}
      \end{equation}
      \begin{equation}
       {\phi(Z)}=const.={{\phi}_0}=5.70+{\frac{2\pi}{T_0}}({t_0}-2002.12)
      \end{equation}
     $A_0$=1.09\,mas and ${\phi}_0$ also represents the precession phase of 
     individual knots. $\phi$ is assumed to be independent of Z and  
     the motion of the knots is approximately ballistic. $T_0$=7.30\,yr. 
     The modeled distribution  of trajectory of superluminal components 
    produced by the jet-nozzle precession of jet-B is shown in Figure 13. We 
    found that most of the superluminal knots followed the precessing common
    trajectories predicted by the precessing nozzle scenario for jet-B.\\
    We would like to point out that the distributions
    of the precessing common trajectory for jet-B (Figure 13) and jet-A 
    (Figure 5) are similar: their jet-axes are at respective position angles 
    $-102.0^{\circ}$ and $-97.2^{\circ}$\footnote{Their difference 
    $4.8^{\circ}$ is similar to the difference in their parameters $\psi$.},
     and their cone apertures are
    ${\sim}36.8^{\circ}$ and $\sim{42.5^{\circ}}$,respectively. However, this
    does not imply that jet-A and jet-B overlap in space. Using the
    values given for the parameters defining the helical trajectories for 
    jet-A in Section 4 and jet-B (this Section), we can
    calculate the angle between their jet-axes and their jet-cones in space
    and obtain the following results (at core separations $r_n{\sim}$1\,mas,
    or corresponding spatial distances Z$\sim$30\,mas): (1) the angle between 
    their jet-axes is $\sim6.8^{\circ}$ in space; (2) their jet
    cone-apertures in space are $\sim{1.12}^{\circ}$ (jet-A) and 
    $\sim{0.98^{\circ}}$, respectively.  Thus the two jets (jet-A and 
    jet-B) are widely separated in space. The observed overlap of the
     projections of jet-A and jet-B in the sky-plane should not be 
    misled into thinking that they are overlapped in space.\\
     As for the case of jet-A, the values selected for the model parameters of
     jet-B  and the associated functions are not statistical samples and not
    unique. They are a specific and physically applicable set of working
    ingredients, which have been obtained through trial and error over the
    past few years. Our aim was to search for evidence of jet precession in
    3C345 and we found that they could be applied to analyze the distribution
    of the observed trajectories and kinematics of superluminal components in 
    3C345 on VLBI-scales and disentangled its  possibly existing double jets.
    In this paper we used the methods of multi-parameter model simulations,
    being not able to give statistical errors to the model parameters. But we
    provided a new criterion to judge the validity of the entire model-fitting
    by showing that the observed trajectories of most knots followed their
    precessing common trajectories predicted in terms of the scenario 
    within $\pm{5\%}$ of the precession period.  
     \begin{figure*}
   \centering
   \includegraphics[width=5.5cm,angle=-90]{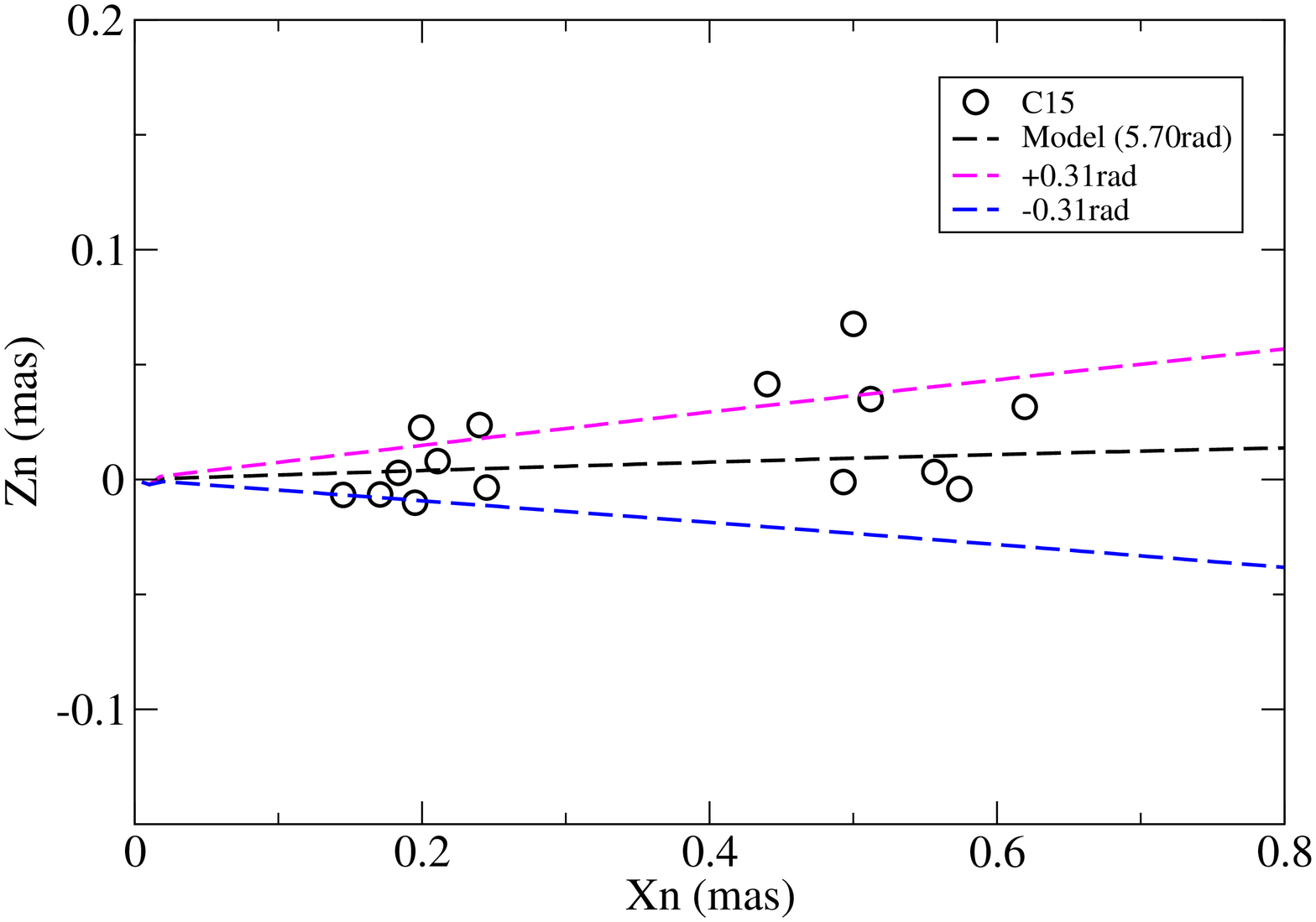}
   \includegraphics[width=5.5cm,angle=-90]{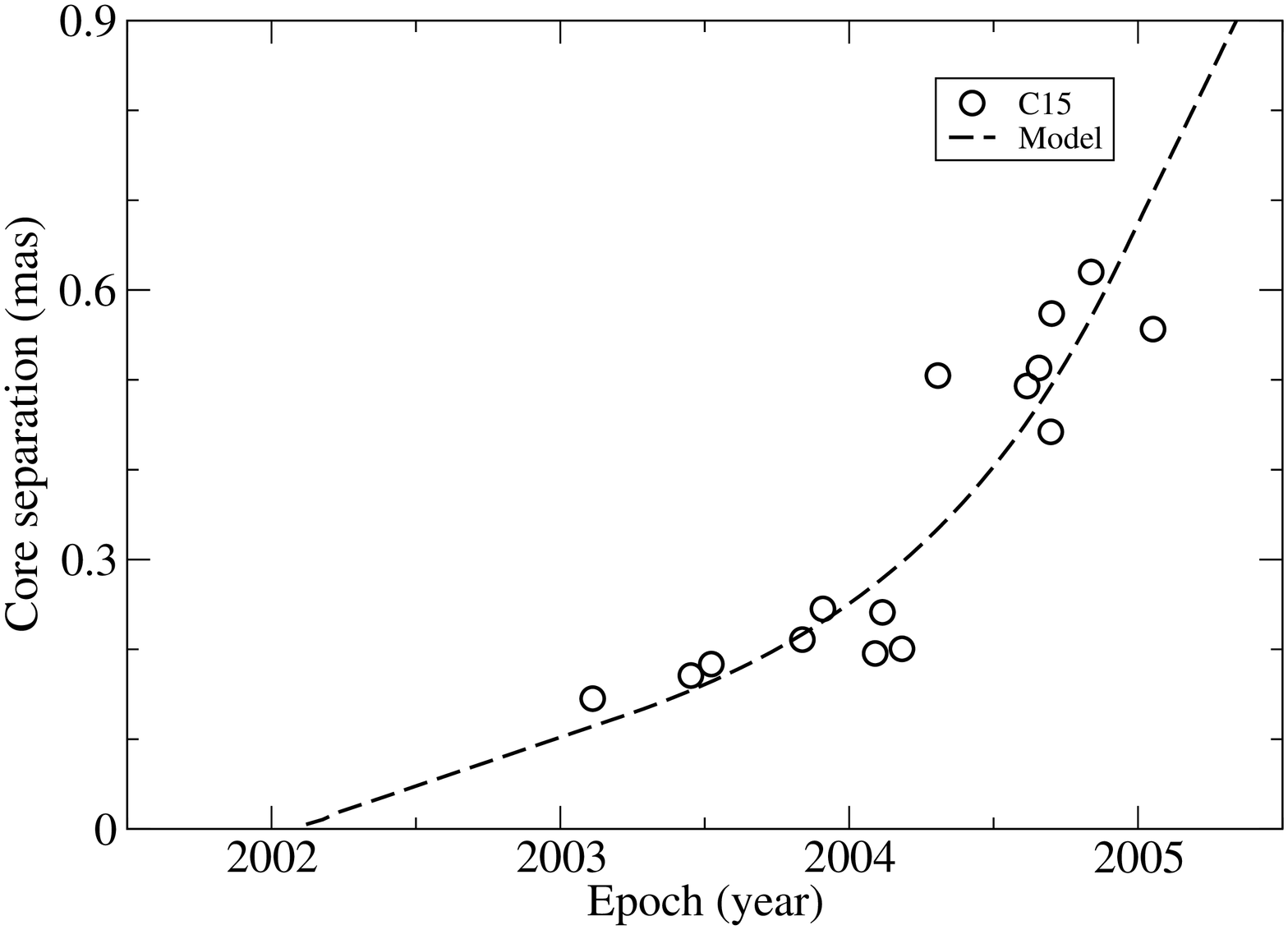}
   \includegraphics[width=5.5cm,angle=-90]{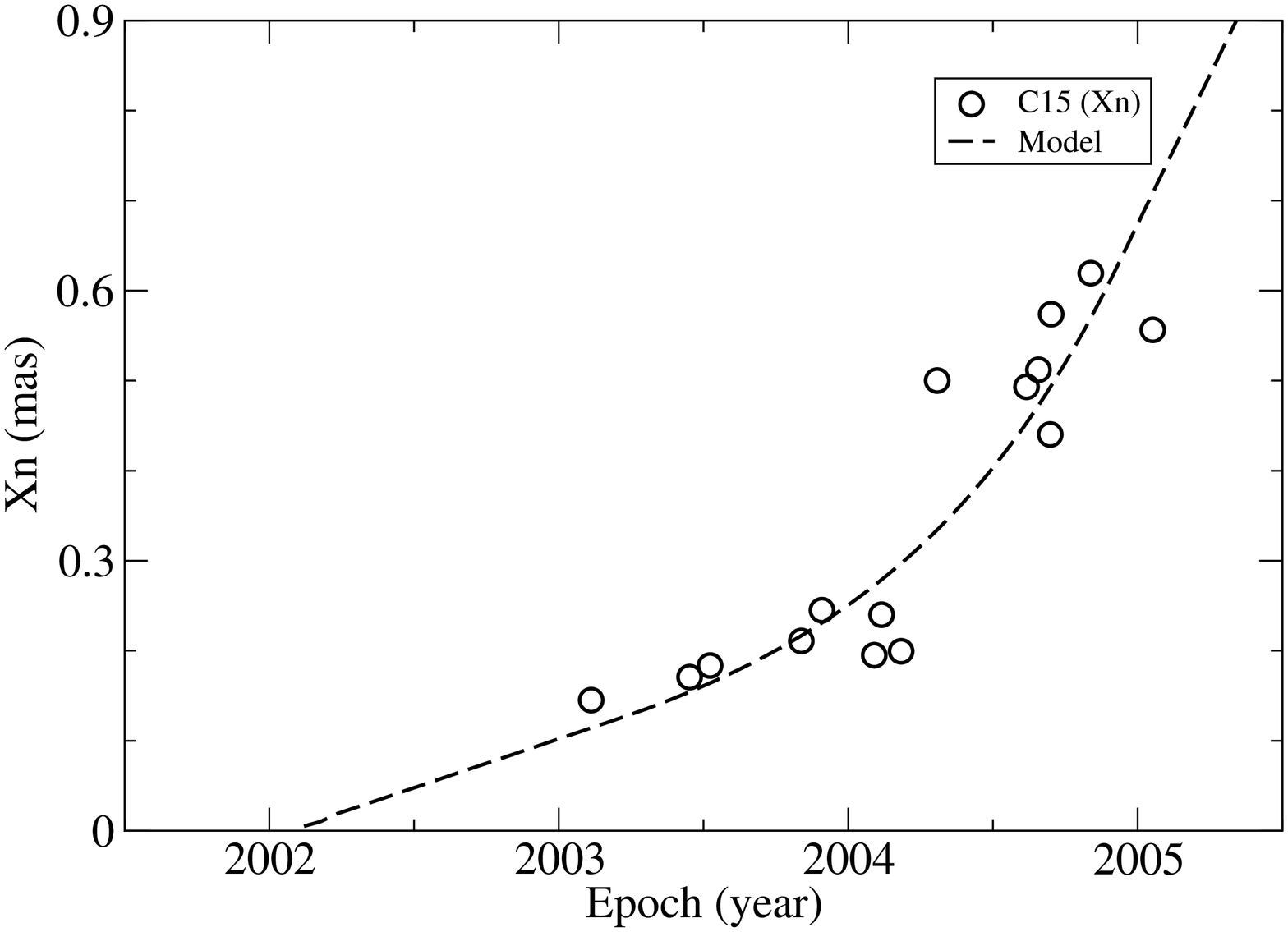}
   \includegraphics[width=5.5cm,angle=-90]{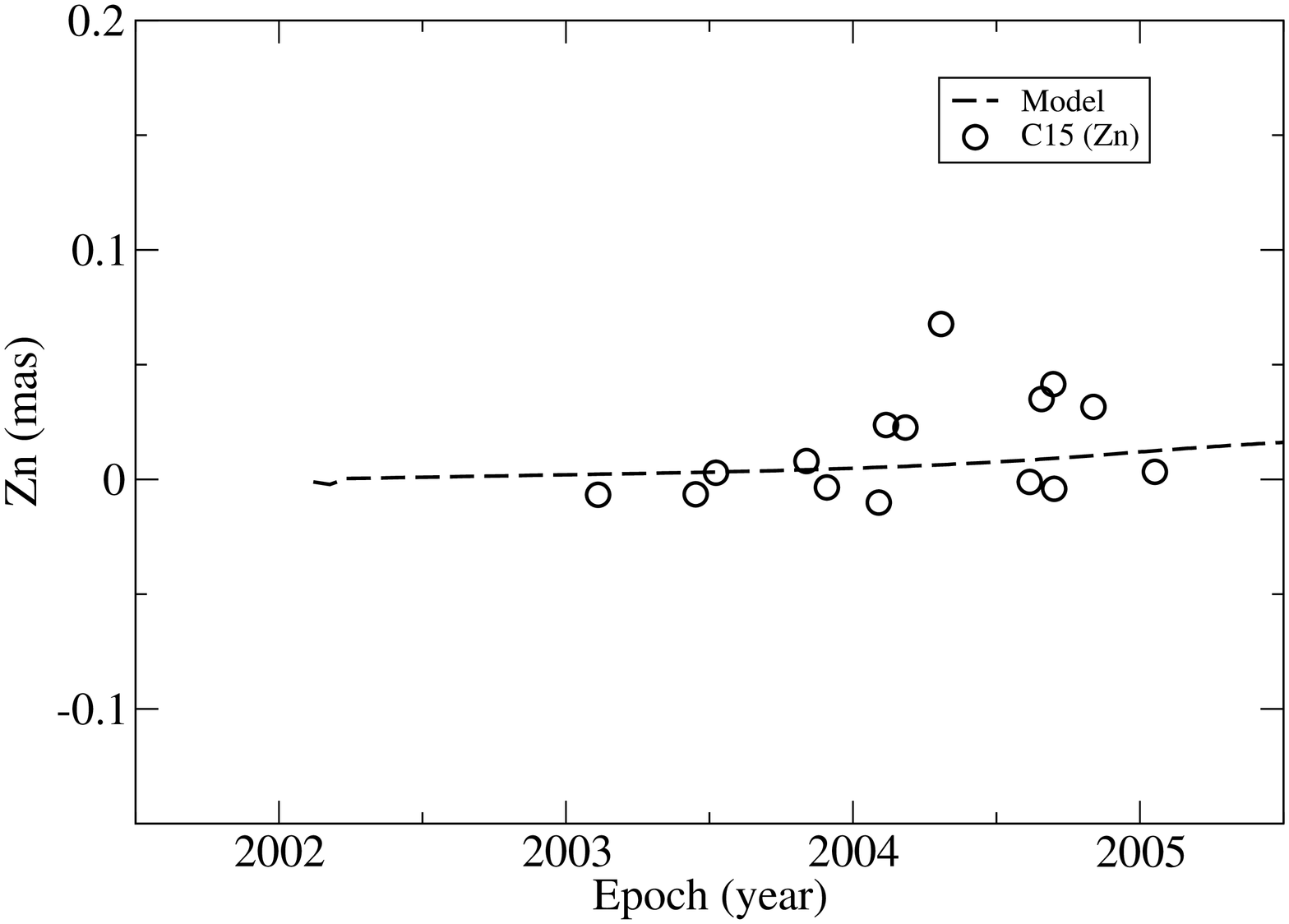}
   \includegraphics[width=5.5cm,angle=-90]{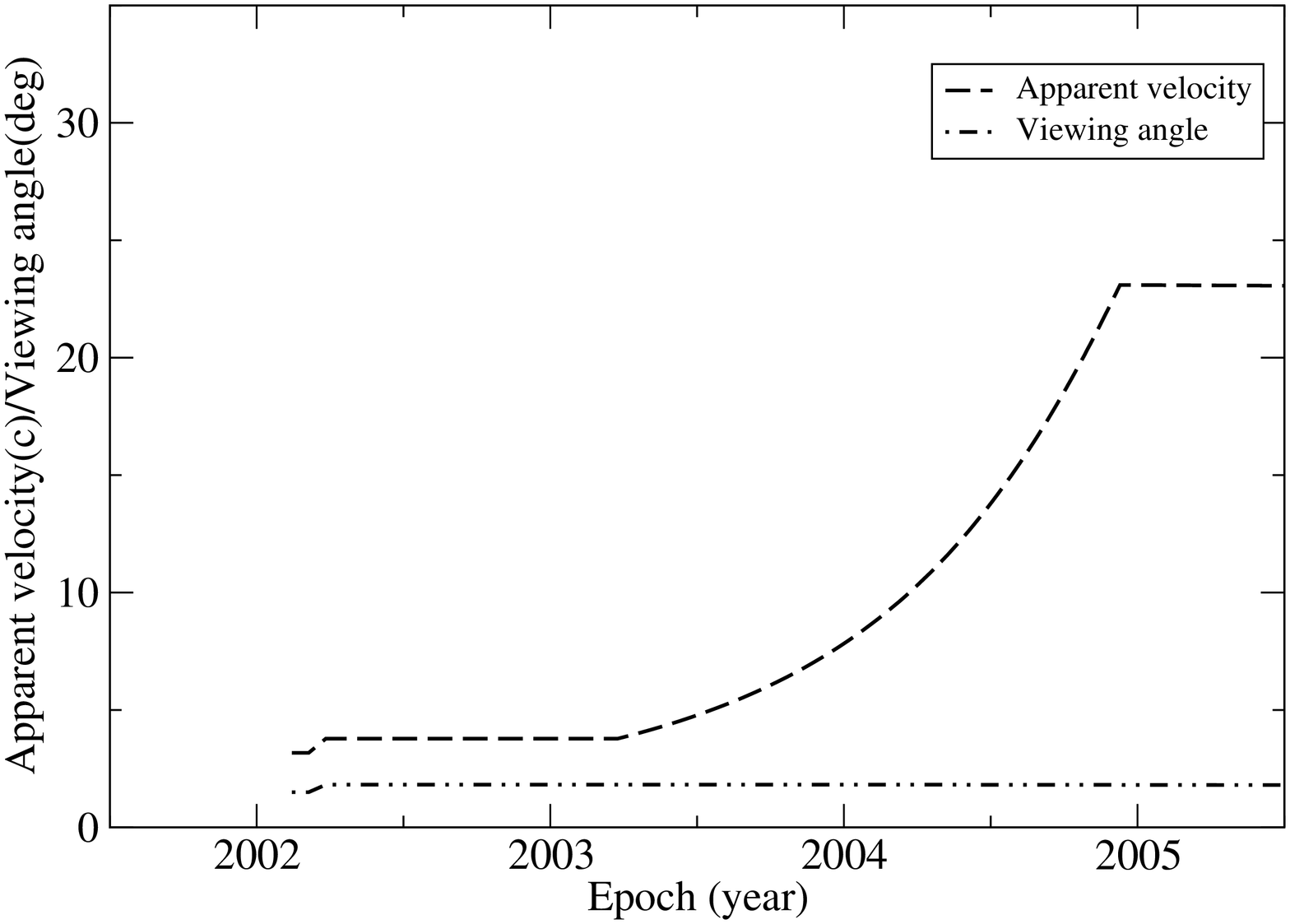}
   \includegraphics[width=5.5cm,angle=-90]{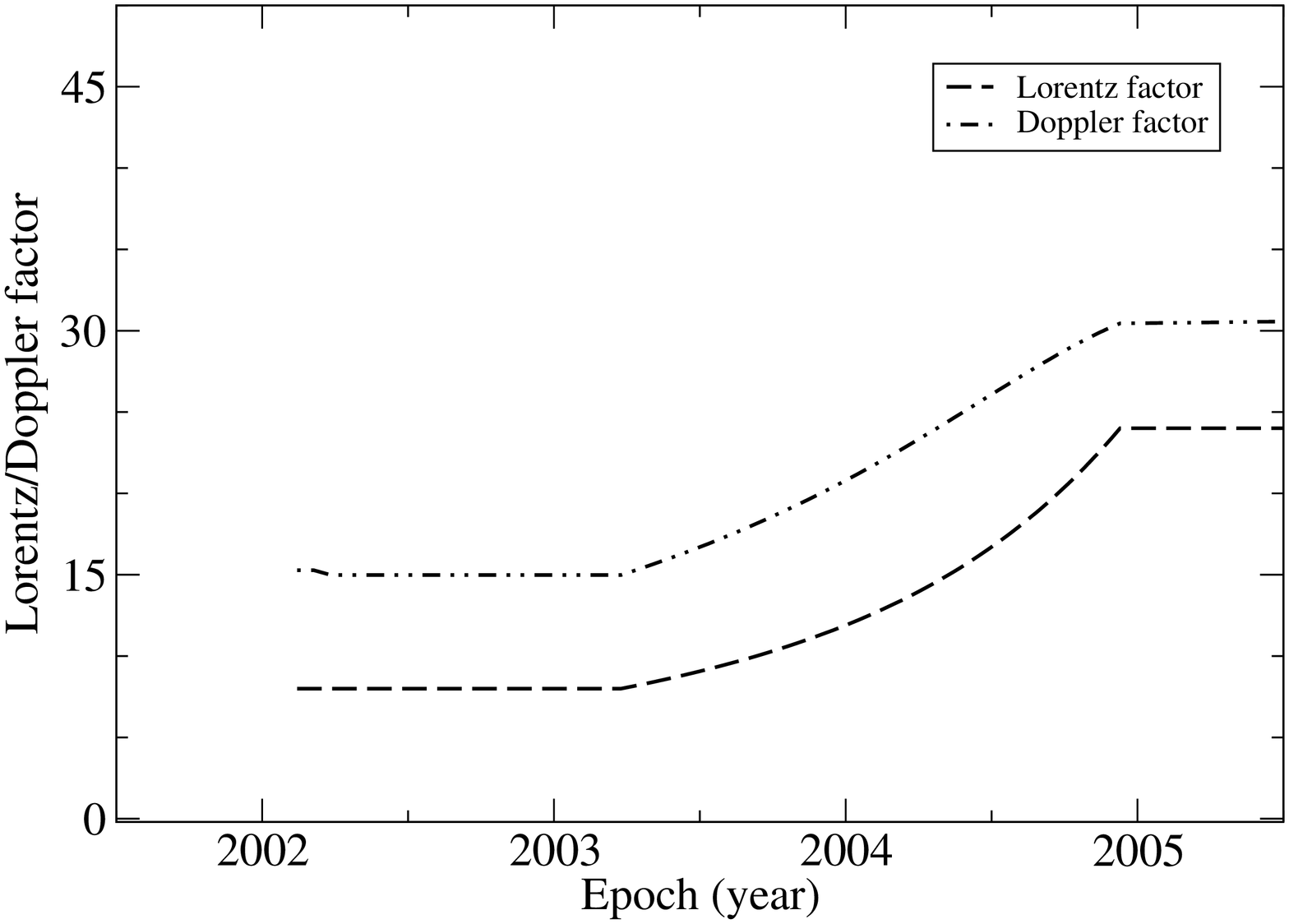}
   \caption{Model fitting of kinematics for knot C15: precession phase 
   $\phi_0$=5.70\,rad, $t_0$=2002.12.}
   \end{figure*}
   \subsection{Model fitting results for knot C15}
   According to the precessing nozzle scenario for jet-B the precession phase
   and  ejection epoch are assumed as : $\phi_0$=5.7\,rad and $t_0$=2002.12.\\
   The model fitting results of its kinematic behavior are shown in
   Figure 14. Obviously, its kinematic behavior (including trajectory,
   core separation and coordinates vs time) can be very well explained
    in terms of the precessing nozzle scenario for jet-B. The apparent
    velocity and bulk Lorentz factor of its motion as functions of time have
    been derived. The viewing angle and Doppler factor vs time are also
    derived.\\ 
     Its observed precessing common trajectory may be assumed to extend to 
    core separation $r_n{\sim}$ 0.7\,mas, corresponding to a spatial distance
     of $Z_{c,m}$$\sim$21.0\,mas or $Z_{c,p}$$\sim$139.7\,pc.
    Its motion was observed to be accelerated and the bulk Lorentz factor
    are modeled as: for Z$\leq$4\,mas $\Gamma$=8; for Z=4--20
    $\Gamma$=8+16(Z-4)/(20-4); for Z$>$20 $\Gamma$=24.\\
    During the period 2003.0-2005.0 its Lorentz factor $\Gamma$, Doppler factor
    $\delta$, apparent velocity $\beta_a$ and viewing angle $\theta$ vary over
    the following respective ranges: [8.0-24.0], [14.9-29.4], [3.9-23.4] 
     and [1.91-1.90].    
   \subsection{Model-fitting results for knot C15a}
    The kinematics of knot C15a has been modeled by using precession
    phase $\phi_0$
   =5.75\,rad and ejection epoch $t_0$=2002.18. The model fitting results are
   shown in Figure A.9. However, its observed precessing common trajectory
    might only extend to core separation $r_n{\sim}$0.17\,mas, corresponding
    to a spatial distance $Z_{c,m}{\sim}$5.60\,mas, or $Z_{c,p}$$\sim$37.2\,pc,
    much smaller than that for knot C15. Thus its outer 
   trajectory have to be fitted by introducing changes in parameters $A_0$ 
   (amplitude) and $\psi$ (rotation of trajectory).  $A_0$ was assumed as:
     for Z$\leq$4\,mas $A_0$=1.09\,mas; for Z=4--10\,mas
    $A_0$=1.09-0.87(Z-4)/(10-4); for Z$>$10\,mas $A_0$=0.218\,mas. And 
   $\psi$ was assumed as: for $\leq$5\,mas $\psi$=0.209\,rad (same as for
    the axis of jet-B);
   for Z=5-10\,mas $\psi$(rad)=0.209+0.146(Z-5)/(10-5)); for Z$>$10\,mas  
    $\psi$=0.355\,rad.\\
    Its accelerated motion could be modeled as: for Z$\leq$5\,mas $\Gamma$=8.5;
    for Z=5--10\,mas $\Gamma$=8.5+1.5(Z-5)/(10-5); for Z$>$10\,mas
    $\Gamma$=10.0.\footnote{Due to lack of data-points within core separation
    $r_n{\sim}$0.15\,mas, the model fitting of its observed precessing common 
    trajectory is marginal.}\\
    During the period 2003-2005.5 its Lorentz factor$\Gamma$, Doppler factor
    $\delta$, apparent velocity $\beta_a$ and viewing angle $\theta$ vary over
    the respective ranges as: [8.5-10.0], [15.8-18.6], [4.2-5.0] and
    [1.80-1.55](deg). 
   \subsection{Model-fitting results for knot C16}
   The model-fitting results for knot C16 are shown in 
   Figure A.10. Its precession phase and ejection epoch are assumed to be
   $\phi_0$=5.80\,rad and $t_0$=2002.24.\\
    It can be seen that its kinematic behavior has been well modeled. 
   Its observed precessing common trajectory may extend to core separation 
    $r_n{\sim}$ 0.8\,mas, corresponding to spatial distance $Z_{c,m}{\sim}$
   24.4\,mas or $Z_{c,p}{\sim}$ 162.2\,pc from the core.\\
   Its accelerated motion is modeled by its bulk Lorentz factor as follows:
    for Z$\leq$6\,mas $\Gamma$=6; for Z=6--20\,mas 
   $\Gamma$=6+12(Z-6)/(20-6); for Z$>$20\,mas $\Gamma$=18.\\
   During the period 2004.0--2008.0 its Lorentz factor $\Gamma$, Doppler factor
   $\delta$, apparent velocity $\beta_a$ and viewing angle $\theta$ vary over 
   the respective ranges: [6.0,18.0], [11.5,26.7], [2.2,15.7] and 
    [1.88,1.87](deg).
   \subsection{Model fitting results for knot C17}
   According to the precessing nozzle scenario the kinematic behavior of knot
    C17 could be modeled by assuming its precession phase 
   $\phi_0$(rad)=1.57+2$\pi$
    and  ejection epoch $t_0$=2004.62. The model fitting results are shown in
   Figure A.11. It can be seen that its kinematic behavior has been well fitted 
    and its observed precessing common trajectory may extend to core separation
   $r_n{\sim}$0.8\,mas, equivalent to spatial distance $Z_{c,m}{\sim}$44.8\,mas
   or  $Z_{c,p}\sim$297.9\,pc from the core.\\
   Its accelerated motion can be modeled by the increase in its bulk Lorentz
    factor as follows: for Z$\leq$2\,mas $\Gamma$=6.8; for Z=2--20\,mas
   $\Gamma$=6.8+13.2(Z-2)/(20-2); for Z$>$20\,mas $\Gamma$=20.0.\\
    During the observed period 2006.5--2009.0 its Lorentz factor $\Gamma$,
   Doppler factor $\delta$, apparent velocity $\beta_a$ and viewing angle
   $\theta$ vary over the respective ranges: [9.9,20.1], [19.1,35.1],
   [3.5,13.1] and [1.01,1.07](deg).
   \subsection{Model fitting results for knot C18}
   The kinematics of knot C18 can be explained in terms of the precessing 
   nozzle scenario for jet-B by assuming its precession phase $\phi_0$(rad)=
   1.62+2$\pi$ and ejection epoch $t_0$=2004.68. The model-fitting
   results of its kinematic behavior are shown in Figure A.12. It can be seen
    that the data-points are well concentrated around the predicted 
   precessing common trajectory. \\
    Its observed precessing common trajectory  may extend to core separation
    $r_n{\sim}$0.6\,mas, corresponding to  spatial distance $Z_{c,m}{\sim}$
    33.8\,mas or $Z_{c,p}{\sim}$224.8\,pc from the core.\\
    The accelerated motion of knot C18 can be modeled by assuming its bulk
   Lorentz factor as follows: for Z$\leq$2\,mas $\Gamma$=6.5; for Z=2--20\,mas
   $\Gamma$=6.5+9.5(Z-2)/(20-2); for Z$>$20 $\Gamma$=16.0.\\
    During the period 2007--2009.5 its Lorentz factor $\Gamma$, Doppler factor
   $\delta$, apparent velocity $\beta_a$ and viewing angle $\theta$ vary over
   the respective ranges: [9.1,16.0], [17.8,29.5], [2.8,8.5] and
    [1.01,1.03](deg).
   \begin{figure*}
   \centering
   \includegraphics[width=5.5cm,angle=-90]{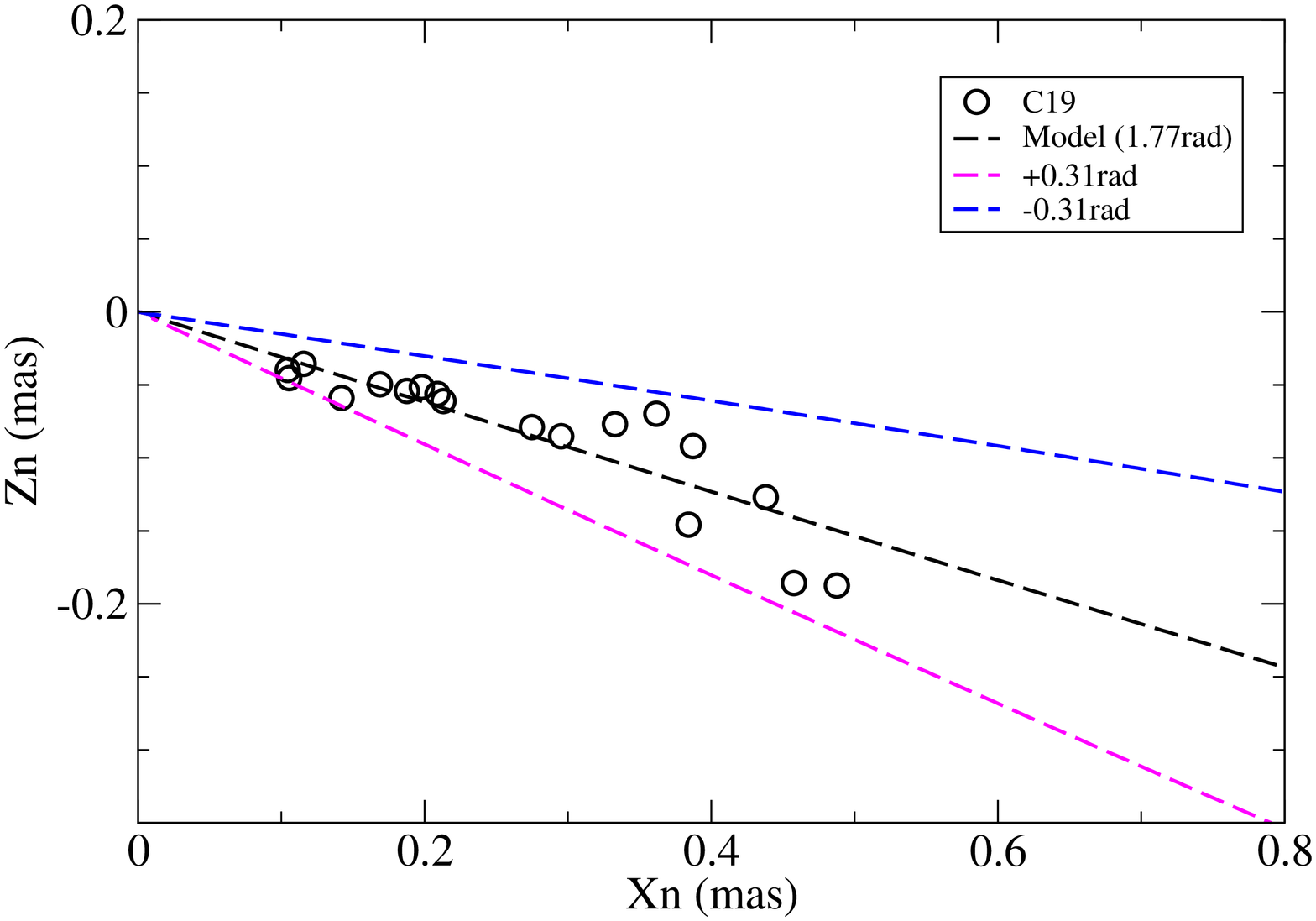}
   \includegraphics[width=5.5cm,angle=-90]{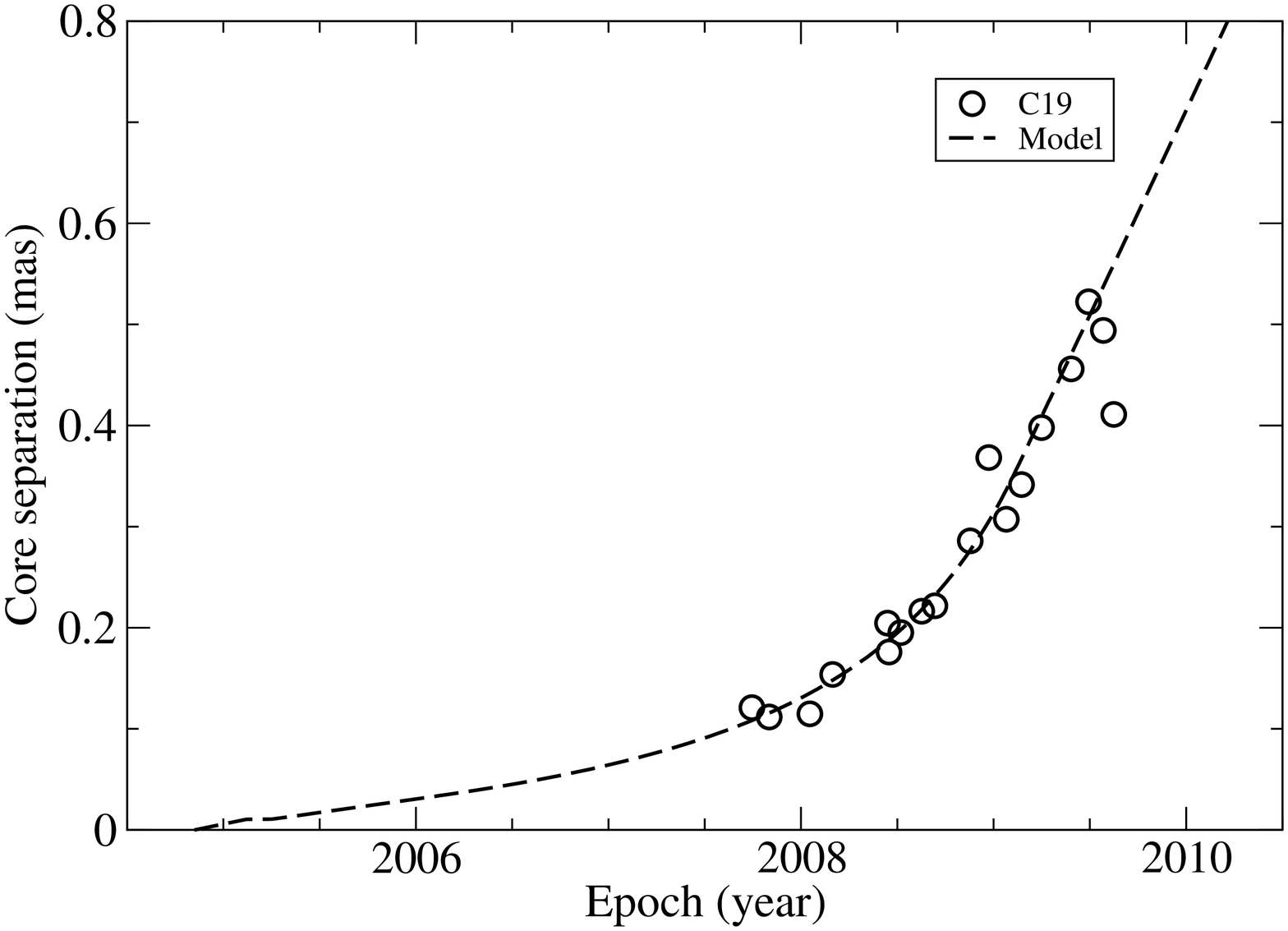}
   \includegraphics[width=5.5cm,angle=-90]{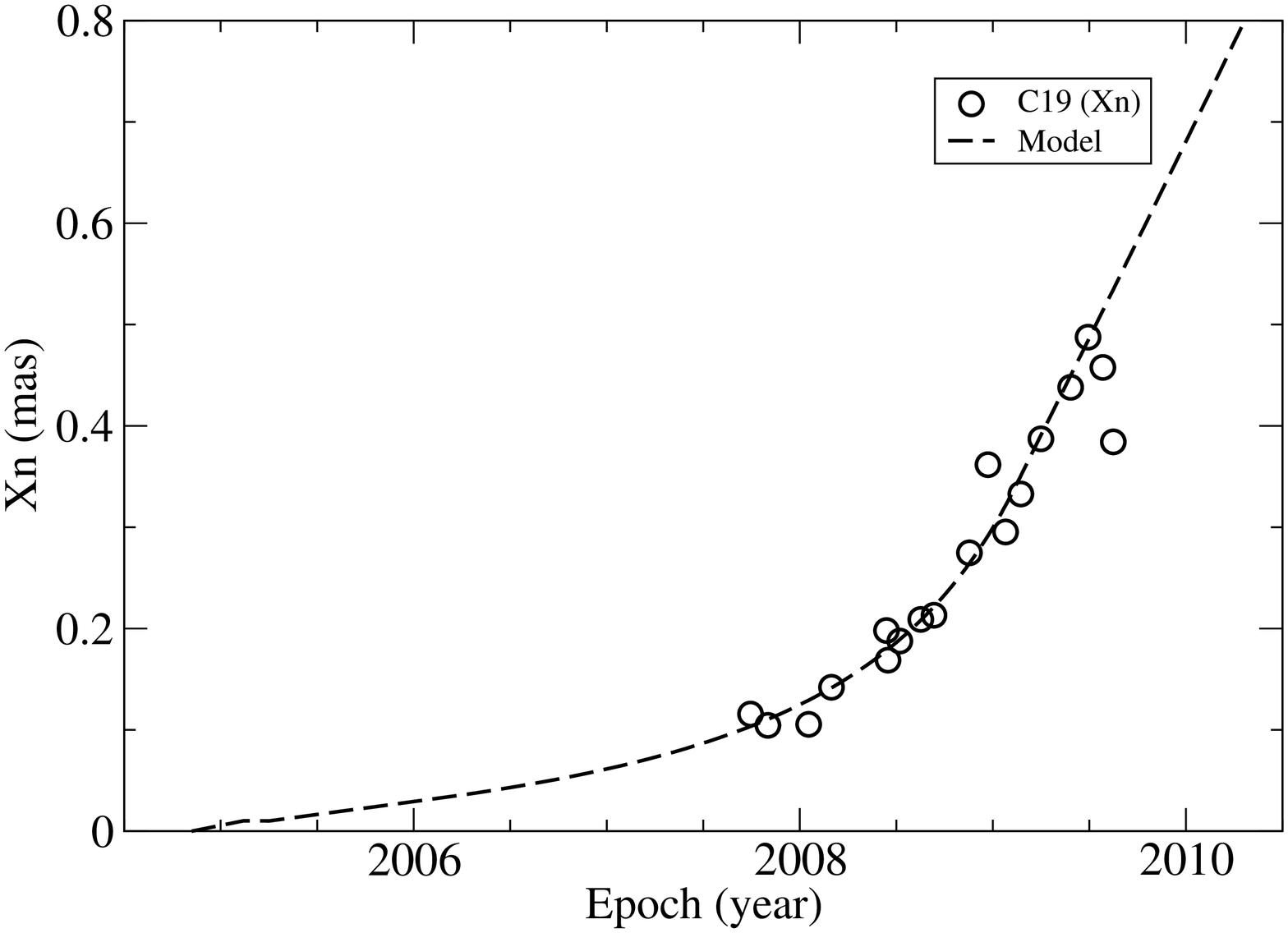}
   \includegraphics[width=5.5cm,angle=-90]{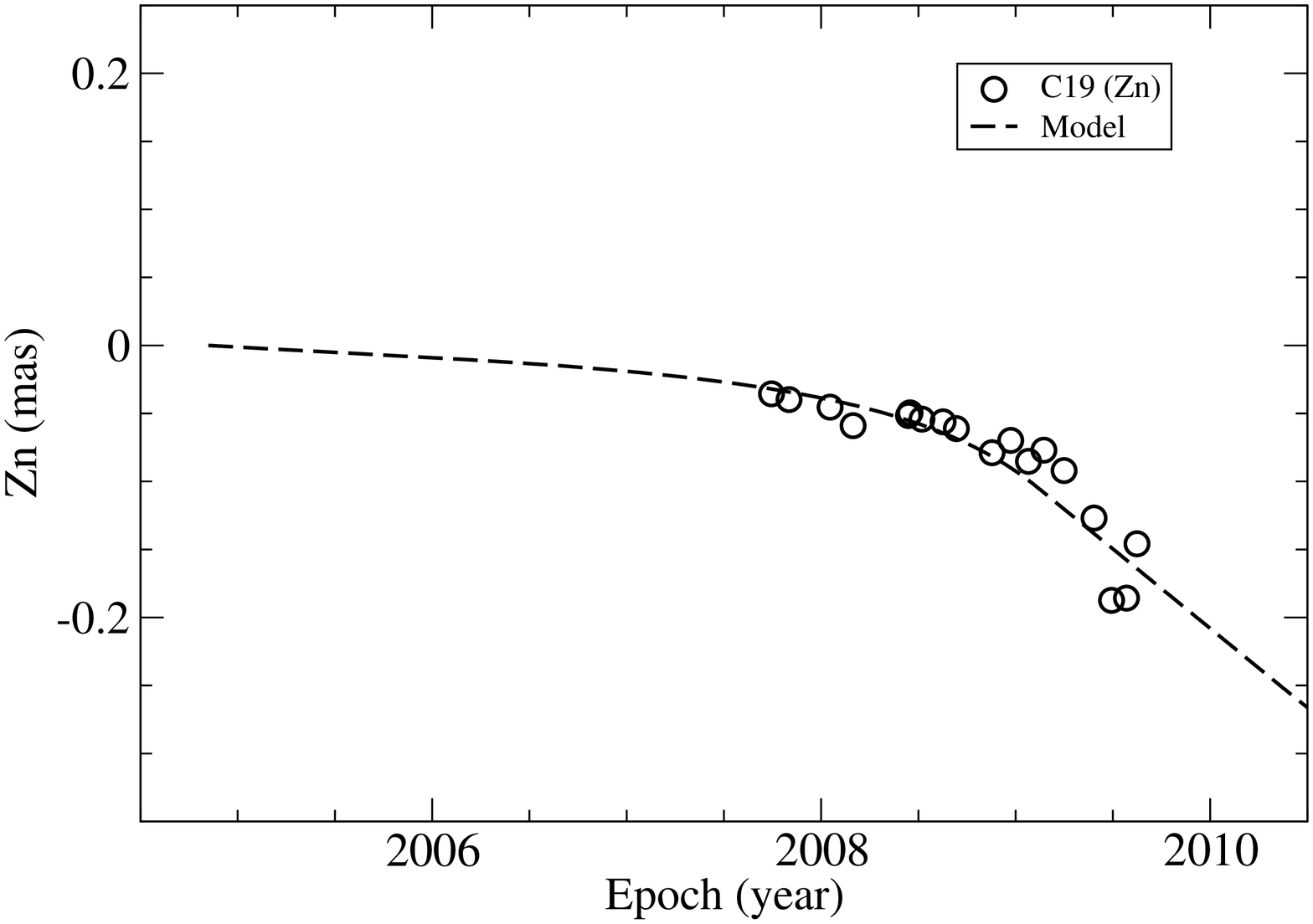}
   \includegraphics[width=5.5cm,angle=-90]{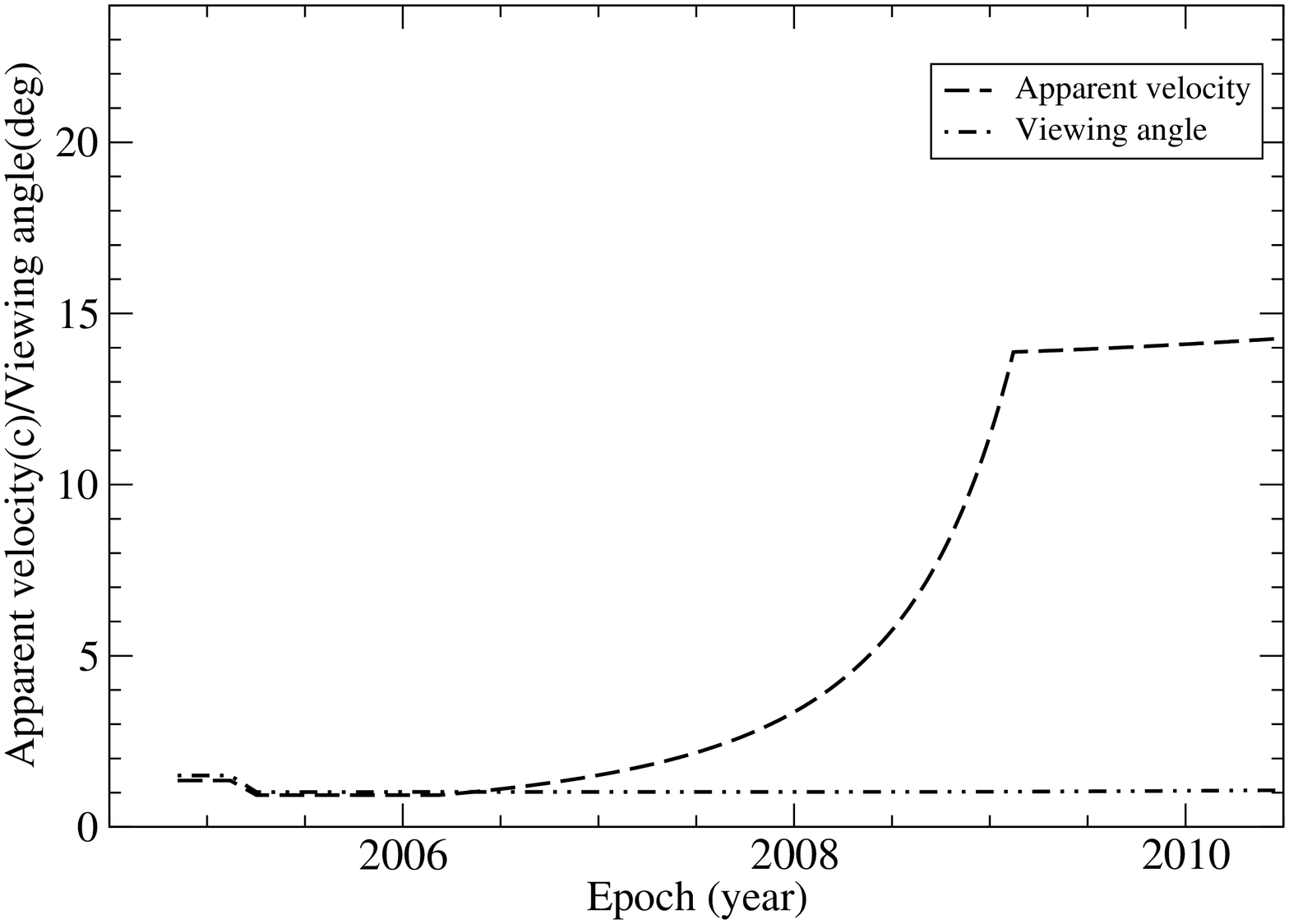}
   \includegraphics[width=5.5cm,angle=-90]{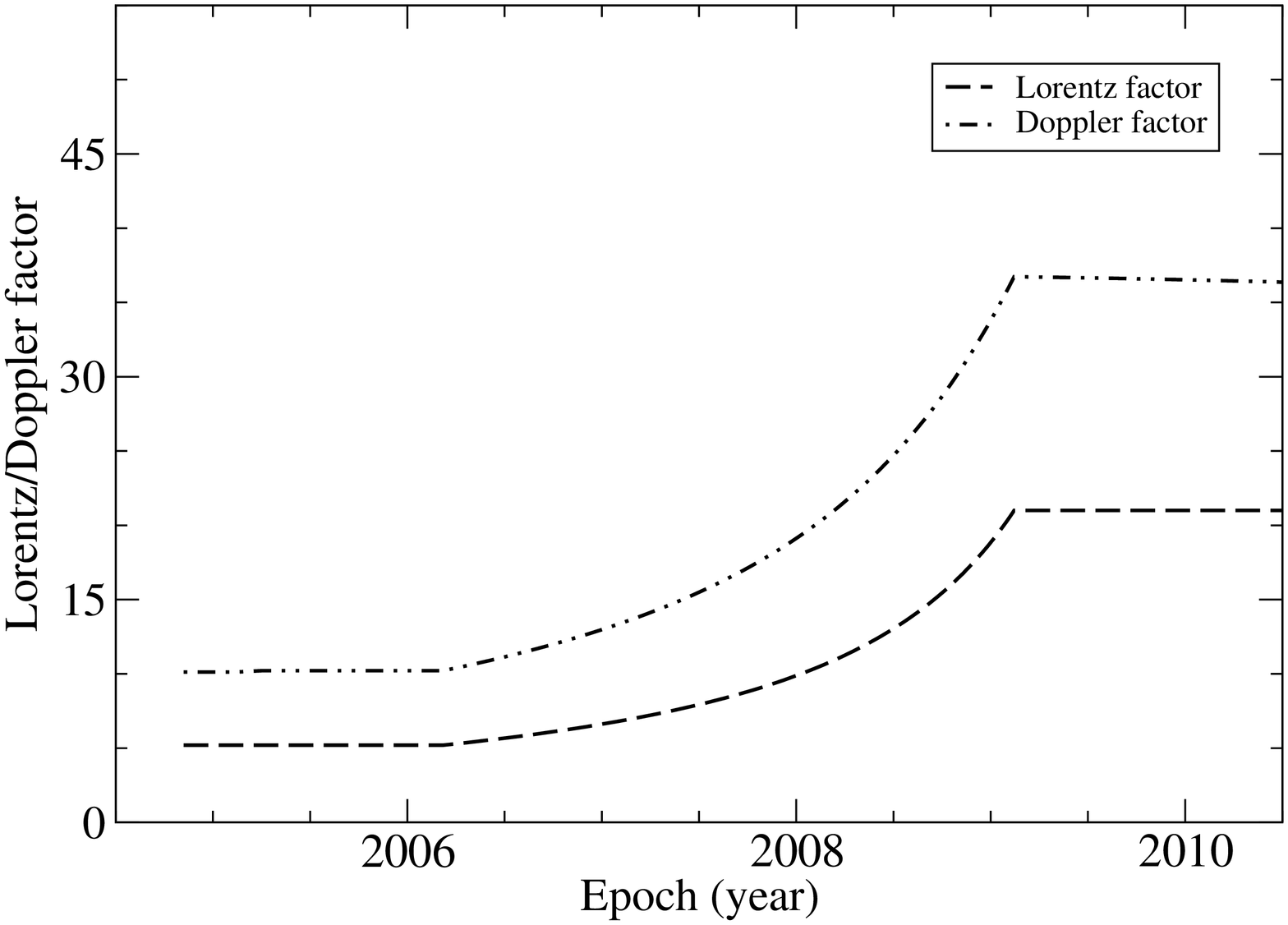}
   \caption{Model fitting of kinematics for knot C19: precession phase
    $\phi_0$=1.77\,rad+2$\pi$. $t_0$=2004.85. Note: the difference between the
    ejection epochs for B6 and C19 is $\sim$7.53\,yr, very close to one 
    precession period (see text).}
   \end{figure*}
   \subsection{Model fitting results for knot C19}
    According to the precessing nozzle scenario for jet-B, the kinematics of
    knot C19 can be explained by assuming its precession phase 
    $\phi_0$(rad)=1.77+2$\pi$ and ejection epoch $t_0$=2004.85.\\
    The model-fitting results of its kinematic behavior are shown in Figure 15.
    Due to plenty of data-points the fitting results are satisfying and all
    the data-points are located within the region defined by precession phases
     $\phi_0$=1.77$\pm$0.31\,rad (corresponding $t_0$=2004.85$\pm$0.36\,yr).\\
     Its observed precessing common trajectory may extend to core separation 
     $r_n{\sim}$0.6\,mas, equivalent to spatial distance 
     $Z_{c,m}{\sim}$33.4\,mas or  $Z_{c,p}$$\sim$222.1\,pc from the core.\\
     Its accelerated motion can be explained by assuming its Lorentz factor
    as follows: for Z$\leq$2 $\Gamma$=5.2; Z=2--20\,mas  $\Gamma$=
    5.2+15.8(Z-2)/(20-2); for Z$>$20 $\Gamma$=21.0.\\
    During the period 2007.5--2010.0 its Lorentz factor $\Gamma$, Doppler factor
    $\delta$, apparent velocity $\beta_a$ and viewing angle 
   $\theta$ vary over the
    respective ranges: [8.0,21.0], [15.6,36.5], [2.2,14.1] and 
    [1.02,1.05](deg).\\
     We emphasize that the difference between the ejection epochs for knot C19
    and knot B6 (see below) is $\sim$7.53\,yr, very close to one precession 
    period.
    \begin{figure*}
   \centering
   \includegraphics[width=5.5cm,angle=-90]{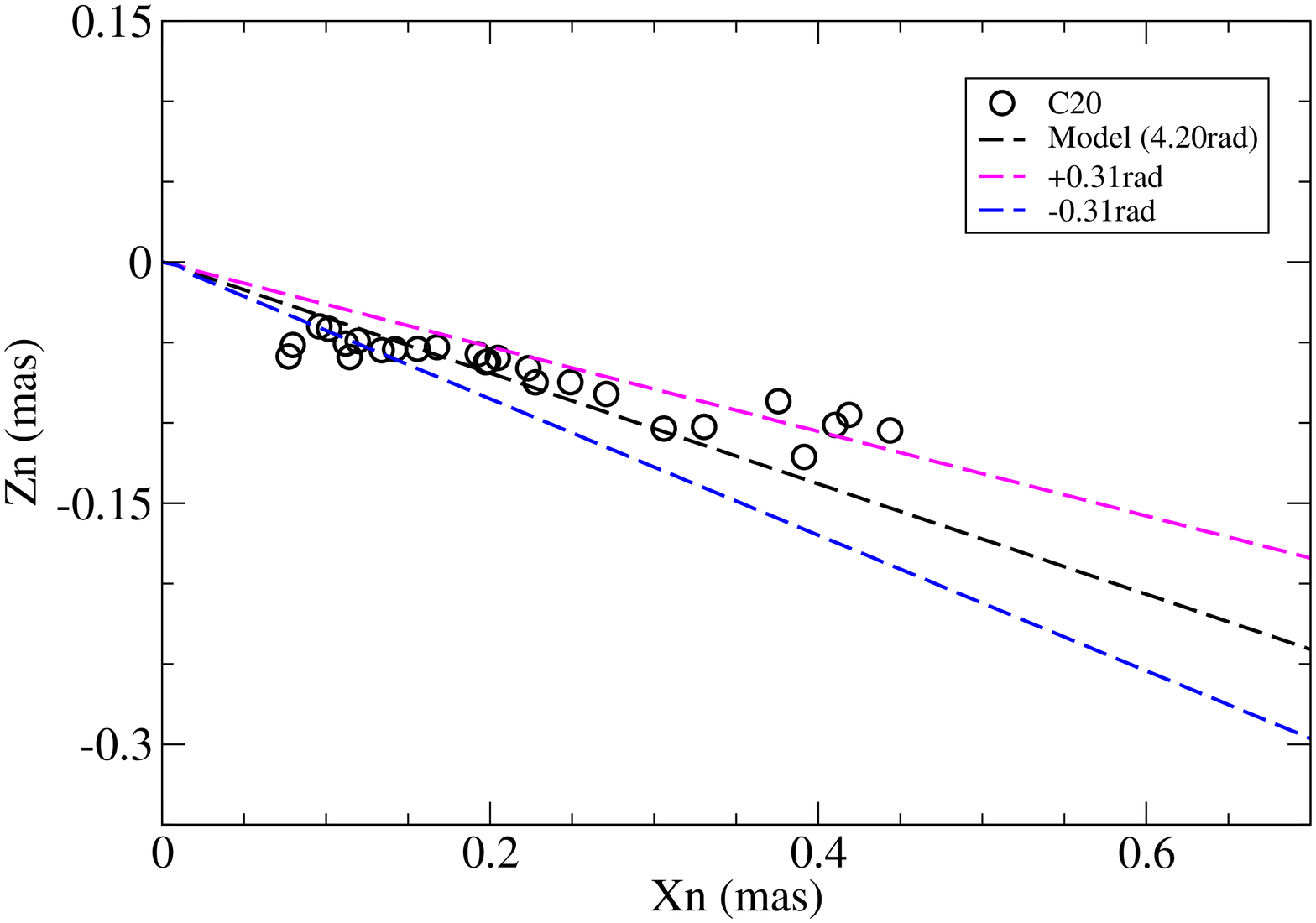}
   \includegraphics[width=5.5cm,angle=-90]{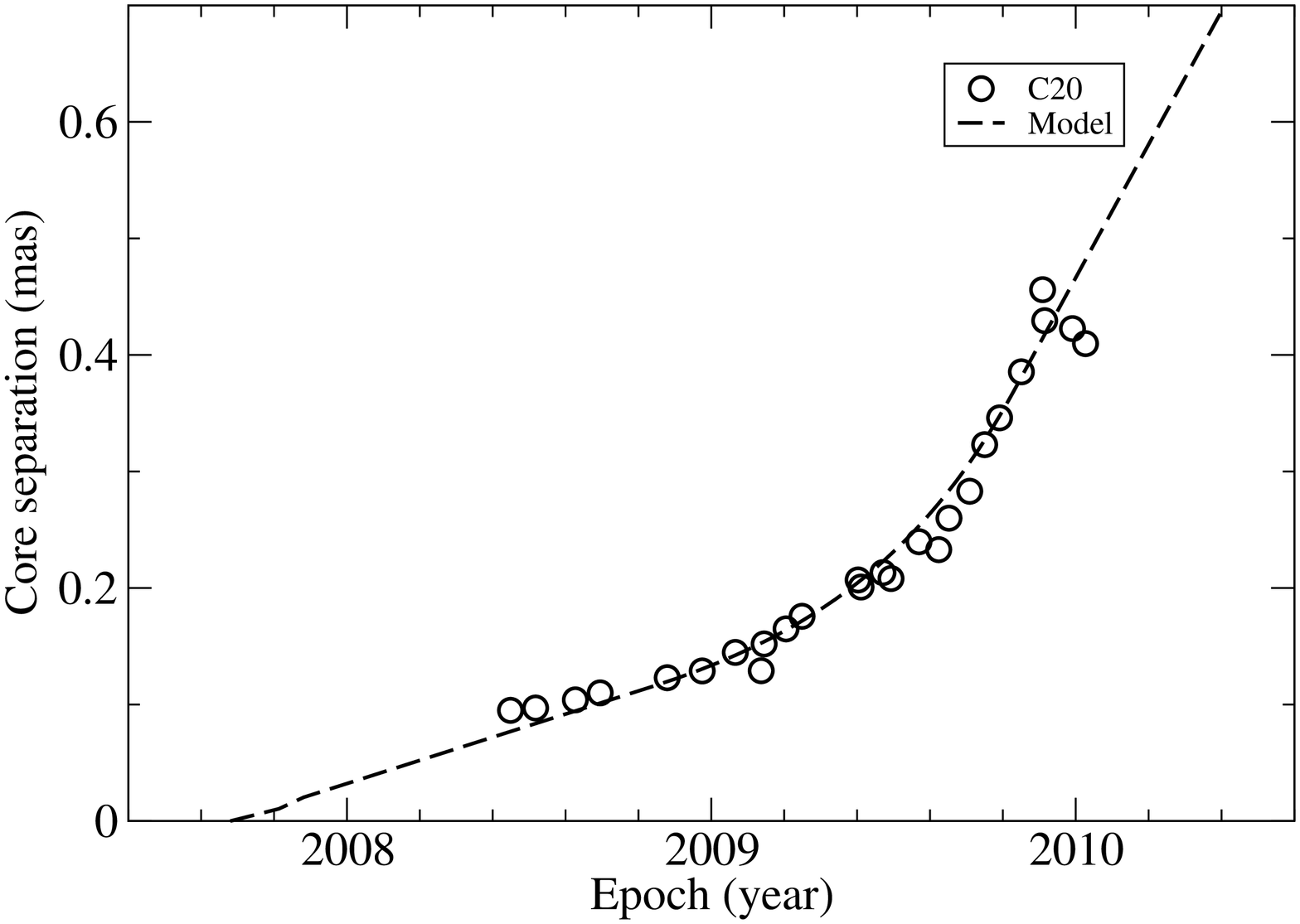}
   \includegraphics[width=5.5cm,angle=-90]{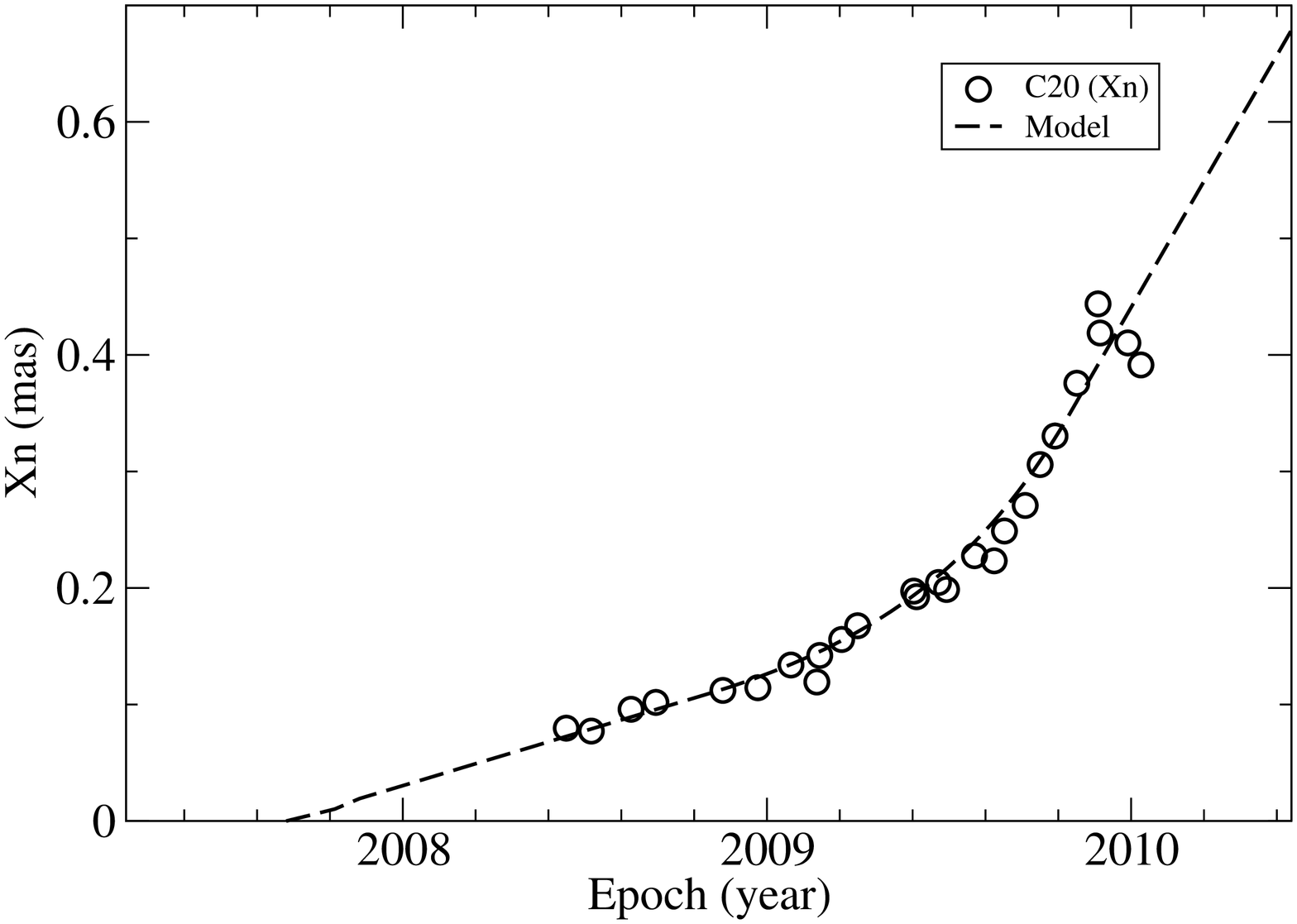}
   \includegraphics[width=5.5cm,angle=-90]{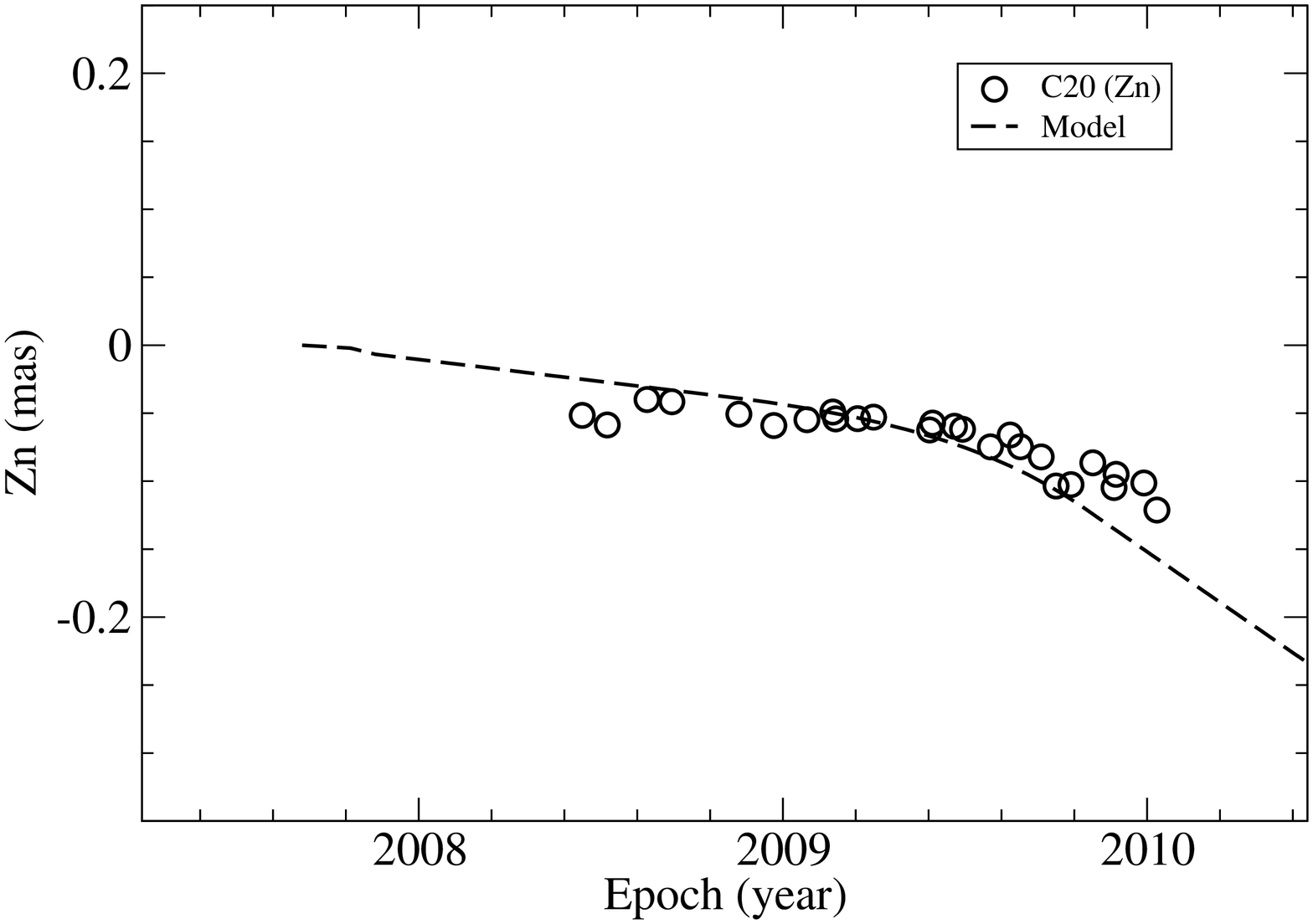}
   \includegraphics[width=5.5cm,angle=-90]{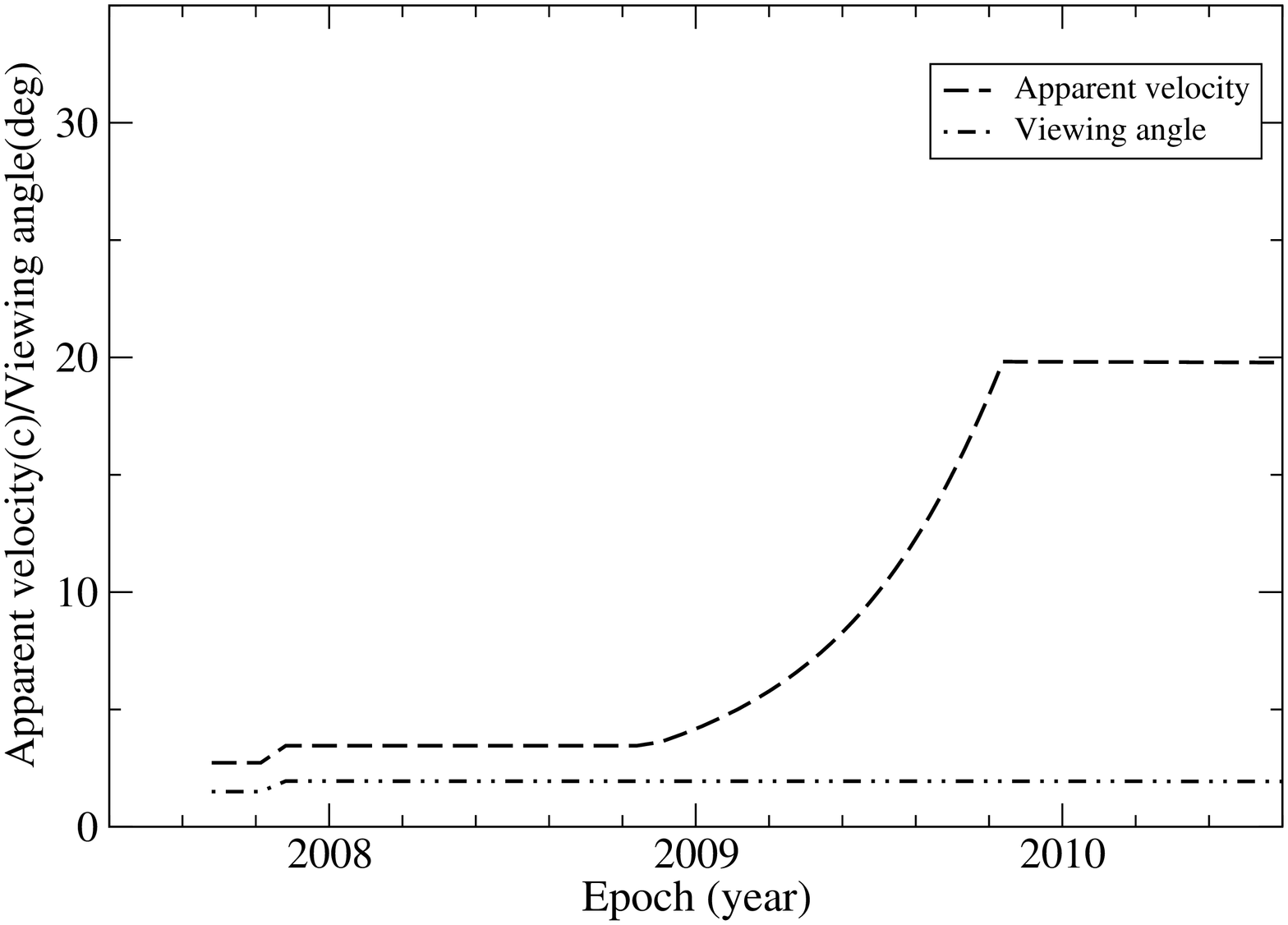}
   \includegraphics[width=5.5cm,angle=-90]{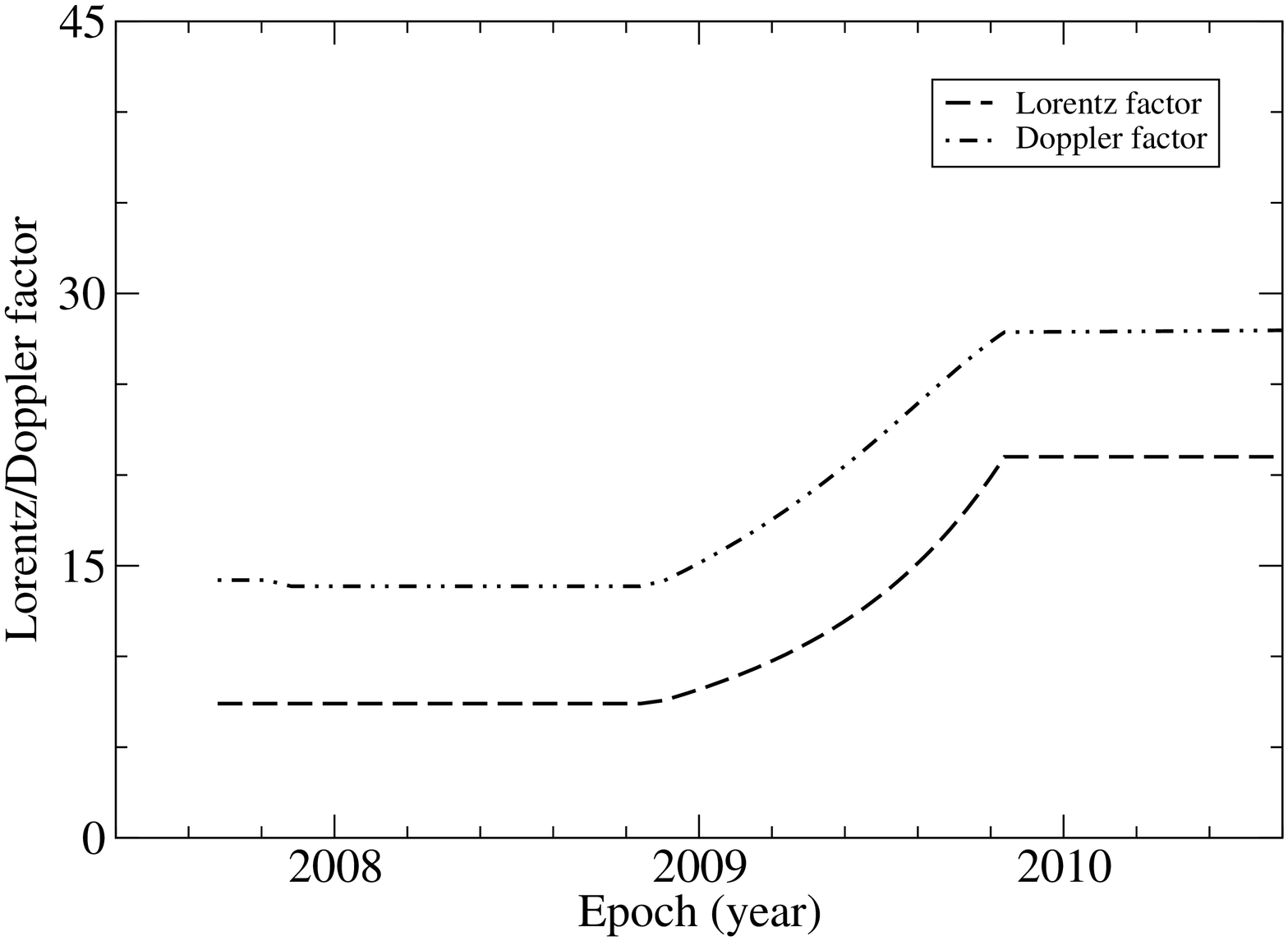}
   \caption{Model fitting of kinematics for knot C20: precession phase
   $\phi_0$=4.20\,rad+2$\pi$, $t_0$=2007.68.}
   \end{figure*}
   \subsection{Model fitting results for knot C20}
    The model-fitting results of the kinematic behavior of knot C20 are shown
    in Figure 16. It can be seen that its kinematics can be very well explained
    by assuming its precession phase $\phi_0$(rad)=4.20+2$\pi$ and
    ejection epoch $t_0$=2007.68. Its observed precessing common trajectory
     may extend to core separation $r_n{\sim}0.45$\,mas, corresponding to
     spatial distance $Z_{c,m}{\sim}$13.2\,mas or $Z_{c,p}{\sim}$87.8\,pc from
    the core. Its accelerated motion can be modeled by assuming its bulk 
    Lorentz factor as follows: for Z$\leq$3.5\,mas  $\Gamma$=7.4; 
    for Z=3.5--11\,mas $\Gamma$=7.4+13.6(Z-3.5)/(11-3.5);
    for Z$>$11\,mas $\Gamma$=21.0.\\
    During period 2008.4--2010.2 its Lorentz factor $\Gamma$, Doppler factor
    $\delta$, apparent velocity $\beta_a$ and viewing angle $\theta$ vary over
    the respective ranges: [7.4,21.0], [13.9,27.9], [3.5,19.8] and 
    [1.95,1.94](deg).\\
    Here we would like to point out that the ejection epoch has been
    determined by linear and polynomial extrapolation 
     methods to be 2007.83 and 2006.91 respectively (Schinzel \cite{Sc12a}).
    Thus the ejection epoch 2007.68 derived for knot C20 by our precessing
    nozzle scenario is closely equal to the averaged epoch obtained by 
    extrapolation methods.\\
    \begin{figure*}
   \centering
   \includegraphics[width=5.5cm,angle=-90]{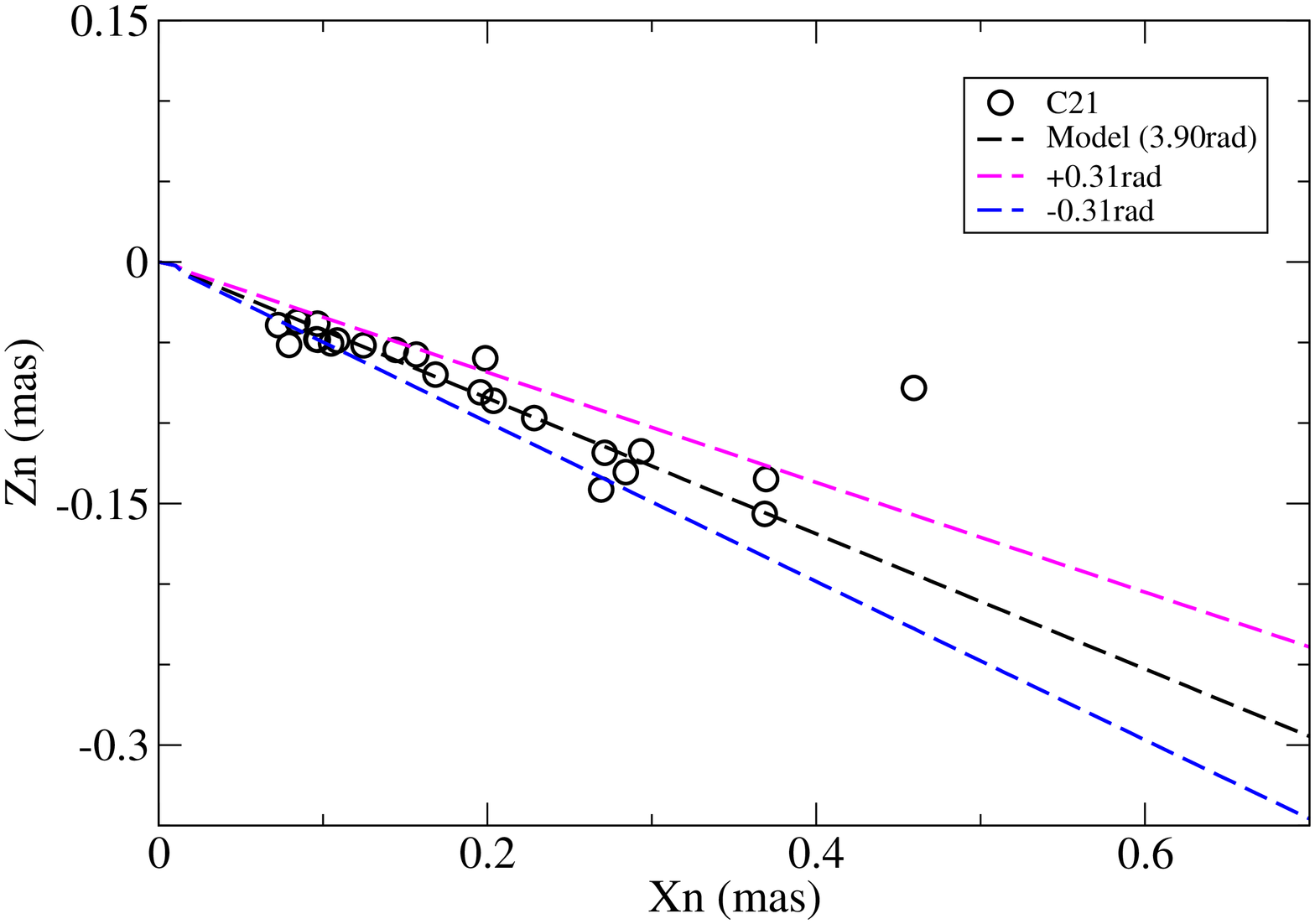}
   \includegraphics[width=5.5cm,angle=-90]{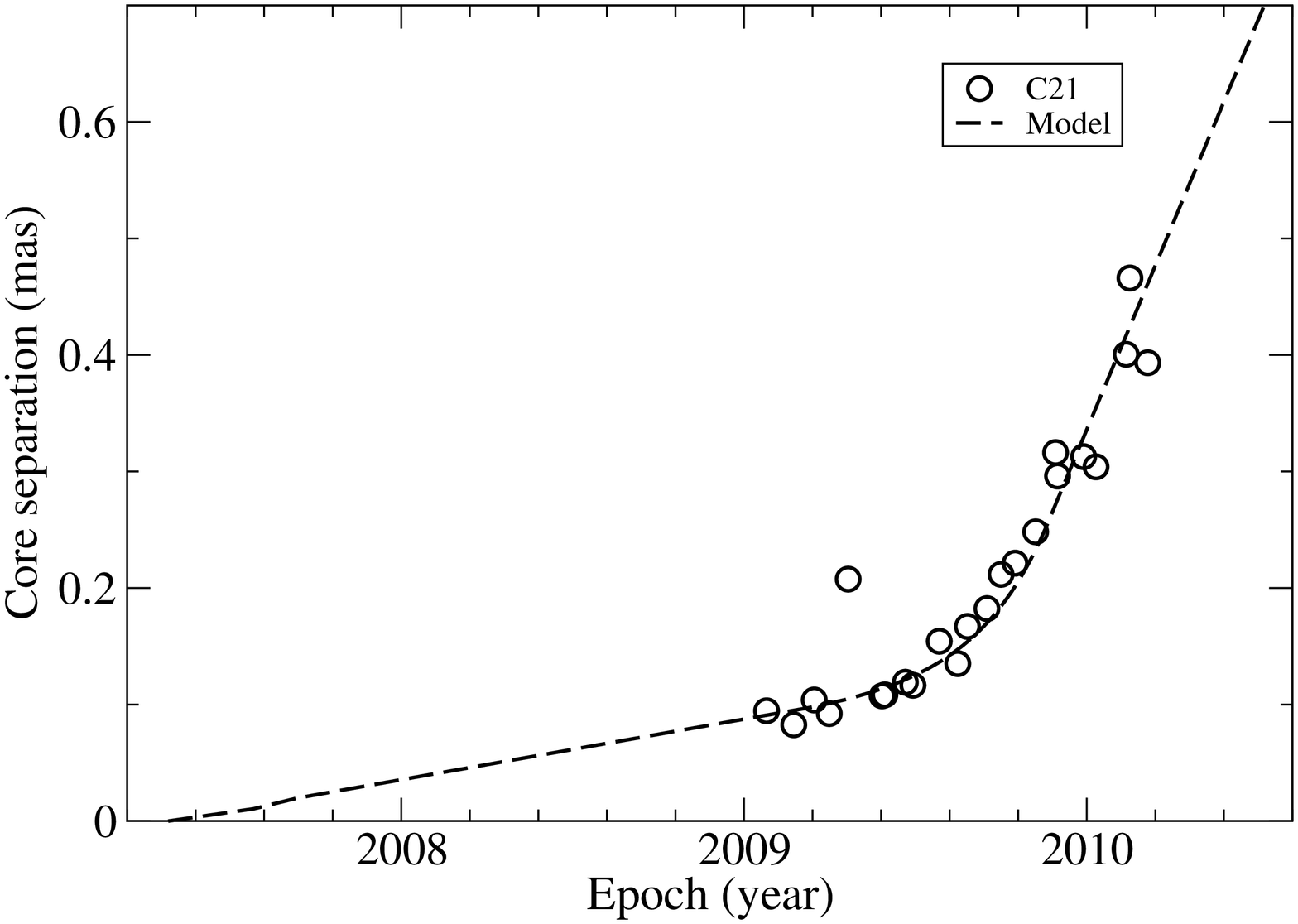}
   \includegraphics[width=5.5cm,angle=-90]{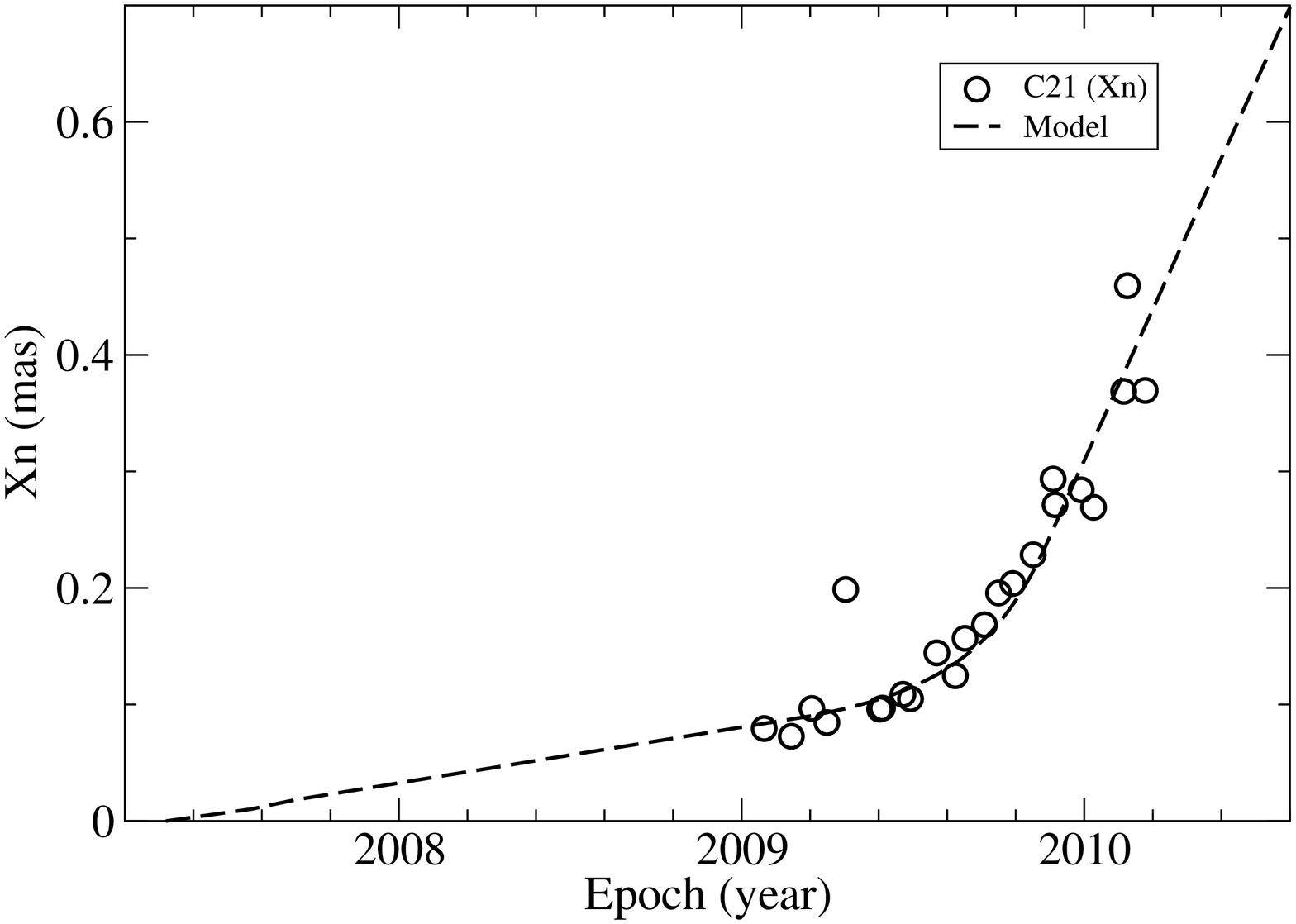}
   \includegraphics[width=5.5cm,angle=-90]{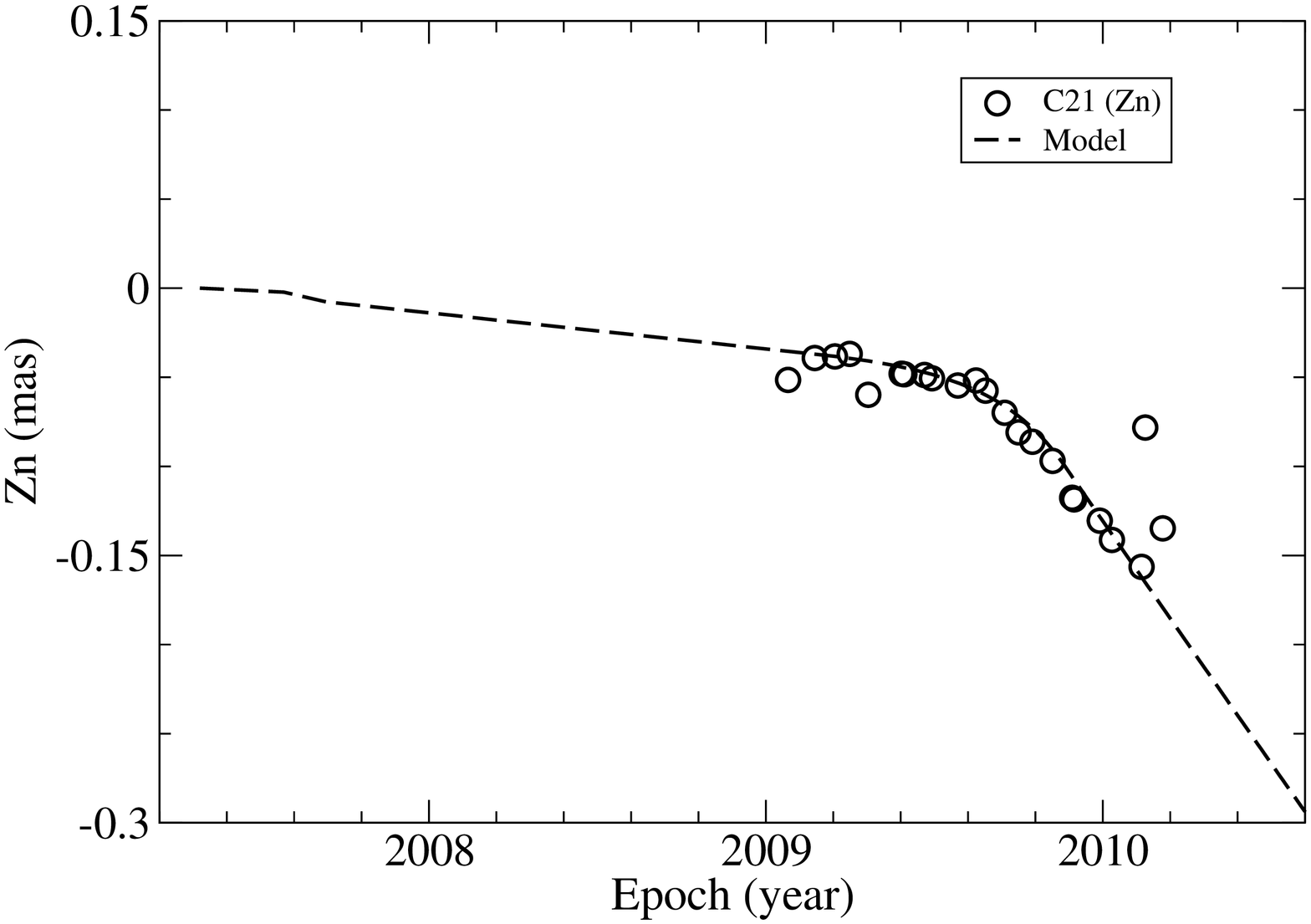}
   \includegraphics[width=5.5cm,angle=-90]{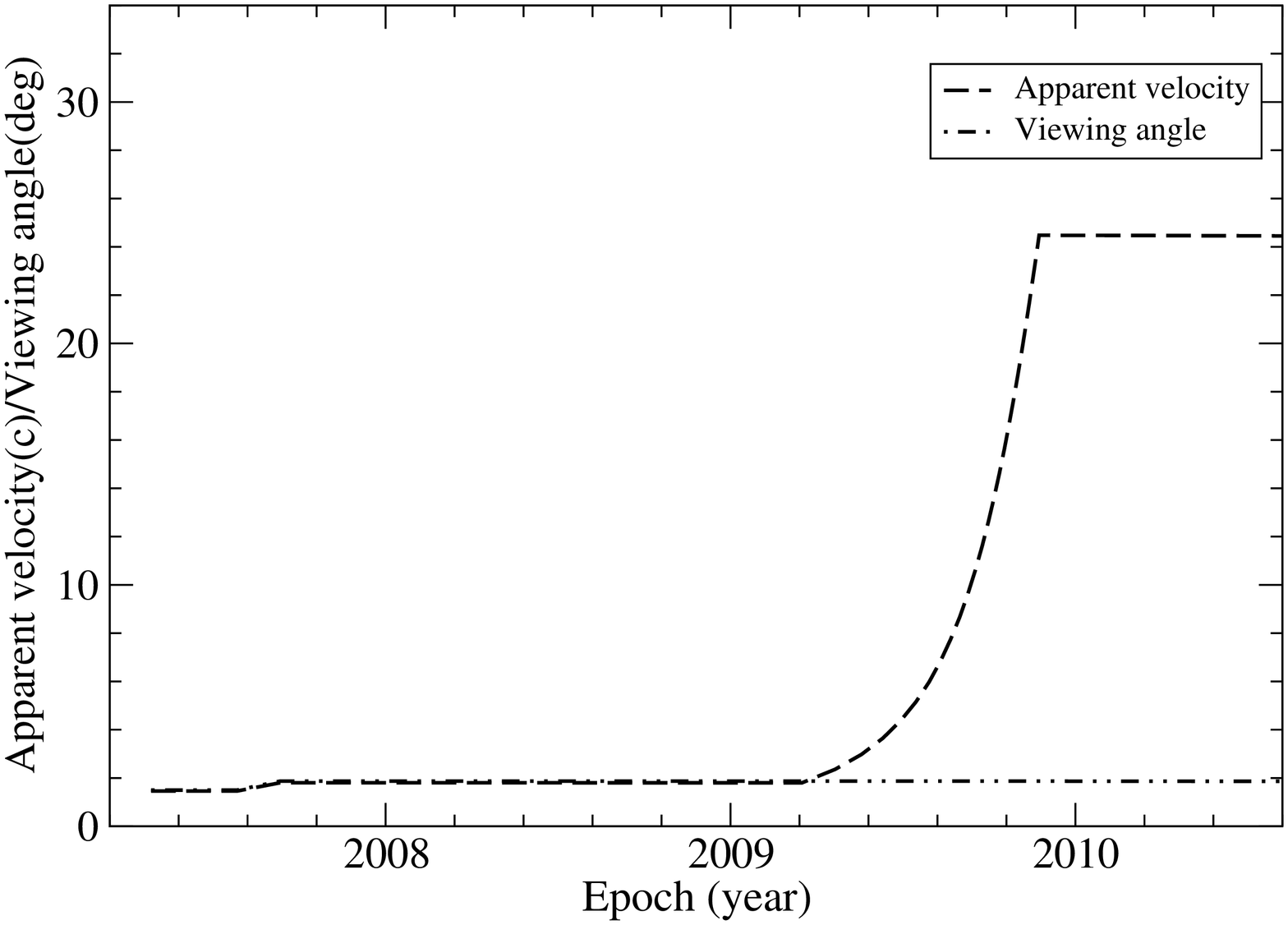}
   \includegraphics[width=5.5cm,angle=-90]{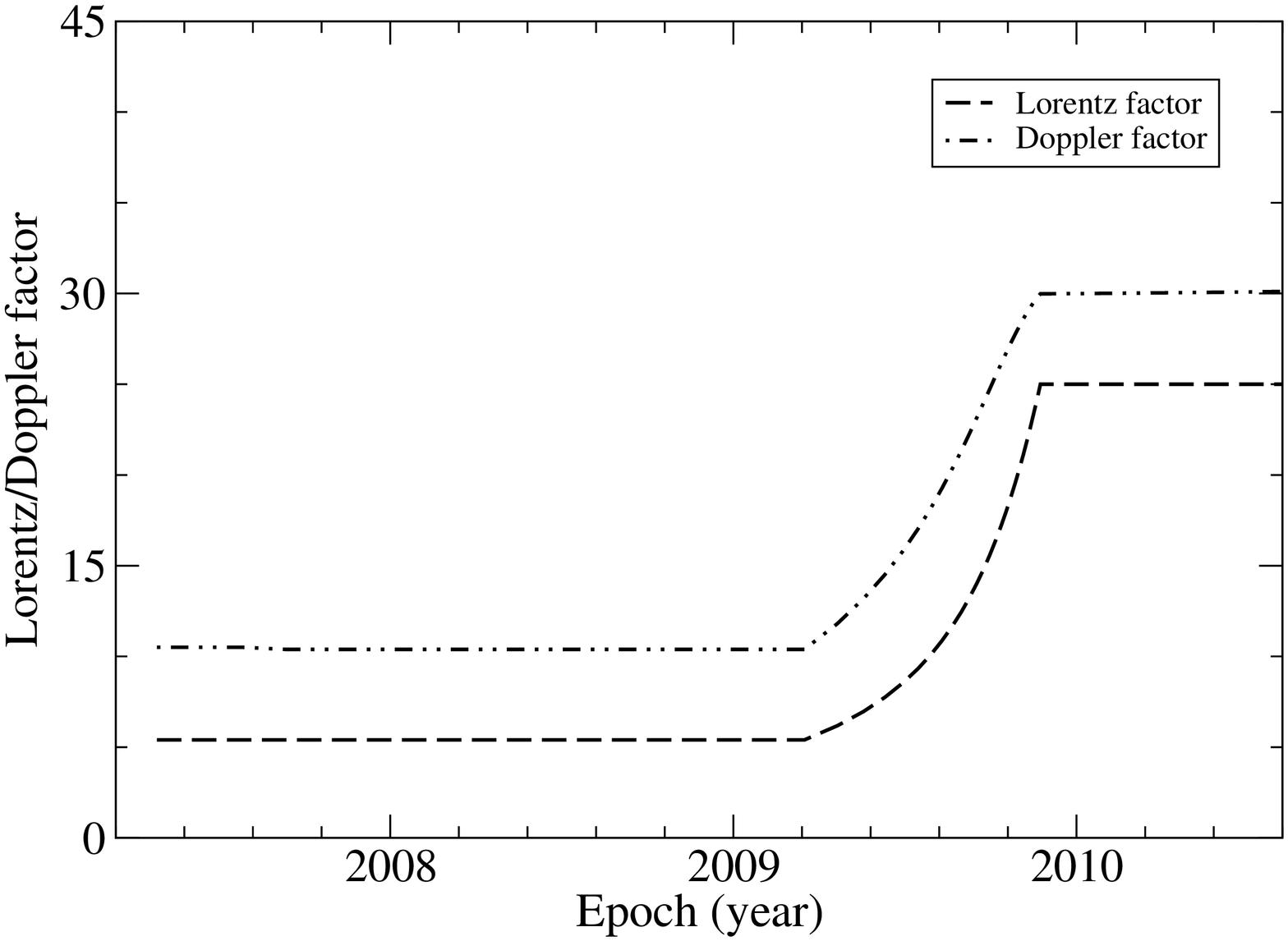}
   \caption{Model fitting of kinematics for knot C21: precession phase 
   $\phi_0$=3.90rad+2$\pi$, $t_0$=2007.32.}
   \end{figure*}
   \subsection{Model fitting results for knot C21}
    The model-fitting results for knot C21  are shown in 
   Figure 17. It can be seen that its kinematic behavior could
    be very well fitted
   by assuming its precession phase $\phi_0$(rad)=3.90+2$\pi$ and ejection
   epoch $t_0$=2007.32.\\
    Its observed precessing common trajectory  may extend to core separation 
   $r_n{\sim}$0.45\,mas, corresponding to spatial distance 
   $Z_{c,m}{\sim}$14.6\,mas or $Z_{c,p}\sim$97.1\,pc.\\
    Its accelerated motion could be modeled by the increase in its Lorentz factor:
    for Z$\leq$3\,mas $\Gamma$=5.4; for Z=3--8\,mas $\Gamma$=
   5.4+19.6(Z-3)/(8-3); for Z$>$8\,mas $\Gamma$=25.0.\\
   During the period 2009.0--2010.4 its Lorentz factor $\Gamma$, 
   Doppler factor $\delta$, apparent velocity $\beta_a$ and viewing angle 
   $\theta$ vary over the following ranges: [5.4,25.0], [10.4,31.4], [1.7,24.1]
   and [1.77,1.76](deg).
   \subsection{Model fitting results for knot B5}
    In the following we discuss the model-fitting of VLBI-kinematics for 
   superluminal components B5--B8, B11 and B12.  Data on these knots
   are kindly provided by S.G.~Jorstad (private communication).\\
    Knot B5 is an interesting case, because  its trajectory was observed 
    nearly at the upper edge of the projected jet-cone. Obviously, it has 
    been curved away northward at $X_n{\stackrel{<}{_\sim}}$0.2\,mas.
    Although its initial trajectory within $X_n{<}$0.2\,mas was not observed,
     it still could be explained in terms of  the precessing nozzle model as 
    shown in Figure A.13 by assuming its precession phase 
    $\phi_0$(rad)=0.33+4$\pi$
    and ejection epoch $t_0$=2010.48. However, in this case  introduction of 
    change in amplitude parameter is required to fit its outer trajectory:
    for Z$\leq$6\,mas $A_0$=1.09\,mas; for Z=6--15\,mas 
    $A_0$=1.09+0.82(Z-6)/(15-6); for Z$>$15\,mas $A_0$=1.91.\\
    Bulk acceleration is also needed: for Z$\leq$15\,mas $\Gamma$=12.0;
    for Z=15--20\,mas $\Gamma$=12+14.5(Z-15)/(20-15); for Z$>$20\,mas 
    $\Gamma$=26.5. \\
    Its precessing common trajectory might be assumed to extend to
     core separation $r_n{\sim}$0.15\,mas, corresponding to a spatial distance
    $Z_{c,m}{\sim}$5.40\,mas or $Z_{c,p}\sim$35.9\,pc.\\ 
     During the period 2011.0-2013.2 its Lorentz factor $\Gamma$, Doppler factor
    $\delta$, apparent velocity $\beta_a$ and viewing angle $\theta$ vary over
    the respective ranges: [12.0,26.5], [21.6,32.7], [7.1,25.7] and
     [1.57,1.70](deg).
   \begin{figure*}
   \centering
   \includegraphics[width=5.5cm,angle=-90]{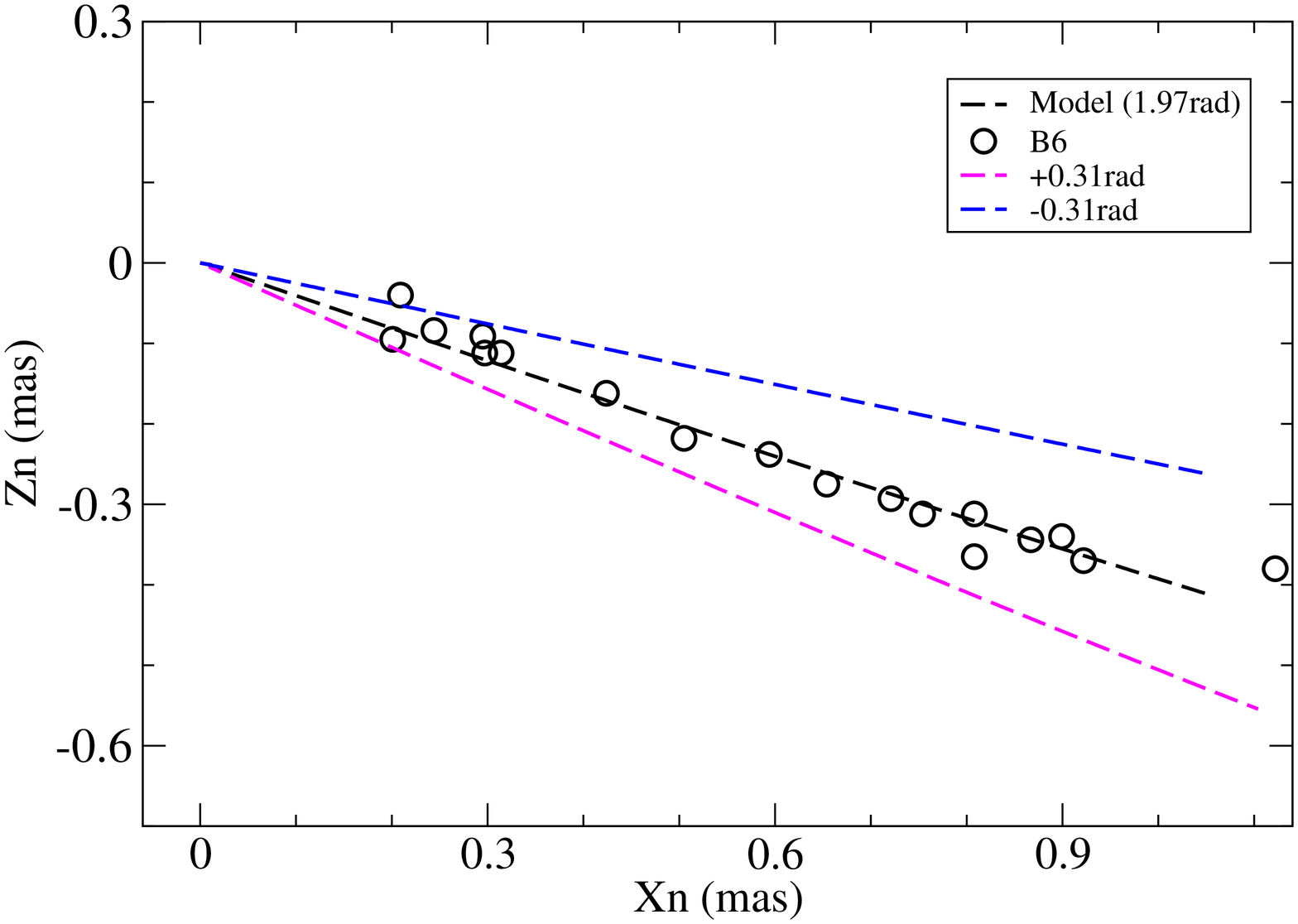}
   \includegraphics[width=5.5cm,angle=-90]{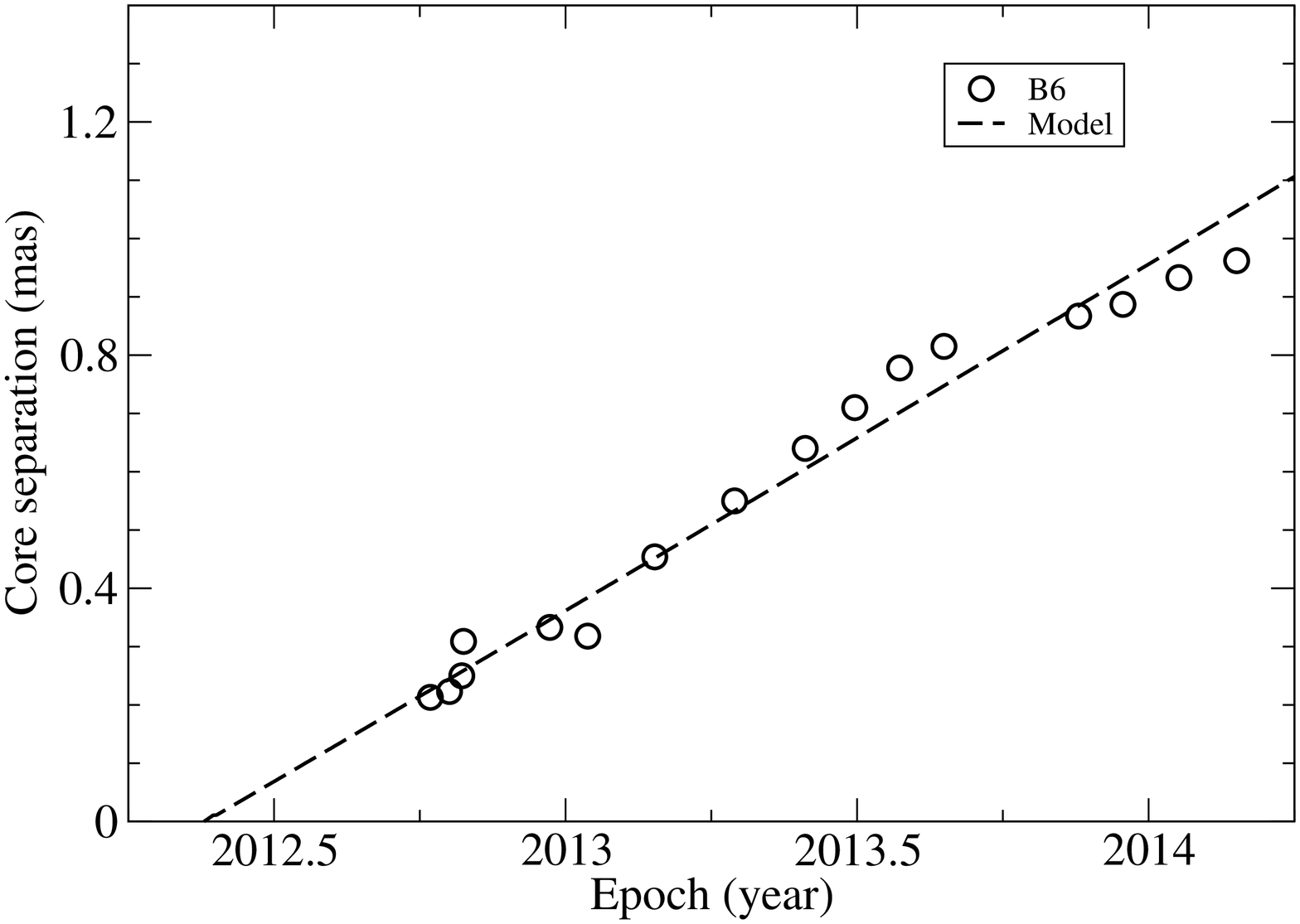}
   \includegraphics[width=5.5cm,angle=-90]{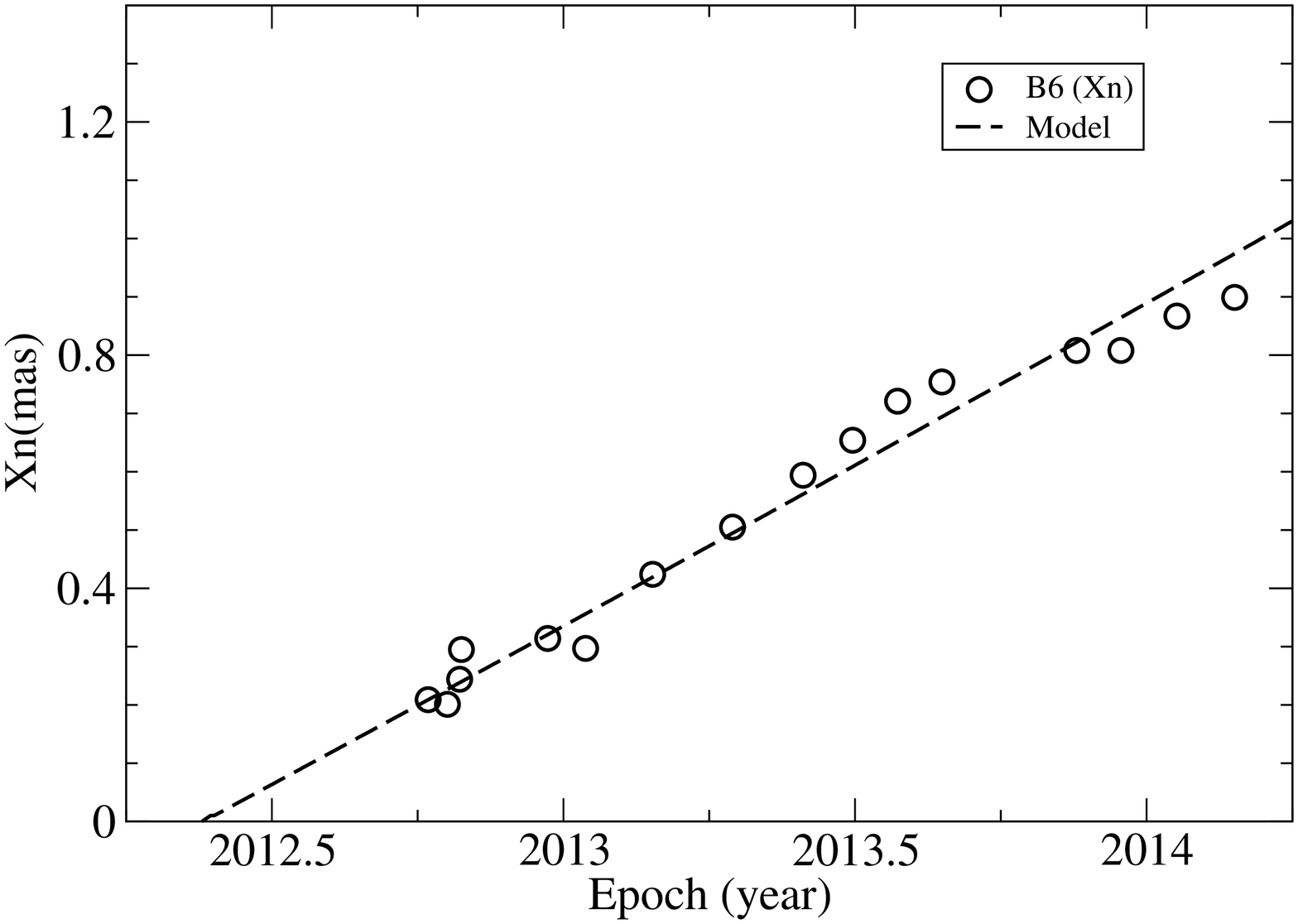}
   \includegraphics[width=5.5cm,angle=-90]{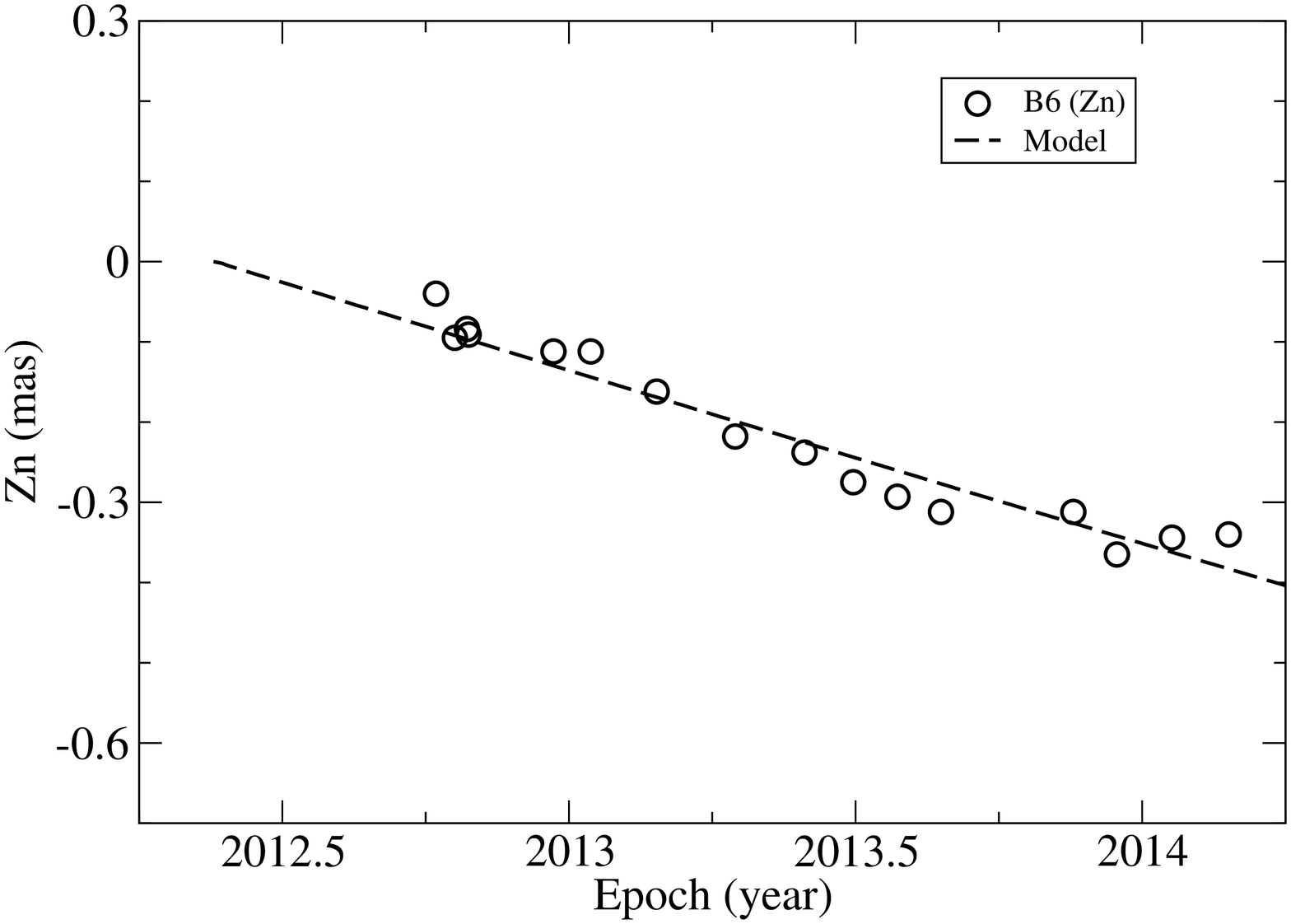}
   \includegraphics[width=5.5cm,angle=-90]{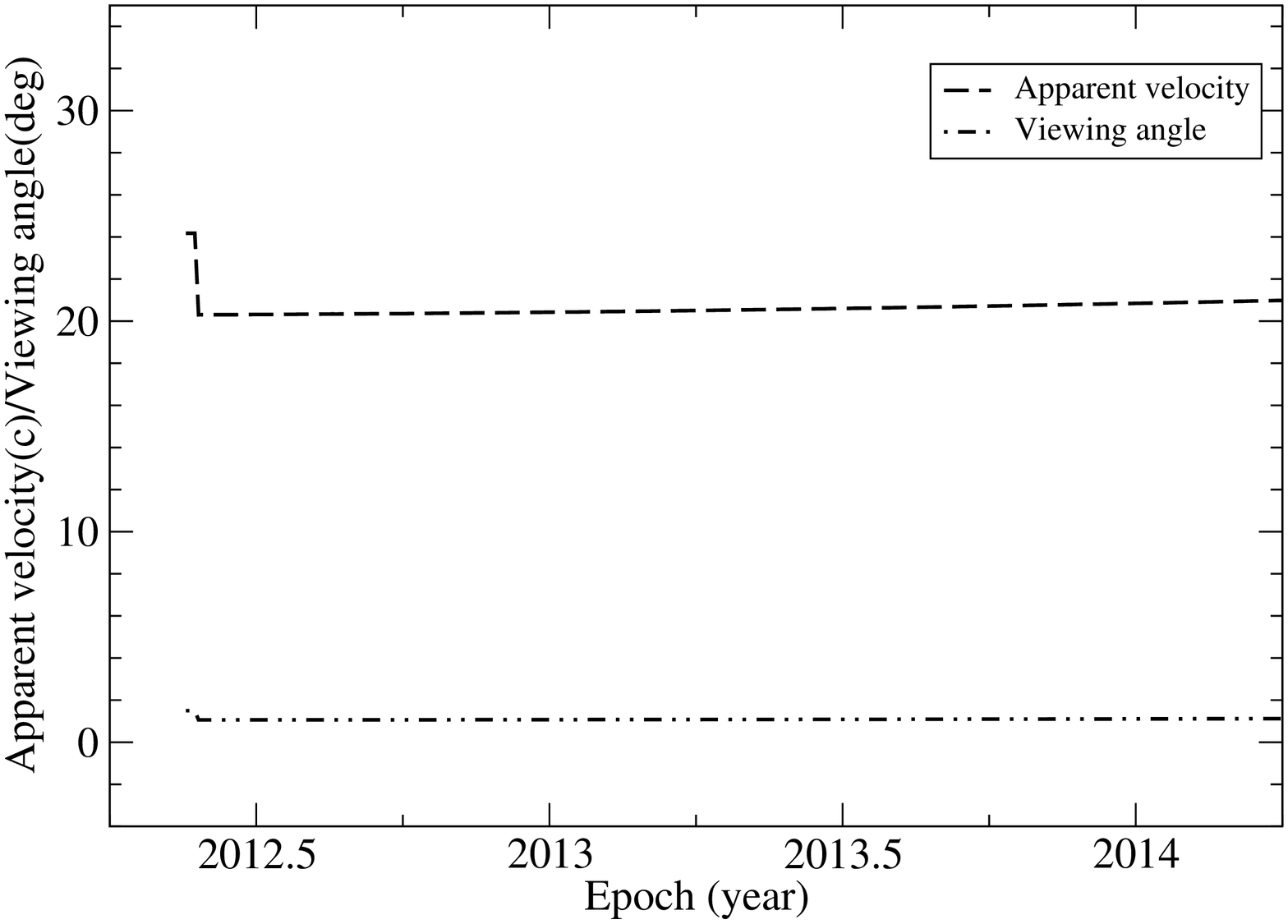}
   \includegraphics[width=5.5cm,angle=-90]{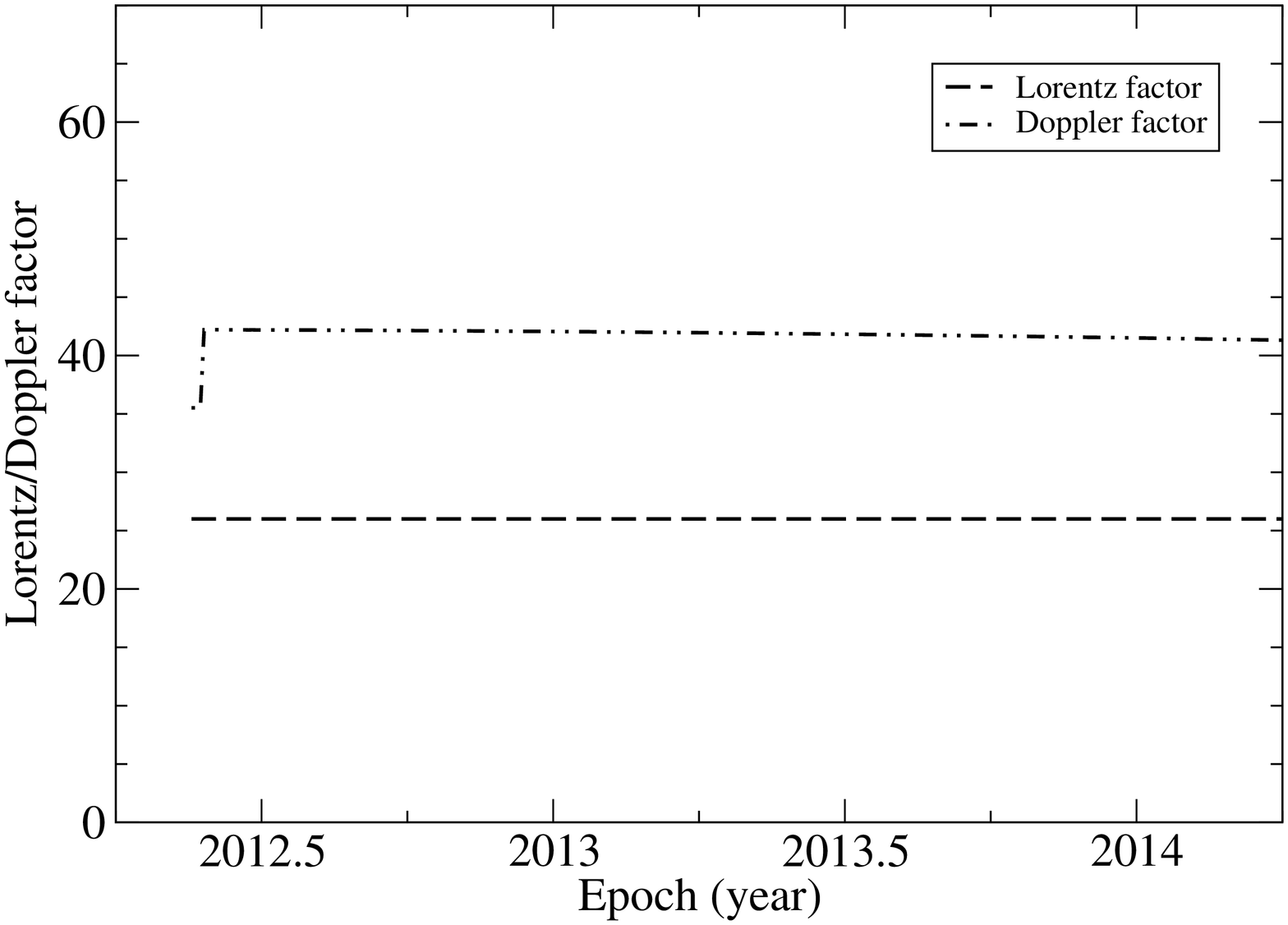}
   \caption{Model fitting of kinematics for knot B6: precession phase
     $\phi_0$=1.97\,rad+4$\pi$, $t_0$=2012.38. Most of the data-points almost
    exactly locate on the predicted precessing common trajectory, justifying
    our precessing nozzle scenario for jet-B.}
   \end{figure*}
   \subsection{Model fitting results for knot B6}
   The model fitting of the kinematics of knot B6 in terms of the precessing 
   nozzle scenario for jet-B is a key ingredient for justifying the double-jet
    scenario for 3C345, because the observational data extends the monitoring 
   time-interval to $\sim$1.4 times the precession period. Its kinematic
   behavior can be very well fitted as shown in Figure 18: most of the 
   data-points almost exactly locate on the predicted precessing common
   trajectory within $X_n{\sim}$1.2\,mas, strongly justifying our 
    precessing nozzle scenario for jet-B. \\
   Its precession phase is modeled as $\phi_0$(rad)=1.97+4$\pi$, corresponding
   to ejection epoch $t_0$=2012.38. In fact the model fitting of the kinematics
   for knot B6 may be regarded as an posterior test of our double-jet scenario,
   because the double precessing jet scenario had been constructed before
   the observational data on knot B6 was collected.\\
   Its observed precessing common trajectory may extend to  core separation 
    $r_n{\sim}$1.2\,mas, corresponding to a spatial distance 
    $Z_{c,m}{\sim}$67.7\,mas or $Z_{c,p}{\sim}$420.3\,pc from the core, the 
   second longest extension of precessing common trajectory in  
   group-B (Table 2).\\
    Its motion is modeled as uniform: Lorentz factor is assumed to be 
    $\Gamma$=26.0. During the period 2012.70--2014.25 its Doppler factor 
    $\delta$, apparent velocity $\beta_a$ and viewing angle $\theta$ vary over
    the following ranges respectively: [42.2,41.3], [20.3,21.0] and 
     [1.06,1.13](deg).
     The successful model-fitting of the kinematic behavior of knot B6 is
    very encouraging, demonstrating our precessing nozzle scenario being 
    fully applicable to blazar 3C345.
    \subsection{Model fitting results for knot B7}
     We would like to note that the following model fitting results for
     knots B7, B8, B11 and B12 were obtained after our finishing the works
     for the thirteen knots (C4 to C23) of jet-A and the ten knots (C5 to B6)
     of jet-B, thus their model-fittings played a role of follow-up
     or afterward verification of the double-jet scenario. The successful 
     model fitting of kinematics for the four 
     knots proved our scenario being valid and effective, and extend the 
     model-fitting of the jet-B  to two precession periods.\\
    In our precessing nozzle scenario, the observed trajectories of knots B7 
    and B8 were distributed earlier than
    that of knot B6, implying their ejection times earlier than that of 
    knot B6 ($t_0$=2012.38, $\phi_0$=1.97\,rad+4$\pi$).
     We assumed $t_0$=2011.60 for knot B7 (equivalent to the precessiom 
    phase $\phi_0$=1.30\,rad+4$\pi$) and 
    $t_0$=2011.95 for knot B8 (equivalent to $\phi_0$=1.60\,rad+4$\pi$).\\
    The model fitting results for knot B7 are shown in Figure A.14. It can be 
    seen that its kinematics can be explained well in terms of our precessing 
    nozzle model. No acceleration in its motion was observed and its
    bulk Lorentz factor was assumed to be $\Gamma$=10.5=const.
     During the period of 2013.0--2017.0, its 
    Doppler factor $\delta$, apparent velocity $\beta_a$, and viewing angle
    $\theta$ vary over the following respective ranges: [20.2-20.2],
    [3.8-3.9], and [1.04-1.06](deg), almost staying constant. 
    Its precessing common trajectory may be regarded as extending to the core
    separation of $r_{n,c}$=0.80\,mas, equivalent to a spatial distance 
    $Z_{c,m}$=43.7\,mas (or $Z_{c,p}$=290.8\,pc).   
    \subsection{Model fitting results for knot B8}
      The model fitting results for knot B8 are shown in Figure A.15. 
    Obviously, its kinematics can be interpreted well in terms of our 
    precessing nozzle scenario. Acceleration in its motion was observed and
    its bulk Lorentz factor was assumed to be as follows: for Z$\leq$20\,mas,
    $\Gamma$=17; for Z=20-30\,mas, $\Gamma$=17+(Z-20)(23-17)/(30-20); and
    for Z$>$30\,mas, $\Gamma$=23.0.\\
    During the period of 2013.0--2015.5, its bulk Lorentz factor
    $\Gamma$, Doppler factor $\delta$, apparent velocity $\beta_a$, and viewing
    angle $\theta$ vary over the following respective ranges: [17.0-23.0],
    [31.2-(39.3)-38.3], [9.4-17.1], and  [1.02-1.11]. 
    Its precessing common trajectory may extend to core separation of 
    $r_{n,c}$=1.20\,mas, equivalent to a spatial distance $Z_{c,p}$=439.8\,pc
    (or $Z_{c,m}$=66.1\,mas). 
    \subsection{Model fitting results for knot B11}
     Knot B11 is an uncertain case.  In our scenario, the trajectory observed
     for knot B11 was distributed later than that of knot B12, but its core
     separation versus time showed that its ejection time should be 
     earlier than B12.
     Thus for explaining the kinematics of B11 some changes in its kinematics
     were needed to take into consideration. We assumed its ejection time 
     $t_0$=2015.67 ($\phi_0$=4.80+4$\pi$). Its bulk Lorentz factor was assumed 
     to follow as: for Z$\leq$10\,mas, $\Gamma$=14.0; for Z=10-15\,mas,
     $\Gamma$=14+(Z-10)(23-14)/(15-10); and for Z$>$15\,mas, $\Gamma$=23.0.
     The changes in its trajectory was described by the change in 
     parameter $\psi$ as follows: for Z$\leq$4\,mas, $\psi$=0.209(rad); for
     Z=4-8\,mas, $\psi$(rad)=0.209-(Z-4)(0.05-0.209)/(8-4); and for 
     Z$>$8\,mas, $\psi$=0.05\,rad.\\
     Its precessing common trajectory might be assumed to extend to 
     $r_{n,c}$=0.12\,mas, equivalent to a spatial distance  $Z_{c,m}$=
     3.40\,mas (or
     $Z_{c,p}$=22.6\,pc). The model fitting results are shown in Figure A.16.
      It can be seen that
     the rotation of its trajectory in the range of  Z=4\,mas to 8\,mas
     made  the kinematics of B11 being well explained within the scenario
     for jet-B.
     Unfortunately, no observational data available to confirm this rotation.
     (in comparison with the case of knot C8).\\
     During the period of 2015.5--2018.0 its bulk Lorentz factor $\Gamma$,
     Doppler factor $\delta$, apparent velocity $\beta_a$, and viewing
     angle $\theta$ vary over the following respective ranges: [14.0-23.0], 
     [22.6-28.4], [11.0-22.3], and [1.99-1.96](deg).
      \begin{figure*}
    \centering
    \includegraphics[width=5.5cm,angle=-90]{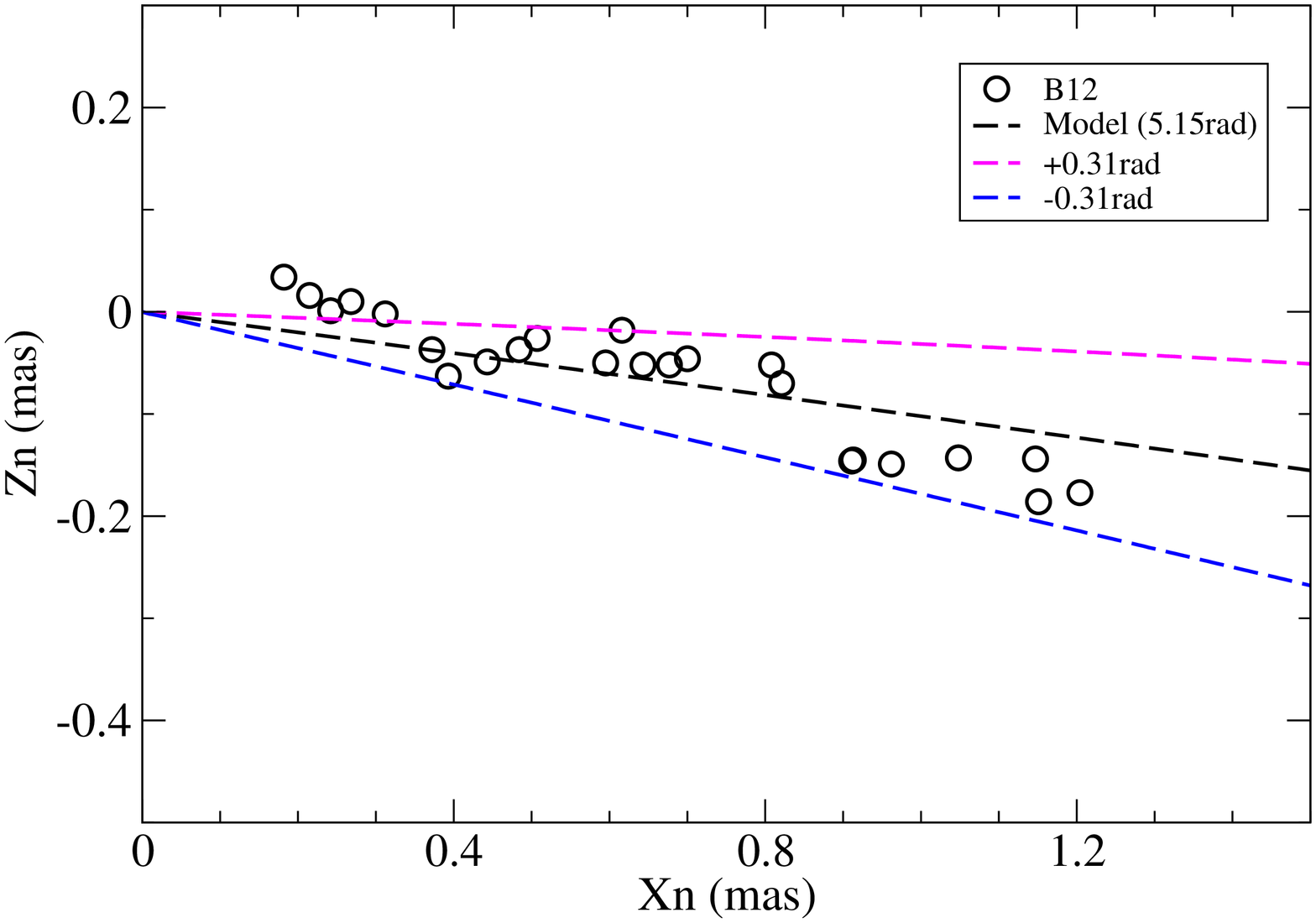}
    \includegraphics[width=5.5cm,angle=-90]{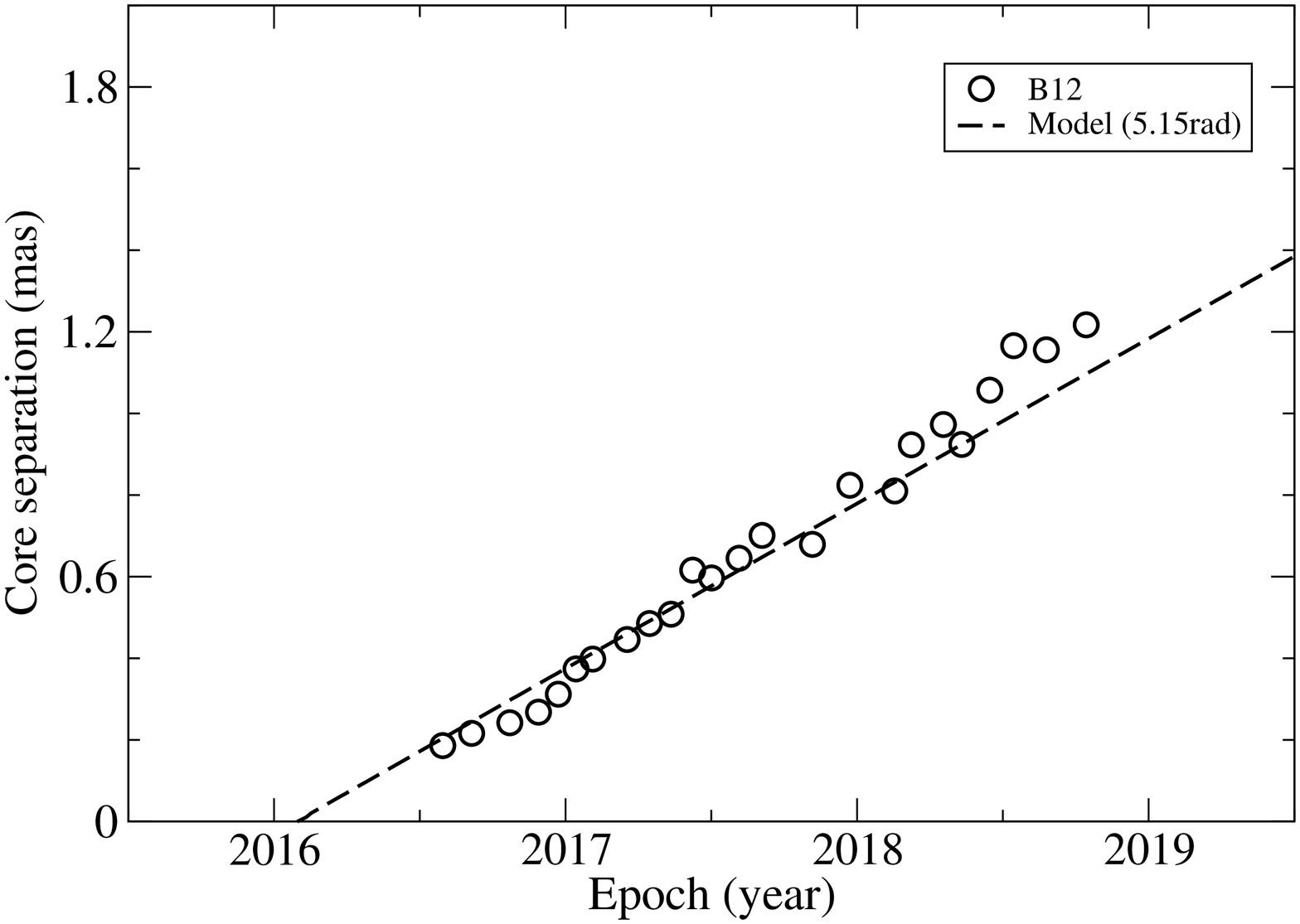}
    \includegraphics[width=5.5cm,angle=-90]{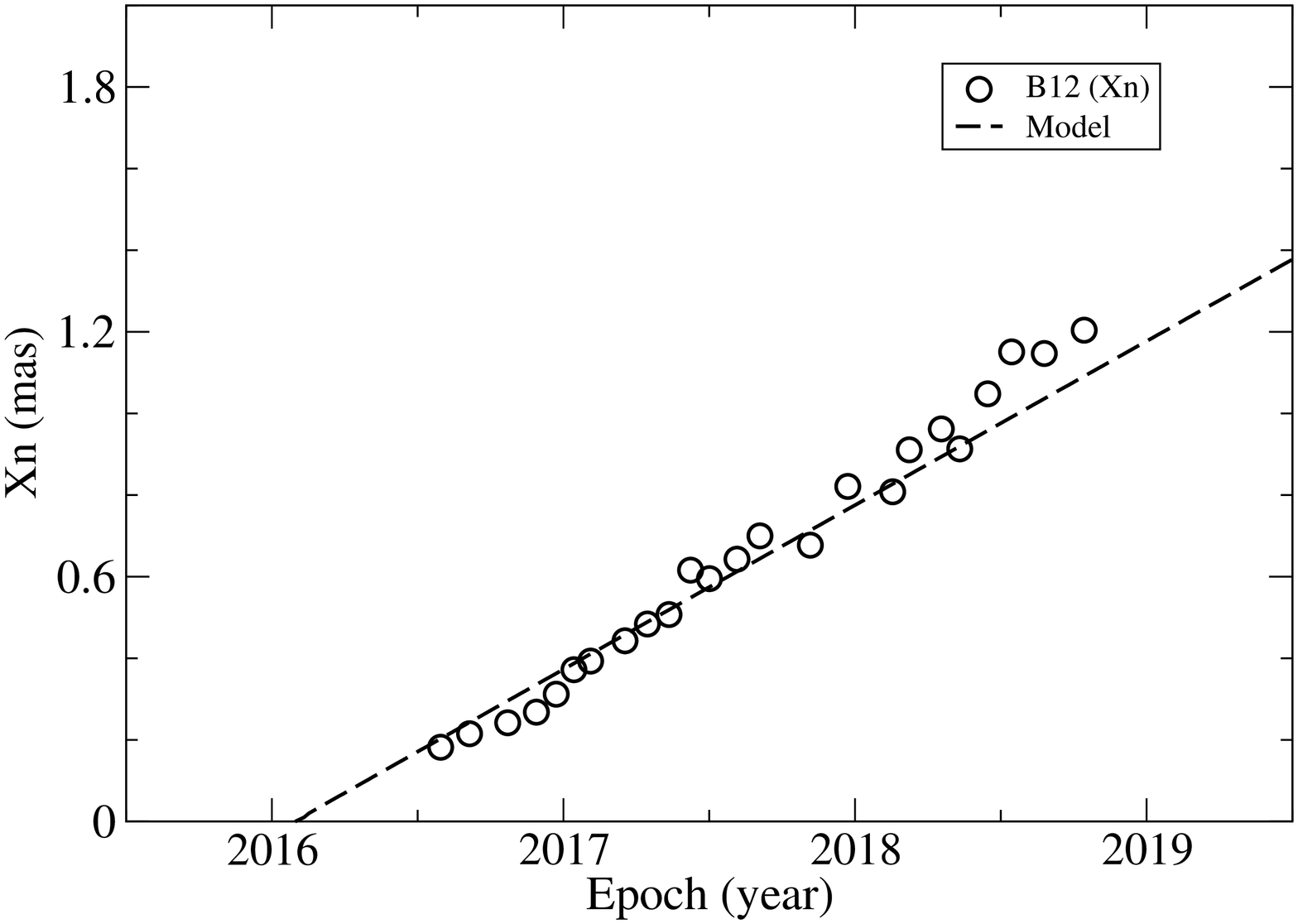}
    \includegraphics[width=5.5cm,angle=-90]{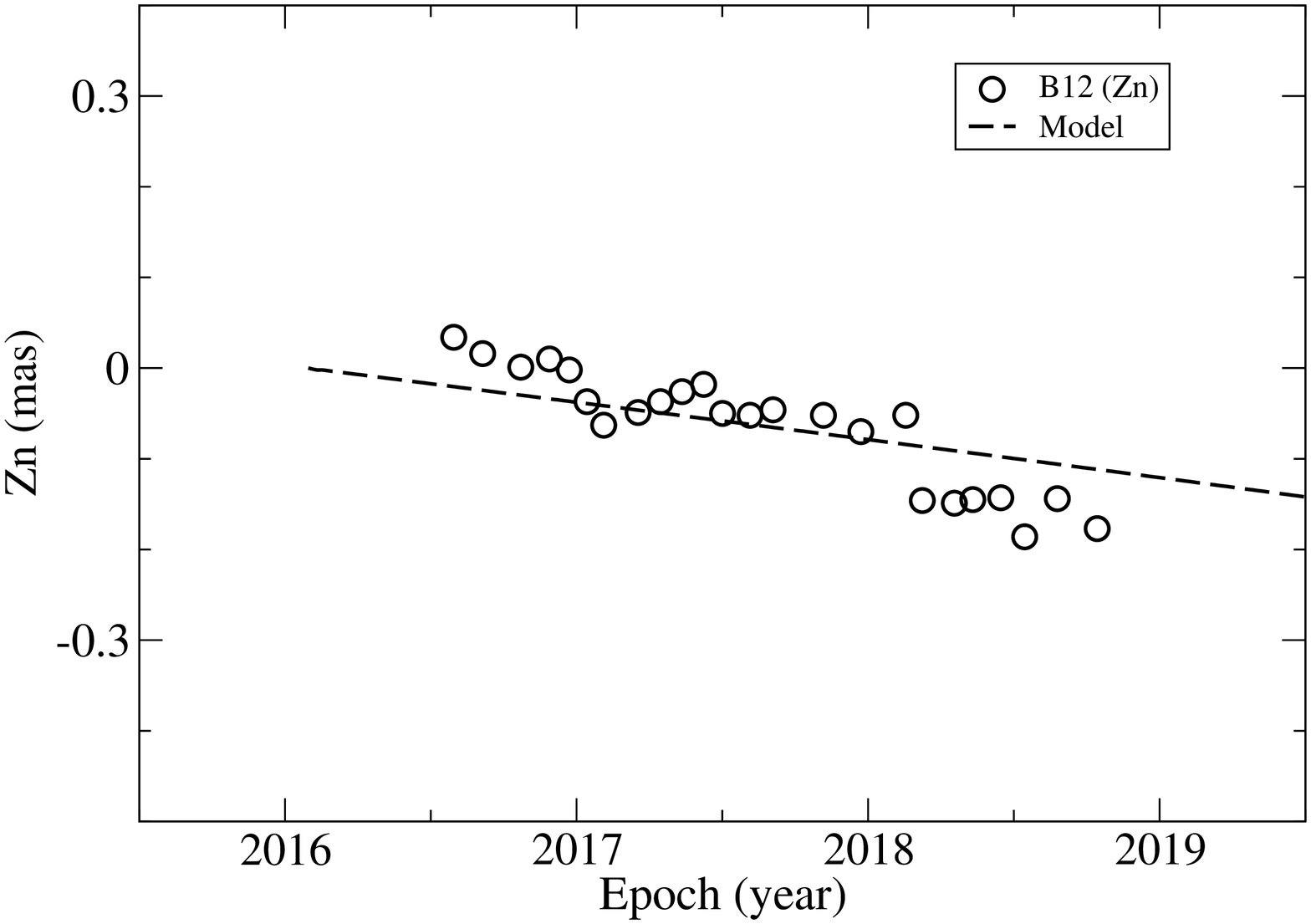}
    \includegraphics[width=5.5cm,angle=-90]{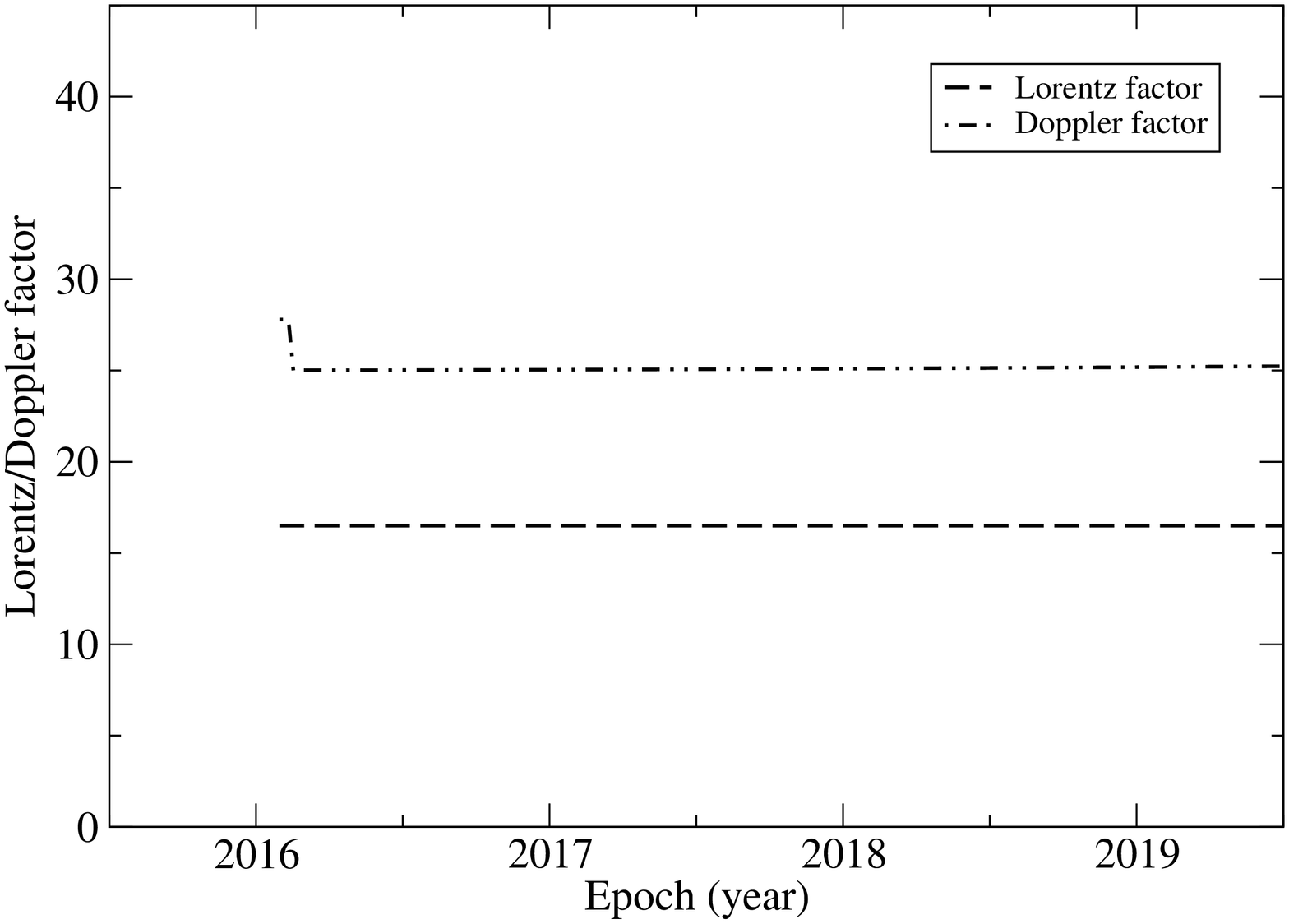}
    \includegraphics[width=5.5cm,angle=-90]{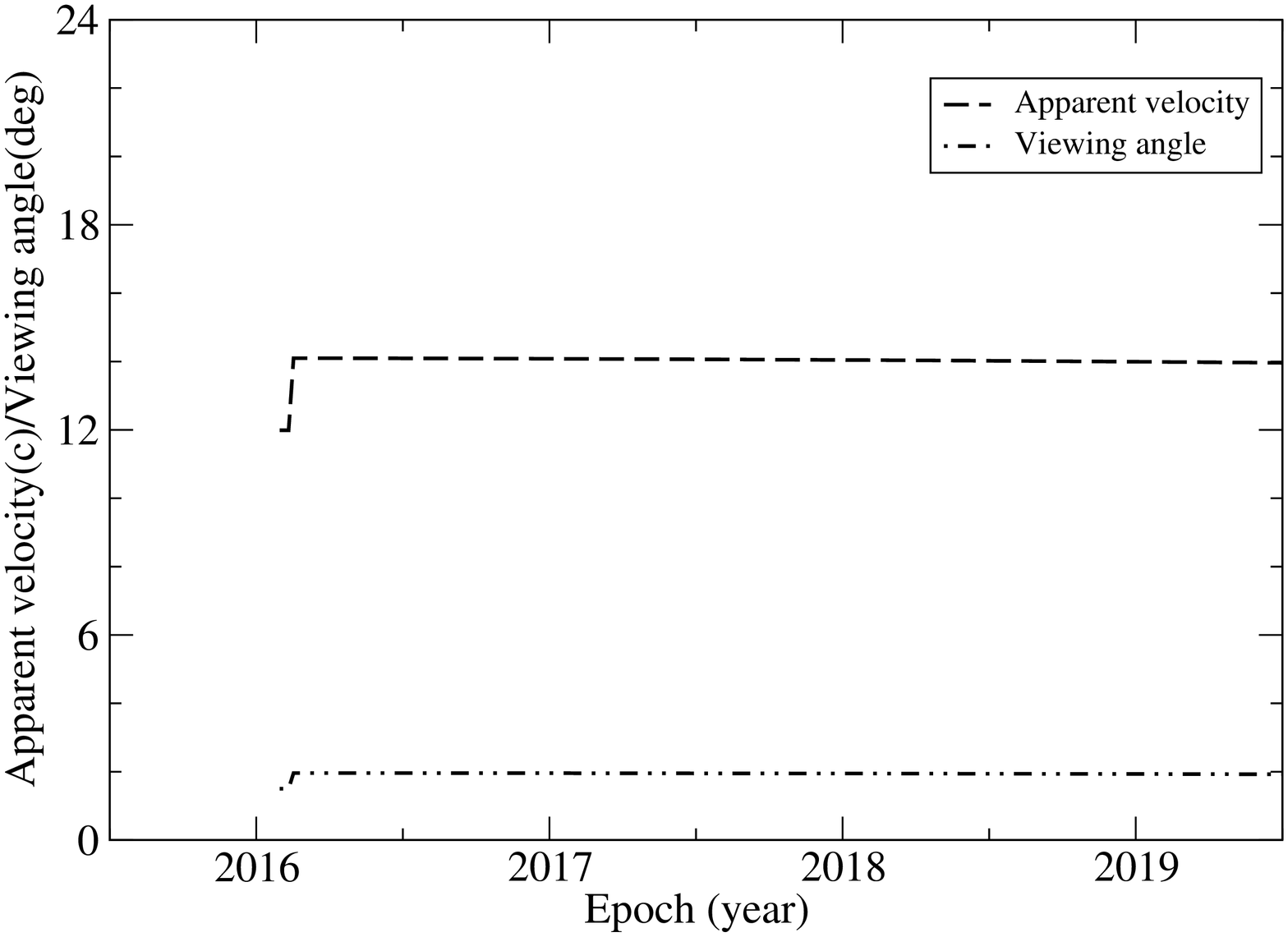}
    \caption{Model fitting results for knot B12. Precession phase
    $\phi_0$(rad)=5.15+4$\pi$ and ejection time $t_0$=2016.08.}
    \end{figure*}
    \subsection{Model fitting results for knot B12}
    Knot B12 is a significant knot for the model fitting of jet-B because
     it regularly followed the precessing common path as the scenario 
    predicted, extending the model fitting time-interval for jet-B to 
    approximately two times the precession period. The model fitting results 
    are shown in Figure 19. Its ejection time was assumed to be $t_0$=2016.08
    (or $\phi_0$=5.15\,rad+4$\pi$). The observational data-points are closely
    distributed around the predicted trajectory within the regions defined by
    lines in magenta and blue.\\
    Its motion was nearly ballistic and no acceleration needs to be modeled.
    Its  bulk Lorentz factor is modeled as $\Gamma$=16.5=const.  During the 
    period of 2016.5--2019.0, its Doppler factor $\delta$, apparent
    velocity $\beta_a$, and viewing angle $\theta$ vary over the 
    following respective ranges:  [25.0-25.2], [14.1-14.0], and
    [1.96-1.93](deg). They almost stay constant. Its precessing common 
    trajectory may be regarded as extending to core separation ${r_n}{\sim}$
    1.20\,mas, corresponding to a spatial distance $Z_{c,m}$=35.2\,mas
    (or $Z_{c,p}$=234.1\,pc).\\
     \begin{figure*}
     \centering
     \includegraphics[width=5.5cm,angle=-90]{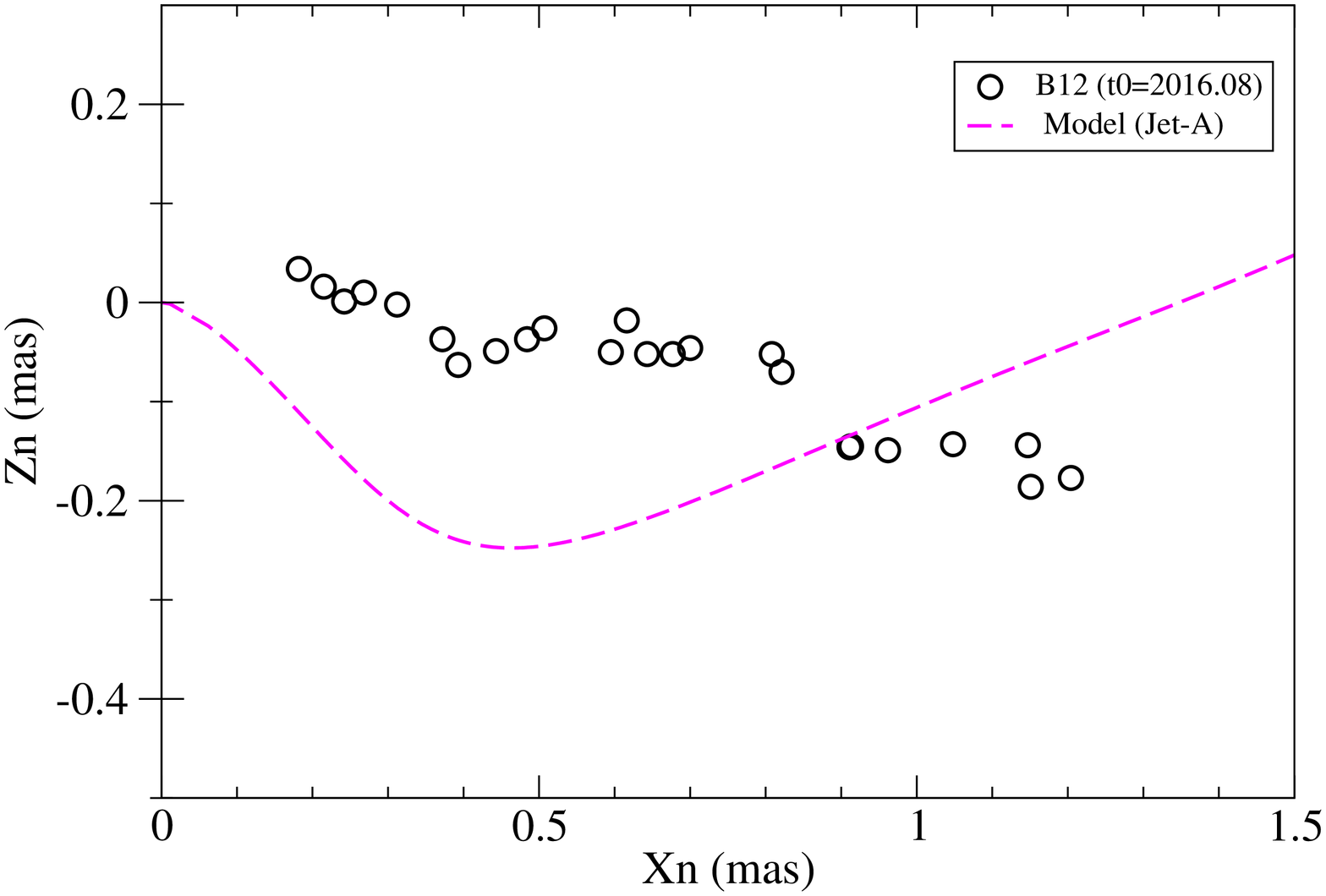}
     \includegraphics[width=5.5cm,angle=-90]{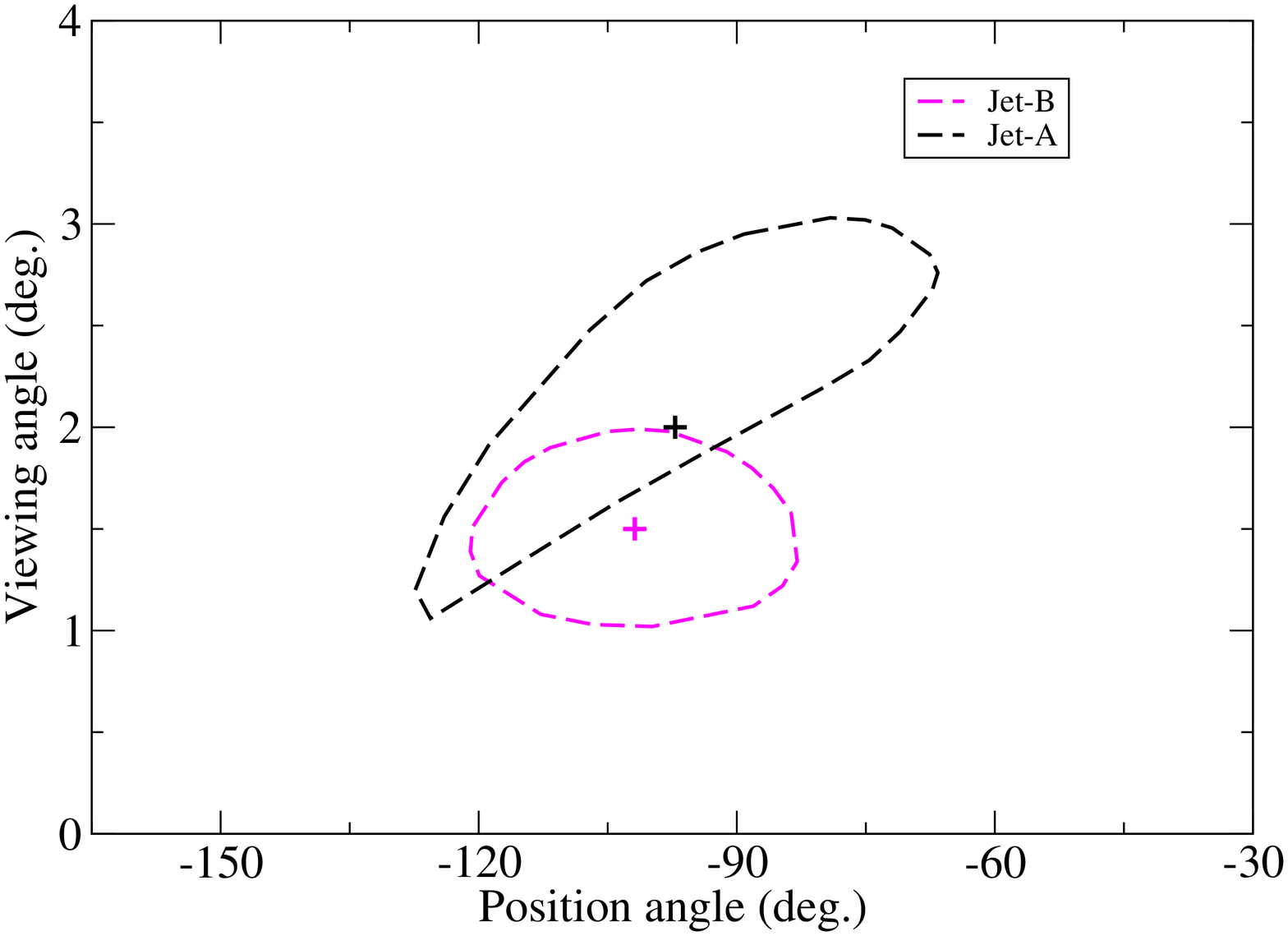}
     \caption{Left panel: Model fitting of the observed trajectory of
       knot B12 in the framework of the scenario for jet-A, showing
       it unlikely being ascribed to jet-A (dashed line represents the 
       predicted trajectory for B12 in the jet-A framework). Right panel:
       The modeled jet-cone patterns (relation between position angle 
       and viewing angle for jet-A (in black) and jet-B (in magenta))
       at core separation 0.3\,mas. Symbol + represents the position of the 
       jet-axis at core separation 0.3\,mas. }
     \end{figure*}
         In order to show that knot B12 should be ascribed to jet-B (not
    to jet-A), its observed trajectory has also been modeled
    in the framework of the scenario for jet-A (taking 
    $t_0$=2016.08 and corresponding precession phase $\phi_0$=4.78+10$\pi$),
     which is shown in Figure 20 (left panel).
    It can be seen that the trajectory of 
    B12 is very different from that predicted by the jet-A modeling scenario,
    but is  well fitted by the jet-B modeling scenario, favoring it associated
    with jet-B.  In Figure 20 (right panel) we present  the modeled relation
    between position angle and viewing angle at core separation 0.3\,mas for
    jet-A and jet-B, which also demonstrates jet-A having a different relation
    from that of jet-B, favoring a double jet structure. \\
    We would like to note that in this paper the superluminal knots
    observed in 3C345 were separated into two groups (group-A and group-B) 
    according to the characteristic features of their kinematic behaviors:
    (1) Knots of group-A moved along curved trajectories while knots of
    group-B moved approximately along straight trajectories;
     (2) Knots of group-A 
    and group-B construct different distributions of precessing common 
    trajectories (i.e., different forms of the function $\phi_0(t_0)$) 
     although they had the same precession period of 7.3\,yr; (3) The axis of 
     jet-A and jet-B have different directions in space: their spatial
     viewing angles differ by ${\sim}6.8^{\circ}$, although their 
     projections in  the sky-plane have closely similar position angles;
     (4) Jet-A and jet-B have different degrees
     of  activity during their respective time-intervals.
     \begin{table*}
    \centering
    \caption{Model parameters for the fourteen superluminal knots of jet-B:
    ejection epoch $t_0$, precession phase $\phi_0$, extension of  
    precessing common trajectory $r_{n,c}$ from the core, corresponding 
    spatial distance  $Z_{c,m}$ (mas) and $Z_{c,p}$ (pc), bulk Lorentz
    factor $\Gamma$, Doppler factor $\delta$, apparent velocity $\beta_a$,
    viewing angle(deg) and status. The superluminal components were ejected
    during a time-interval of about two precession periods. Column 
     'status' denotes the quality
    of trajectory model-fits: + for well-fitted cases and $-$ for not 
    fitted cases.}
   \begin{flushleft}
   \centering
   \begin{tabular}{lllllllllll}
   \hline
   Knot & $t_0$ & ${\phi}_0$ & $r_{n,c}$ & $Z_{c,m}$ & $Z_{c,p}$ & $\Gamma$ &
              $\delta$ & $\beta_a$  & $\theta$ & status\\
   \hline
   C15 & 2002.12 & 5.70 & 0.70  & 21.0 & 139.7 & 8-24 &
           14.9-29.4   & 3.9-23.4  & 1.91-1.90 & +\\
   C15a & 2002.18  & 5.75 & 0.17 &  5.60  & 37.2 & 8.5-10 &
           15.8-18.6   & 4.2-5.0   & 1.80-1.55 & +\\
   C16 & 2002.24 & 5.80 & 0.80 & 24.4 & 162.2 & 6.0-18.0 &
           11.5-26.7   & 2.2-15.7  & 1.88-1.87 &  +\\
   C17 &  2004.62 & 1.57+2$\pi$  & 0.80 & 44.8 & 297.9 & 9.9-20.1 &
           19.1-35.1 & 3.5-13.1   & 1.01-1.07 & +\\
   C18  & 2004.68 & 1.62+2$\pi$   & 0.60  & 33.8 & 224.8 & 9.1-16.0 &
           17.8-29.5  & 2.8-8.5  & 1.01-1.03 & +\\
   C19  & 2004.85 & 1.77+2$\pi$   & 0.60 & 33.4 & 222.1 & 8.0-21.0 &
           15.6-36.5 & 2.2-14.1  & 1.02-1.05 & +\\
   C20  & 2007.68 & 4.20+2$\pi$  & 0.45  & 13.2 & 87.8 & 7.4-21.0  &
           13.9-27.9  & 3.5-19.8  & 1.95-1.94 & +\\
   C21  & 2007.32 & 3.90+2$\pi$ & 0.45 & 14.6 & 97.1 & 5.4-25.0 &
           10.4-31.4   & 1.7-24.1   & 1.77-1.76 & +\\
   B5   & 2010.48  & 0.33+4$\pi$ & 0.15 & 5.40 & 35.9 & 12.0-26.5 &
          21.6-32.7  & 7.1-25.7 & 1.57-1.70 & --\\
   B6  & 2012.38  & 1.97+4$\pi$  & 1.20 & 67.7 & 420.3 &  26.0=const.  &
          42.2-41.3 & 20.3-21.0  & 1.06-1.13 & +\\
   B7  & 2011.60  & 1.30+4$\pi$  & 0.80 & 43.7 & 290.8  & 10.5=const.  &    
       20.2=const. & 3.8-3.9 & 1.04-1.06 & +\\
   B8  & 2011.95  & 1.60+4$\pi$  & 1.20 & 66.1 & 439.8 & 17.0-23.0 &
       31.2-(39.3)-38.3 & 9.4-17.1  & 1.02-1.11 & + \\
   B11  & 2015.67  & 4.80+4$\pi$ & 0.12 & 3.40 & 22.6 & 14.0-23.0  &
       22.6-28.4 & 11.0-22.3  & 1.99-1.96 & -- \\
   B12  & 2016.08  & 5.15+4$\pi$ & 1.20 & 35.2 & 234.1  & 16.5=const.  &
       25.0-25.2 &  14.1-14.0   &  1.96-1.93 & + \\
   \hline
   \end{tabular}
   \end{flushleft}
   \end{table*}  
   \subsection{A brief summary for jet-B}
    We briefly summarize the model-fitting results for jet-B.
   \begin{itemize}
    \item The kinematics of the fourteen superluminal knots (C15--C21,
     B5--B8, B11 and B12)
    can be consistently and well modeled in the precessing nozzle scenario
    for jet-B.
    \item Its precession period is just the same with that for jet-A 
    (7.30\,yr) and the sense of precession is also similar to that for jet-A: 
     counterclockwise seen along the line of sight.
    \item As shown in Table 2 the time-interval for our model-fitting is
    $\sim$14\,yr, corresponding to $\sim$1.5 times the precession period.
    \item Combined with the model-fitting results for the 13 superluminal
     knots of jet-A, we would be able to tentatively conclude that QSO 3C345
    could contain a double jet system formed by a putative binary supermassive
     black hole.
    \item The jet cone described by the precessing nozzle of jet-B was 
    distinctly different from that of jet-A and this could be tentatively 
    regarded as a significant clue to the double-jet structure in 3C345.
    \item As argued above, the knots of jet-B approximately moved 
     along straight-lines (ballistically), while the knots of jet-A along
     curved trajectories. This might imply that the knots of jet-B
     were observed at larger core distances. In fact, there were seven knots
     of jet-B were observed till spatial core distances greater than 200\,pc
     (see parameter $Z_{c,p}$ in Table 2), while only three knots for jet-A
     (see Table 1). This might imply that the knots of jet-B could move 
     along helical trajectories with much smaller pitch angles.  
   \end{itemize}
    \section{Discussion}
      We have investigated  and explained the kinematic behavior on pc-scales
    of twenty-seven superluminal components in QSO 3C345. It was  tentatively
    found that they could be divided into two groups (group-A and group-B;
     see Table 1 and Table 2, respectively) and  the superluminal components 
    could be ejected from a double-jet system consisting of jet-A and jet-B,
    which produce the superluminal components of group-A and group-B,
     respectively.\\
    Based on detailed  model fitting of the kinematics for the knots 
    of group-A and group-B in terms of our precessing nozzle scenario,
    it was found that both jet-A and jet-B could  precess with a 
    precession period of 7.30$\pm$0.36\,yr. This is fully consistent with 
    the analysis in Qian et al. (\cite{Qi09}), where a precession period 
    of $\sim$7.36\,yr was derived from model-fitting of the position angle
    swings observed for seven knots (C4--C10) at 
   different core separations (0.10, 0.15 and 0.20\,mas). \\
     For 25 out of the 27 superluminal knots the  model-fitting of their 
    kinematic behavior showed that bulk acceleration are required for 
    explaining their motion. The modeled Lorentz factor varies over a 
    range of  $\sim$4--30. \\
      Based on the results obtained in this paper we would point out 
    the following.
    \begin{itemize}
    \item Our model-fitting results reveal 
    that the acceleration zones could extend
   to spatial distances of $\sim$300--400\,pc from the cores 
    (for knots C4, C5, C9,
    C17 and B6, see Table 1 and 2). This result is consistent with MHD
    theories for jet formation. For example, Vlahakis \& K\"onigl (\cite{Vl04})
      suggested that relativistic MHD models of jet formation could provide 
     extended acceleration on kilo-parsec scales and proposed a specific 
     MHD model for 
     the jet in 3C345 and predicted an acceleration to $\Gamma{\sim}$10 at
     a radial distance Z$\sim$20\,pc. Interestingly, this model also predicted
     an acceleration to $\Gamma{\sim}$20 at a radial distance Z$\sim$300\,pc.
     Thus the Vlahakis/K\"onigl's radial self-similar solution of MHD 
     jet-formation mechanism with their specifically selected  parameters
      may be already a successful theory. Similar intrinsic acceleration 
      observed in other blazars may also be understood within this framework.
      Moreover stronger and more extended bulk acceleration could be expected
      if magnetic collimation occur in  cylindrical-type magnetic surface
       structures.\\
      Radial scales of $\sim$300\,pc approximately 
   correspond to $\sim{10^6}{R_s}$, 
     if the mass of central supermassive black hole is
       $\sim{3\times{10^9}}{M_{\odot}}$
      (Woo \& Urry \cite{Wo02}; $R_s$--Schwarzschild radius).\\
      The long-extended acceleration/collimation of relativistic jets 
     might also be helpful to explain why jets in giant radio galaxies 
      could extend to Mpc-scales 
     (e.g. in giant radio galaxy DA240, see Tsien (\cite{Ts82a}),
     Tsien \& Saunders (\cite{Ts82b}),
      Willis et al.(\cite{Wi74})). As Qian et al.
     (\cite{Qi17}) suggested that except precession of jet-nozzles, other 
      mechanisms may also cause jet precession on different time scales,
      e.g., geodetic precession, Newtonian-driven precession, Lense-Thirring
     effect (Begelman et al. \cite{Be80}, Katz \cite{Ka97}, Lense \& Thirring
     \cite{Len18}).  Especially, geodetic precession  may occur on timescales 
    of ${\sim}10^{4-5}$ years. This kind of jet precession has been observed 
    in the inverse symmetric distribution of lobes and hotspots in FRII 
    radio galaxies (e.g., in Cygnus A and NGC326, see  Ekers et al. 
    \cite{Ek81} and  Ekers \cite{Ek82}; also referring to
    Hargrave \& Ryle \cite{Hag74}; Perley et al. \cite{Pe84}; Oort \cite{Oo82};
    Fanaroff \& Riley \cite{Fa74}; Tsien \cite{Ts82a}, Tsien \& Saunders
    \cite{Ts82b}, Tsien \& Duffet-Smith \cite{Ts82c}). In blazars  different
    jet-precession mechanisms may cause different periodicities in their
    optical/radio light-curves.\\
    \item Previously through model-fitting of the kinematics of 
    superluminal components 
    in blazars in terms of the precessing nozzle scenario we have tentatively
    found three blazars, where double-jet systems might exist in their nucleus:
     3C279 (Qian et al. \cite{Qi19a}), OJ287 (Qian \cite{Qi18a}), 3C454.3
     (Qian et al. \cite{Qi21}). Here we add a new one: 3C345. 
      These double-jet systems may be produced by binary 
    supermassive black hole systems and thus our precessing nozzle scenario
    would be useful to investigate the characteristics of binary black holes,
    e.g., determining orbital period, mass of black holes, precession of 
    jet-nozzle, mechanism of precession, etc.
   It might be expected that more blazars could be found to possibly house
    binary supermassive  black hole systems, if  observational data are plenty 
     enough to make  model-simulation of their  VLBI-kinematics.\\
    \item  As previously suggested, in our precessing nozzle scenario for
     3C345 and other blazars we assumed that  precessing nozzles not only 
    eject superlumianl knots (relativistic shocks or plasmons), but also eject
    rotating magnetized plasmas. Thus the whole jets (or jet-bodies) comprise 
    multiple superluminal knots (ejected at different times, distributing 
    at different positions and moving along different helical trajectories)
    plus the rotating plasmas associated with the superluminal knots. The 
    accelerated motion and the increase in Lorentz factors along helical
     trajectories are one of the most important features of these knots, which
    can  well be interpreted  in terms of MHD theories for 
    jet-formation/collimation/acceleration mechanisms (e.g., Vlahakis \& 
    K\"onigl \cite{Vl03}, \cite{Vl04}; Blandford \& Znajek \cite{Bl77};
      Meier \& Nakamura \cite{Me06}). The apparently superluminal components
    can be interpreted as relativistic shocks traveling along helical
    trajectories toward us in the work-frame of electromagnetic mechanisms.\\
    The results obtained in this work for 3C345, especially those for its 
    knot C9, firmly support these viewpoints on the nature of superluminal
    components (Jorstad et al. \cite{Jo05}, \cite{Jo13}).\\
     In contrast, some authors suggested that the apparent trajectories of
    superluminal components could result from the underlying jet  
    structure pattern (formed by Kelvin-Helmholtz instabilities) lit up 
    by passages of plasma condensations ejected during nuclear flares. 
    The close correlation between flux evolution and Doppler boosting effect
    (found for knot C9 in this paper) 
    \footnote{Such kind of close correlation between flux evolution and 
    Doppler boosting will be presented elsewhere for more superluminal 
    components in 3C345 (Qian, in preparation).} obviously do not support such interpretations.\\
    Since 3-dimensional models were used for simulating the trajectory 
    and kinematic behavior of knots in 3C345, we could separate their intrinsic
    bulk acceleration from the effects of trajectory curvature. 
    Thus their kinematic parameters (ejection
    time; bulk Lorentz factor, viewing angle, Doppler factor and apparent 
    velocity vs time) and the location of bulk acceleration zone could be 
     consistently modeled. Generally, the modeled parameters are 
    consistent with those derived in other works by using different methods 
    (e.g., Jorstad et al.\cite{Jo05}, \cite{Jo13}, \cite{Jo17}; Klare
    \cite{Kl03}; Schinzel \cite{Sc11a}): for example,  our modeled Doppler 
    factors could be compared with those 
    derived from variability time scales of optical/radio outbursts.\\
    \item  Our model-fitting results for 3C345 demonstrate that within the 
    collimation/acceleration zone its superluminal knots ejected from the 
    precessing nozzle could have a common inner trajectory pattern, which 
    precesses to produce the observed inner trajectories of the knots.
    However, beyond this zone their outer motion would follow different
    individual trajectories. This precessing common inner trajectory could 
    extend to different distances for different knots (see Table 1 and 
    Table 2). We have made use of the precessing nozzle scenario 
    to model fitting
    of the inner-jet kinematics of superluminal knots in 3C345.  The concept 
    of precessing common trajectory may be essential for doing this study.
     Obviously, the strong magnetic fields of the magnetosphere in its nucleus
    may play determinative role to form the steady common trajectory pattern,
    which precesses to produce the trajectories of the knots ejected at 
   different times. The model-fitting of the inner trajectories of the knots
    in terms of the precessing nozzle scenario naturally explain 
   the position angle swings observed in 3C345. In other blazars and QSOs 
   (e.g., 3C279, OJ287, 3C454.3, B1308+326, PG1302-102, NRAO 150) similar 
   phenomena were discovered and model-fitted (Qian et al. \cite{Qi19a},
   Qian \cite{Qi18a} and references therein). We would like to point out that
   the precessing nozzle scenarios for these blazars and QSOs are not only 
   based on the model-fittings of the available observational data-sets, but
   also are confirmed by other observations. For example, Hodgson et al.
   (\cite{Hod17}) have found some evidence for the possible presence of double
   ejection directions. Particularly, in the case of 3C279 one of the double
    jet (jet B) which was predicted and searched for quite a long time
    had already been  observed in much earlier years (Pauliny-Toth et al. 
   \cite{Pau87}, Pauliny-Toth \cite{Pau98}; de Pater \& Perley \cite{Pat83};
   Cheung \cite{Che02}). For blazar OJ287 the possibility of double jet 
   structure has been suggested (Villata et al. \cite{Vi98}, 
   Qian \cite{Qi18b}). Thus the assumptions in our precessing nozzle
   scenarios seem  valid for some blazars.\\
    \item In the process of model-simulations bulk Lorentz factor $\Gamma$
    as function of time was derived for each superlumianl components.
    Interestingly, most of the components (for both jet-A and jet-B) are 
    modeled as accelerated and their Lorentz factors
    varied over quite large ranges, e.g. knot C10 (jet-A) in [4.5,29] (Table 1).
    These intrinsic acceleration should be produced by the strong 
    torsion of toroidal magnetic fields in the collimation/acceleration 
    zones (e.g., Vlahakis \& K\"onigl \cite{Vl04}), which could extend as far
    as $\sim$400\,pc from the radio core (e.g. for knot C9; Table 1). 
    Generally, when
    the Doppler factor versus time of a knot was derived as in this paper, one
    can investigate the intrinsic radio light-curves of the knot and 
    model-simulating the intrinsic evolution of its electron density/magnetic
     field (e.g., Qian et al. \cite{Qi96}). However,
    the 15GHz and 43GHz light-curves of knot C9 could almost completely be 
    interpreted in terms of its Doppler boosting effect during  the main flare 
    (during $\sim$1997.50--2000.25), implying that the 
     intrinsic radio emission of 
    this superluminal knot was very stable or the relativistic shock
     responsible for producing the knot having very stable physical properties
    (see Sec.4.4.6). This result is important, because it is for the first 
    time to find the radio light-curves at both 15GHz and 43GHz being
    closely coincident with the Doppler boosting profile and strongly
   justify our precessing nozzle scenario. The common viewpoint  (or scenario)
   for superluminal knots participating relativistic motion and 
   magnetohydrodynamic acceleration mechanisms are firmly supported.
   \footnote{Only within our 
   precessing nozzle scenario, where Doppler factor curve $\delta(t)$ (and
   Doppler boosting profile $[\delta(t)]^{3+\alpha}$) can be obtained for 
   interpreting flux evolution of superluminal knots. Usual analysis of 
   VLBI-observations obtains only averaged values for bulk Lorentz 
   factor or Doppler factor, which could not be used to study flux evolution of
    superluminal knots.}
    \begin{table*}
    \centering
    \caption{Parameters of the binary black holes suggested in blazar 3C345:
     total mass M+m (in units of 
    ${10^8}{M_{\odot}}$), separation between the holes $r$ (in units of parsecs),
     post-Newtonian  parameter ($\epsilon_n$) and 
    gravitational radiation lifetime $t_{gr}$ (\,yr)
    calculated for different masses M of the primary black hole.}
    \begin{flushleft}
    \centering
    \begin{tabular}{lllll}
    \hline
     M & M+m & $r_{pc}$ & $\epsilon_n$ & $t_{gr}$ \\
    \hline
    \hline
     1.0 & 1.87 & $8.2{\times}10^{-3}$ & $1.09{\times}10^{-3}$  &
                   $1.58{\times}10^6$\\
     1.5 & 2.81 & $9.37{\times}10^{-3}$ & $1.43{\times}10^{-3}$ &
               $7.96{\times}10^5$ \\
     2.0 & 3.74 & $1.03{\times}10^{-2}$ & $1.74{\times}10^{-3}$  &
               $4.93{\times}10^5$\\
     3.0 & 5.60 & $1.18{\times}10^{-2}$ & $2.28{\times}10^{-3}$ &
                $2.53{\times}10^5$\\
     4.0 & 7.48 & $1.30{\times}10^{-2}$ & $2.76{\times}10^{-3}$  &
                $1.56{\times}10^5$\\
     5.0 & 9.35 & $1.40{\times}10^{-2}$ & $3.21{\times}10^{-3}$  &
                $1.08{\times}10^5$\\
    \hline
    \end{tabular}
    \end{flushleft}
    \end{table*}
   \begin{table*}
   \centering
   \caption{Orbital period $P_{orb}$ (yr, in source frame) and mass ratio q
    for the four blazars: 3C345, 3C454.3, OJ287 and 3C279. z--redshift.}
   \begin{flushleft}
   \centering
   \begin{tabular}{llll}
   \hline
   source & z & $P_{orb}$ & q \\
   \hline
   3C345 & 0.595 &  4.6 & 0.87 \\
   3C454.3 &  0.859 & 5.6 & 0.30 \\
   OJ287 & 0.306 &  9.2 & 0.30 \\
   3C279 & 0.538 & 16.3 & 0.50 \\
   \hline
   \end{tabular}
   \end{flushleft}
   \end{table*}
   \begin{table*}
   \centering
   \caption{Parameters of the binary black holes suggested in  blazars
   3C345, 3C454.3, OJ287 and 3C279: Mass M and m of the primary and secondary
    holes (in units of $10^8{M_{\odot}}$), post-Newtonian parameter 
   $\epsilon_n$, gravitational radiation lifetime $t_{gr}$ (\,yr), 
   $r/R_g$ ($R_g$--gravitational radius of the primary hole) and $r/r_g$ 
   ($r_g$--gravitational radius of the secondary hole) calculated for 
   different orbital separations $r$ in circular motion (in units of parsecs).}
   \begin{flushleft}
   \centering
   \begin{tabular}{lllllll}
    \hline
   Source & parameter & $r_{pc}$=0.01 & $r_{pc}$=0.02 & $r_{pc}$=0.03 &
                $r_{pc}$=0.04 & $r_{pc}$=0.05 \\
   \hline
   \hline
    3C345 &  $M_8$ & 1.83 &  14.7 & 49.5  & 1.17$\times{10^2}$  & 
                2.29$\times{10^2}$ \\
         &  $m_8$ & 1.59 & 12.8  & 43.0  & 1.02$\times{10^2}$  & 
               1.99$\times{10^2}$ \\
         & $\epsilon_n$ & 1.64$\times{10^{-3}}$ & 6.58$\times{10^{-3}}$ &
       4.44$\times{10^{-2}}$  & 2.63$\times{10^{-2}}$  & 4.11$\times{10^{-2}}$\\
          & $t_{gr}$  & 5.73$\times{10^5}$ & 1.78$\times{10^4}$ & 
         2.34$\times{10^3}$ & 5.56$\times{10^2}$ & 1.83$\times{10^2}$ \\
          & $r/R_g$ & 1140 & 284 & 127 & 71.6 & 45.7 \\
          & $r/r_g$ & 1320 & 328 & 146 & 122 & 52.6 \\
   \hline
   \hline
    3C454.3 & $M_8$ & 1.78 & 14.2 & 48.0 & 1.14$\times{10^2}$ & 
           2.23${\times}10^2$  \\
            & $m_8$ & 0.53 & 4.26 & 14.4 & 34.1 & 66.7  \\
            & $\epsilon_n$ & 1.1${\times}10^{-3}$ & 4.43$\times{10^{-3}}$  &
             9.99$\times{10^{-3}}$ & 1.78$\times{10^{-2}}$ &  
             2.78$\times{10^{-2}}$\\
            & $t_{gr}$ & 2.59$\times{10^6}$ & 8.22$\times{10^4}$ &
             1.07$\times{10^4}$ &  2.54$\times{10^3}$  & 8.27$\times{10^2}$\\
          & $r/R_g$  & 1180 & 295 & 131 & 73.5 & 47.0 \\
          & $r/r_g$  & 3950 & 984 & 437 & 246 & 157 \\
    \hline
    \hline
    OJ287 & $M_8$ & 0.66 & 5.27  &  17.8 &  42.2 &  82.5   \\
          & $m_8$ & 0.20 & 1.58  &  5.34  &  12.7  &    24.8   \\
          & $\epsilon_n$ & 4.13$\times{10^{-4}}$ & 1.64$\times{10^{-3}}$  &
         3.70$\times{10^{-3}}$  & 6.59$\times{10^{-3}}$  &
          1.03$\times{10^{-2}}$ \\
          & $t_{gr}$ & 5.02$\times{10^7}$ & 1.60$\times{10^6}$ & 
           2.10$\times{10^5}$ & 4.96$\times{10^4}$  &  1.62$\times{10^4}$ \\
           & $r/R_g$ & 3170 & 795 & 354 & 198 &  127 \\
          & $r/r_g$ & 10500 & 2650 & 1180 & 660 & 422 \\
    \hline
    \hline
    3C279 & $M_8$ & 0.18 & 1.45  & 4.91  & 11.6  &   22.7 \\
          & $m_8$ & 0.09 & 0.73   & 2.45   & 5.80    &   11.4 \\
          & $\epsilon_n$ & 1.3$\times{10^{-4}}$ & 5.23$\times{10^{-4}}$  &
        1.18$\times{10^{-3}}$ & 2.09$\times{10^{-3}}$ &
           3.27$\times{10^{-3}}$ \\
          & $t_{gr}$ & 1.27$\times{10^9}$  & 3.95$\times{10^7}$   & 
          5.21$\times{10^6}$  & 1.24$\times{10^6}$  & 4.04$\times{10^5}$  \\
         & $r/R_g$ & 11650 & 2880 & 1280 & 721 & 461 \\
         & $r/r_g$  & 23300 & 5740 & 2560 & 1450 & 917 \\
    \hline
    \hline
    \end{tabular}
    \end{flushleft}
    \end{table*}
     \item Due to both jet nozzles (nozzle-A and nozzle-B of 3C345) precessing 
     with the  same period (4.6\,yr in source frame), the  precession of the
     nozzles could be caused by the orbital motion of the binary holes,
     rather than other mechanisms of precession (for example, geodetic 
      precession and Newtonian-driven precession; referring to 
     Qian et al. \cite{Qi17}, \cite{Qi18a}; Britzen et al. \cite{Br17})
     \footnote{Our works have shown 
     that the double-jet structures revealed in the four blazars (3C279, 
     3C454.3, 3C345 and OJ287) have similar (common) features: both jets 
     precess in the same direction with the same period. Such kind of jet
     precession could not be caused by hydrodynamical instabilities induced
     by the interactions between the jets and the surrounding media.}.
     If this interpretation is
     valid, the mass-ratio q=m/M between the primary and secondary
     holes could be approximately estimated to be equal to the ratio 
     between the jet apertures: q$\sim$0.87 \footnote{This value is estimated
     at core separation $\sim$0.5\,mas. In the case of knots moving along 
     curved trajectories the jet-cone apertures critically depend on core
     separations and thus the value q=0.87 is only a rough estimation.}.
     The orbital period and mass ratio obtained for 3C345 here could 
    tentatively provide
     some useful constraints on the total mass and gravitational radiation 
     lifetime of the putative supermassive black hole binary  in 3C345.
     Some results are presented in Table 3 to show the relation between 
     the mass of the primary black hole and the parameters of the binary 
     (total mass, orbital separation, post-Newtonian parameter and 
     gravitational radiation lifetime), showing that the total mass 
     should be less than  9.5$\times{10^8}{M_{\odot}}$ if its gravitational
      radiation lifetime $t_{gr}{>}{10^5}$\,yr.
    \item Interestingly and 
       to our own surprise, the VLBI-kinematics of the four blazars (3C345, 
    3C454.3, 3C279 and OJ287) could have been interpreted in terms of the
     precessing jet-nozzle scenario with binary black hole systems,
      and their precession periods and mass ratios could have been 
      tentatively derived 
    (Qian \cite{Qi18b}, Qian et al. \cite{Qi18a}, \cite{Qi21} and this paper).
   Thus we would have to investigate whether there are physically reasonable 
     parameters to describe these binary black hole systems which have
    the orbital period $P_{orb}$ and mass ratio q as listed in Table 4. 
    For the four blazars (or the four putative binary black hole systems),
    their parameters (mass $M_8$ and $m_8$, post-Newtonian parameter 
    $\epsilon_n$, gravitational radiation lifetime $t_{gr}$, ratios $r/R_g$
    and $r/r_g$ between orbital separation and  gravitational radius 
    calculated for different orbital separations $r$ (0.01\,pc--
    0.05\,pc) are given in Table 5, taking the $P_{orb}$ and q listed in 
    Table 4 into account\footnote{Formulas for calculating these parameters
     can be found in Qian et al. (\cite{Qi17}).
      Values for $r$=0.014 and 0.015\,pc
    were also used , but not listed in Table 5. For the four blazars 0.01\,pc
    corresponds to angular distance of $\sim$1-2\,$\mu$as.}. Based on Table 5,
    we can search for appropriate 
    ranges for the masses and separations $r$ for the four binary systems, 
    if requiring their gravitational radiation lifetime limited in the range
    $10^5$--$10^6$\,yr.\\
    It was found that: (a) For 3C279, the ranges [$r$=0.03--0.04\,pc] and
     [$t_{gr}$=5.21\,$10^6$--1.24\,$10^6$]\,yr correspond to mass ranges
     [$M_8$=4.91, $m_8$=2.45] and [$M_8$=11.6, $m_8$=5.80], respectively.
     These masses are well consistent with the values 
     (3--8\,${10^8}{M_{\odot}}$) given in Woo \& Urry (\cite{Wo02}),
     Wang et al. 
     (\cite{Wa04}),  Gu et al. (\cite{Gu01}) and Nilsson et al. (\cite{Ni09}). 
     This is a very good case, indicating that the related parameters of the 
     binary system derived for 3C279 by our precessing jet-nozzle scenario
     seem to be physically sound;\\
     (b) For 3C454.3, the ranges [$r$=0.015--0.02\,pc] and
     [$t_{gr}$=3.41\,$10^5$--8.22\,$10^4$]\,yr correspond to the mass ranges
     [$M_8$=6.01, $m_8$=1.80] and [$M_8=$14.2, $m_8$=4.26]. These masses
     are well consistent with the values of 1.3--1.5\,$10^9{M_{\odot}}$
     given in Woo \& Urry (\cite{Wo02}) and Wang et al. (\cite{Wa04}); \\
     (c) For 3C345, as mentioned in the previous item, its total mass should be
     less than 9.5\,$10^8{M_{\odot}}$ if requiring $t_{gr}$$>{10^5}$\,yr.
     According to Woo \& Urry its black hole mass=2.63\,$10^9{M_{\odot}}$,
     but Lobanov \& Roland (\cite{Lo05}) adopted a mass
     1.42\,$10^9{M_{\odot}}$ for their binary black hole model. Here
      we found that the ranges [$r$=0.01--0.014\,pc] and 
     [$t_{gr}$=5.73\,$10^5$--1.06\,$10^5$]\,yr correspond to the mass ranges
     of [$M_8$=1.83--5.02] and [$m_8$=1.59--4.37], which seem to be
     reasonable values.  According to Wang et al. (\cite{Wa04}), 3C345 is
     distinctly different from 3C279 and 3C454.3 in the
     relation between its kinematic luminosity ($L_{kin}$) and broad-line 
     region luminosity ($L_{BLR}$), possibly implying some jet-disk 
     connection which could result in its measured mass higher than that
     of the black hole itself. In addition, Xie et al. (\cite{Xi05}) measured
     a mass of 2.57\,$10^8{M_{\odot}}$, similar to that we derived;\\
     (d) For OJ287, the ranges [$r$=0.02--0.03]\,pc and [$t_{gr}$=
     1.60\,$10^6$--2.10\,$10^5$]\,yr correspond to the mass ranges 
     [$M_8$=5.27--17.8] and [$m_8$=1.58--5.34]. These values are broadly
     consistent with the values (mostly $\leq{10^9{M_{\odot}}}$) 
     adopted in several works (e.g., Valtaoja et al. \cite{Va00}, Liu \& Wu
     \cite{Liu02}, Wang et al. \cite{Wa04}, Gupta et al.
     \cite{Gup12}, Katz \cite{Ka97}). However, in the disk-impact scenario
     proposed by Lehto \& Valtonen (\cite{Le96}; also Dey et al. 
    \cite{De19}, Valtonen et al. \cite{Val12}), the masses of the binary holes
     are modeled to be $\sim{1.5}{\times}{10^{10}}{M_{\odot}}$ and
      $\sim{1.4}{\times}{10^8}{M_{\odot}}$ with a mass-ratio q$\sim$ 0.008.
     Obviously, the issue on the properties of the black hole binary in
      OJ287 should be further investigated and clarified (Qian \cite{Qi18b},
      \cite{Qi19b}, \cite{Qi19c}, \cite{Qi20}, Britzen et al. \cite{Br18},
      Villata et al. \cite{Vi98}, Sillanp\"a\"a et al. \cite{Si88},
       \cite{Si96}).\\
        \item In Table 5 are also given the ratios ($r/R_g$ and $r/r_g$)
     between orbital separation and gravitational radius for the primary 
    and secondary holes. It can be seen that the reasonable ranges of 
   masses described above correspond to reasonable ranges of the ratios for
  the primary and secondary holes: approximately a few hundreds to
   a few thousands (for example, for 3C279 $r/R_g$=1280-721 and
   $r/r_g$=2560-1450), indicating that the formation of accretion disks and jets
  could proceed without destructive influences by the gravitational 
  interaction  between the two holes. However, the formation of double-jet and
  disks and their instabilities  need to be investigated for understanding 
  the complex  phenomena observed in blazars (including the VLBI-kinematics and
  the multi-wavelength radiation and the connection between low-energy and
  high-energy radiations). In binary black hole systems cavity accretion might 
  restrain the formation of large-scale accretion disks, but spin of the
  holes might play more significant roles in formation of relativistic jets.
  Perhaps the binary black holes in blazars might be rapidly rotating with spin
  parameter j${\geq}$0.5 (j=J/$J_{max}$, $J_{max}$=G$M^2$/c--maximum spin
  angular momentum of a black hole). In addition, for the reasonable
   ranges of the black hole
  masses, the post-Newtonian parameter $\epsilon_n$ given in Table 5 has its
  values in the range of [1-4]${\times}10^{-3}$, implying that
  the keplerian motion of the black hole binaries in the four blazars
  is still non-relativistic. That is, the black hole binaries of the four
   blazars have not been entering the stages of orbital evolution dominated by
  the general relativity effects (Einstein \cite{Ei16}, \cite{Ei18}). The
  results listed in Table 5 for the physical parameters of the black hole
  binaries could only describe the characteristic features of their initial 
  in-spiraling processes.
   \item We would like to point out that our results derived for the putative 
  black hole binaries  in the four blazars (3C279, 3C454.3, OJ287 ans 3C345) 
  were well consistent with the theoretical arguments about close binary 
  systems by Begelman et al. (\cite{Be80}). This was unexpected and confirmed
  posteriorly. They have shown that, when the orbital separation of binaries
  approaches $R_{gr}$ at which gravitational radiation  becomes to
  dominant their orbital evolution, their keplerian motion will have an orbital
  period of $T_{orb}(R_{gr})$=48.4\,yr for a specific model (assuming
  q=0.3, M=$10^9\,{M_{\odot}}$ and corresponding $r_{gr}$=0.067\,pc and 
  $t_{gr}(R_{gr})$=1.8$\times{10^8}$\,yr; referring to Begelman et al.). If 
  the orbital separation $r<{R_{gr}}$, for example, $r=0.32{R_{gr}}$ (0.021\,pc)
  and $r=0.18{R_{gr}}$ (0.012\,pc), corresponding 
  [$t_{gr}=1.8\times{10}^6$\,yr, $T_{orb}$=8.6\,yr] 
   and [$t_{gr}=1.8\times{10^5}$\,yr, $T_{orb}$=3.7\,yr], respectively. These
  values are closely similar to 
   the gravitational radiation lifetimes and orbital 
  periods derived by our works for the four blazars (especially for 3C279; 
  Tables 3--5). The double jet-nozzle precession we tentatively found through
  analysis of the VLBI-kinematics of the four blazars could be the direct
  consequences of their orbital motion. Therefore both theoretical
  investigations and VLBI-observations seem consistently to approach the same
  conclusion: our investigations and analyzes of the VLBI-kinematics of
  superluminal components in the four blazars (3C345, 3C454.3, 3C279 and OJ287)
  might have revealed the keplerian motion of the black hole binaries 
  putatively suggested to be existing in their nucleus.\footnote{Double-jet
  structure in blazar OJ287 has tentatively been suggested by Qian \cite{Qi18b}
  through model-simulation of its superlunimal components in terms of our 
  precessing nozzle scenario.  The secondary jet ejected by its secondary
  black hole could be observed in near future (Dey et al. \cite{De21})}\\
  In fact, only in blazars  one could possibly discern (or detect) the 
  precession of double jets caused by the keplerian motion through analyzing 
  their VLBI-kinematic behaviors, since blazars are observed at very small 
  viewing angles ($\sim{1-5^{\circ}}$ for the four blazars discussed here),
  which assure the precession cones of the double jets projected in the plane
  of the sky being sufficiently wide  and separated to be resolved by 
  VLBI-observations, and the superluminal components can radiate strongly to be
   measured on VLBI-scales due to relativistic beaming effects. Search for 
  evidence of possible keplerian orbital motion of black hole binaries in 
  blazars may be a severe challenge for future VLBI-observations. 
    \item Relativistic jets in blazars (and generally in AGNs) are believed 
    to be formed through MHD processes in the magnetosphere of 
    black-hole/accretion disk systems in the nuclei of host galaxies (e.g.,
    Blandford \& Payne (\cite{Bl82}), Blandford \& Znajek (\cite{Bl77}),
     Camenzind (\cite{Ca86}, \cite{Ca87},\cite{Ca90}), 
   Lovelace et al. (\cite{Lo86}), Li et al. (\cite{Li92}), 
    Vlahakis \& K\"onigl (\cite{Vl03}, \cite{Vl04}). In these scenarios
    magnetic fields dominate the processes: magnetic pressure gradient
   accelerate the jets and magnetic pinch effects of the toroidal fields 
   collimate the plasma flows. MHD theoretical scenarios
    can also explain the extended (parsec-scale) acceleration observed in
    blazars. It is shown by Blandford\& Znajek (\cite{Bl77}), Li et al.
   (\cite{Li92}) and Vlahakis \& K\"onigl (\cite{Vl04}) that there are
     radially self-similar solutions for stationary axisymmetric MHD
    flows in the magnetosphere of black-hole/accretion-disk systems which 
   allow extended acceleration after the flow passes the classical 
    fast-magnetosonic point approaching the modified fast-magnetosonic point
    through "magnetic nozzle mechanism". Vlahakis \& K\"onigl (2004)
     nicely explained the
     accelerating component C7 in blazar 3C345 (observed by Unwin et al.
   \cite{Un97}) in terms of their  radially self-similar MHD solution 
    for a proton-electron jet. The acceleration zone  $\sim$ (30-300\,pc)
    and the range of bulk Lorentz factor ($\sim$4--20) derived in 
   our model fittings can be understood within the Vlahakis-K\"onigl's 
    radially self-similar MHD model. Thus available
    MHD jet-formation  theories are already effective to explain
    the extended  acceleration observed in blazars.\\
     We would like
    to point out that the cavity-accretion models 
    may be very helpful to understand the accretion process, jet-formation
    and ejection of superluminal components and alternative
     quasi-periodic activity in blazars with double-jets, hosting binary 
    black holes (e.g., Shi et al. \cite{Sh12}, Shi \& Krolik \cite{Sh15};
    Artymovicz \& Lubow  \cite{Ar96}, Artymovicz \cite{Ar98}). Specifically,
    Shi \& Krolik (\cite{Sh15}) argued that in MHD scenarios cavity-accretion 
    rates can be raised to the level that both black holes can produce a jet,
    forming a double jet system.
    One distinct feature of our results is the precession of a common
     trajectory pattern around the  jet axis to produce the observed inner 
    trajectories of the  knots and the 7.30\,yr precession period for 3C345.
    It is worth emphasizing that for jet-A its nozzle-precession has been 
    observed over about four precession periods ($\sim$1979--2009) and for
     jet-B over about two precession periods ($\sim$2002--2016), 
    implying that jet-A and jet-B  have been active during respective periods
    with respective levels of activity\footnote{Thus it seems uncertain whether
    the cores of jet-A and jet-B could be concurrently observed.}.
    Precessing nozzles could possibly exist in the putative binary black hole
    system in 3C345, which needs to be tested by more observations. 
   \item Based on the assumptions about the possible existence of a  precessing
    jet-nozzle and precessing common trajectory in 3C345, we have analyzed the
    kinematic behavior of  27 superluminal components 
    in terms of the precessing nozzle scenario and model-fitted their
    trajectory $Z_n(X_n)$, core separation $r_n$(t), coordinates
    $X_n$(t) and $Z_n$(t) and apparent velocity $\beta_a$(t) versus time during 
     a time-interval of $\sim$38 years (1979--2016). The double jet structure
    was disentangled and their precession periods (7.30\,yr) were derived. 
     The derived bulk Lorentz factor $\Gamma$, Doppler factor $\delta$ and
     viewing angle $\theta$ as functions of time may be very useful for
     studying the intrinsic emission and  physical properties of these knots,
     and the connection  between the radio, optical and $\gamma$-ray emitting
     regions. The double-jet structure and their precession might be useful
     for investigating the putative binary black hole system, providing some
     constraints on its physical properties.\\
      We have used a new method, which is different from the ordinary methods
     for analyzing the observational data obtained on VLBI-scales. 
     The assumptions we suggested should be tested by future observations of
      the kinematics for superluminal components
    within core separations $r_n{<}$0.05--0.1\,mas, where the superluminal
    knots might follow helical trajectories with large pitch angles (or strong
    toroidal field-lines; referring to Qian (\cite{Qi19c})) and
     might be more difficult to determine their 
    precessing behaviors\footnote{At large core-separations helical
    magnetic fields may have much smaller pitch angles and thus the motion
    of superluminal knots becomes to be approximately ballistic.}. 
   The assumption of precessing common trajectory
     would be confronted with future VLBI-observations with higher-resolutions 
    in order of $\sim$10$\mu$as. \\
    In fact, there have been different results by analyzing the kinematic 
    behaviors of superluminal components  for 3C345 and quite a lot of
    different suggestions were proposed in literature. For example, 
    Klare (\cite{Kl03}) claimed the presence of jet precession of 
    $\sim$8-10\,yr, but Schinzel (\cite{Sc11a}) claimed no clear evidence for
    periodic trends in the kinematic behavior of superluminal components;
     Lobanov \& Zensus (\cite{Lo99}) suggested a $\sim$8-10\,yr period; 
     Lobanov \& Roland (\cite{Lo05}) suggested a binary  hole scenario and 
    9.5\,yr precession period, and Qian et al. (\cite{Qi09}) suggested
    a 7.36\,yr precession period, etc.\footnote{In fact, all 
    these determinations of precession period were derived by using the
    observational data on the knots of jet-A only. One could not find any 
    precession by using observational data on the knots of both jet-A 
    and jet-B.} 
     Our results in this paper may be regarded  as one of the
    alternative explanations. Using our new methods for analyzing the 
    VLBI-kinematics in 3C345 much more information on its kinematic properties
     and physical implications could be obtained.\\
      However, this work was established on the assumption that jets in blazars
     should precess with certain regular periods. At present it still
     remains unsettled as a question: whether blazar
     3C345 has a single-jet structure or double-jet structure with or without
     jet-nozzle precession. More observations and investigations are needed
     to solve this issue.
   \item Similar issues are present for  other blazars, for example, for blazar
    3C279. Recently, performing VLBI-observations at 
    mm-wavelengths by using Event-Horizon-Telescope, Kim et al. (\cite{Ki20}) 
    found similar position angles in 2011 and 2017, suggesting a
    precession period of $\sim$6\,yr  and claiming to exclude the 25\,yr 
    precession period which was derived through analyzing the kinematic
    behavior of $\sim$30 superluminal knots observed during $\sim$ 30
    years (Qian et al. \cite{Qi19a}).
    \footnote{Similar position angles observed
    at two different epochs is inadequate to determine the precession
    period, because it still needs to confirm that the two 
    epochs correspond to a difference in precession phase of 2$\pi$. In our
    precessing jet-nozzle scenarios suggested for the blazars (3C345, OJ287,
    3C345 and 3C454.3) recurrences of the curved trajectories observed at 
    different epochs corresponding to differences in precession phases 
    of $\sim$2$\pi$  and the distribution of the modeled precessing common
    trajectory were very helpful for determining their precession
     periods.}.
    Here we would like to propose an alternative interpretation: 
    Kim's finding might not necessarily be contradictory to the 25\,yr
     period. It could  possibly be compatible with  the
    precession period of $\sim$25\,yr derived in Qian et al.(\cite{Qi19a})
    because the 25\,yr precession period permits similar position angles
    appearing at two epochs different within $\sim$12.5\,yr (a half of 
    the precession period of  25\,yr). Such kind of similar position angles
    at two times could be observed  at the either side of the projected 
    jet boundaries (or the "jet edges"). An
    appropriate example may be: through model-fitting of the 
    VLBI-kinematics  for OJ287 (Qian \cite{Qi18b}), it was found that its 
    knots C11 and C12 were observed at similar position angles
    with a difference in their ejection times of $\sim$3.8\,yr. The two knots
    were ejected at the either side of the southern edge of its southern jet
    with a difference in their precession phases of $\sim$2.0\,rad,
    corresponding to a time-interval smaller than $\sim$6\,yr 
    (half of the precession period 12\,yr).\\
    The presence of double-jet structure, jet-nozzle precession and periodicity
    in ejection of superluminal components in 3C345 and other blazars are still
    issues in debate and need to be further investigated. Recently,
    Event-Horizon-Telescope (EHT) Collaboration  has begun monitoring 
    observations for a few blazars at 230\,GHz with resolutions of 
    $\sim$20$\mu$as (EHT-Collaboration et al. \cite{EHT19}). As suggested by
    Qian (\cite{Qi18b}, \cite{Qi20}) that the non-thermal radiation
     emitted by blazar OJ287 might originate from its double jets with 
     precessing jet-nozzles. EHT-monitoring observations in the near future
     would be helpful to reveal its double jet structure, searching
    for the second jet ejected from the secondary black hole in OJ287  
     (Qian \cite{Qi18b}, Villata et al. \cite{Vi98}, Dey et al.
    \cite{De21}, \cite{De19}). Similar search for double-jet structure 
    in blazars 3C279, 3C454.3 and 3C345 should be tried, and
    EHT-monitoring observations will be very helpful. All kinds of 
    scenarios proposed to interpret the kinematic phenomena in blazars
    would experience severe tests.
     \end{itemize}
    \section{Concluding remarks}
    In this work the kinematics of 27 superluminal components in blazar 3C345
    were model-simulated  in terms of  our precessing nozzle scenario. 
    Double jet-nozzle structure was tentatively suggested consisting of 
    jet-A and jet-B. Interestingly, taking the model-simulation of kinematic
    behavior of knot C9 as an example, its trajectory was extremely well
    fitted by the precessing common trajectory at precession phase 5.54+4$\pi$
    corresponding to its ejection time 1995.06, with a very high  accuracy
    of $\pm$5\% precession period (Fig.3). Its observed kinematic properties
    (trajectory $Z_n(X_n)$, core separation $r_n(t)$, coordinates 
    ($X_n(t)$ and $Z_n(t)$) and apparent velocity ($\beta_{app}(t)$) were  well
     fitted within core separation $\sim$1.8\,mas, equivalent to spatial
     distance from the core of $\sim$400\,pc (Fig.8). Moreover, its radio light
    curves were confirmed to be extremely well coincident with its Doppler
    factor profile (Figs.9 and 10), thus demonstrating its bulk Lorentz
     factor and viewing 
    angle as functions of time being correctly derived and superluminal knot C9
    being a relativistic shock moving toward us with acceleration. 
    This was unexpected \footnote{Unfortunately, for the other three blazars
    3C279, 3C454.3 ad OJ287 the relation between flux evolution of superluminal
    knots and their Doppler boosting effect have not been studied until know.}
    and for the first time both the kinematic properties and Doppler boosting
    effect in  the flux evolution of a superluminal knot can be studied in
    detail. Thus the model-simulation for the kinematic and physical properties
    in terms of our precessing nozzle scenario was almost perfect.\\ 
    Model-simulation results of both kinematic behavior and Doppler boosting 
    effect in flux evolution for more superluminal knots in 3C345 are
    in preparation, showing their kinematic and physical properties similar to
     that for knot C9. All these results certainly 
    indicate that our precessing nozzle 
    scenario  may be an effective one and appropriate to study the kinematic 
    and physical properties of blazars.\\
      As generally suggested, the swing of ejection position angle observed 
    for superluminal components in QSOs or blazars has been assumed to be
    related to the jet precession in the sources. For blazar 3C345
    precession periods of $\sim$7-10\,yr were really identified (e.g.,Lobanov 
    \& Zensus \cite{Lo99}, Klare \cite{Kl03}, Qian et al. \cite{Qi09}).
    Interestingly, these studies only used the earlier VLBI-observational data
     (equivalent to the data for jet-A identified in this paper). However,
    by using the entire dataset till recent years (e.g., to $\sim$2010, 
   including the dataset for jet-A and jet-B both) no trend of jet precession
     was identified  (this work, Schinzel et al. \cite{Sc11c}). As shown
     by the detailed analysis of
    the kinematics of all superluminal components in 3C345, the source could
    possibly comprise of two groups of superluminal components, ascribed to
    two jets (jet-A and jet-B). In this case both  jets could be modeled as
    having their own precession cone pattern, but having similar periods and 
   in the same direction. This kind of precession of double-jet could be 
   produced in close black hole binaries through the modulation of jet-flow by 
    keplerian orbital motion (Artymovitz \cite{Ar98}, Qian et al. \cite{Qi21},
    \cite{Qi17}, \cite{Qi18b}, \cite{Qi19a}, Roos et al. \cite{Ro93}). Thus
    in this paper we tentatively adopted the following assumptions:(1)
    3C345 has double-jet structure with double precessing nozzles for jet-A
     and jet-B; (2) both jet-A and jet-B precess with the same period 
    $\sim$7.30\,yr; (3) the two precessing jets have different jet directions
    and precession cone patterns. Based on these assumption the kinematic
    behavior of superluminal components ascribed to jet-A and jet-B were 
    successfully model-simulated respectively and the relation between the
    flux evolution and Doppler-boosting effect for some superluminal components
    were studied and interpreted.\\
     3C345 is the fourth blazar which was 
    model-simulated in terms of our precessing nozzle scenario within the
    framework of double jet-nozzle structure. Interestingly, it was found 
    consistently that for all the four blazars, only under the assumption of
    double nozzle structures, the precession of both jet-nozzles
     could be  model-simulated. These double jet structures naturally 
    led to black hole binaries putatively existing in the nuclei
    of the four blazars. Double jet-nozzle scenario may not necessarily
    imply concurrent ejection  of superluminal components:
    the two nozzles might be active alternatively or one jet could  exist
    temporarily.  At present, few theoretical 
    investigations of the formation of double relativistic jets in black hole
    binaries. Obviously, formation of double jets, jet precession, their 
    stability and  short orbital period, etc. involve new theoretical fields
    in binary black hole physics and should be intensively investigated for 
    providing deeper understanding of blazar phenomena.\\
     Although under the assumption of double-jet structure for 3C345, 
   significant and unprecedented results and physical information about
   its kinematic behavior and physical nature were derived,
   single-jet structure scenario can not be excluded. The existence 
    of a double-jet structure or single-jet structure in 
   blazar 3C345 is still a matter in debate.
   However blazar OJ287 seems likely to have a double-jet structure in its 
   nucleus (Qian \cite{Qi18b}, Villata et al. \cite{Vi98}, Dey et al.
    \cite{De21}).  At present, for blazars 3C345, 3C454.3 and 3C279, double
   jet-nozzle scenarios may be regarded as some 
   hypothesized working-frames which need to be tested by future observations. 
   \acknowledgements{Qian thanks Drs. S.G.~Jorstad  and Z.~Weaver 
   (Boston university, USA) for kindly providing 43GHz VLBA-data on 
   superluminal components measured during the period 2007-2018, which are
   from the the VLBA-BU Blazar Monitoring Program, funded by NASA through 
   the Fermi Guest Investigator Program.  
   Qian wishes to thank
    Dr. T.K.~Krichbaum (MPIfR) for helpful discussions and introduction of 
   observational data on 3C345. Qian is most grateful to 
   Prof. Xiao-Zhong Chen for his advice on plasma physics and 
   magnetohydrodynamics over years.}
  
  \begin{appendix}
   \section{Figs. A.1--A.8 for knots C5,C6,C7,C10,C11,C12,C13 and C14 of jet-A
    and Figs. A.9--A.16 for knots C15a,C16,C17,C18,B5,B7,B8 and B11 of jet-B}
   \begin{figure*}
   \centering
   \includegraphics[width=6cm,angle=-90]{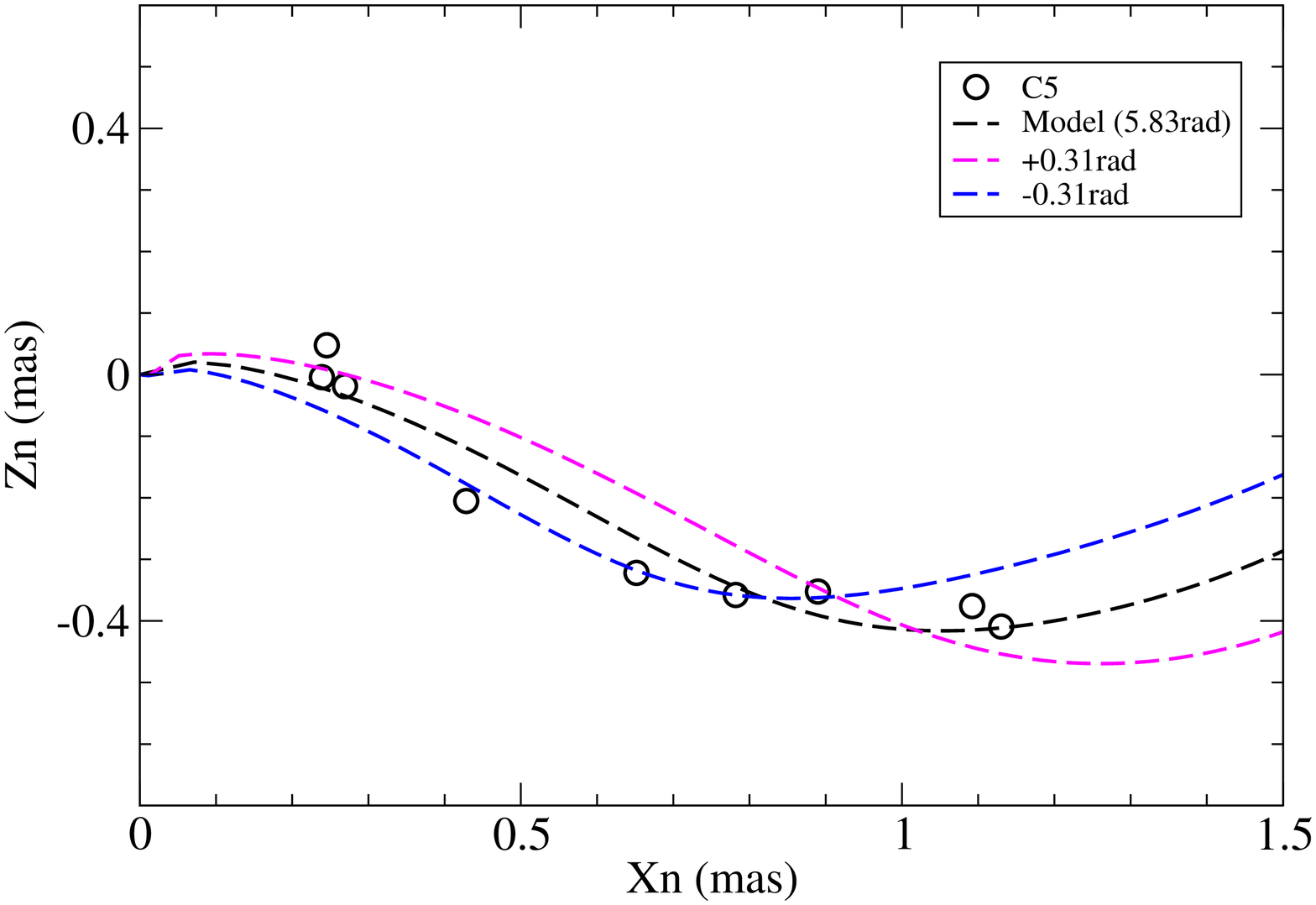}
   \includegraphics[width=6cm,angle=-90]{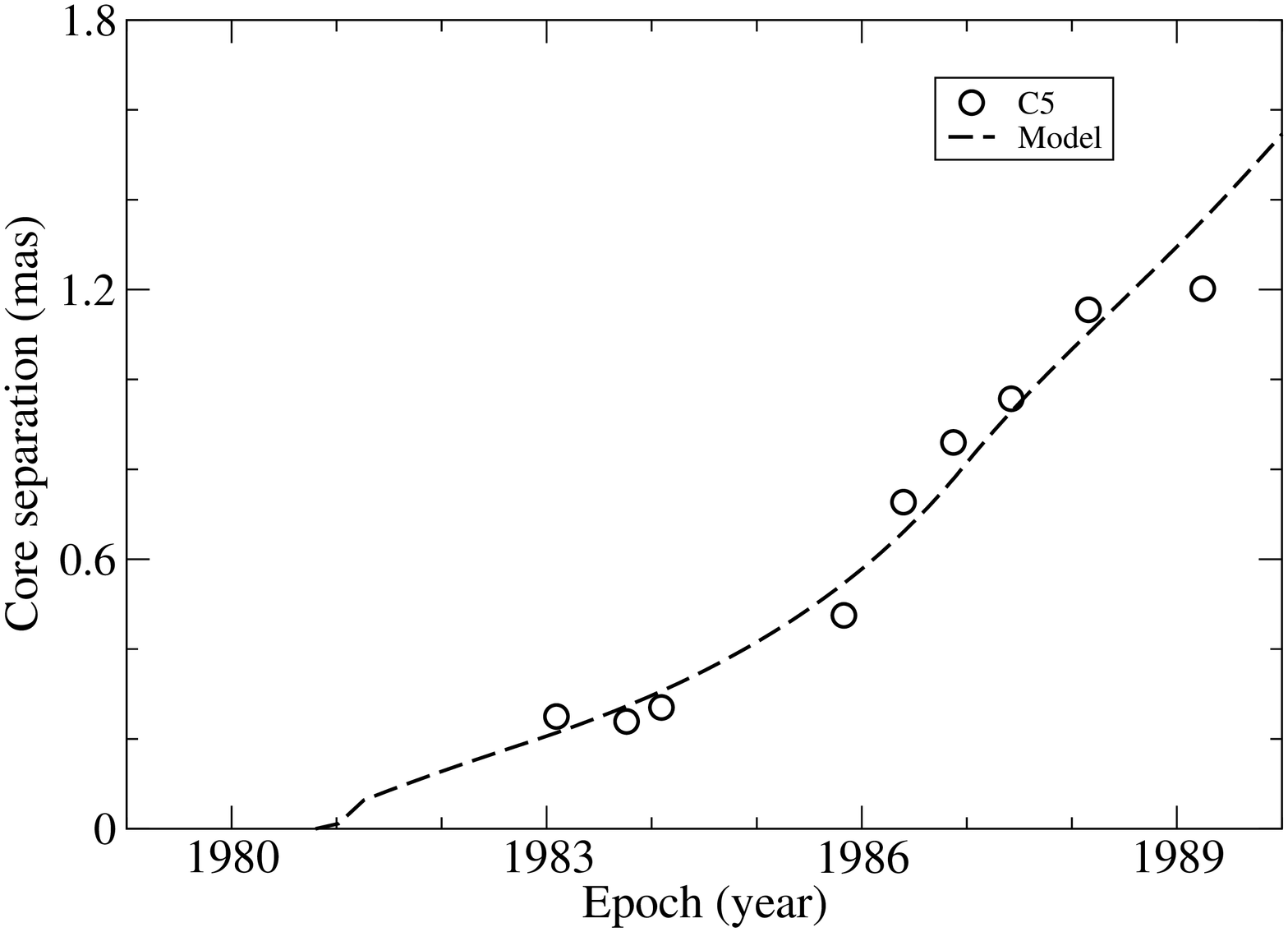}
   \includegraphics[width=6cm,angle=-90]{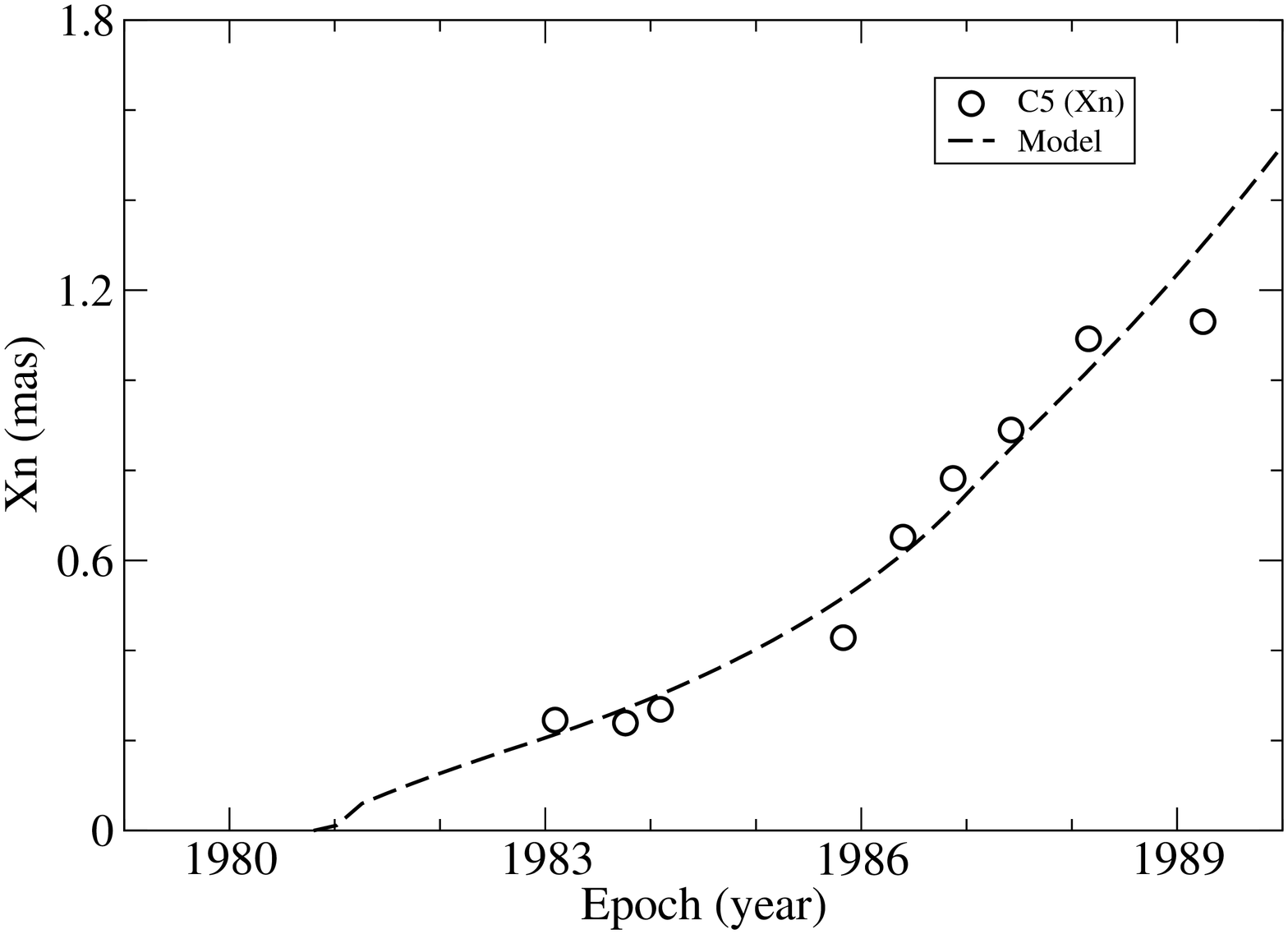}
   \includegraphics[width=6cm,angle=-90]{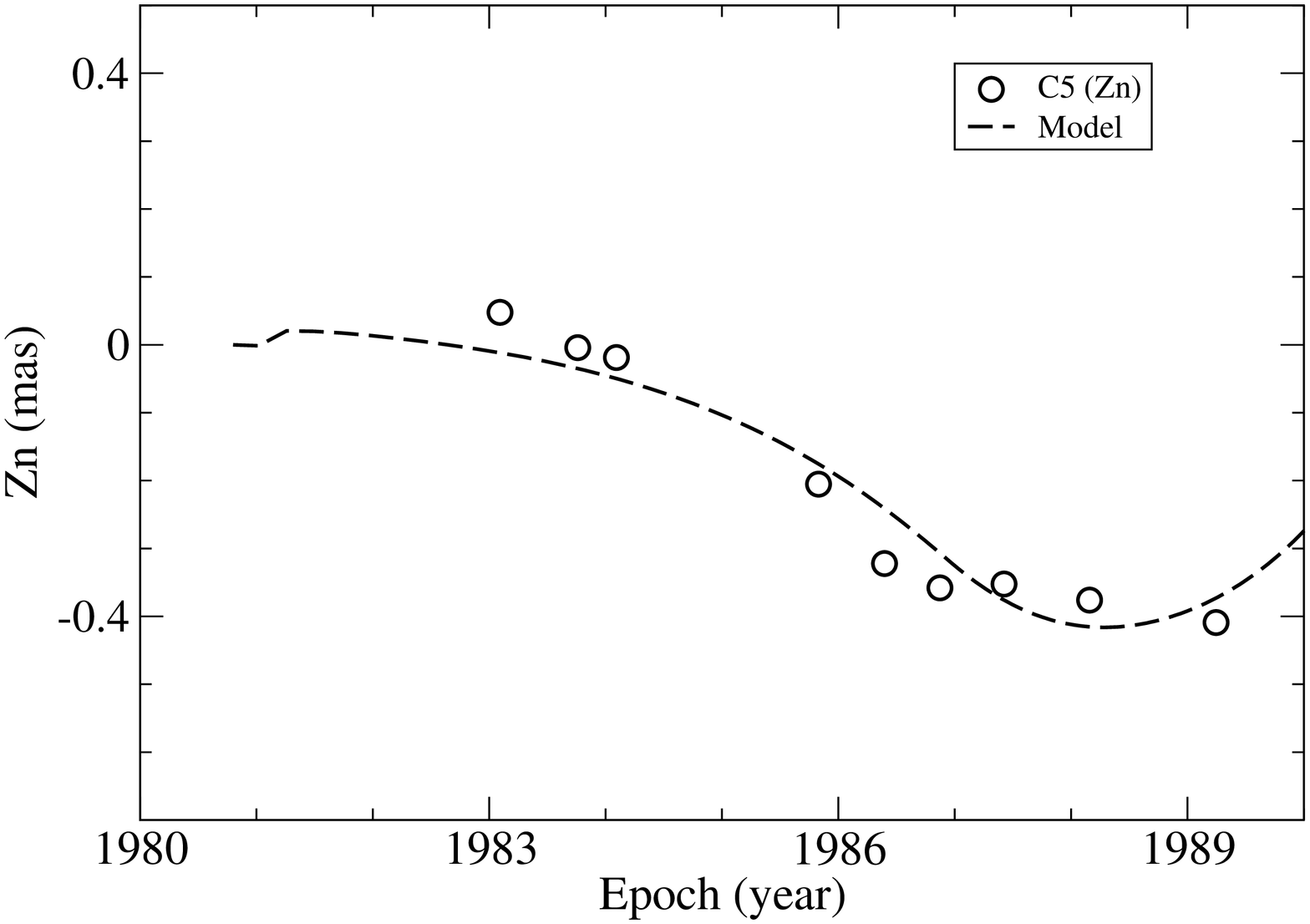}
   \includegraphics[width=6cm,angle=-90]{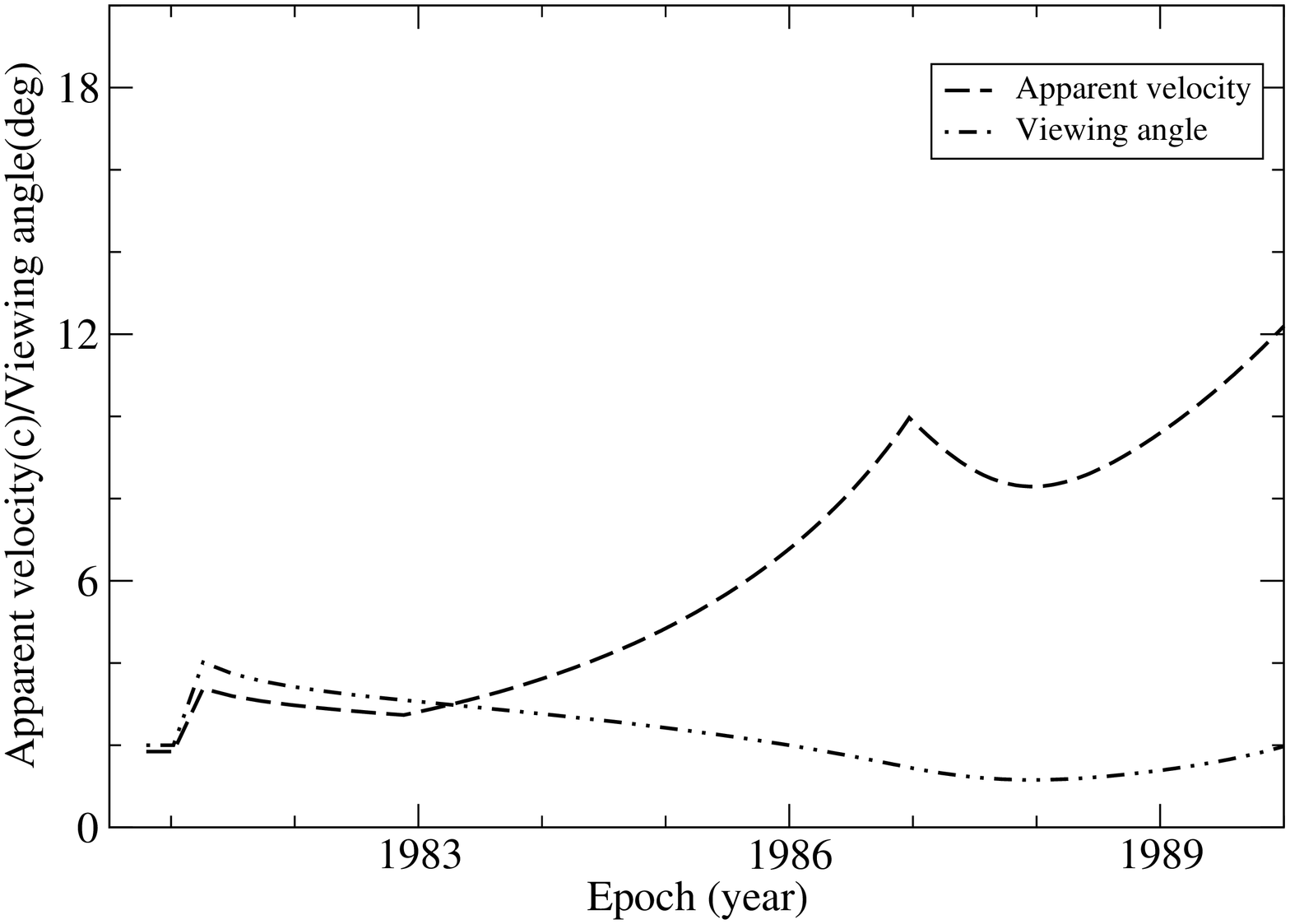}
   \includegraphics[width=6cm,angle=-90]{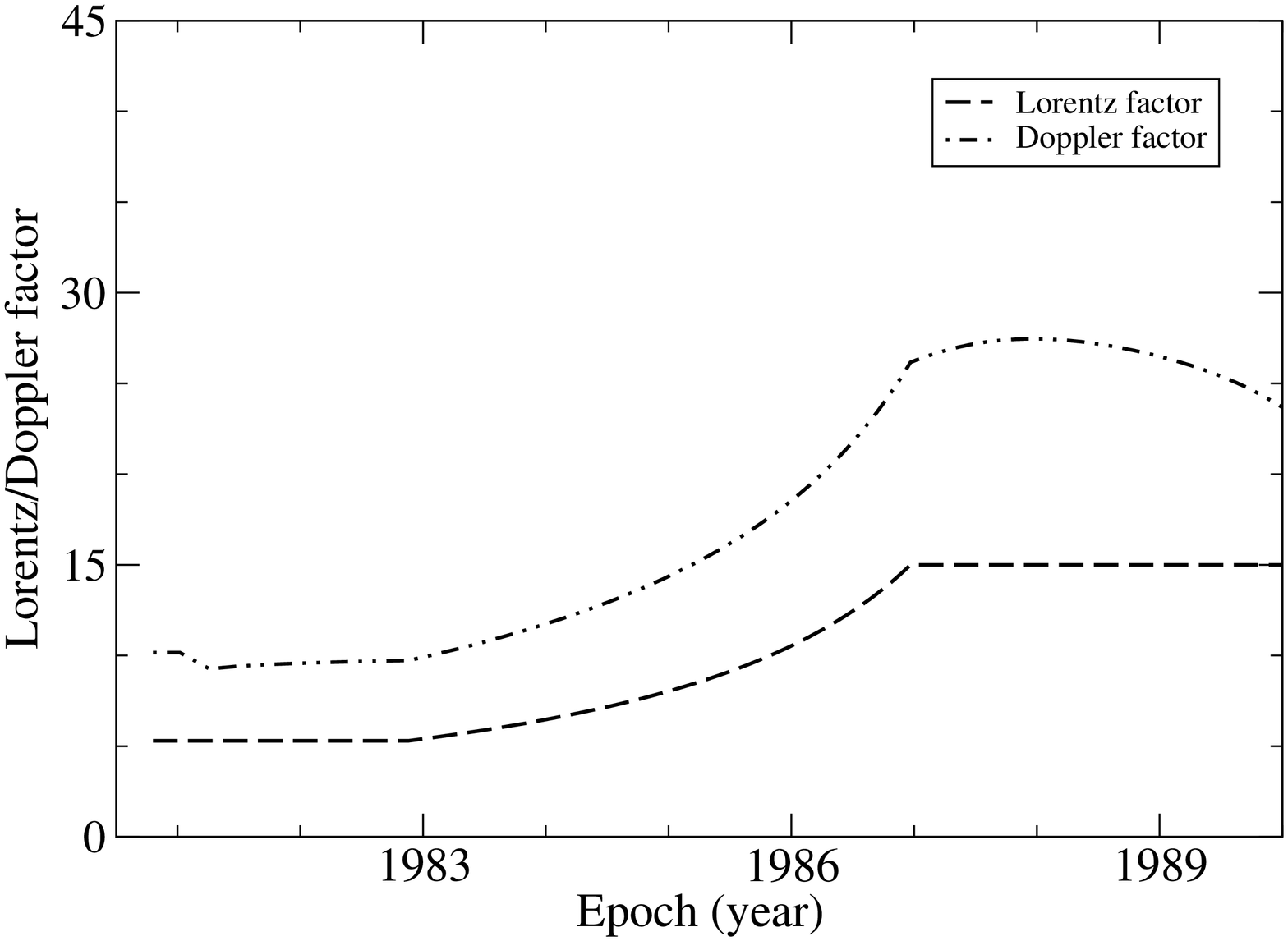}
   \caption{Knot C5: Precession phase $\phi_0$=5.83\,rad and ejection time 
   $t_0$=1980.80. Model-fitting results: trajectory $Z_n(X_n)$, coordinates
   $X_n(t)$ and $Z_n(t)$, core separation $r_n(t)$, modeled apparent velocity
   $\beta_a$(t) and viewing angle $\theta$(t), bulk Lorentz factor $\Gamma$(t)
   and Doppler factor $\delta$(t).}
   \end{figure*}
   \begin{figure*}
   \centering
   \includegraphics[width=6cm,angle=-90]{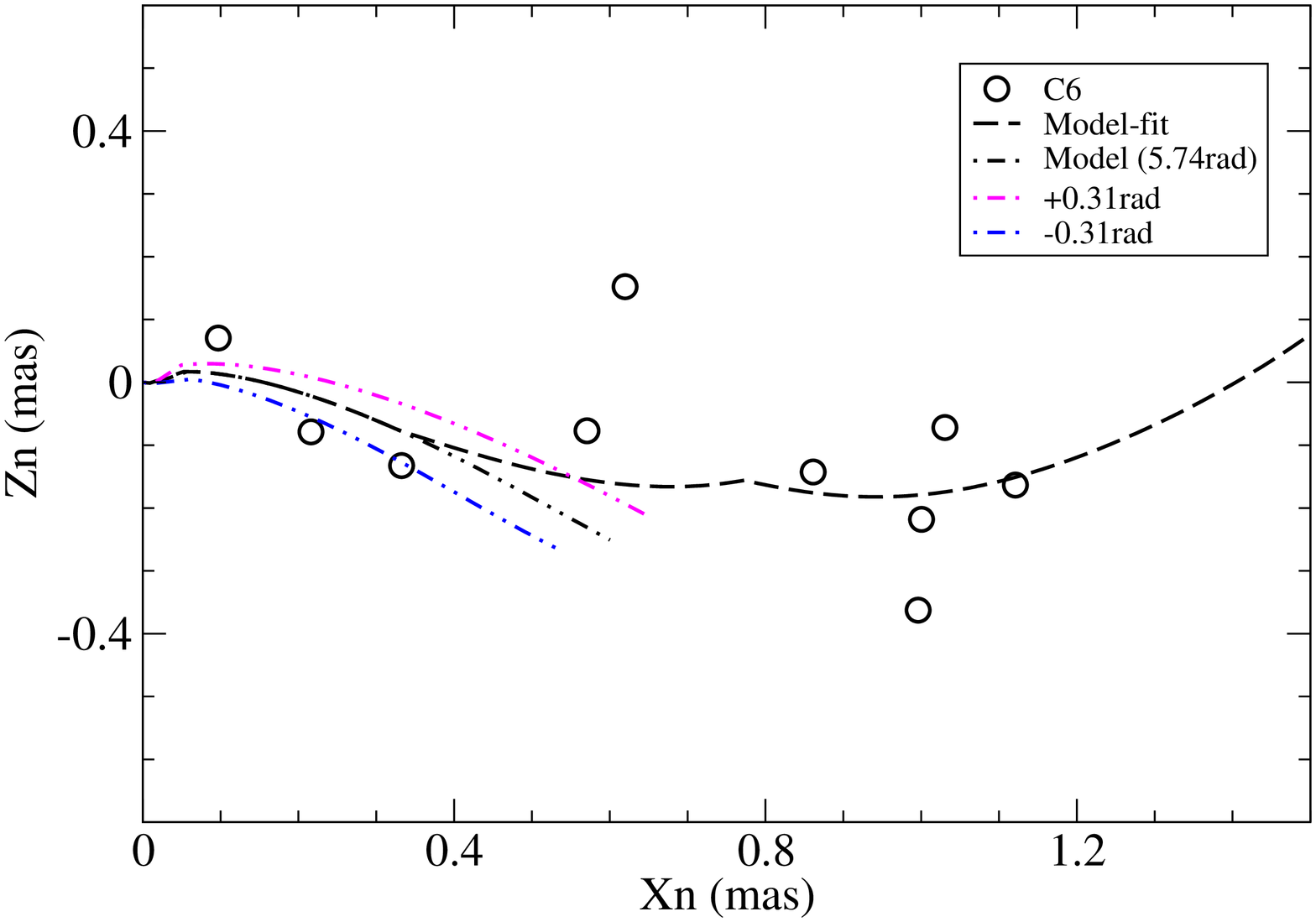}
   \includegraphics[width=6cm,angle=-90]{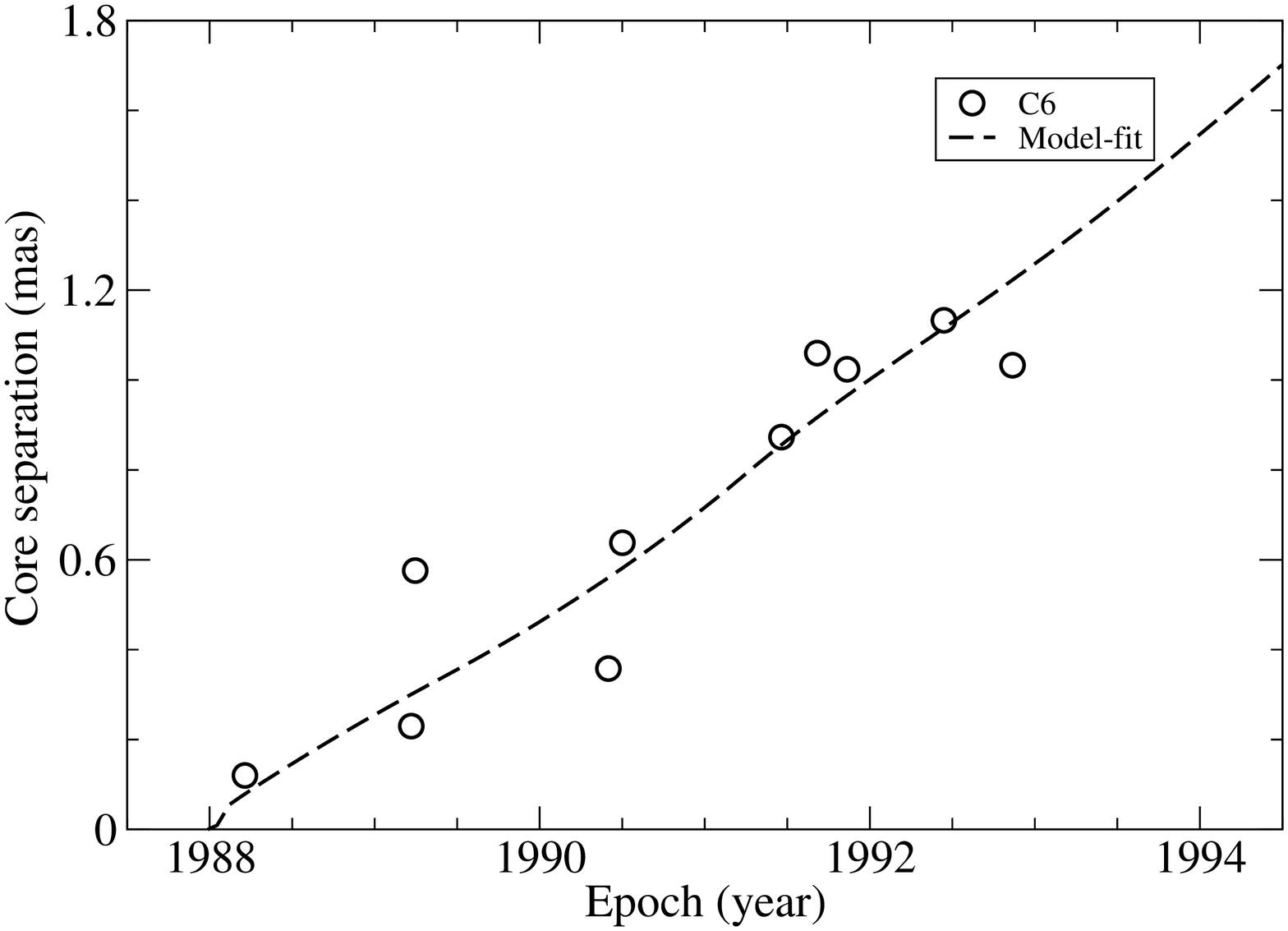}
   \includegraphics[width=6cm,angle=-90]{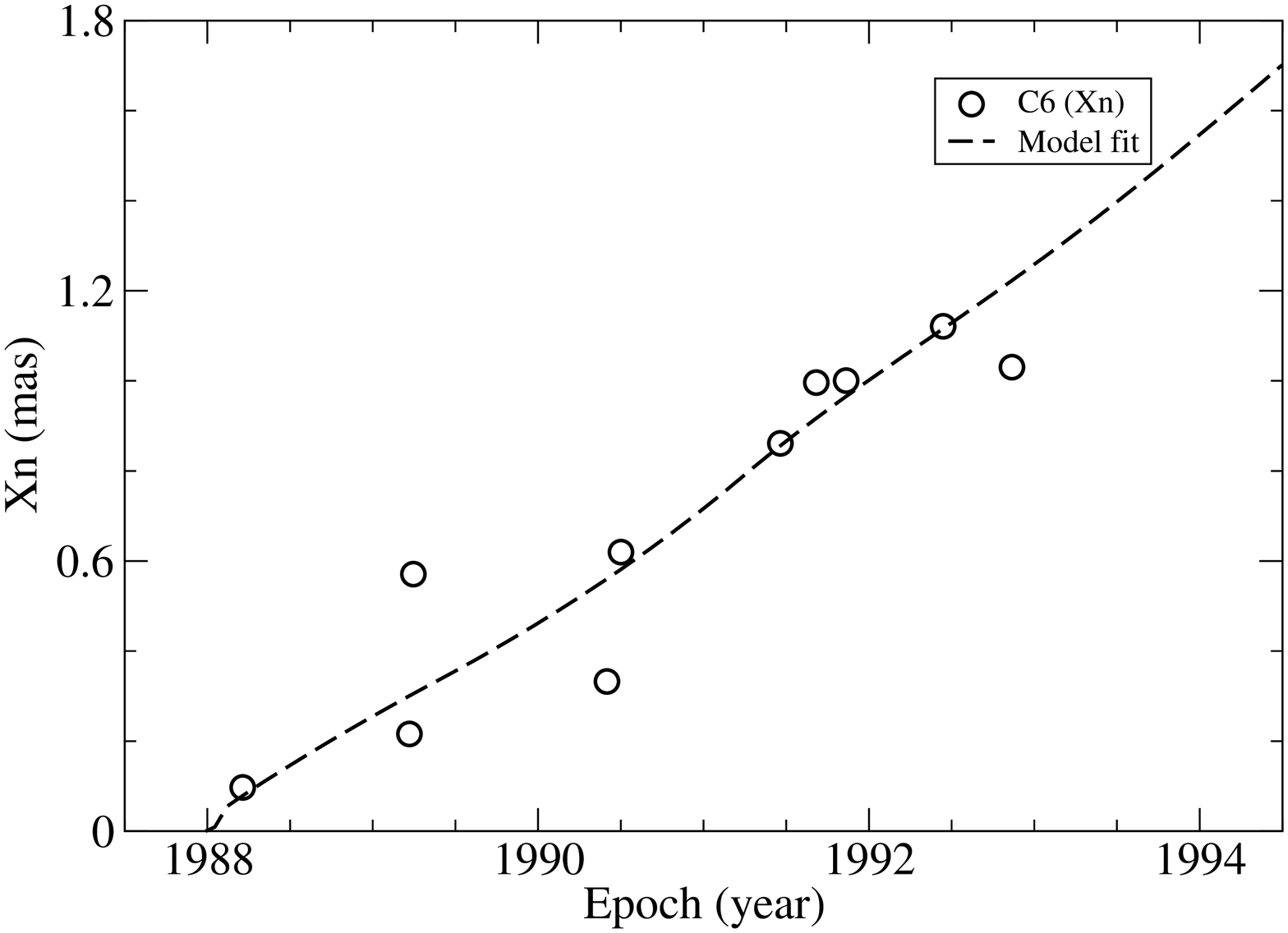}
   \includegraphics[width=6cm,angle=-90]{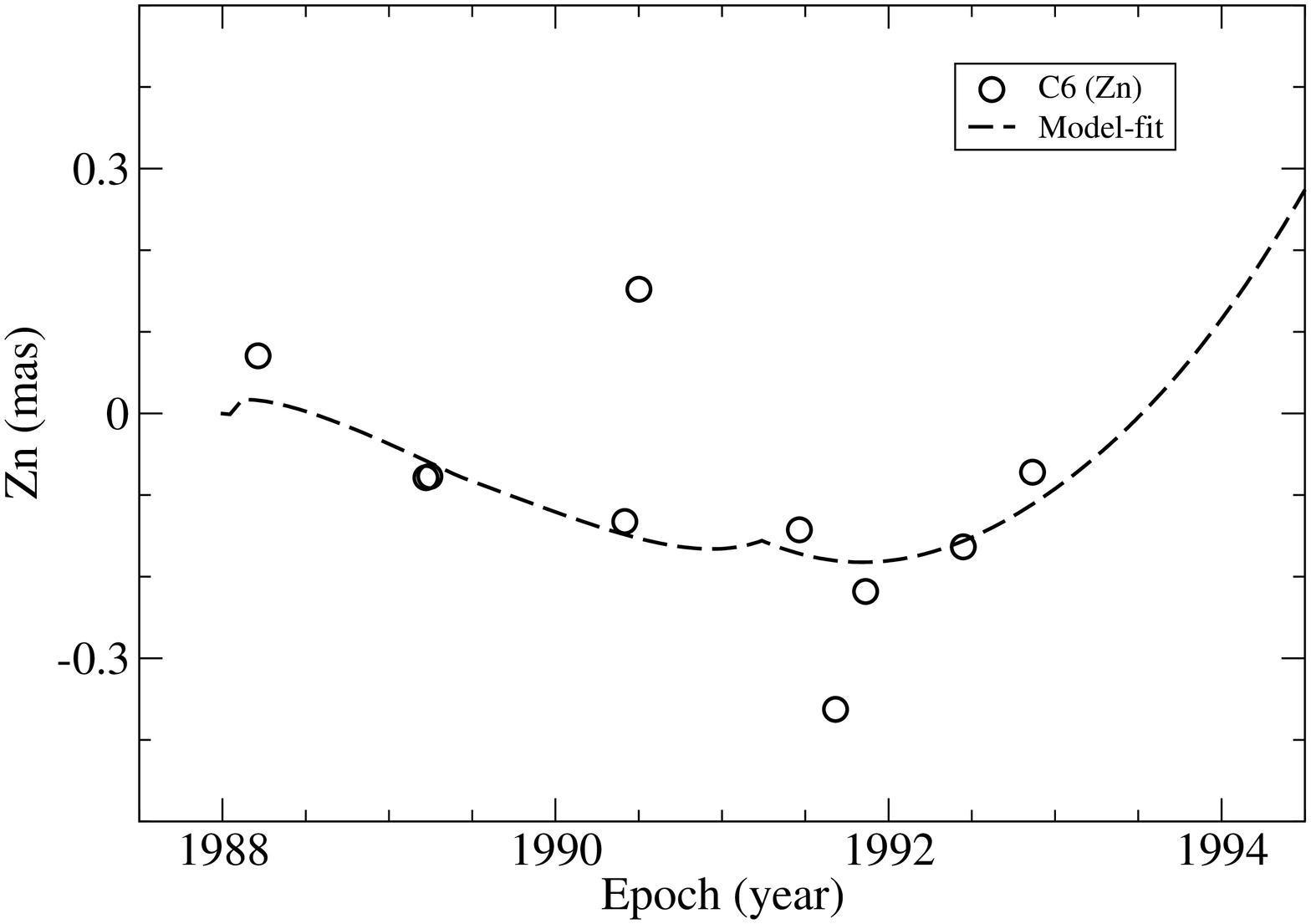}
   \includegraphics[width=6cm,angle=-90]{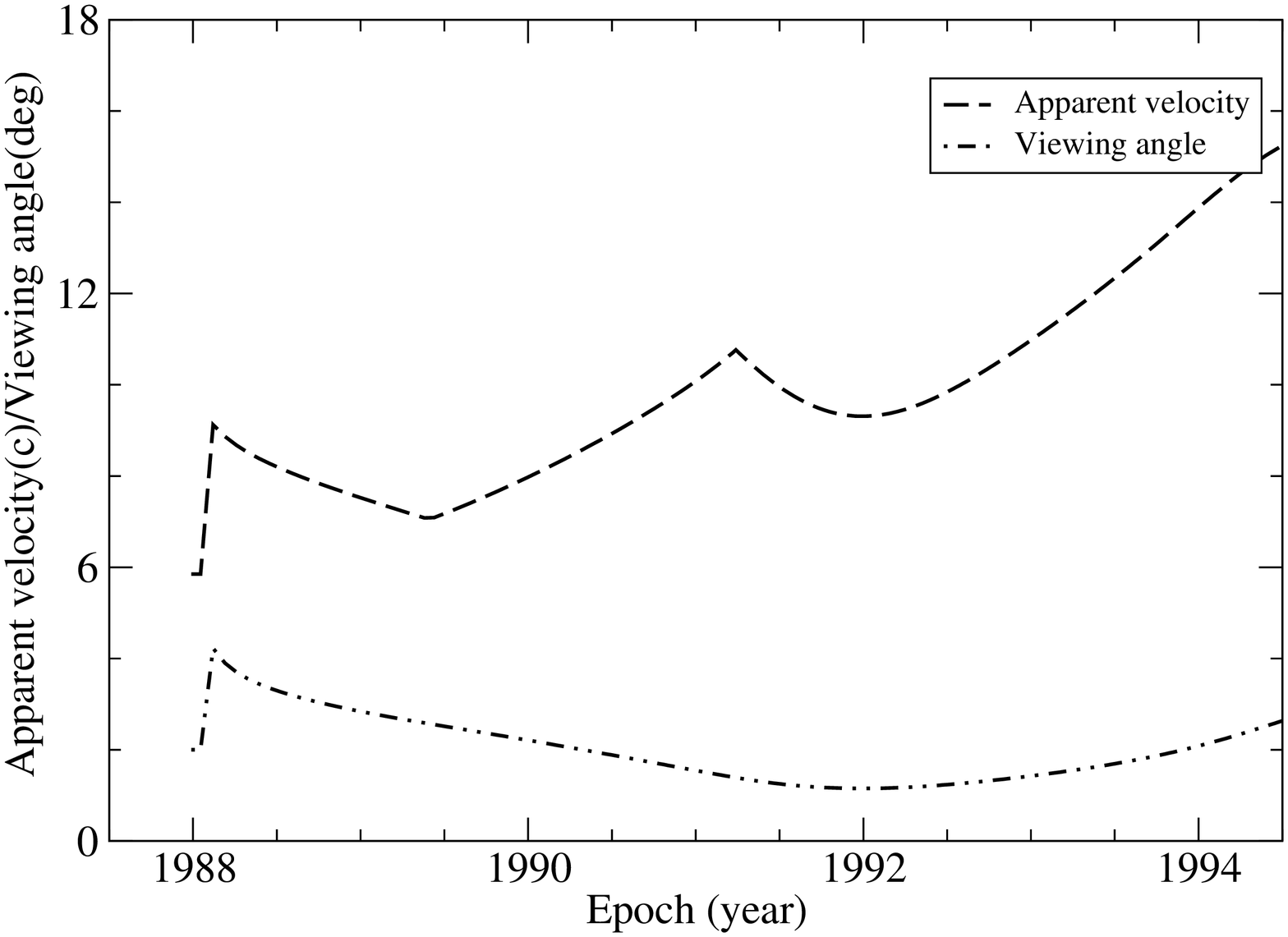}
   \includegraphics[width=6cm,angle=-90]{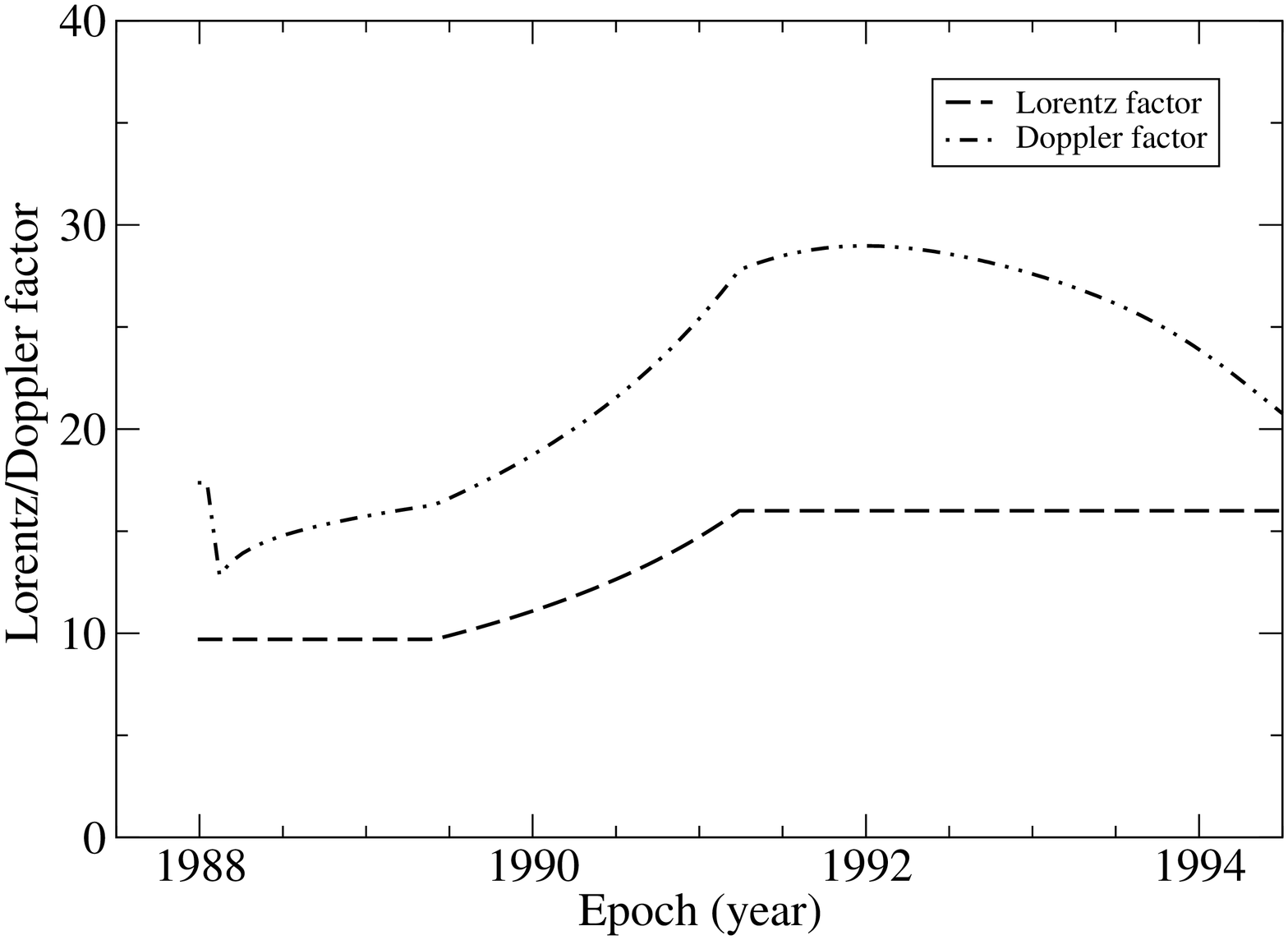}
   \caption{Knot C6: precession phase $\phi_0$(rad)=5.74+2$\pi$ and ejection 
   time $t_0$=1987.99. Mode-fitting results: trajectory $Z_n(X_n)$, coordinates
   $X_n$(t) and $Z_n$(t), core separation $r_n$(t), modeled apparent velocity
    $\beta_a$(t) and viewing angle $\theta$(t), bulk Lorentz factor $\Gamma$(t)
   and Doppler factor $\delta$(t). Kinematics for $r_n{>}$0.4\,mas has 
    been fitted by introducing changes in parameter $\psi$ (see text).}
   \end{figure*}
   \begin{figure*}
   \centering
   \includegraphics[width=6cm,angle=-90]{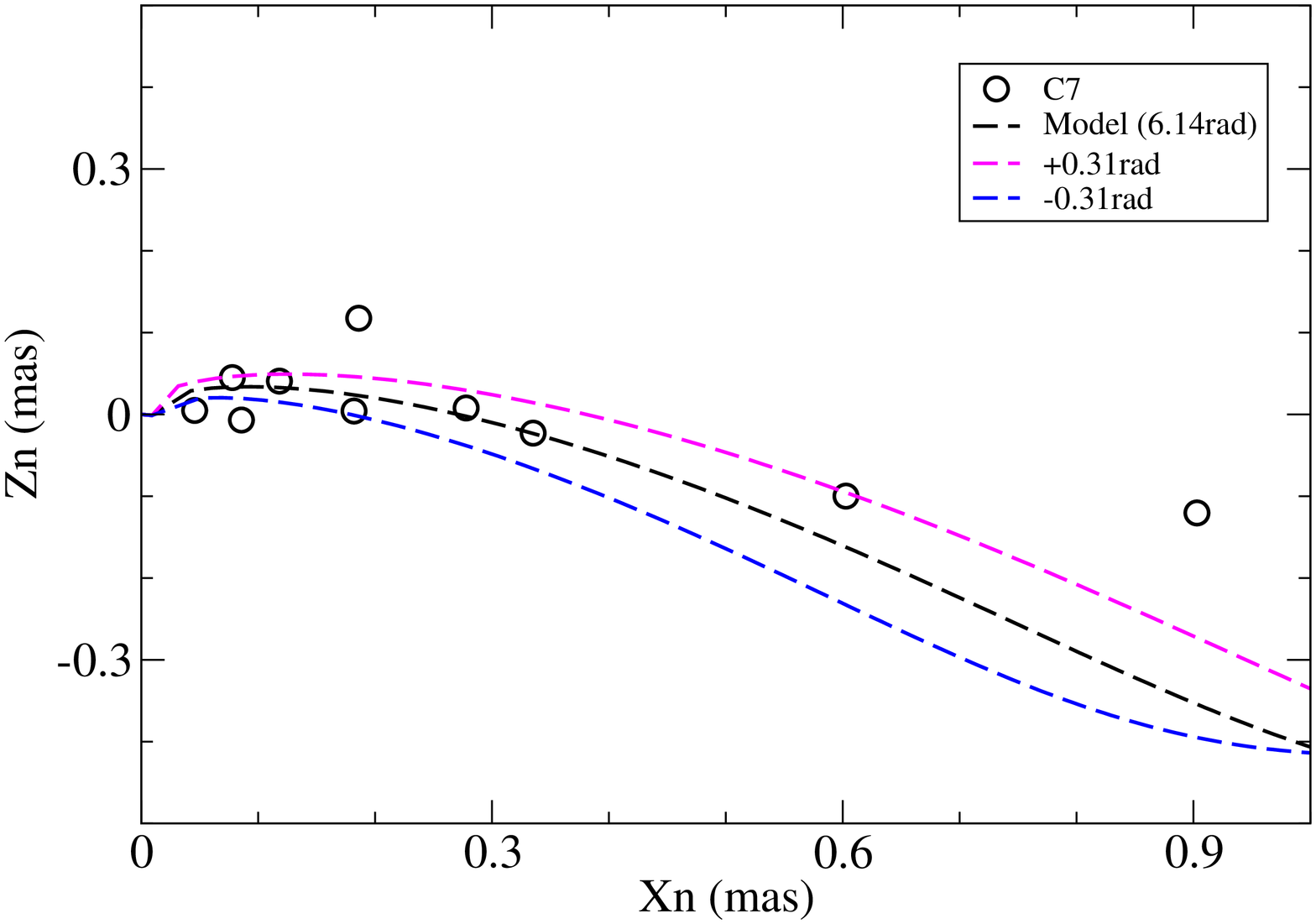}
   \includegraphics[width=6cm,angle=-90]{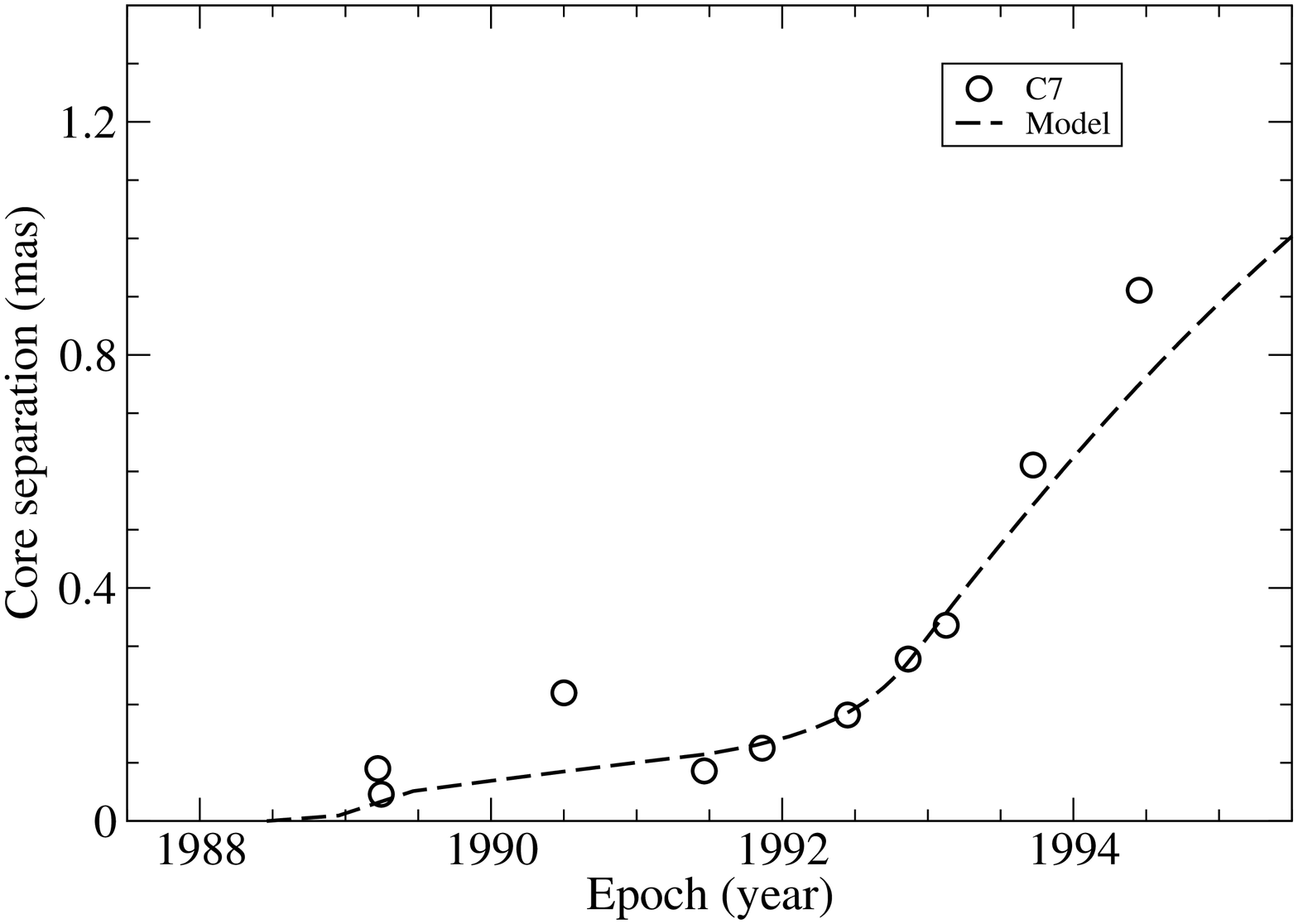}
   \includegraphics[width=6cm,angle=-90]{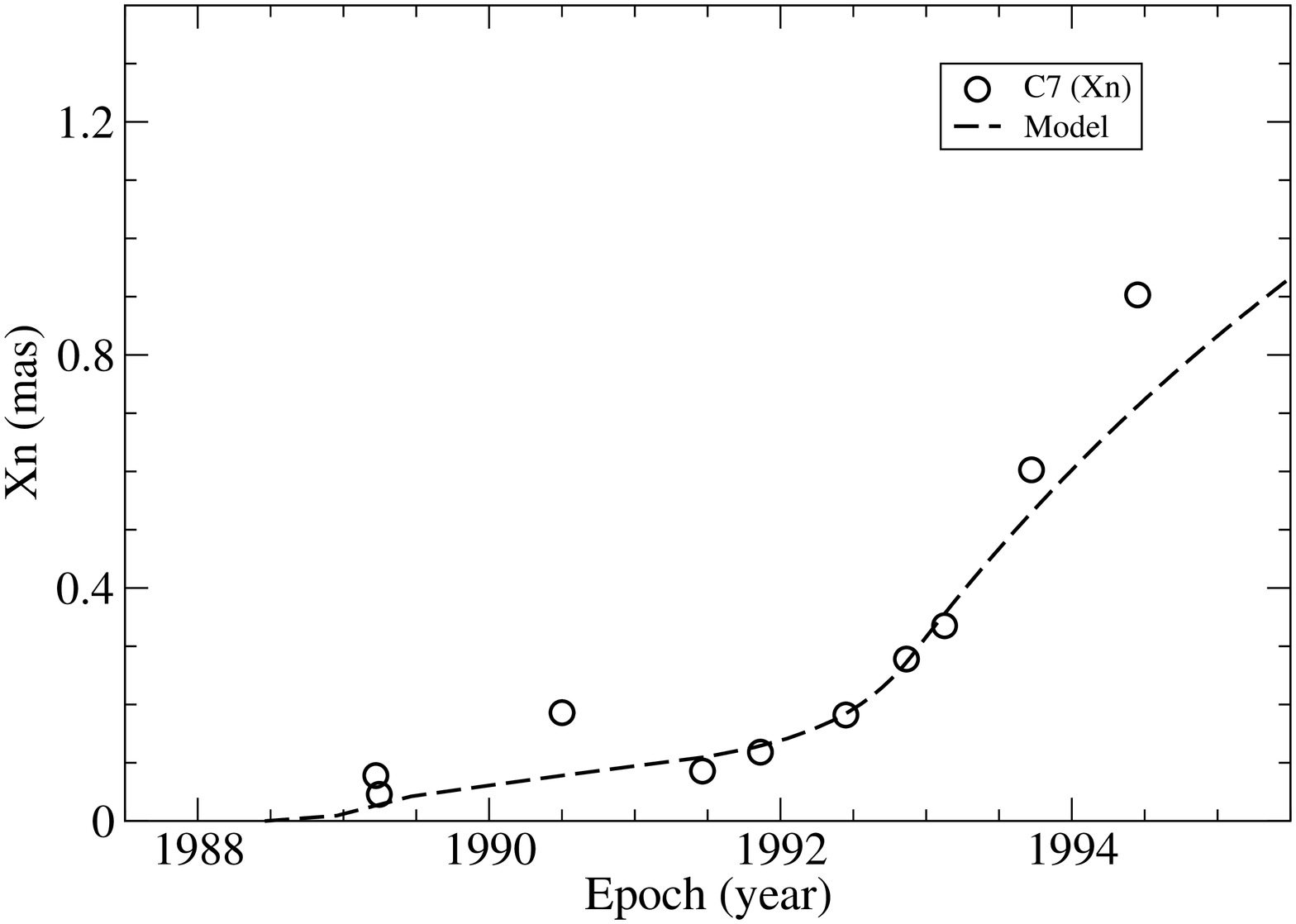}
   \includegraphics[width=6cm,angle=-90]{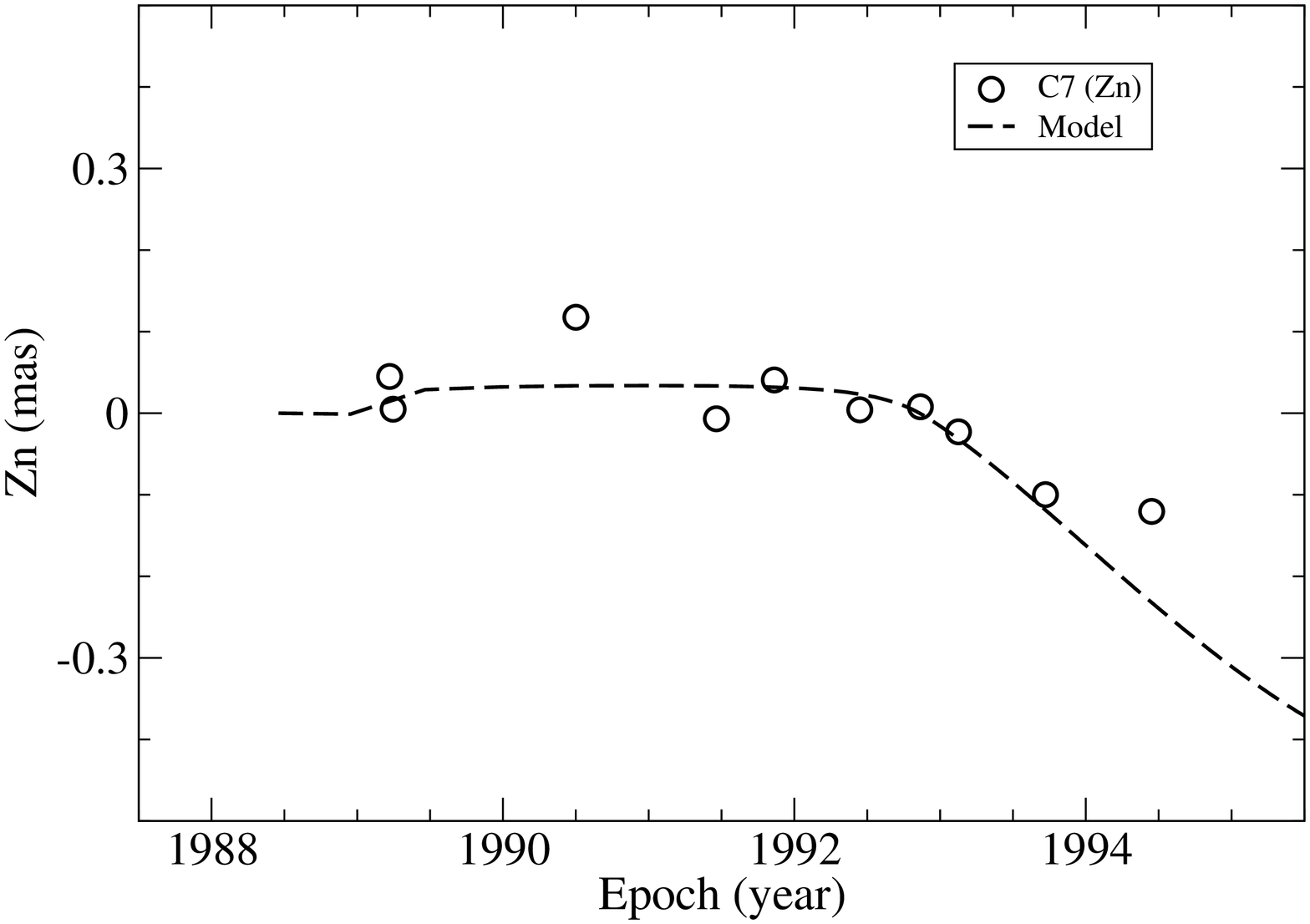}
   \includegraphics[width=6cm,angle=-90]{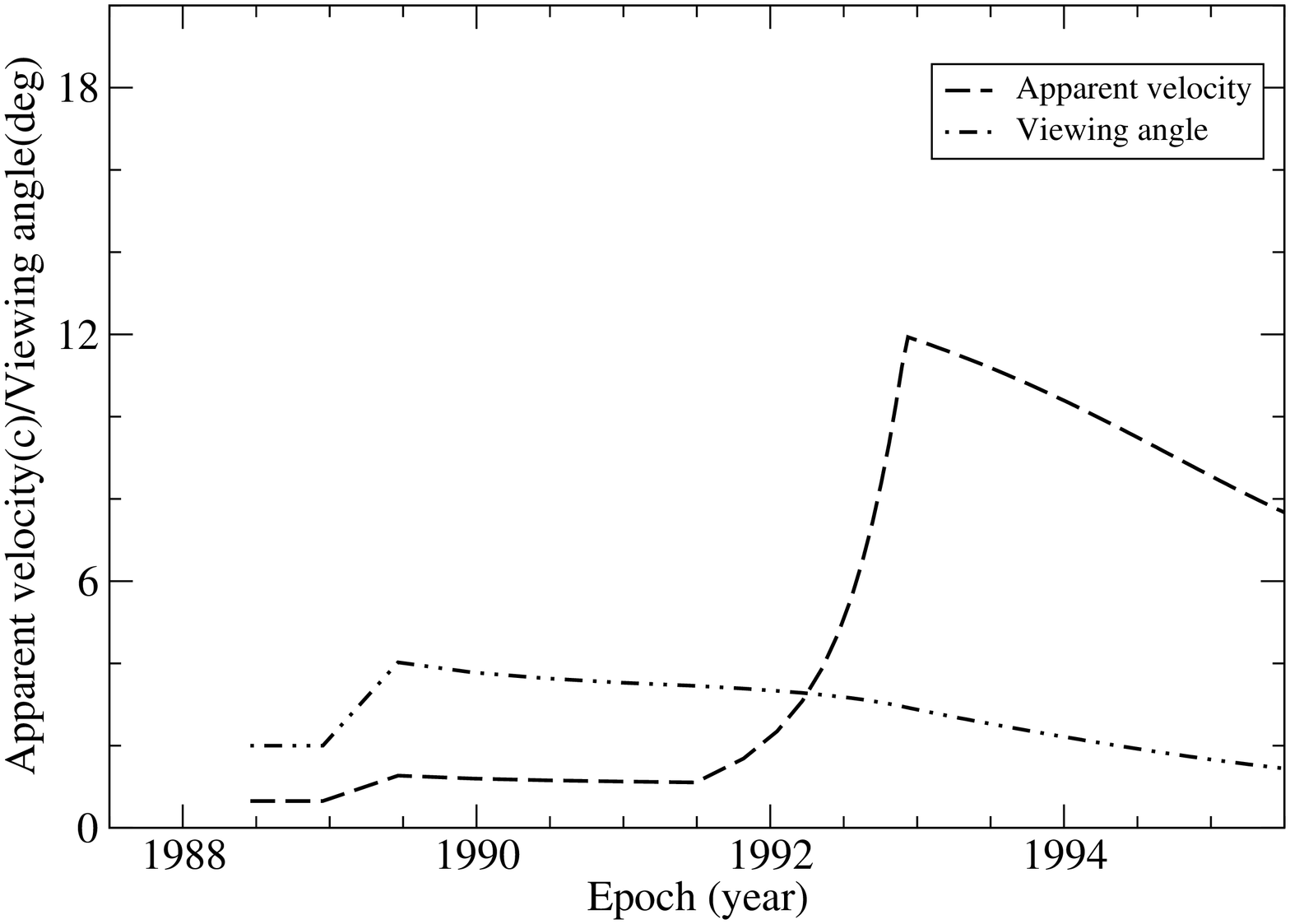}
   \includegraphics[width=6cm,angle=-90]{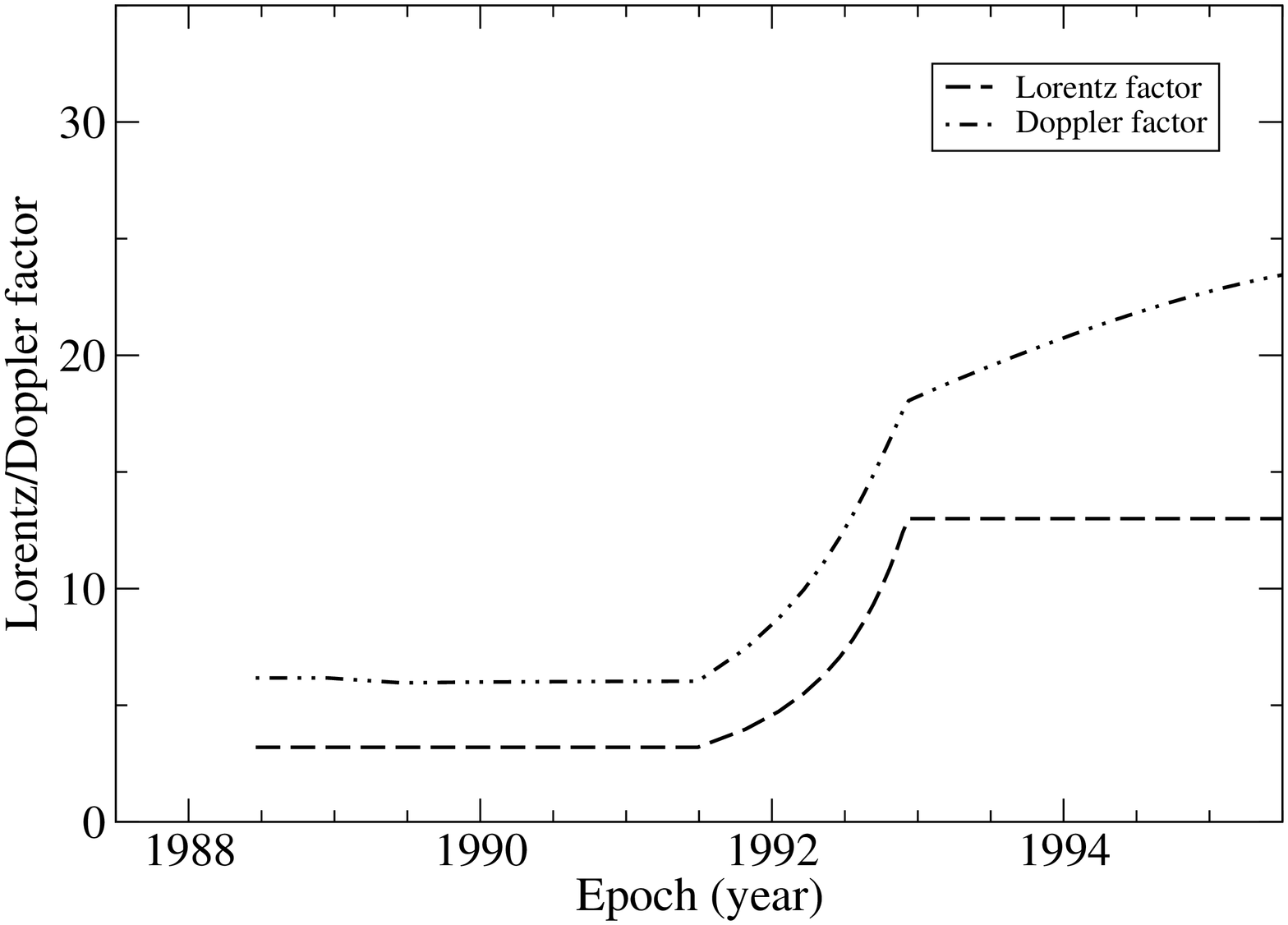}
   \caption{Knot C7: Precession phase $\phi_0$(rad)=6.14+2$\pi$ and ejection
   time $t_0$=1988.46. Model-fitting results: trajectory $Z_n(X_n)$, 
   coordinates $X_n(t)$ and $Z_n(t)$, core separation $r_n(t)$, modeled
   apparent velocity $\beta_a(t)$ and viewing angle $\theta(t)$, bulk Lorentz
   factor $\Gamma(t)$ and Doppler factor $\delta(t)$. Its kinematics within 
    $r_n{\sim}$0.7\,mas can be well fitted in terms of the precessing nozzle
   scenario, or its observed precessing commom trajectory extends to core
   separation $\sim$0.7\,mas.}
   \end{figure*} 
   \begin{figure*}
   \centering
   \includegraphics[width=6cm,angle=-90]{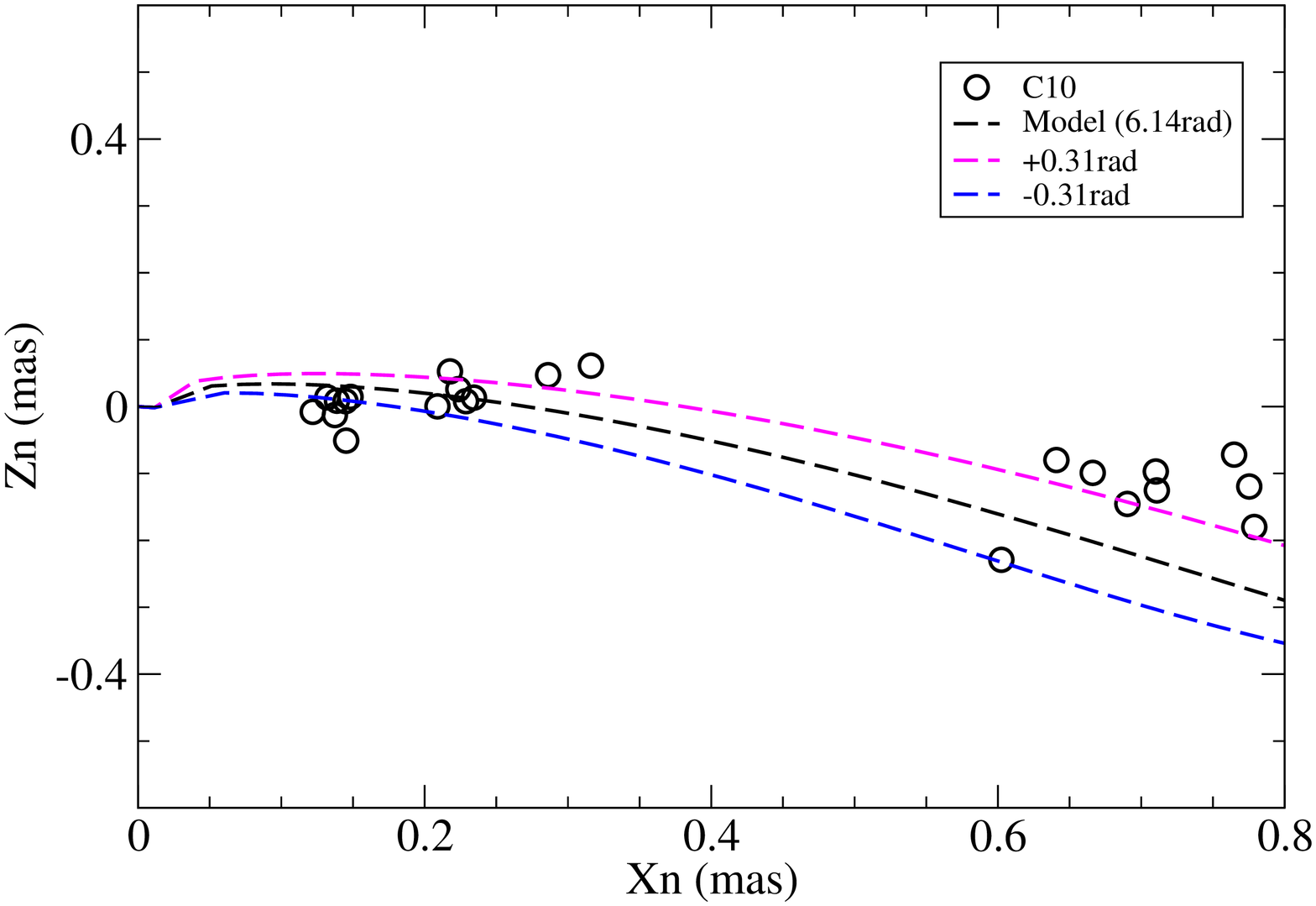}
   \includegraphics[width=6cm,angle=-90]{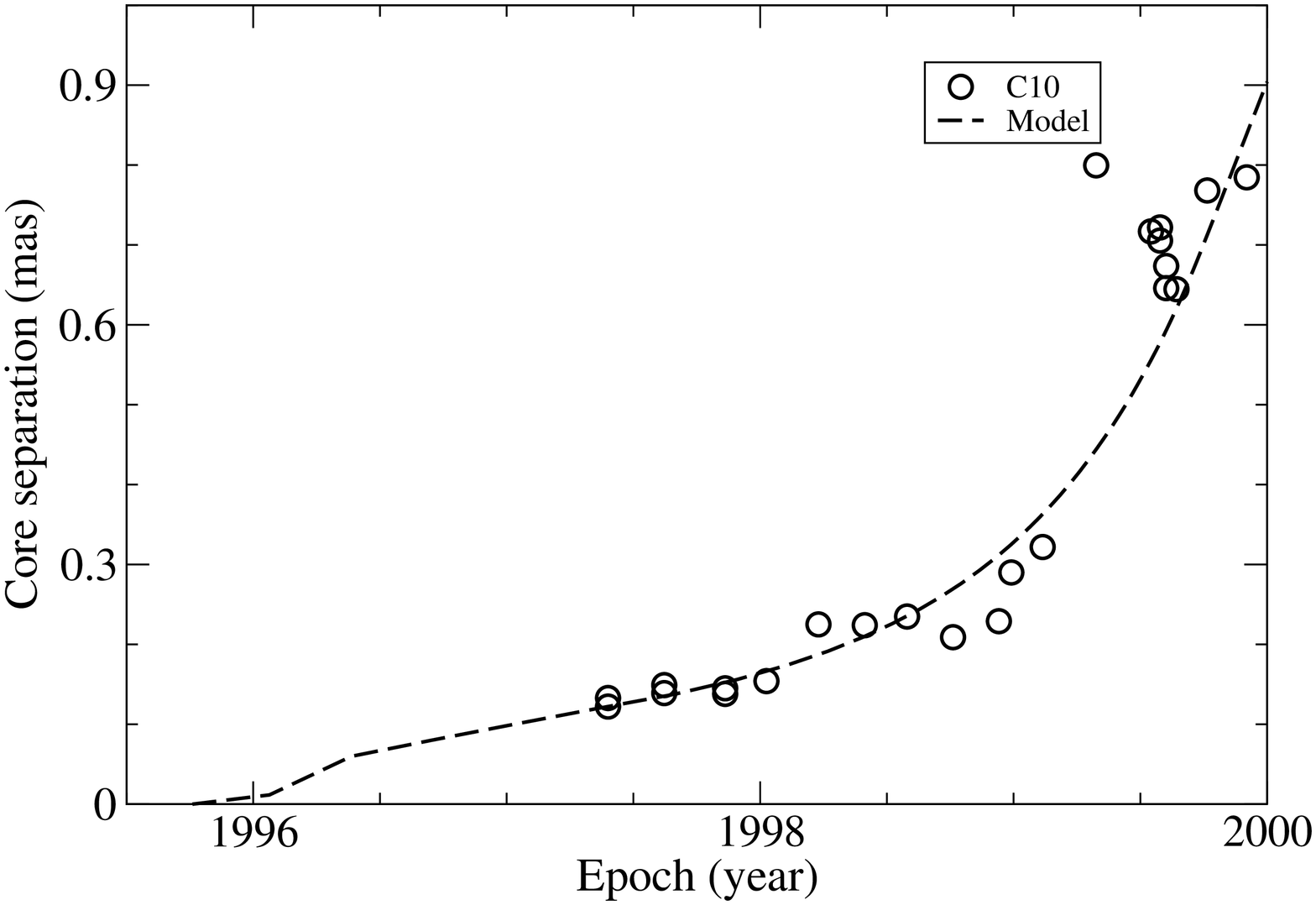}
   \includegraphics[width=6cm,angle=-90]{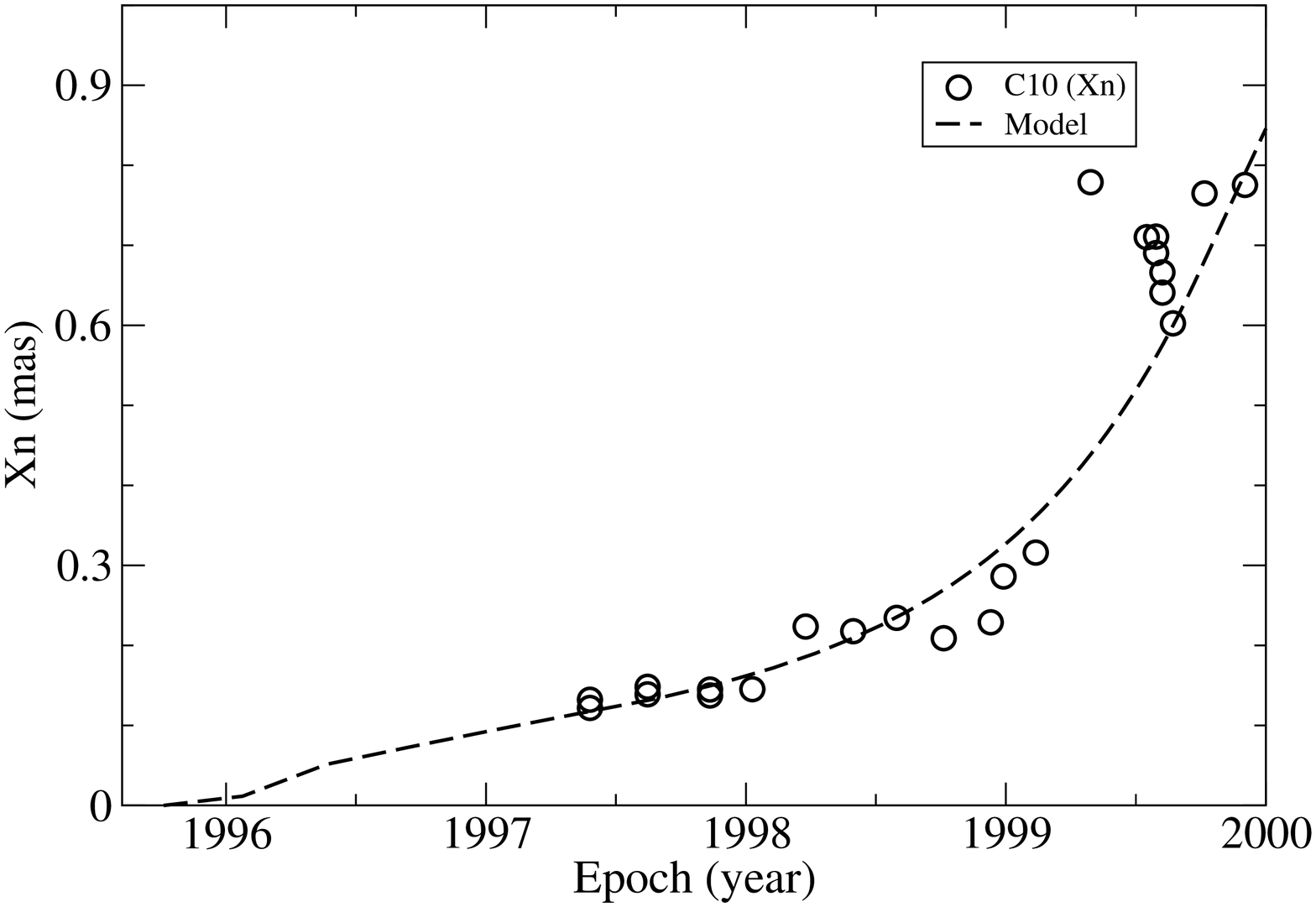}
   \includegraphics[width=6cm,angle=-90]{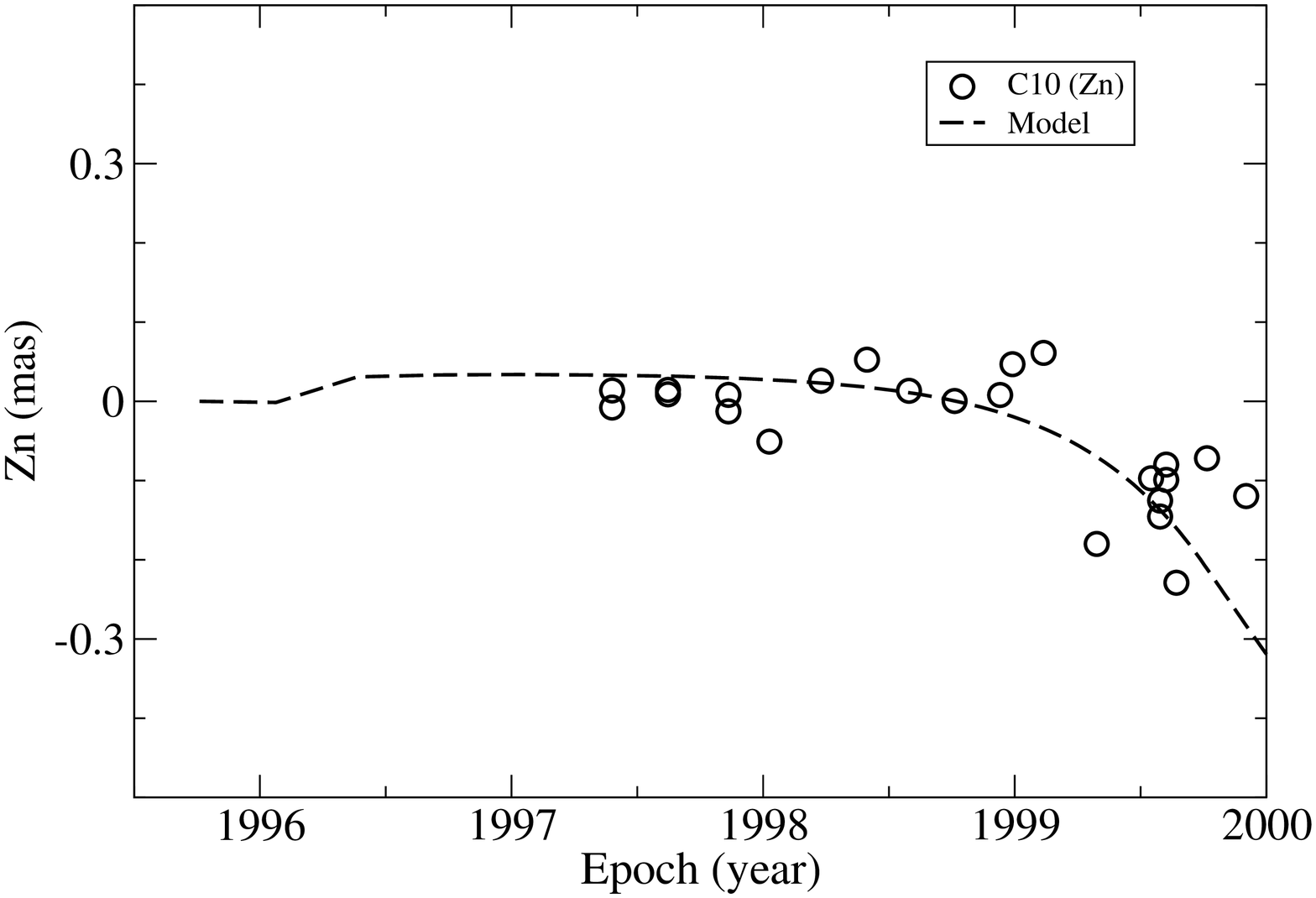}
   \includegraphics[width=6cm,angle=-90]{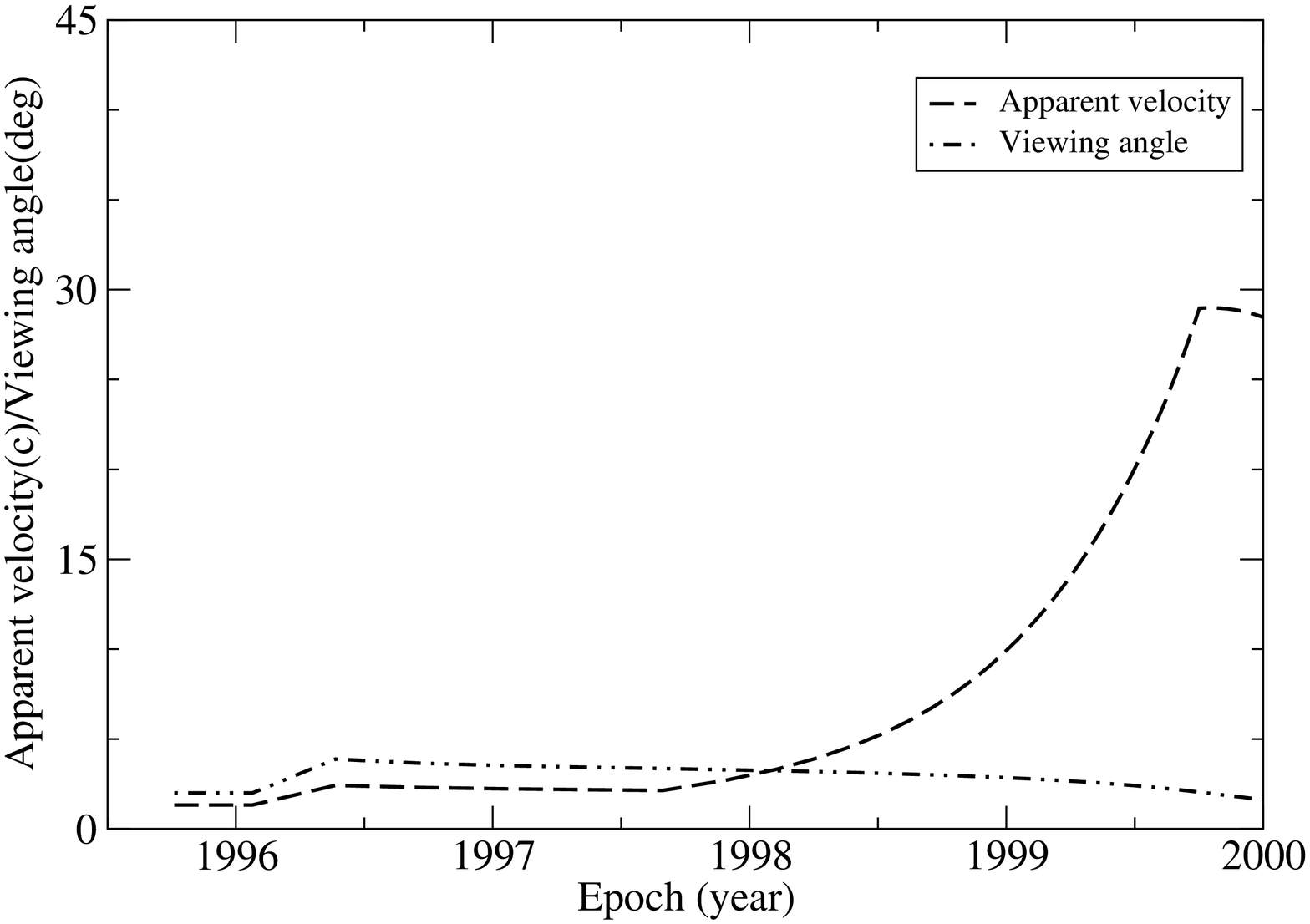}
   \includegraphics[width=6cm,angle=-90]{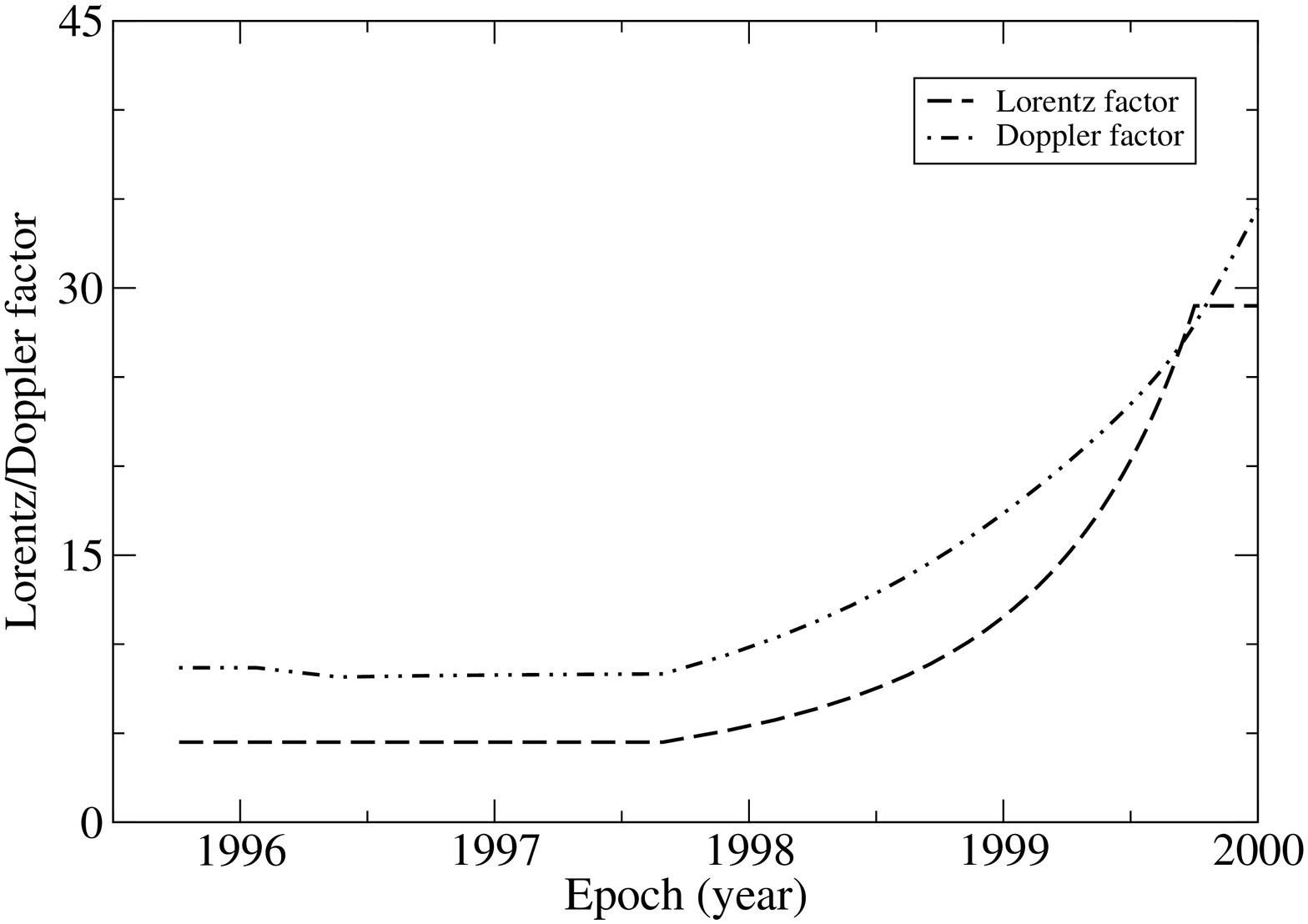}
   \caption{Knot C10: precession phase $\phi_0$(rad)=6.14+4$\pi$ and ejection
   time $t_0$=1995.76. Model-fitting results: trajectory $Z_n(X_n)$, 
   coordinates $X_n(t)$ and $Z_n(t)$, core separation $r_n(t)$, modeled
   apparent velocity $\beta_a(t)$ and viewing angle $\theta(t)$, bulk
   Lorentz factor $\Gamma(t)$ and Doppler factor $\delta(t)$. Its observed
   precessing common trajectory is assumed to extend to $r_n{\sim}$0.8\,mas.}
   \end{figure*} 
   \begin{figure*}
   \centering
   \includegraphics[width=6cm,angle=-90]{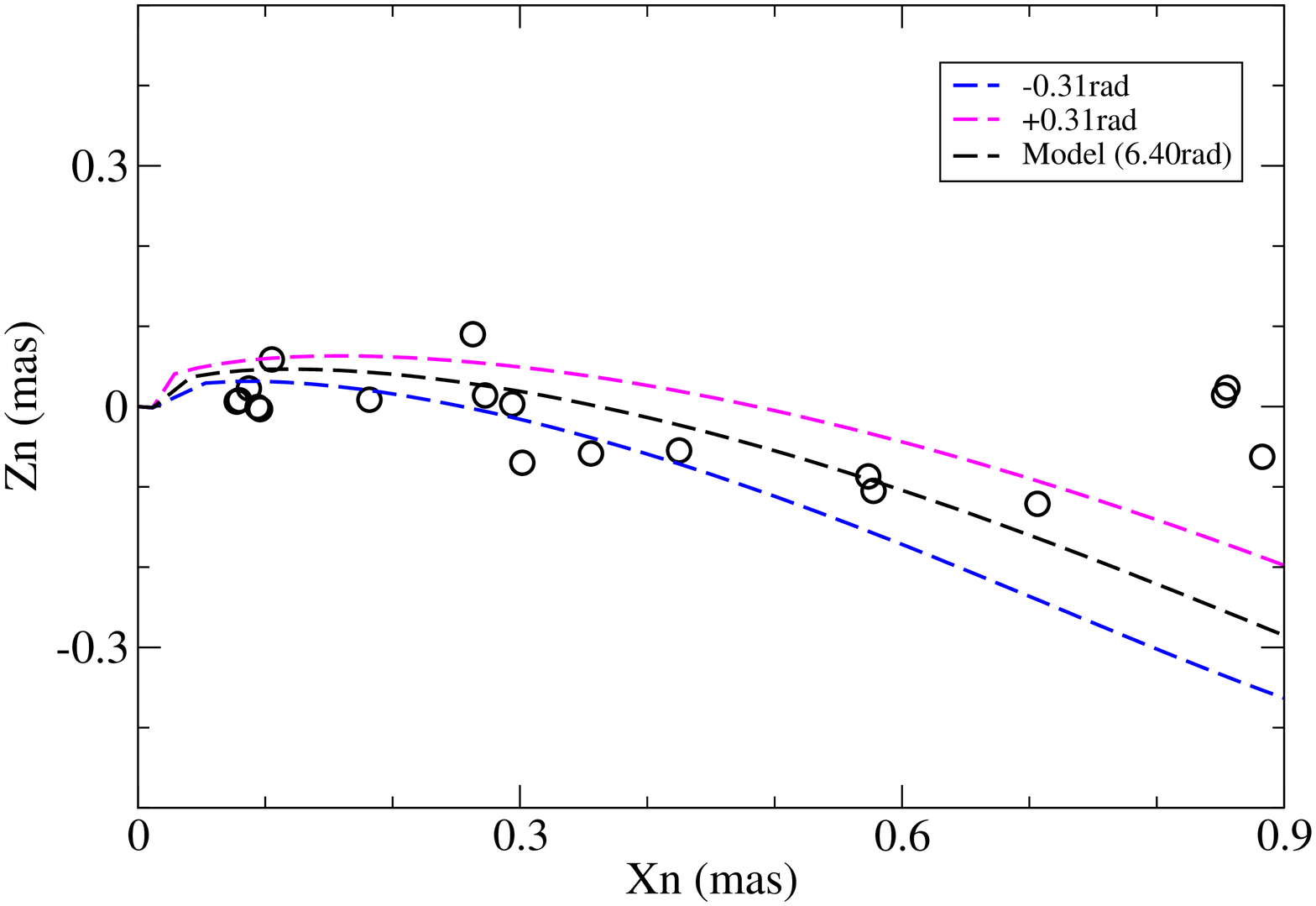}
   \includegraphics[width=6cm,angle=-90]{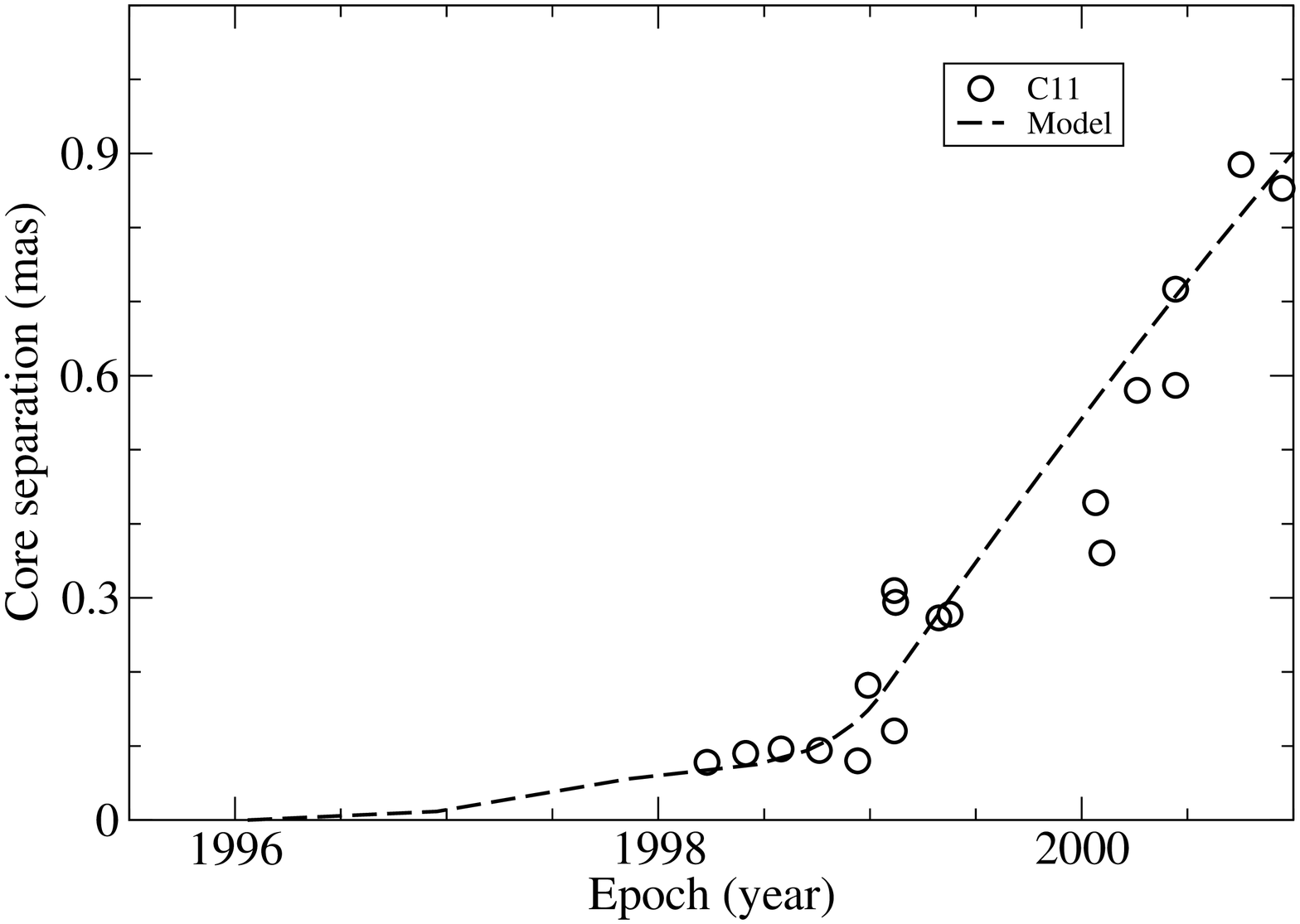}
   \includegraphics[width=6cm,angle=-90]{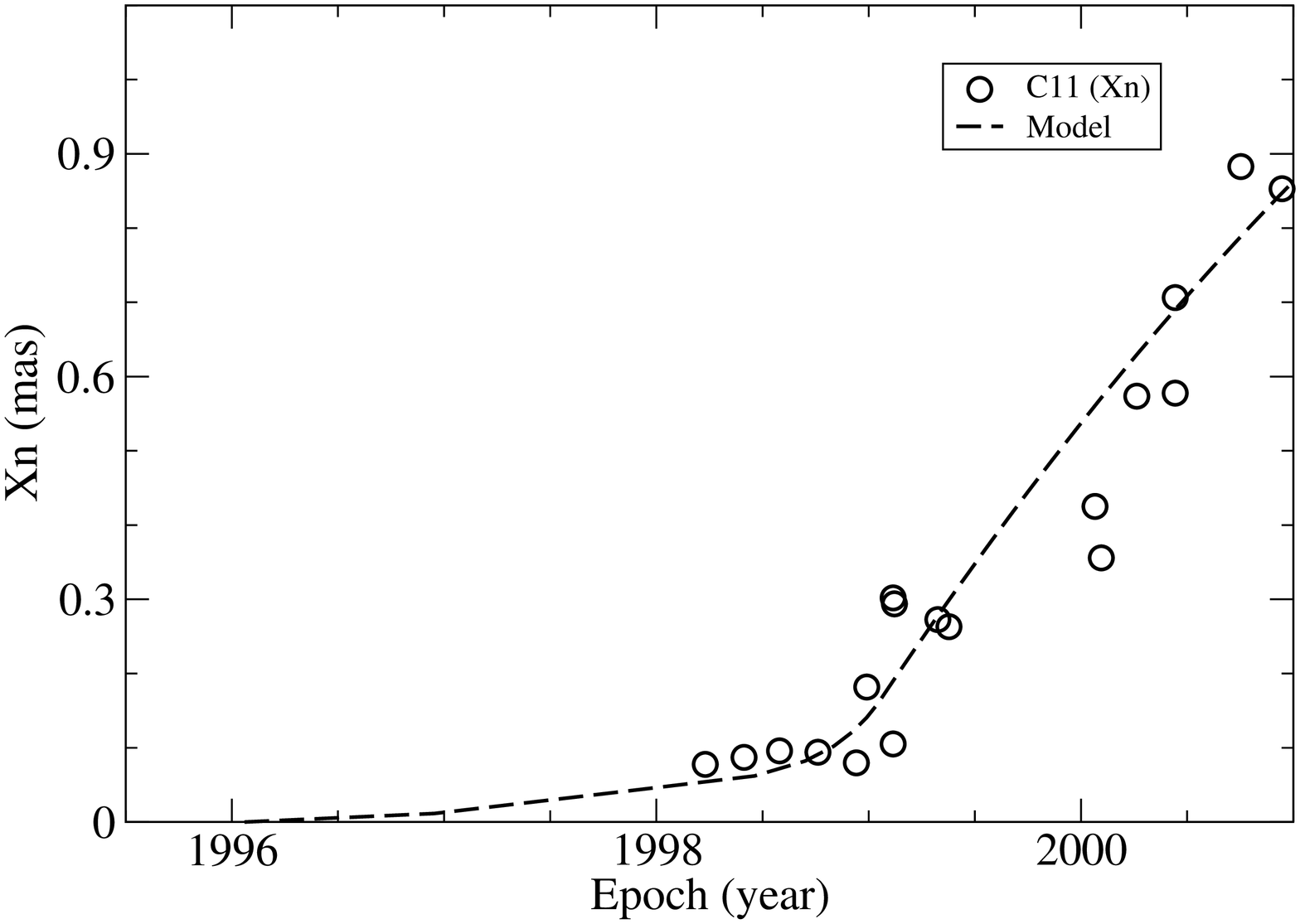}
   \includegraphics[width=6cm,angle=-90]{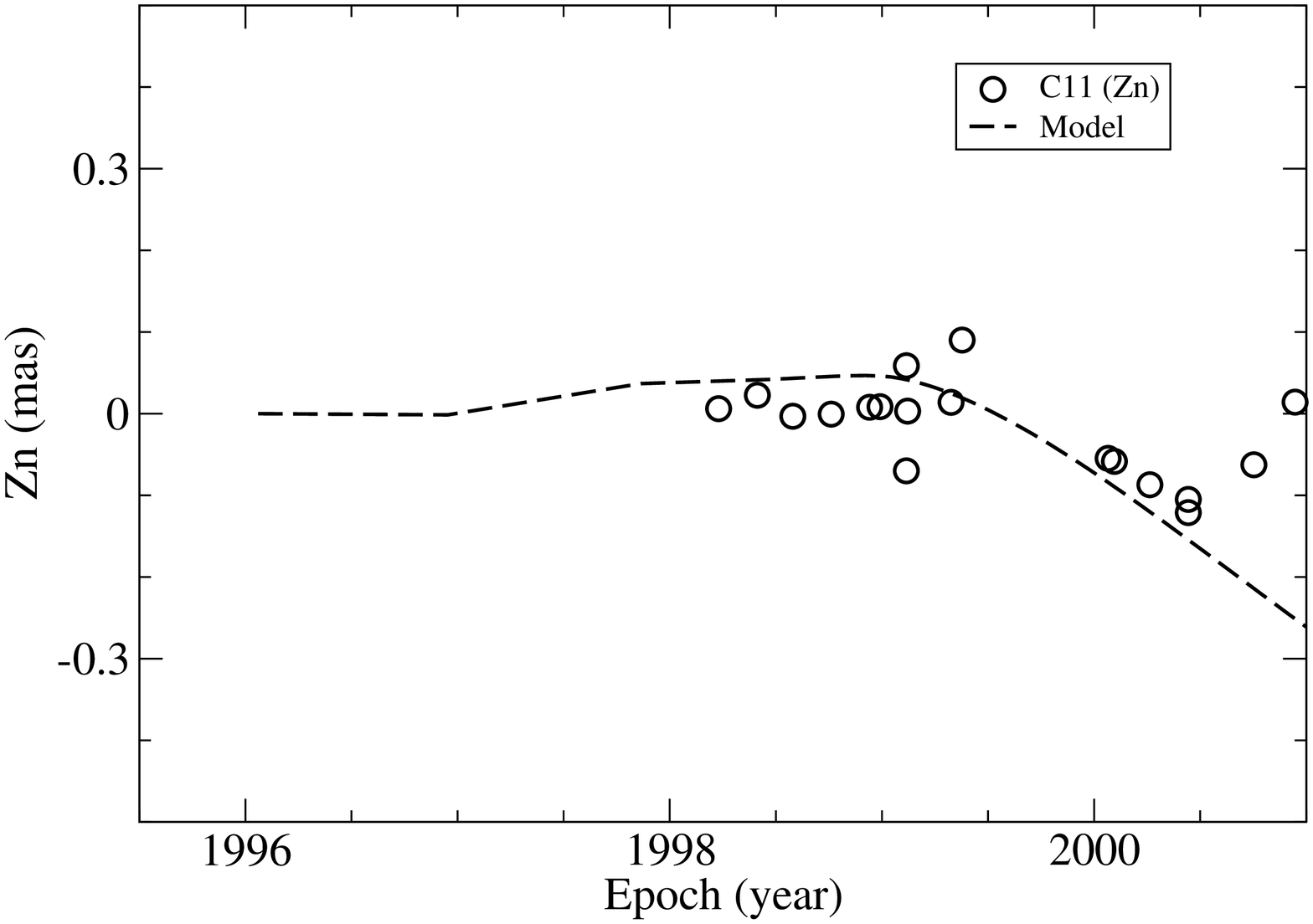}
   \includegraphics[width=6cm,angle=-90]{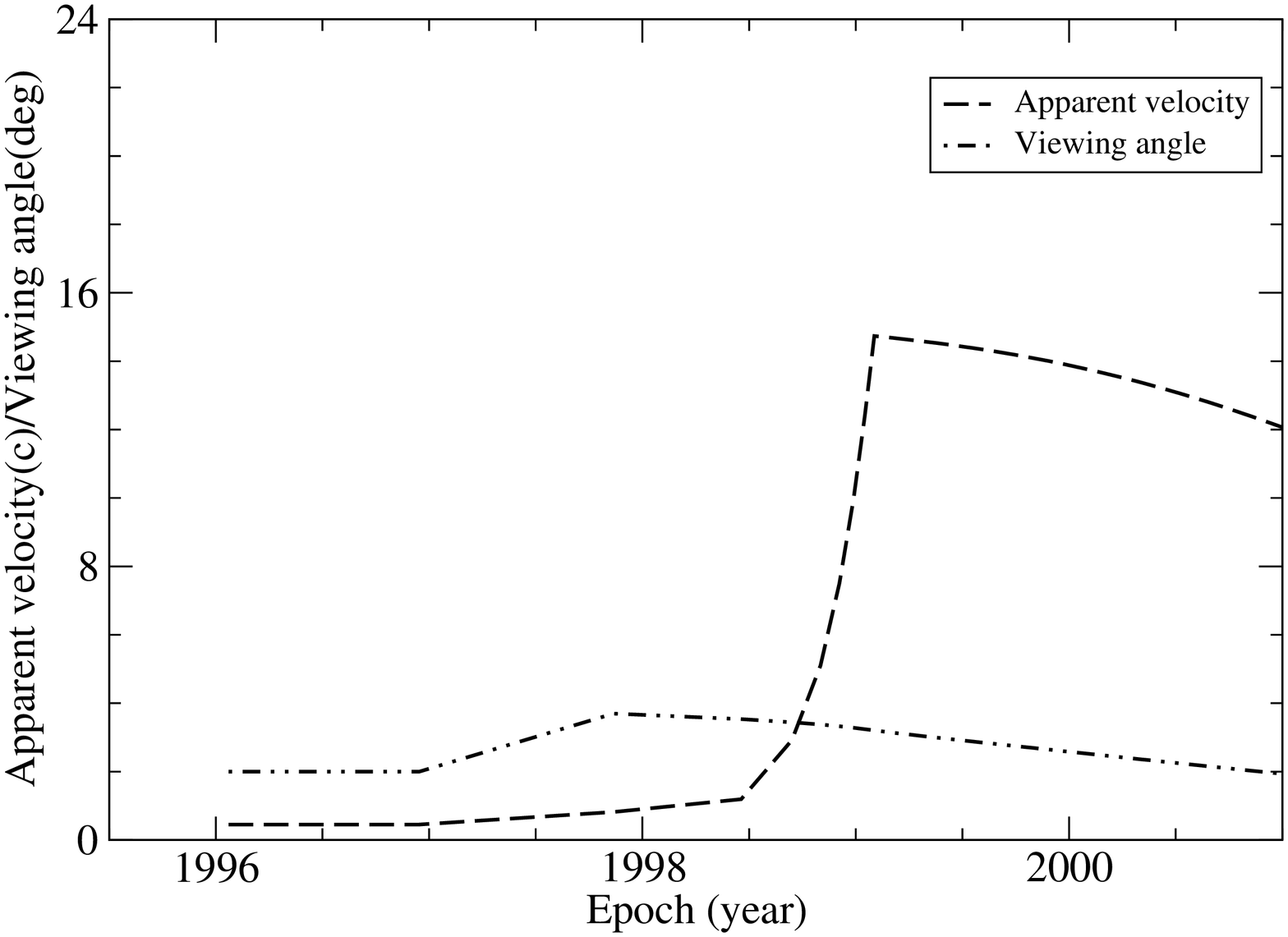}
   \includegraphics[width=6cm,angle=-90]{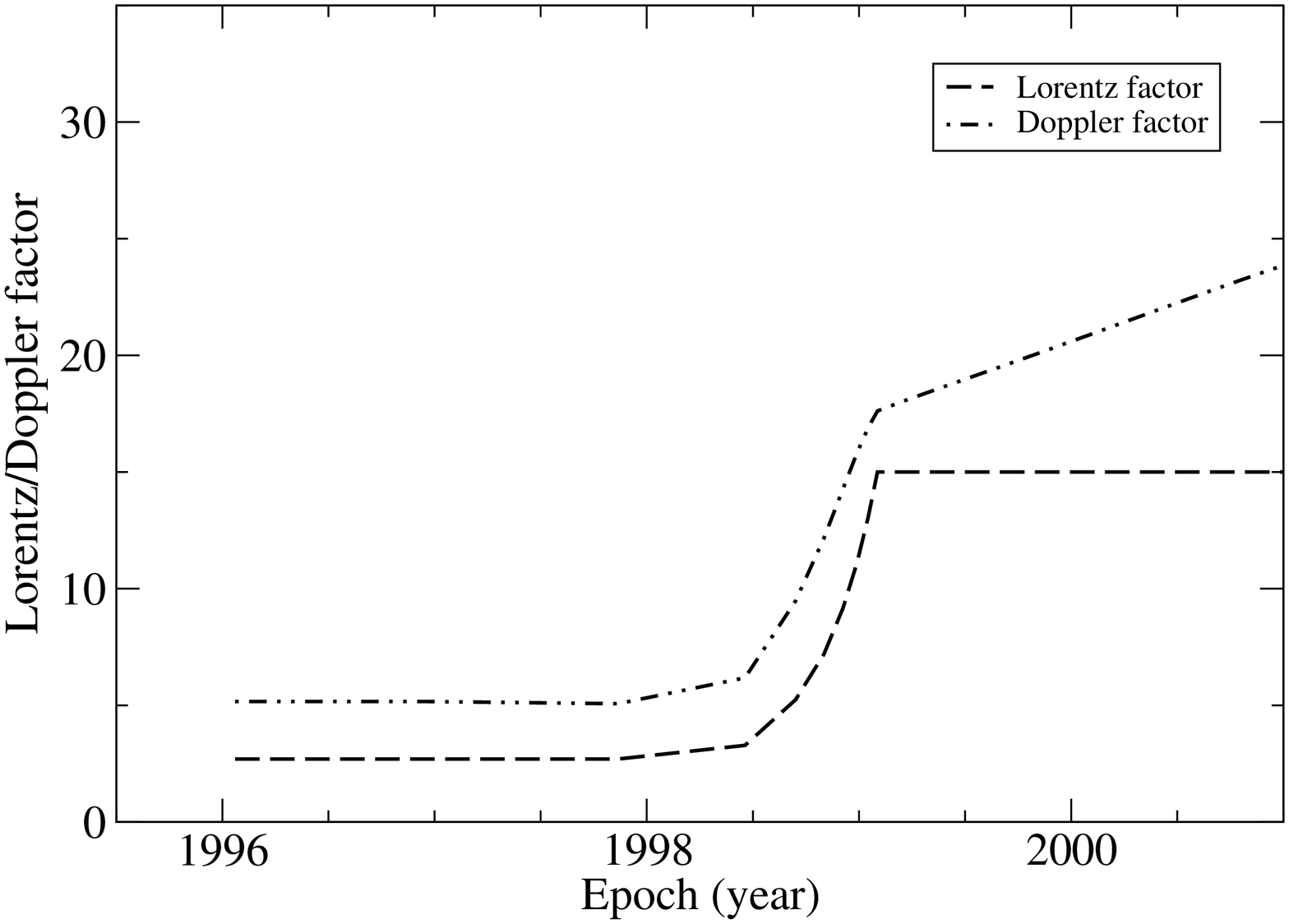}
   \caption{Knot C11: precession phase $\phi_0$(rad)=5.88+4$\pi$ and ejection 
   time $t_0$=1995.46. Model-fitting results: trajectory $Z_n(X_n)$, 
   coordinates $X_n(t)$ and $Z_n(t)$, core separation $r_n(t)$, modeled
   apparent velocity $\beta_a(t)$ and viewing angle $\theta(t)$, bulk Lorentz
   factor $\Gamma(t)$ and Doppler factor $\delta(t)$. Its kinematics within
   $r_n{\sim}$0.75\,mas can be well fitted in terms of the precessing 
   nozzle scenario.}
   \end{figure*}
   \begin{figure*}
   \centering
   \includegraphics[width=6cm,angle=-90]{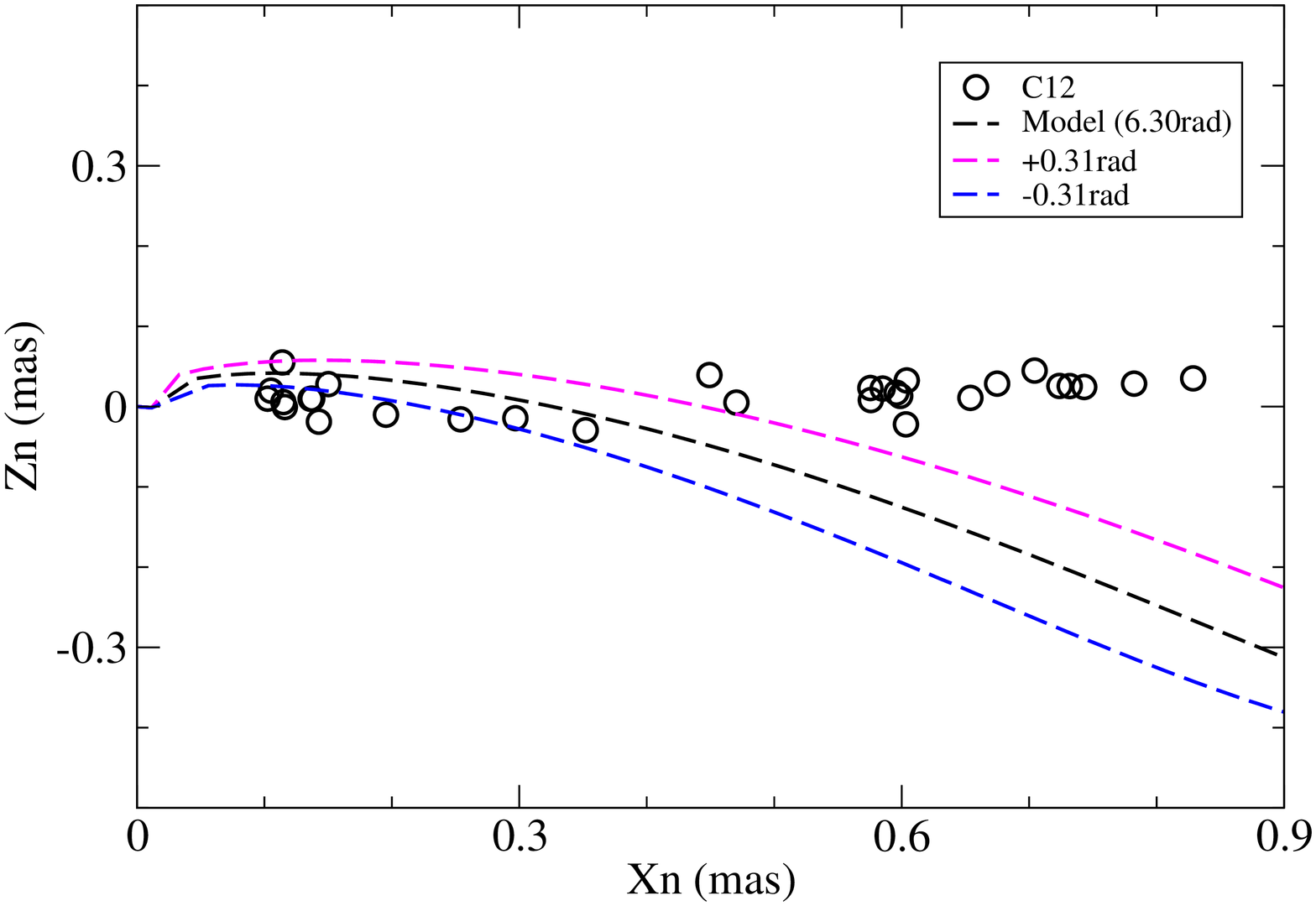}
   \includegraphics[width=6cm,angle=-90]{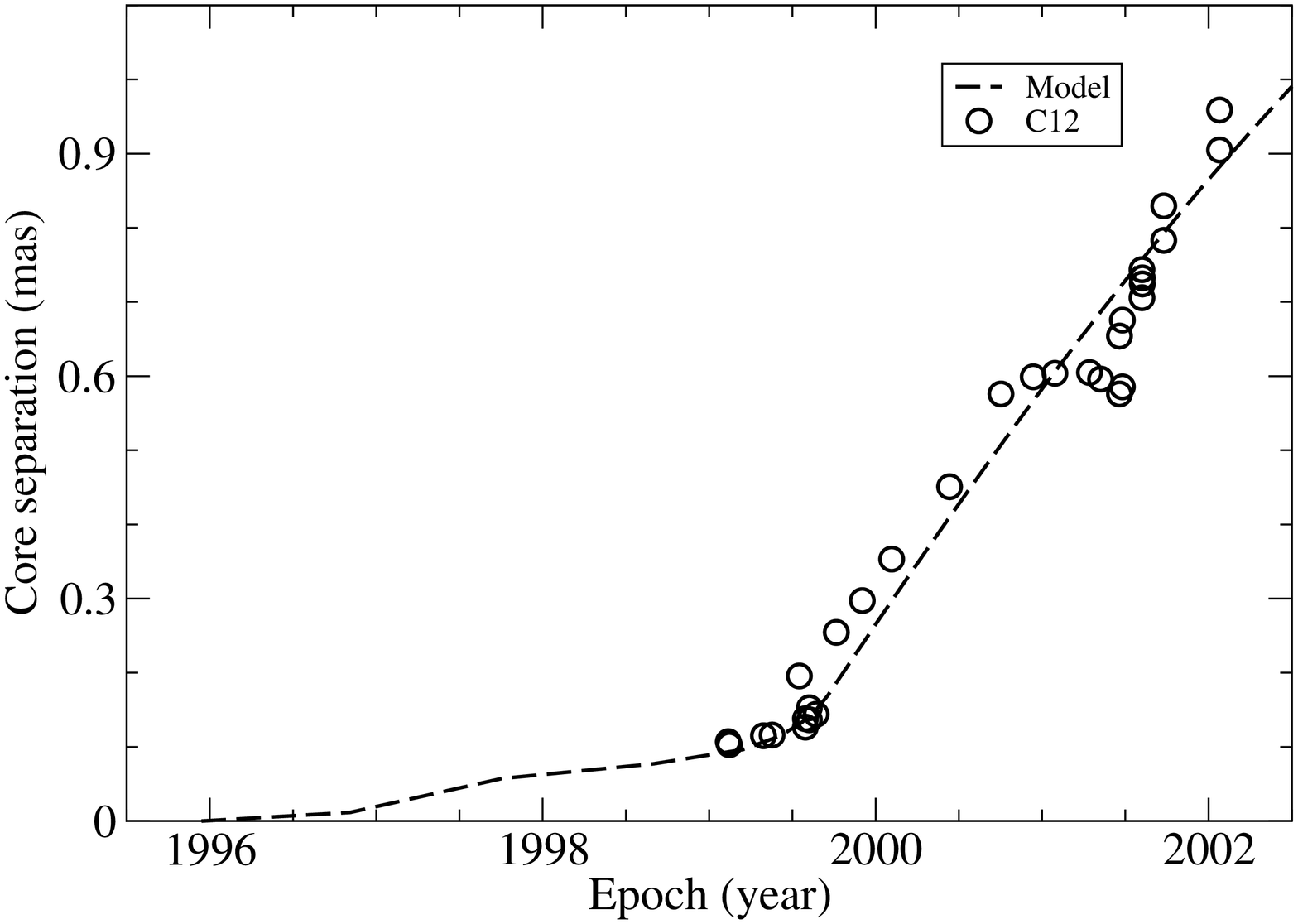}
   \includegraphics[width=6cm,angle=-90]{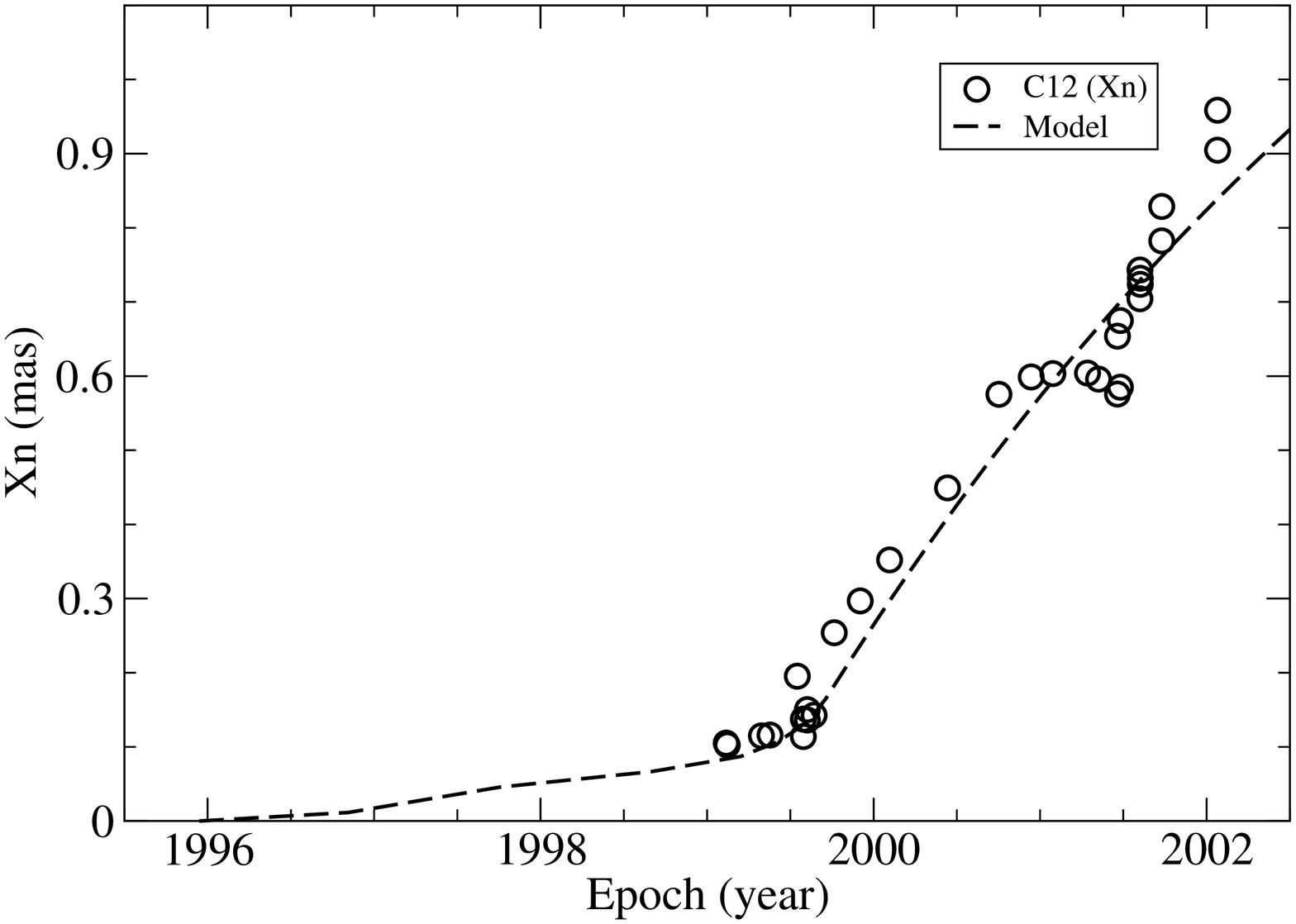}
   \includegraphics[width=6cm,angle=-90]{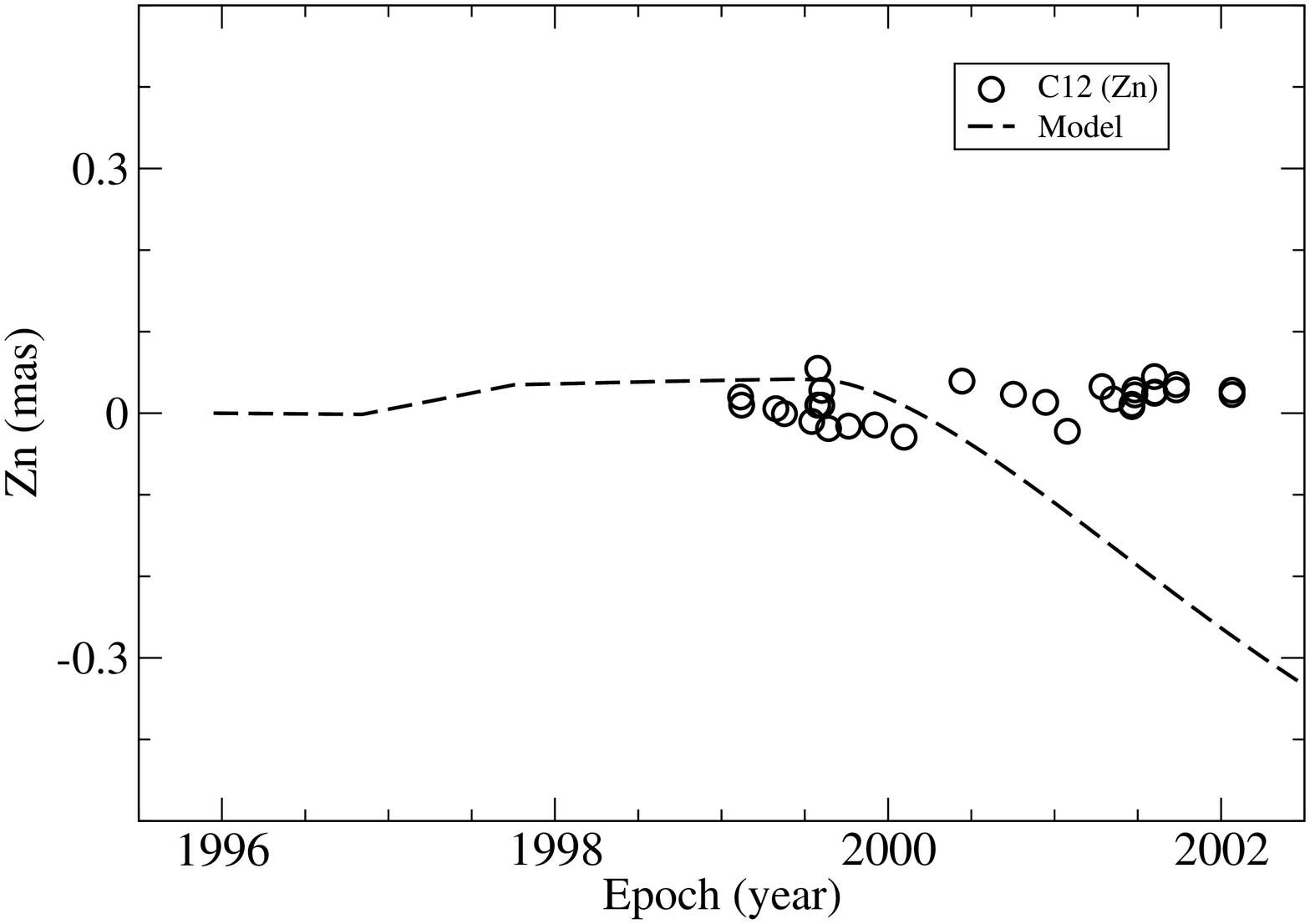}
   \includegraphics[width=6cm,angle=-90]{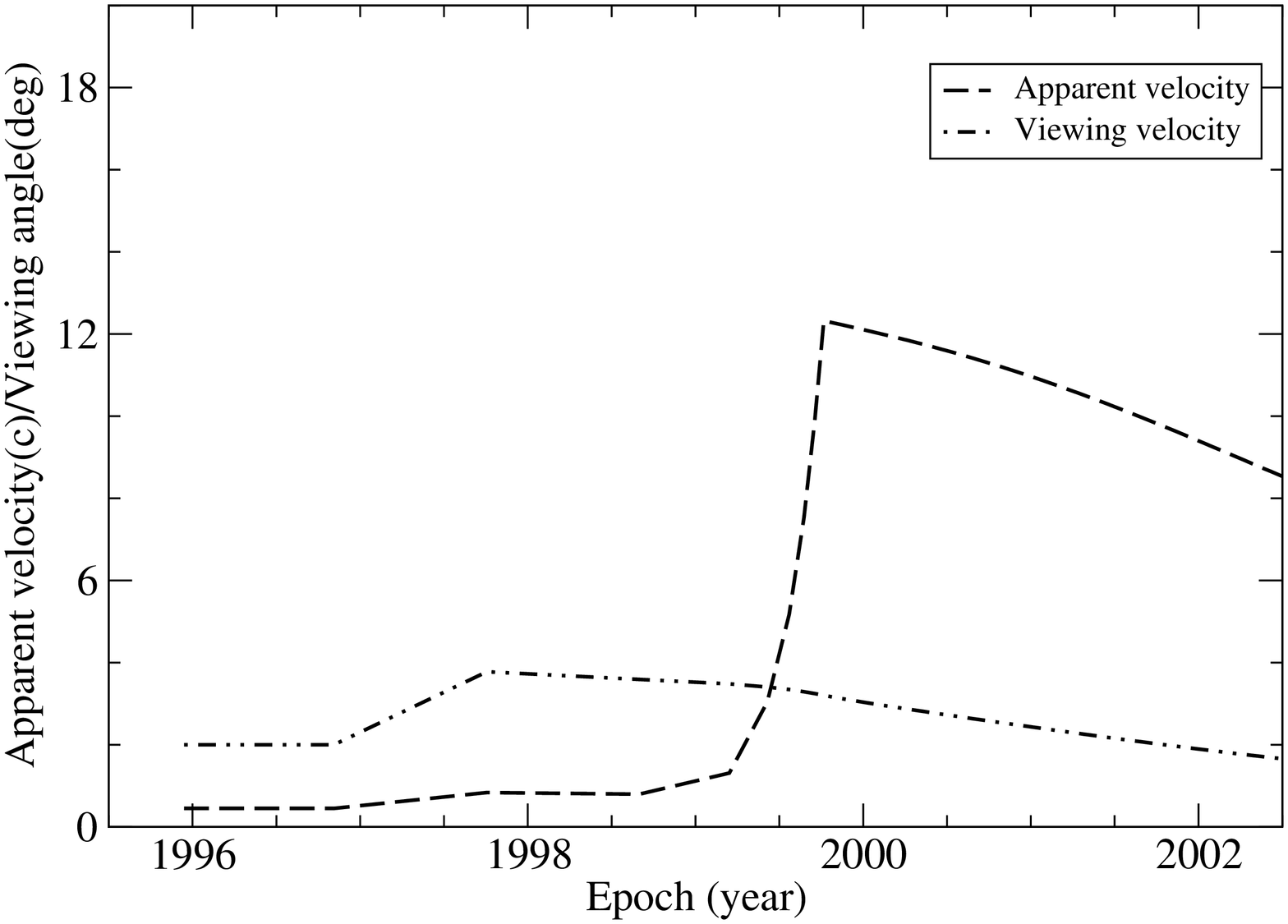}
   \includegraphics[width=6cm,angle=-90]{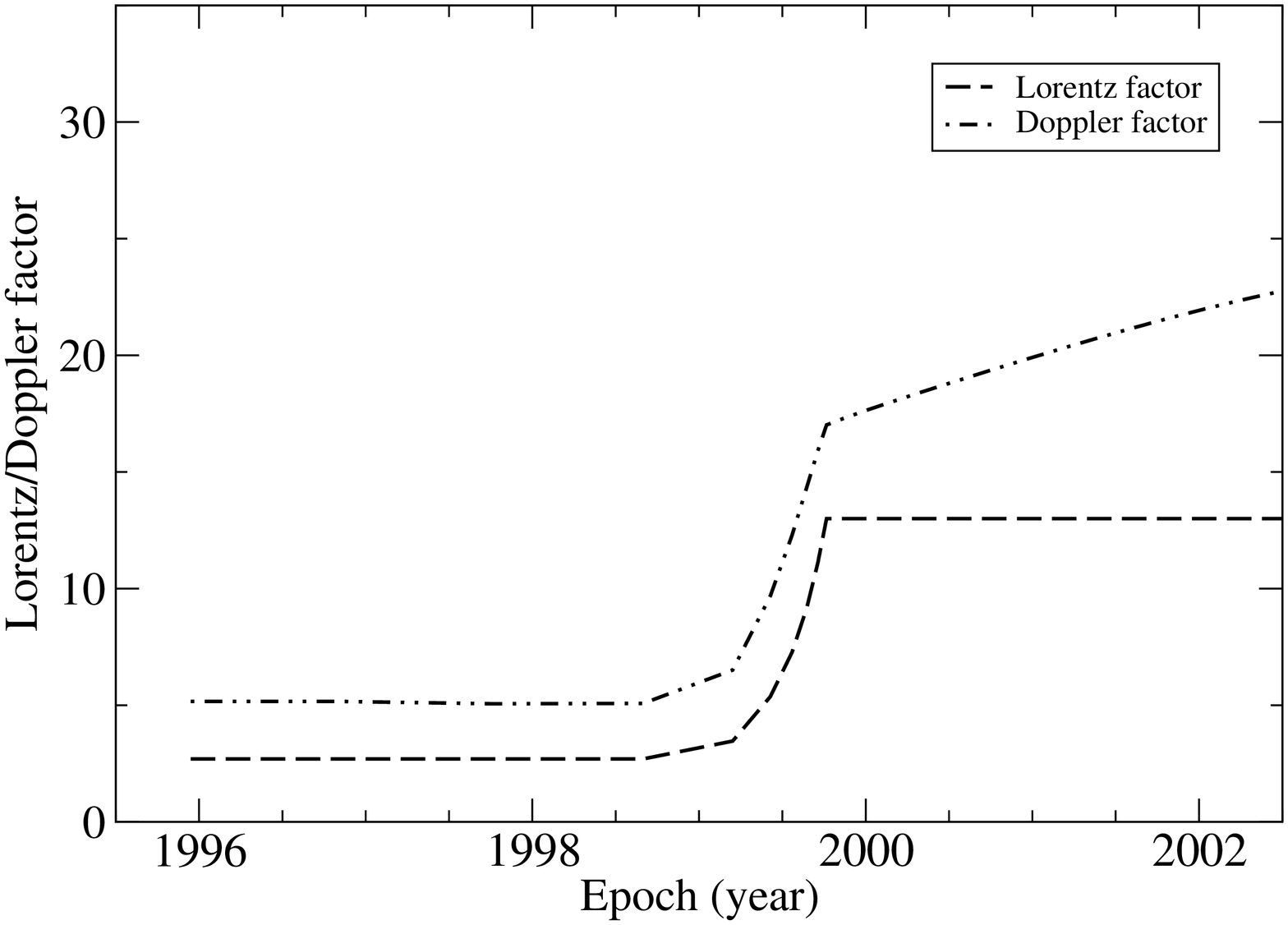}
   \caption{Knot C12: precession phase $\phi_0$(rad)=6.30+4$\pi$ and ejection
   time $t_0$=1995.95. Model-fitting results: trajectory $Z_n(X_n)$,
   coordinates $X_n(t)$ and $Z_n(t)$, core separation $r_n(t)$, modeled
   apparent velocity $\beta_a(t)$ and viewing angle $\theta$(t), bulk Lorentz
   factor $\Gamma(t)$ and Doppler factor $\delta(t)$. Its kinematics within
    $r_n{\sim}$0.5\,mas could be well fitted by the precessing nozzle model.}
   \end{figure*}
   \begin{figure*}
   \centering
   \includegraphics[width=6cm,angle=-90]{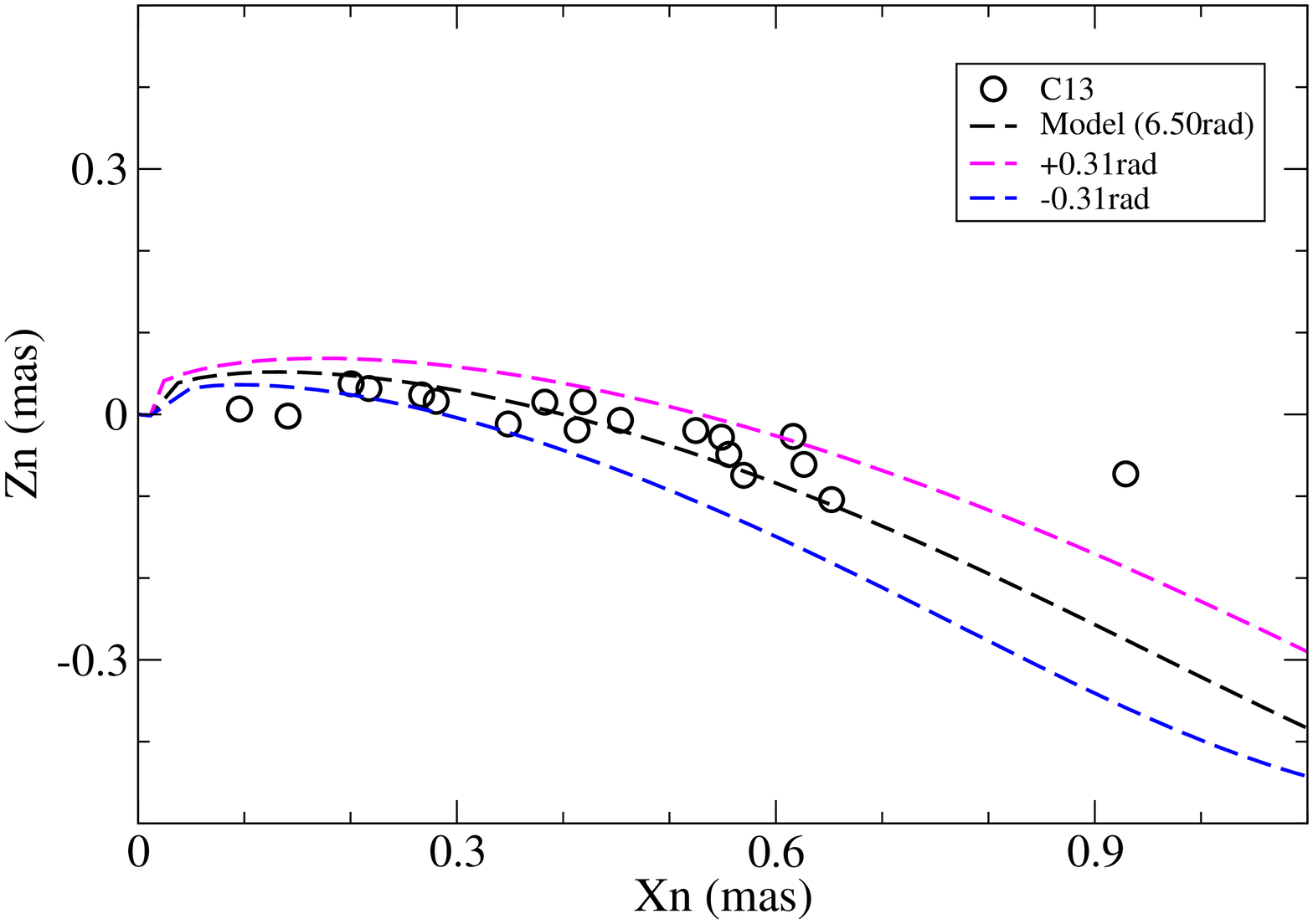}
   \includegraphics[width=6cm,angle=-90]{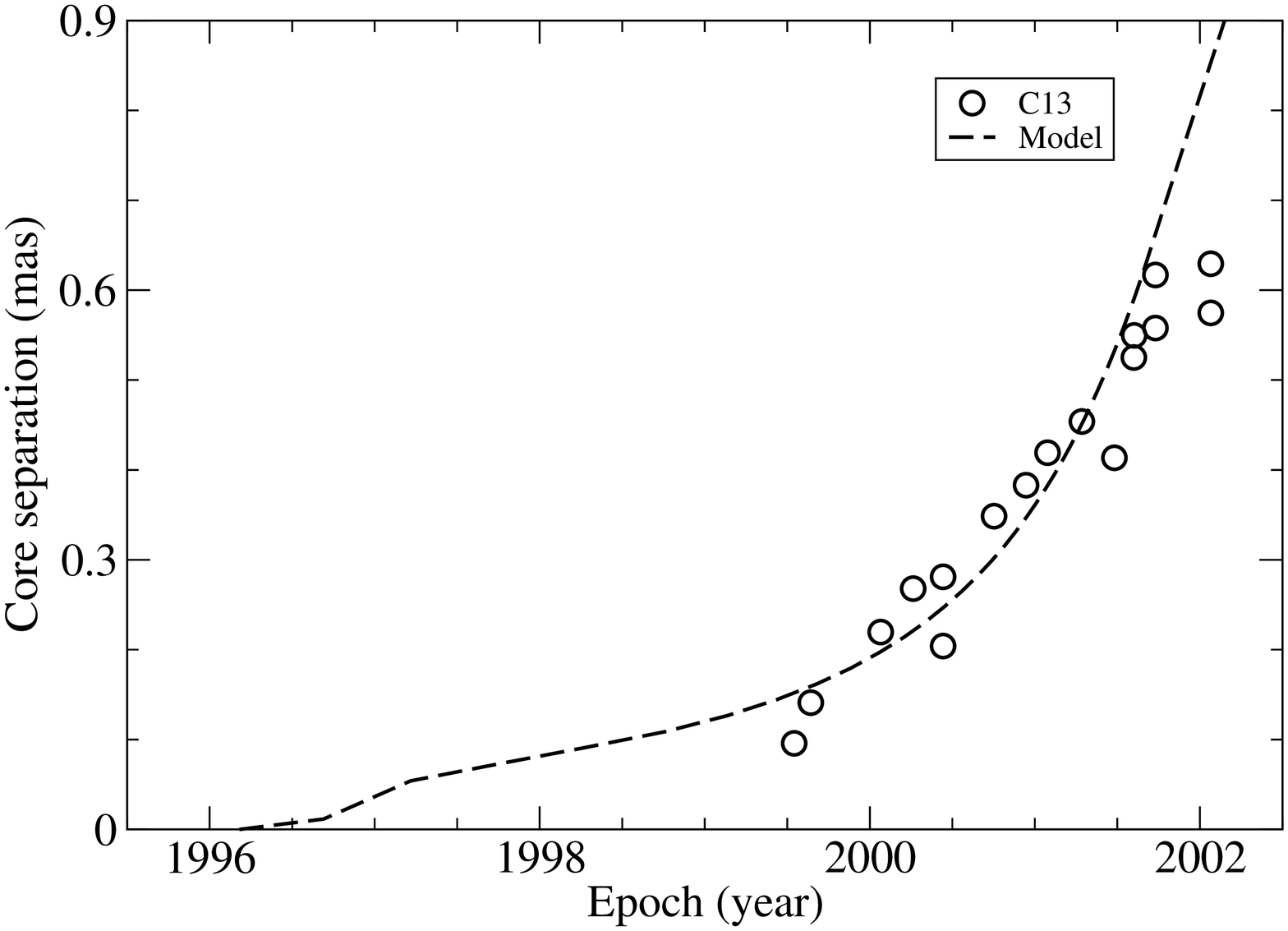}
   \includegraphics[width=6cm,angle=-90]{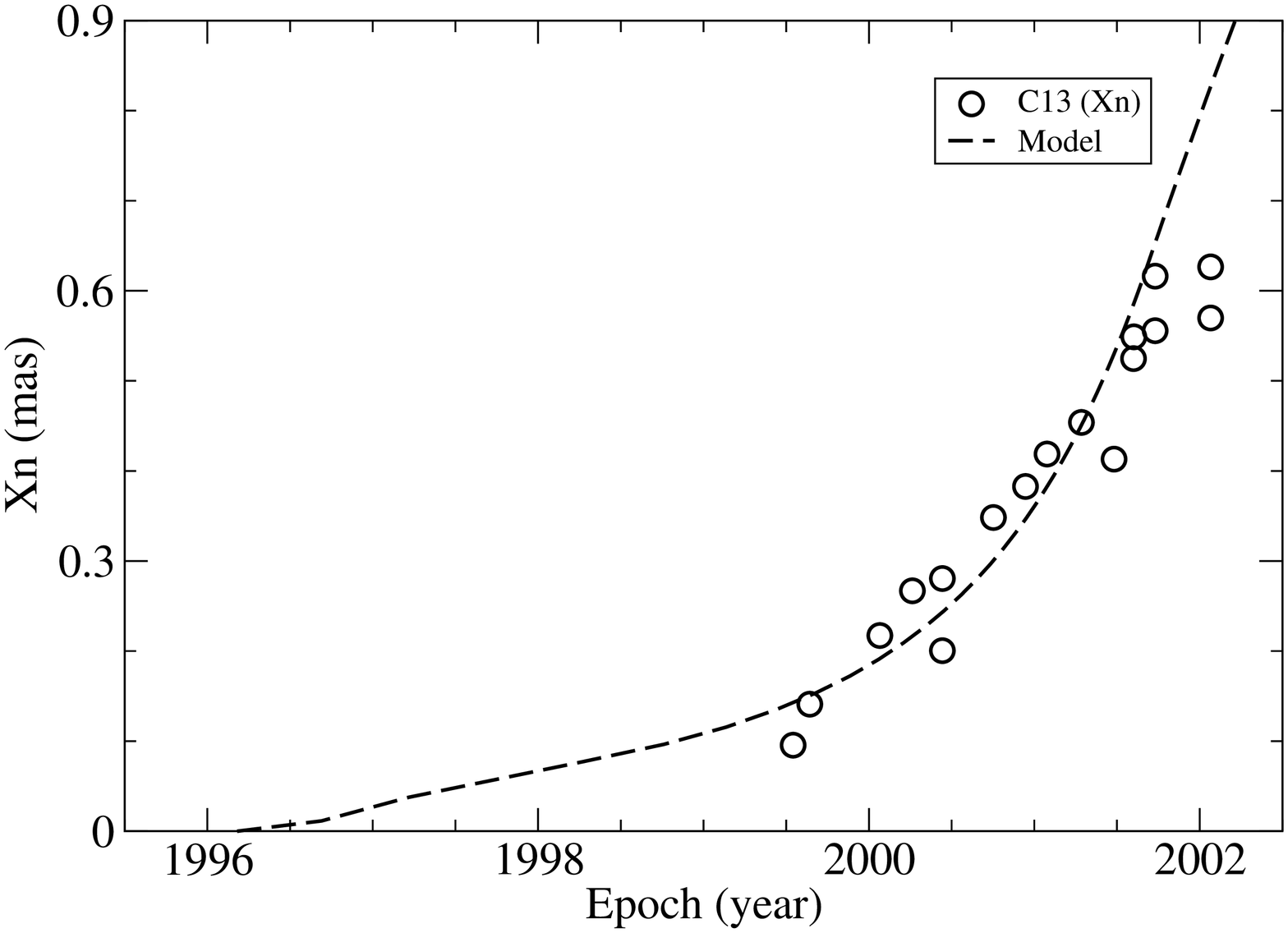}
   \includegraphics[width=6cm,angle=-90]{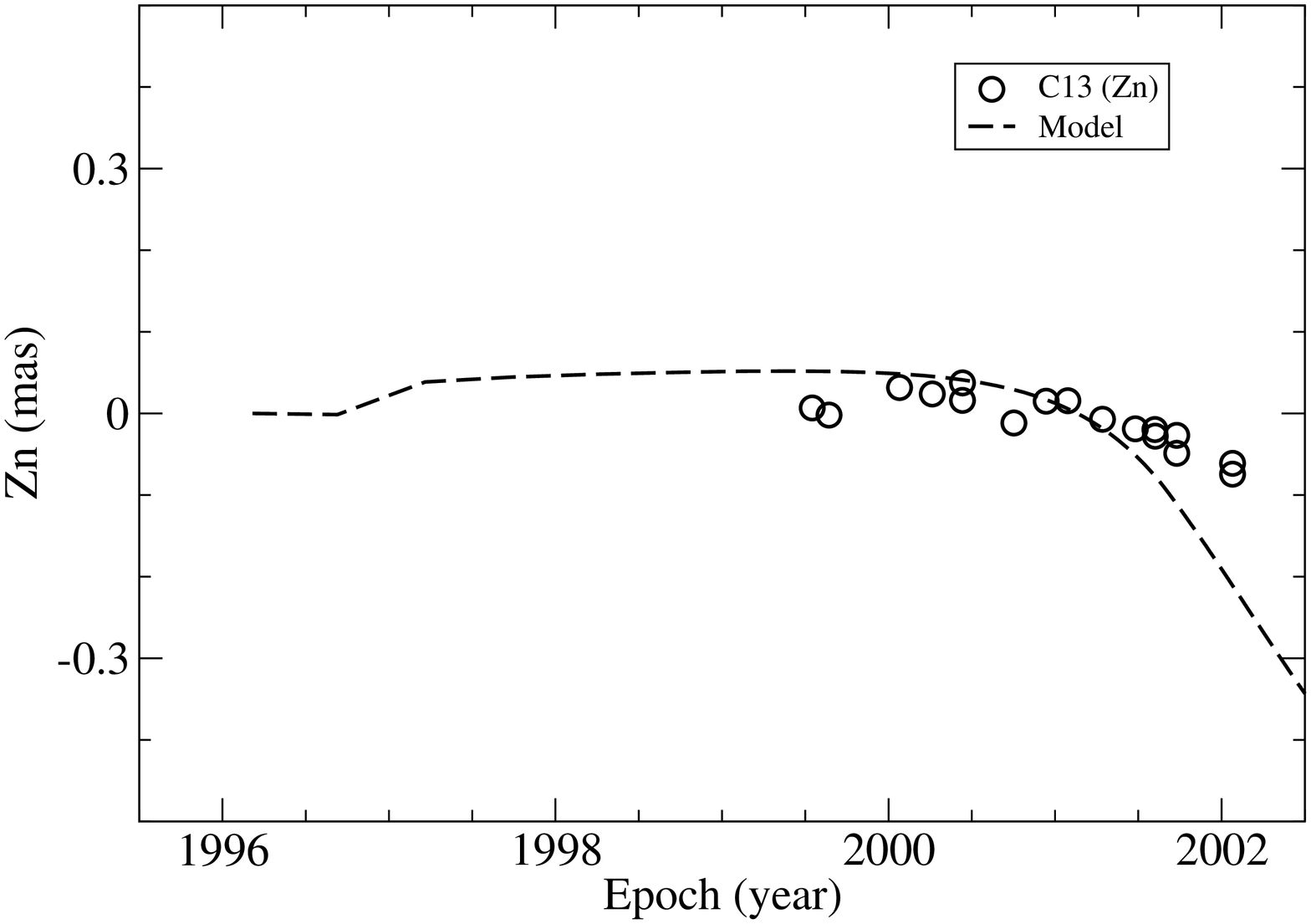}
   \includegraphics[width=6cm,angle=-90]{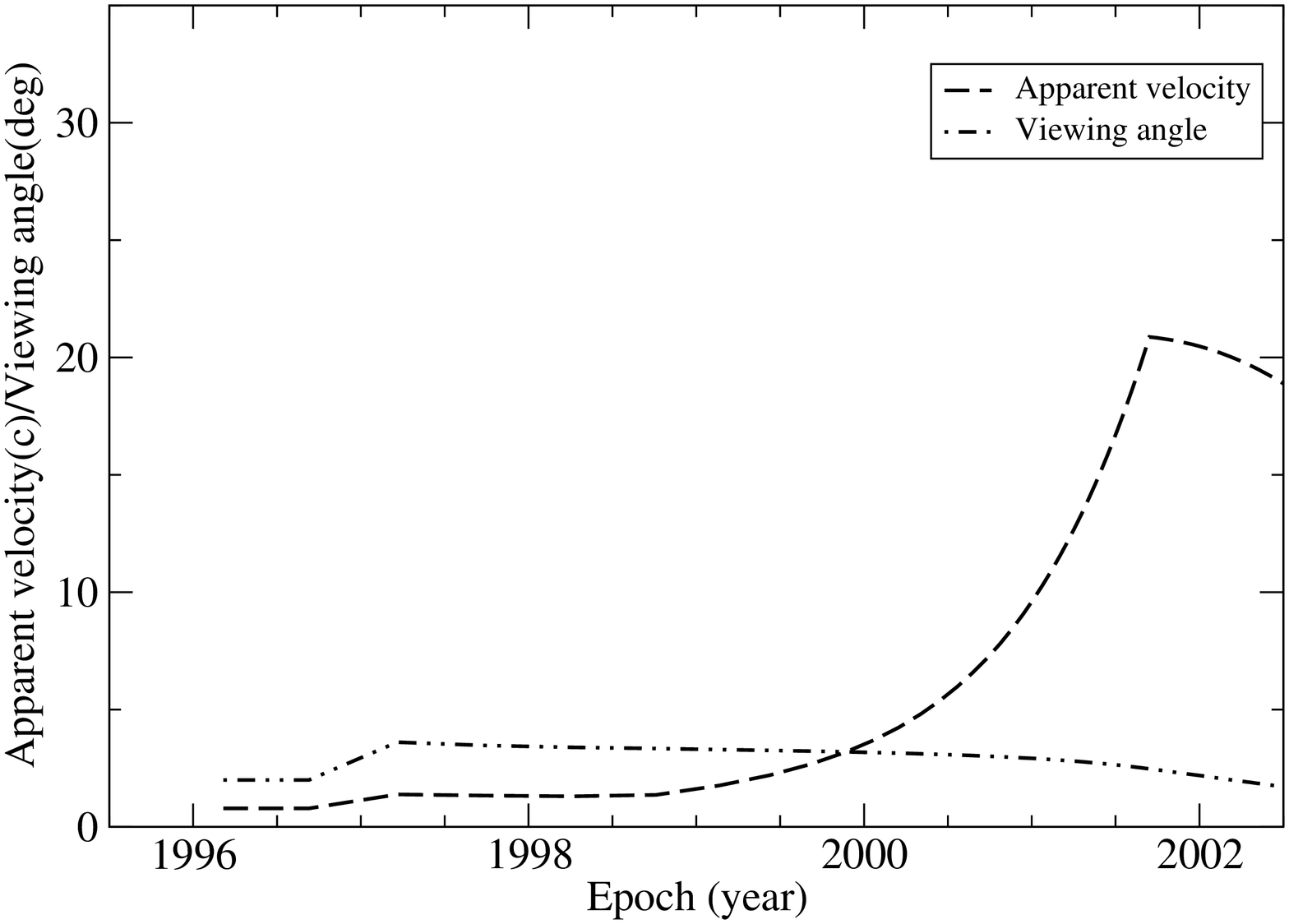}
   \includegraphics[width=6cm,angle=-90]{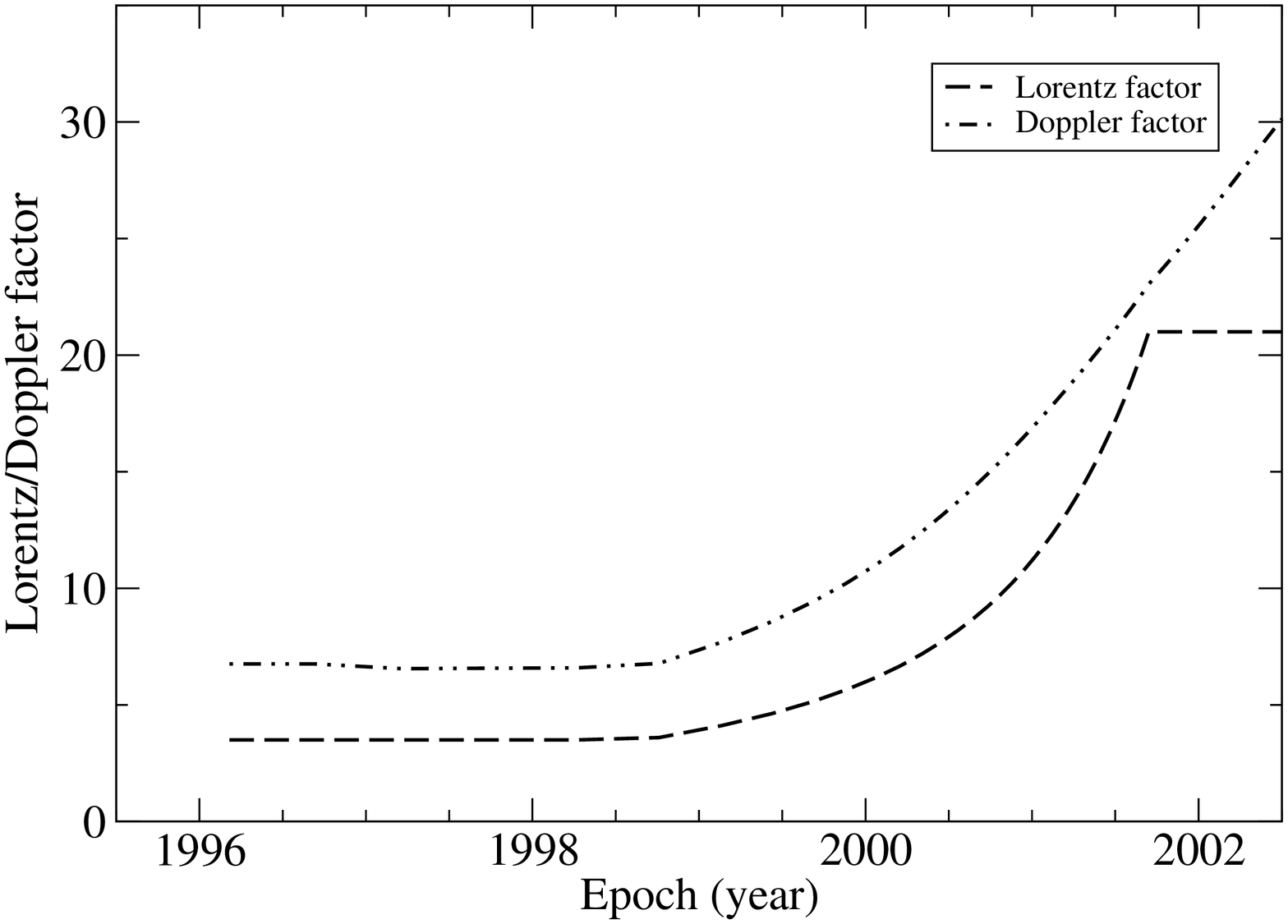}
   \caption{Knot C13: precession phase $\phi_0$(rad)=6.50+4$\pi$ and ejection
     time $t_0$=1996.18. Mode-fitting results: trajectory $Z_n(X_n)$,
   coordinates $X_n(t)$ and $Z_n(t)$, core separation ($r_n(t)$), modeled
   apparent velocity $\beta_a(t)$ and viewing angle $\theta(t)$, bulk Lorentz
   factor $\Gamma$(t) and Doppler factor $\delta(t)$. Its kinematics within
   $r_n{\sim}$0.70\,mas could be very well model-simulated in terms of
    the precessing nozzle scenario.}
   \end{figure*}
   \begin{figure*}
   \centering
   \includegraphics[width=6cm,angle=-90]{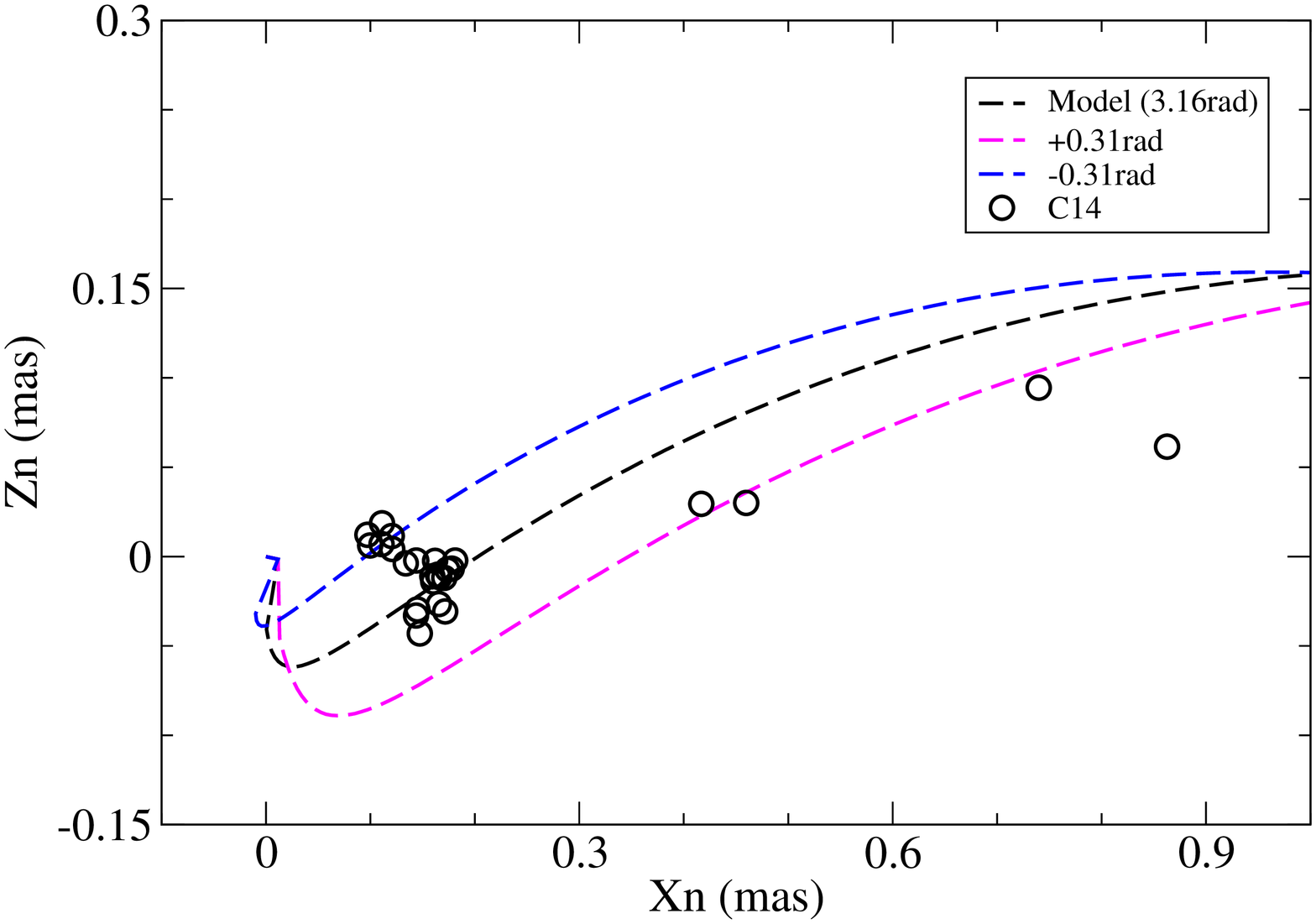}
   \includegraphics[width=6cm,angle=-90]{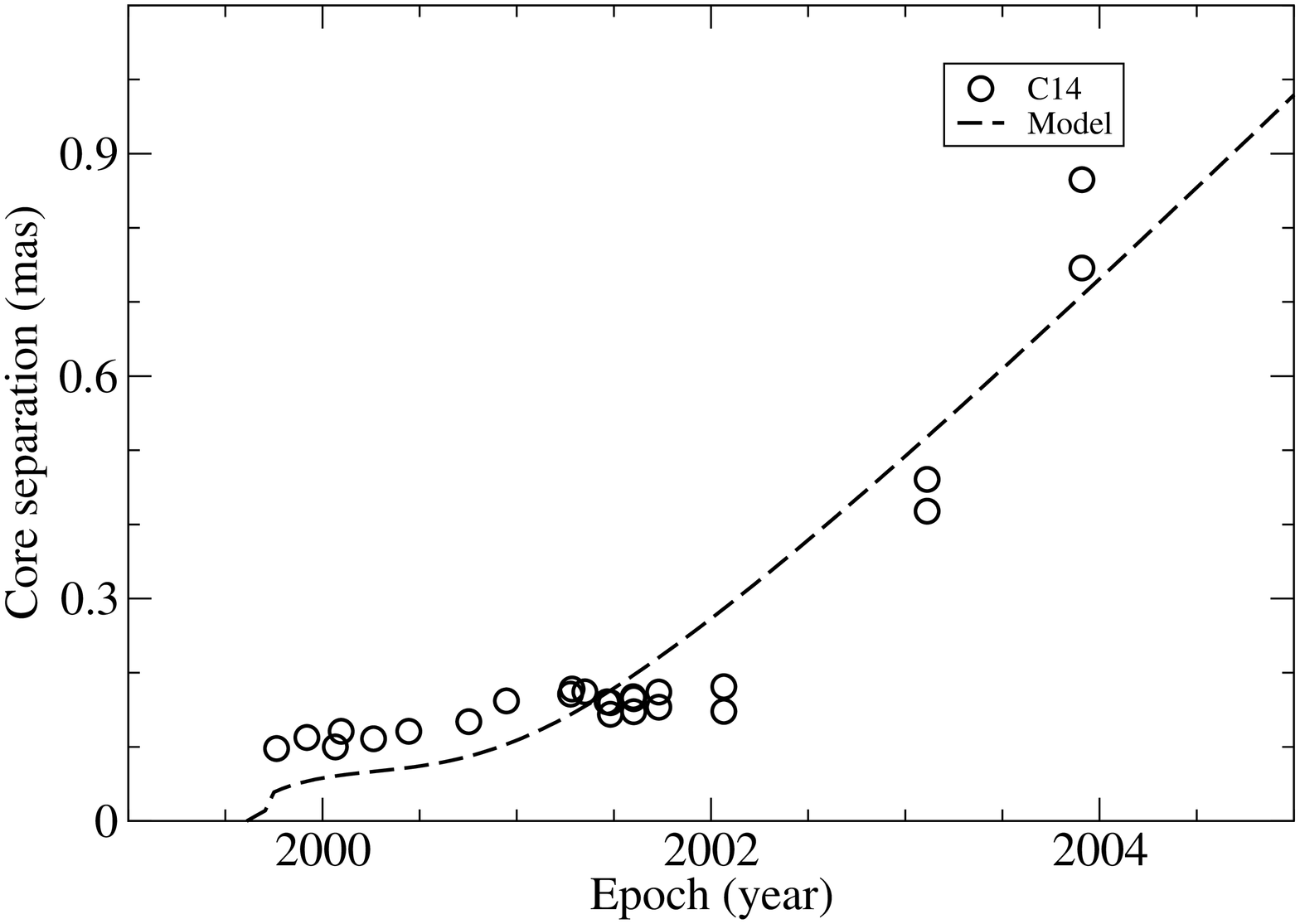}
   \includegraphics[width=6cm,angle=-90]{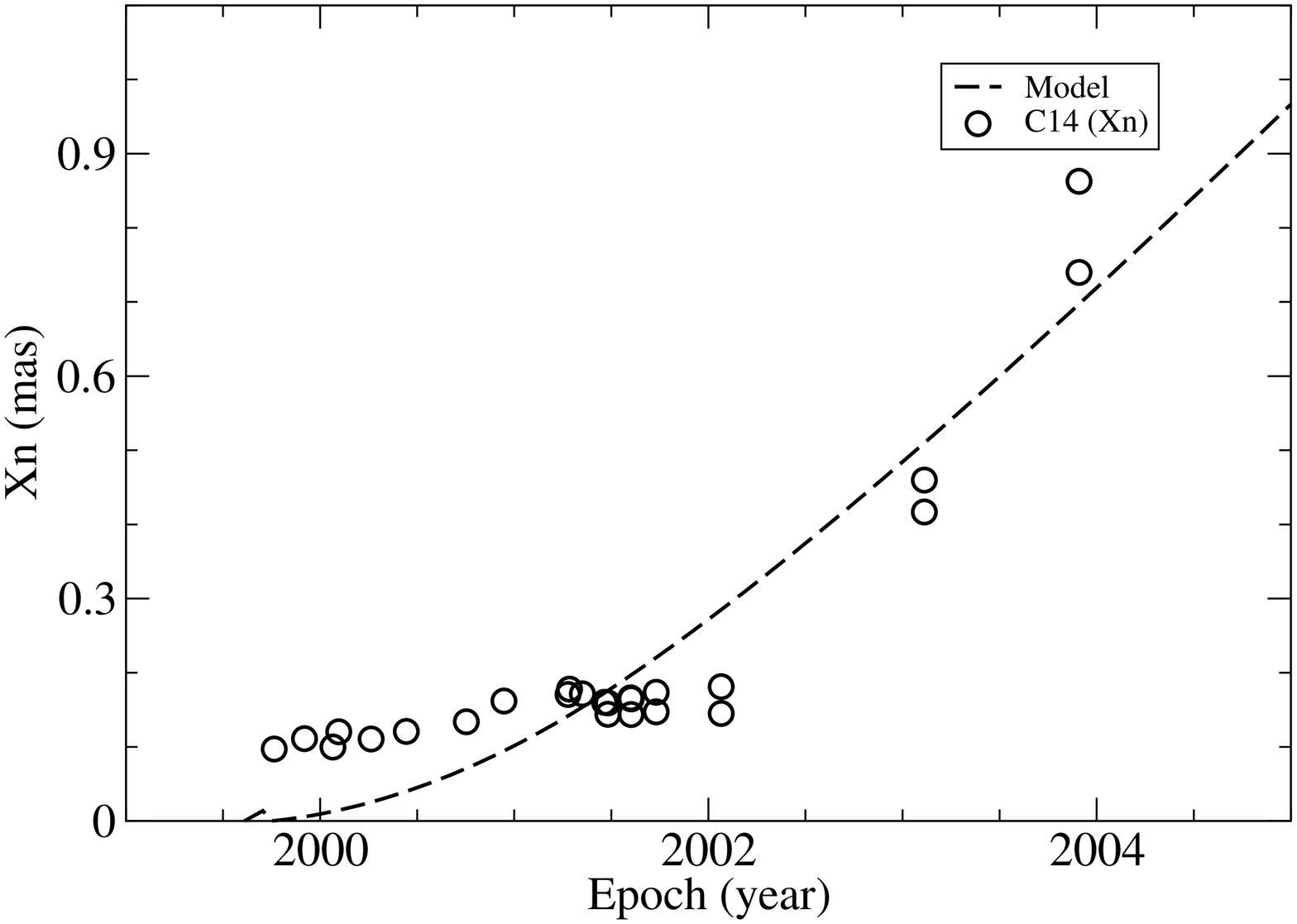}
   \includegraphics[width=6cm,angle=-90]{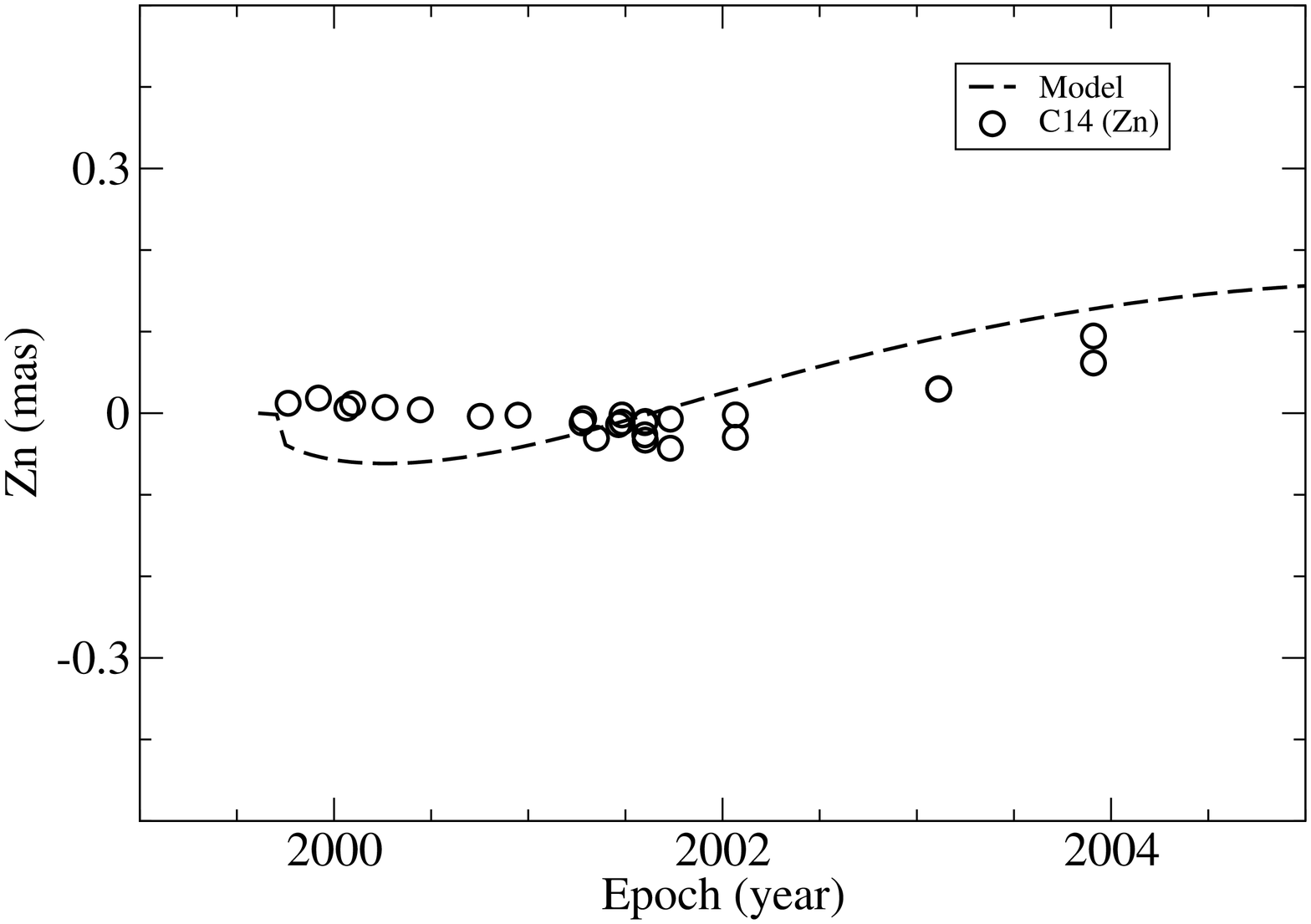}
   \includegraphics[width=6cm,angle=-90]{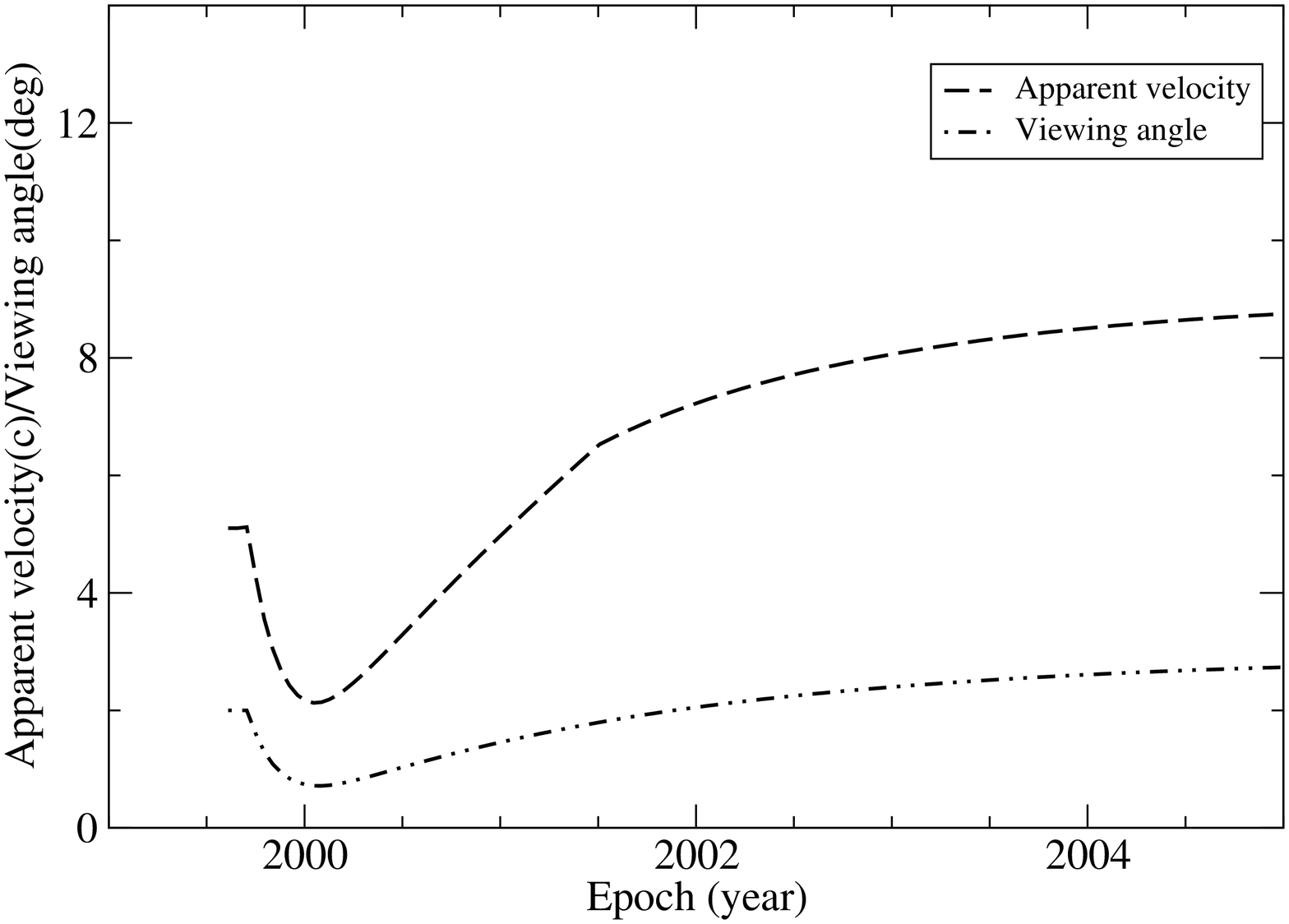}
   \includegraphics[width=6cm,angle=-90]{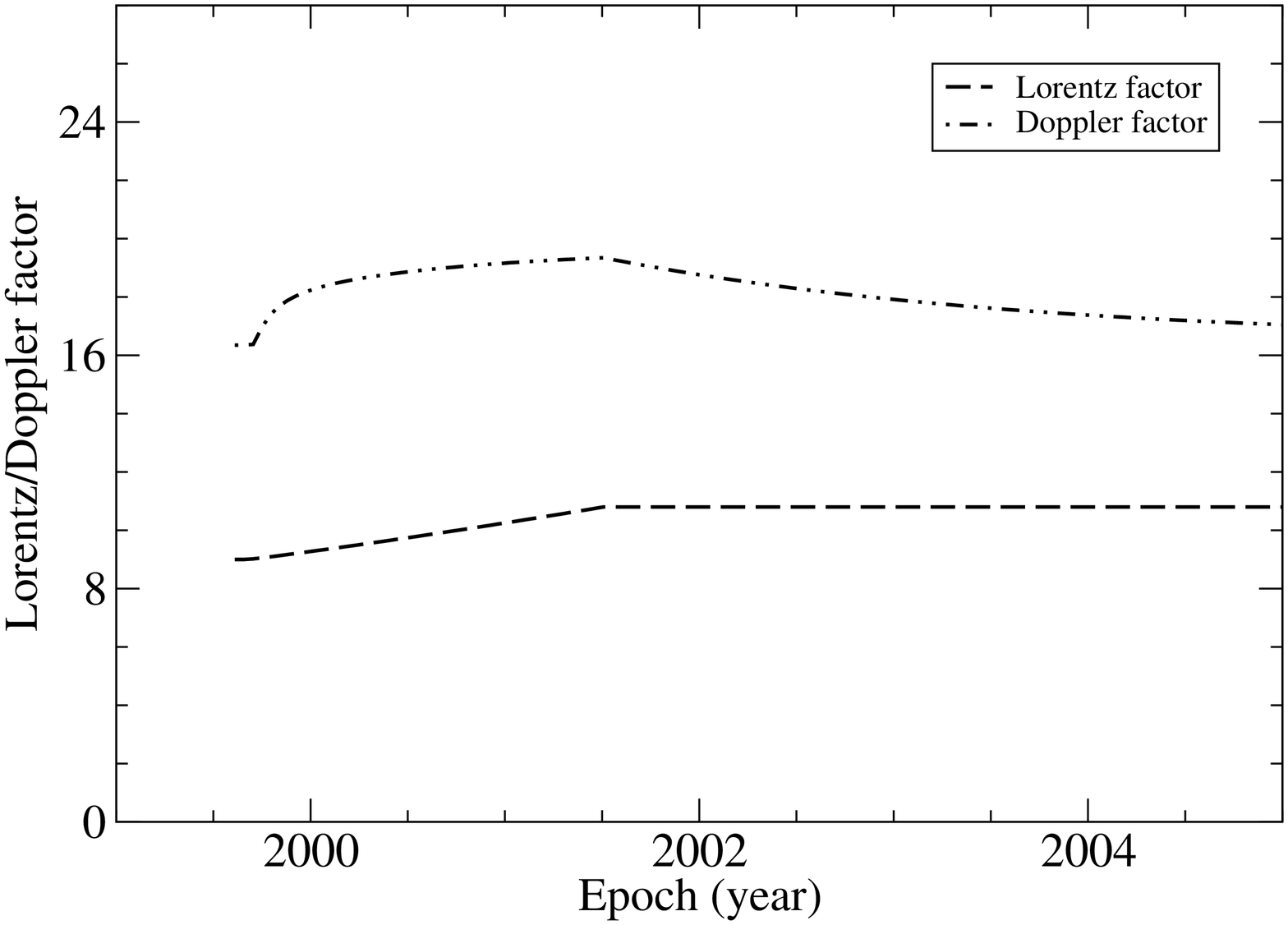}
   \caption{Knot C14: precession phase $\phi_0$(rad)=3.16+6$\pi$ and ejection
   time $t_0$=1999.61. Model-fitting results: trajectory $Z_n(X_n)$, 
   coordinates $X_n(t)$ and $Z_n(t)$, core separation $r_n(t)$, modeled
   apparent velocity $\beta_a(t)$ and viewing angle $\theta(t)$, bulk Lorentz
   factor $\Gamma$(t) and Doppler factor $\delta(t)$. Its kinematics within 
   $r_n{\sim}$0.50\,mas could be well fitted in terms of the precessing nozzle
   mdel.}
   \end{figure*}
   \begin{figure*}
   \centering
   \includegraphics[width=6cm,angle=-90]{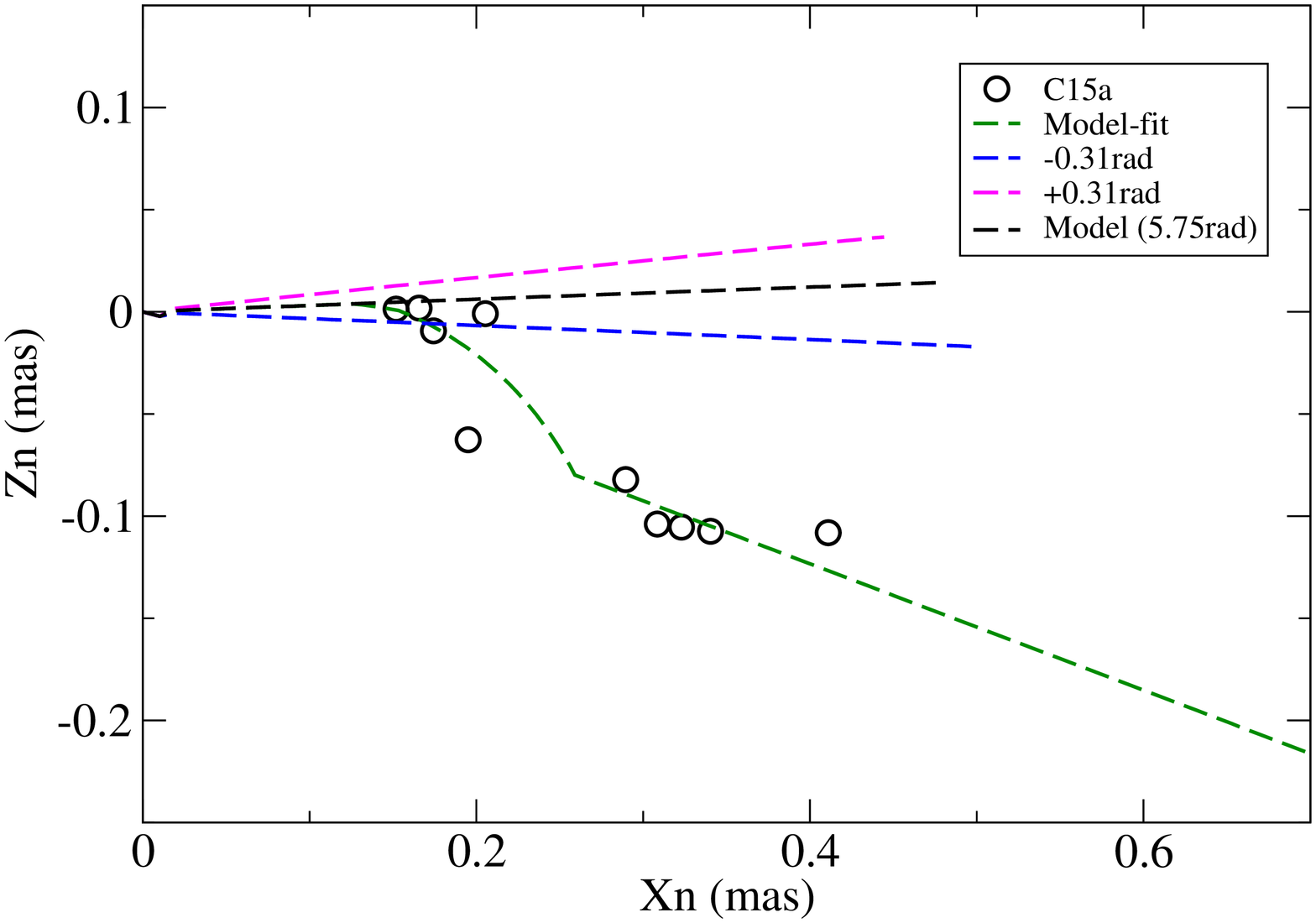}
   \includegraphics[width=6cm,angle=-90]{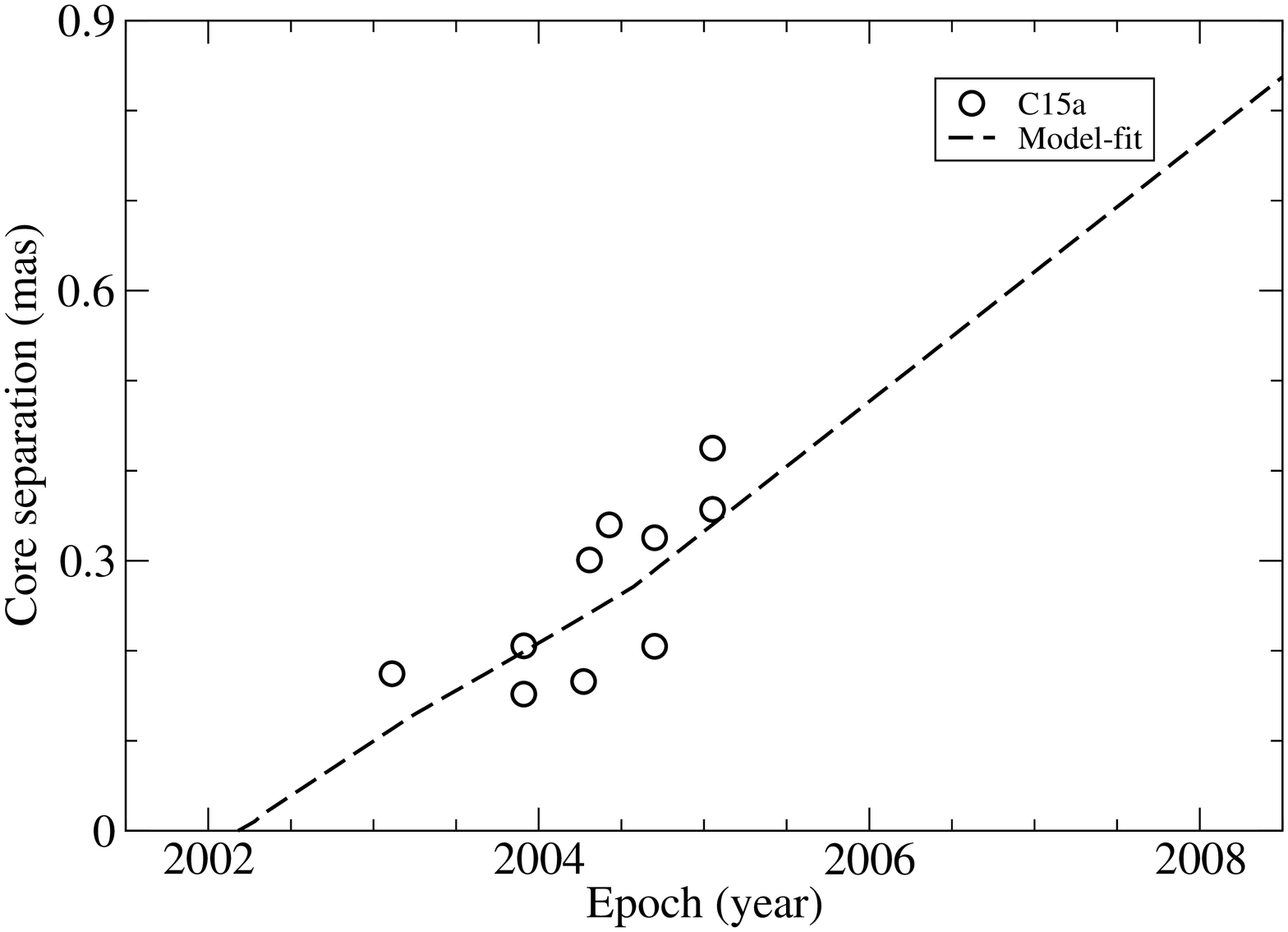}
   \includegraphics[width=6cm,angle=-90]{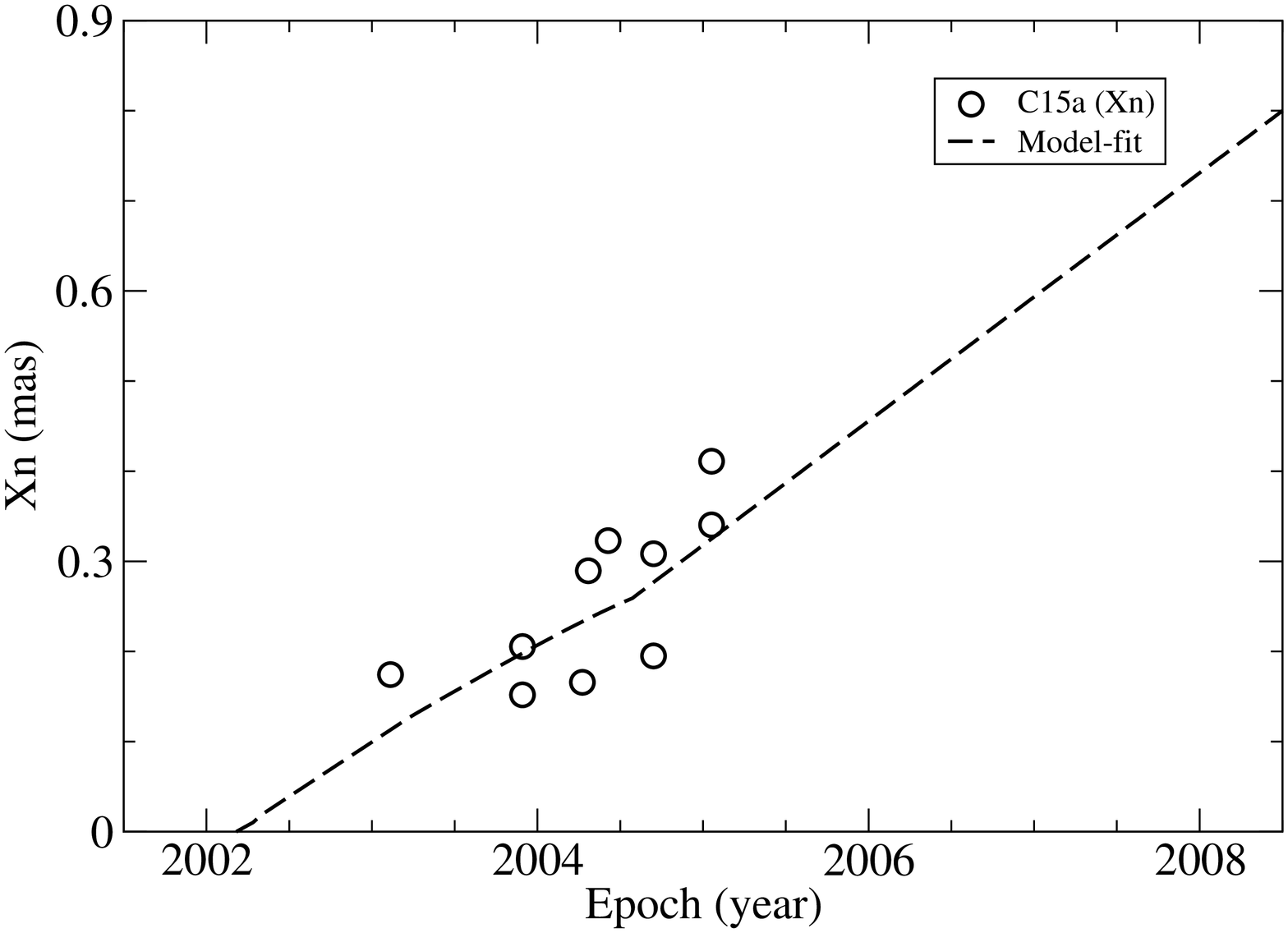}
   \includegraphics[width=6cm,angle=-90]{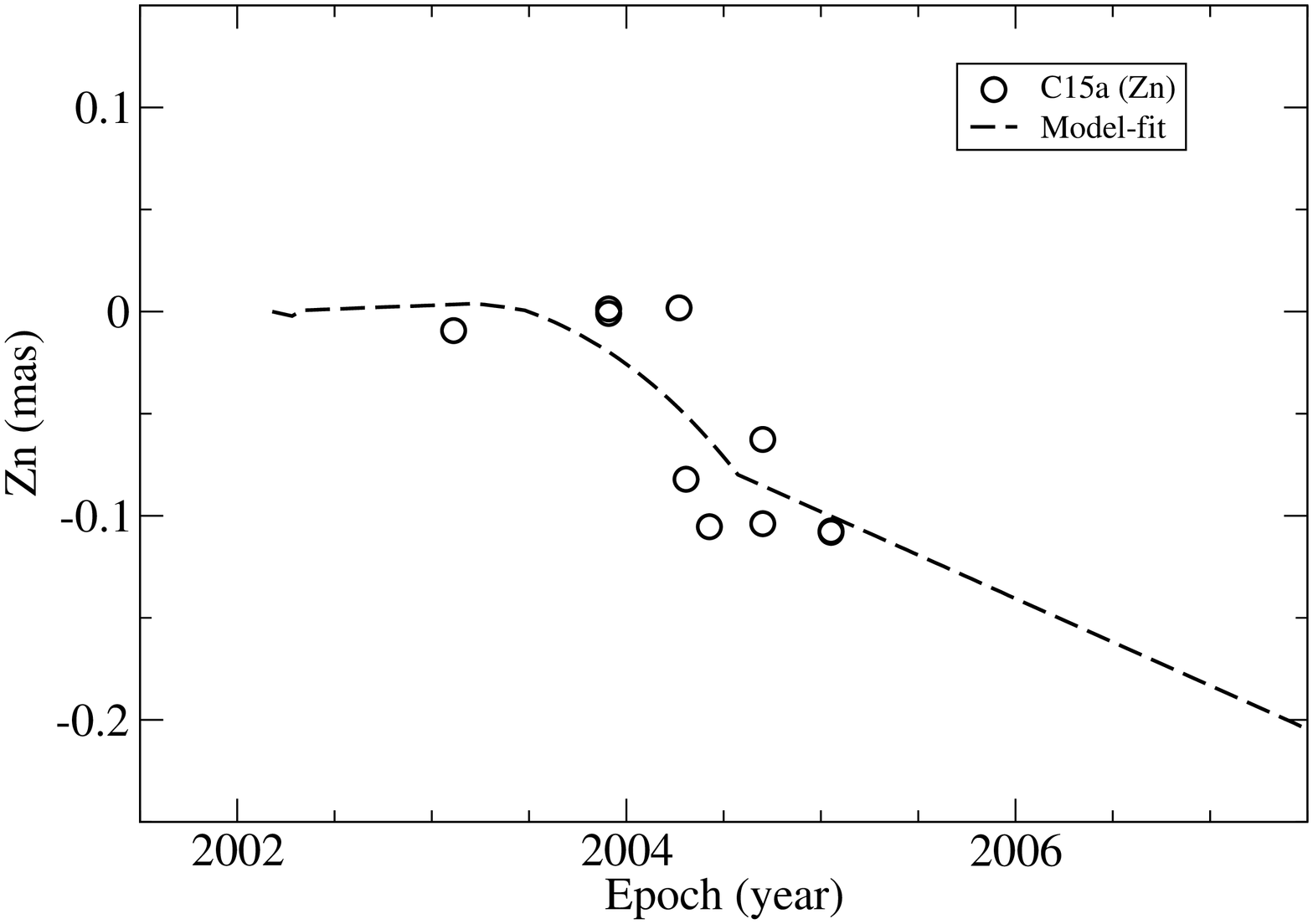}
   \includegraphics[width=6cm,angle=-90]{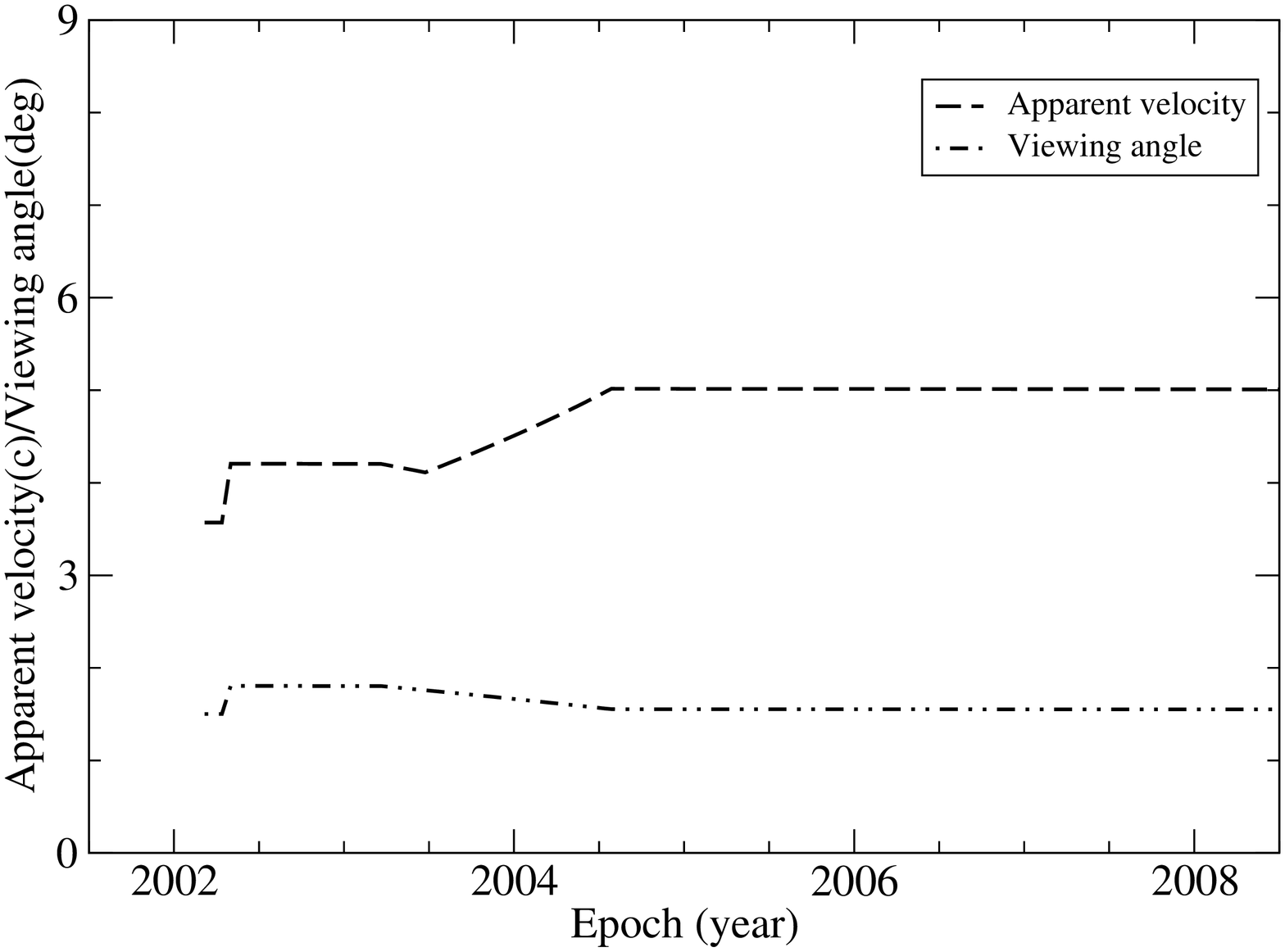}
   \includegraphics[width=6cm,angle=-90]{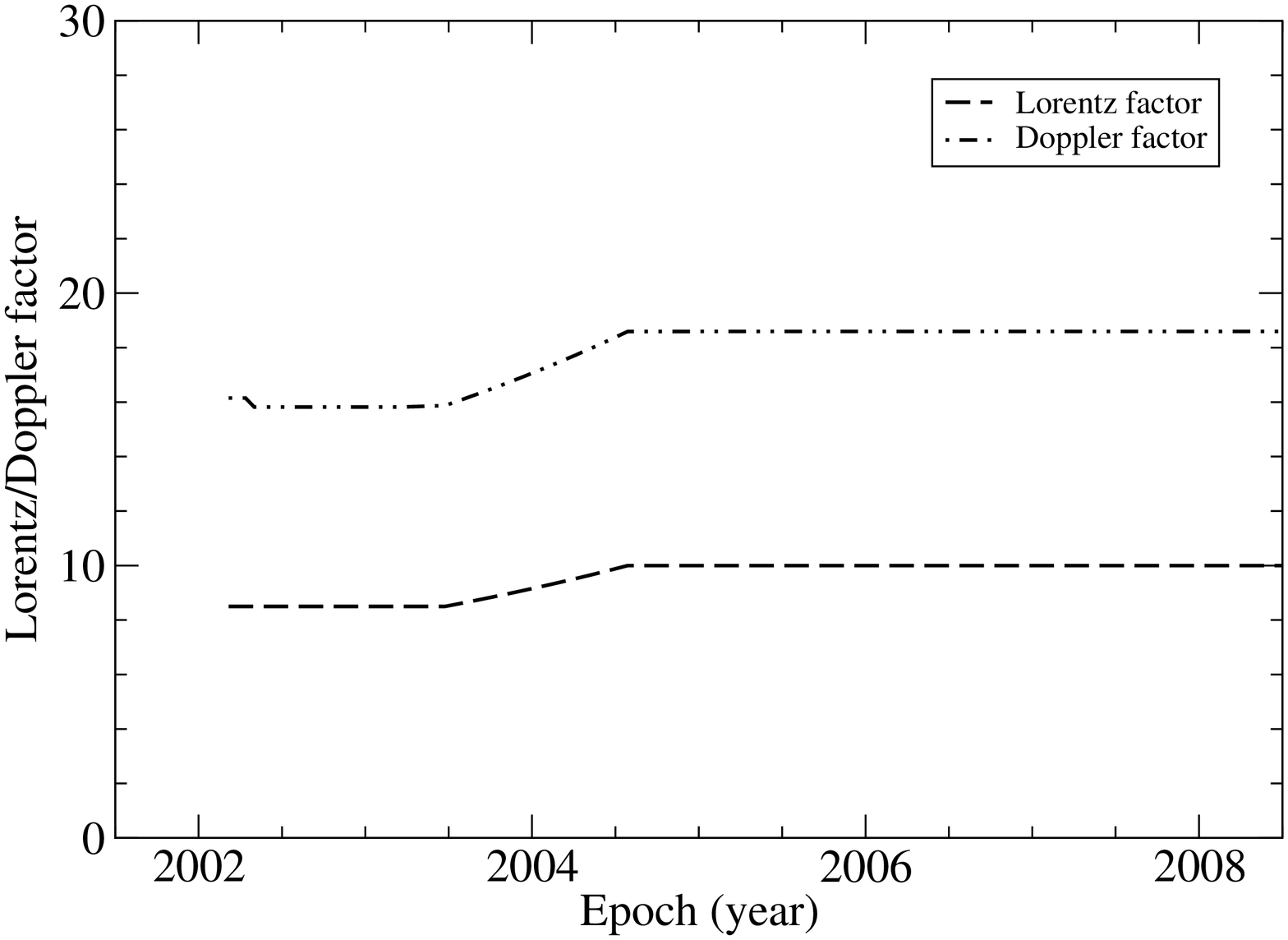}
   \caption{Knot C15a: precession phase $\phi_0$=5.75\,rad and ejection time
   $t_0$=2002.18. Model-fitting results: trajectory $Z_n(X_n)$, coordinates
   $X_n(t)$ and $Z_n(t)$, coordinates $X_n(t)$ and $Z_n(t)$, core separation
   $r_n(t)$, modeled apparent velocity $\beta_a$(t) and viewing angle
    $\theta$(t), bulk Lorentz factor $\Gamma$(t) and Doppler factor 
   $\delta(t)$. Its observed precessing common trajectory was assumed to extend
    to core separation $r_n{\sim}$0.17\,mas.}
   \end{figure*}
   \begin{figure*}
   \centering
   \includegraphics[width=6cm,angle=-90]{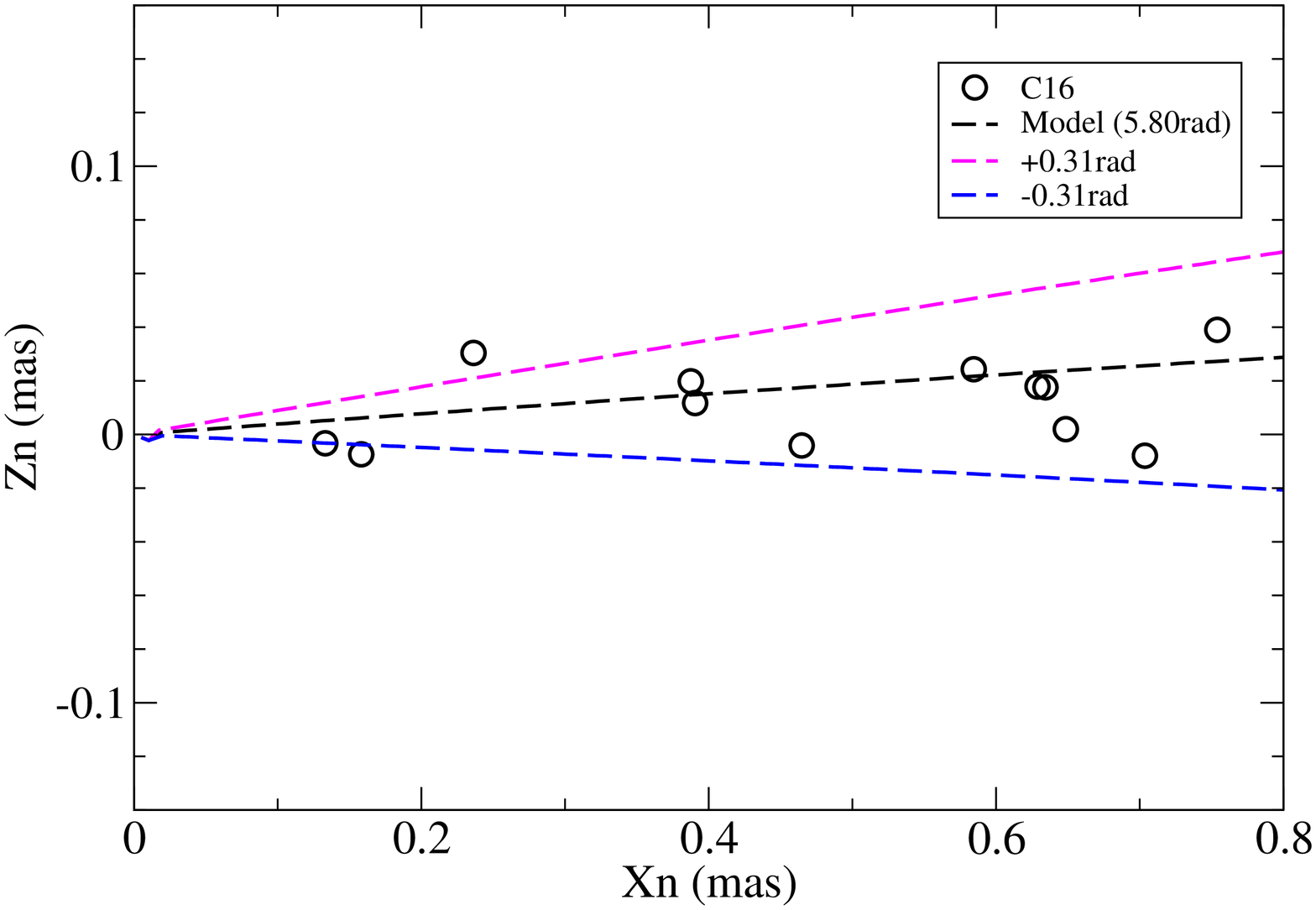} 
   \includegraphics[width=6cm,angle=-90]{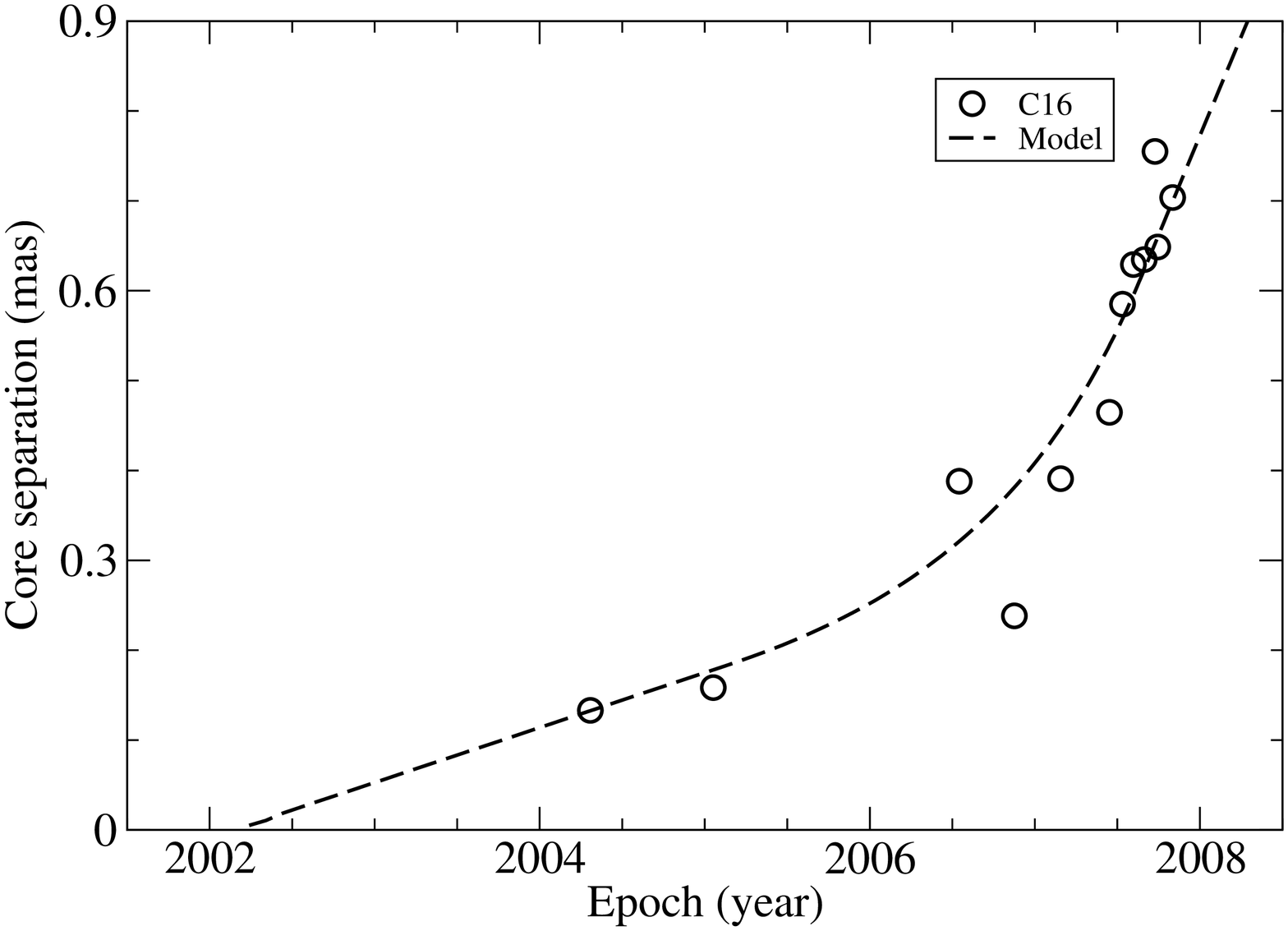}
   \includegraphics[width=6cm,angle=-90]{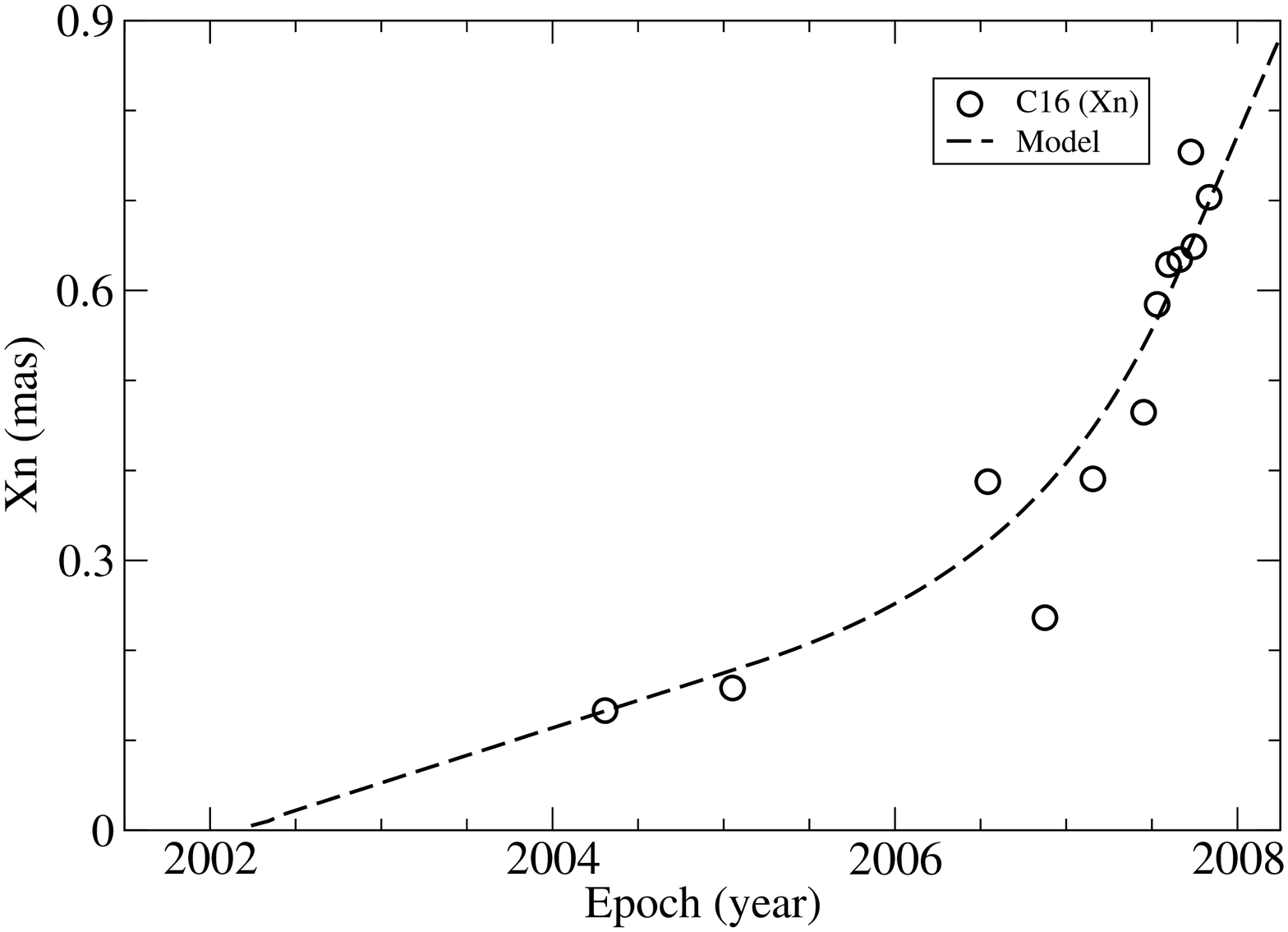}
   \includegraphics[width=6cm,angle=-90]{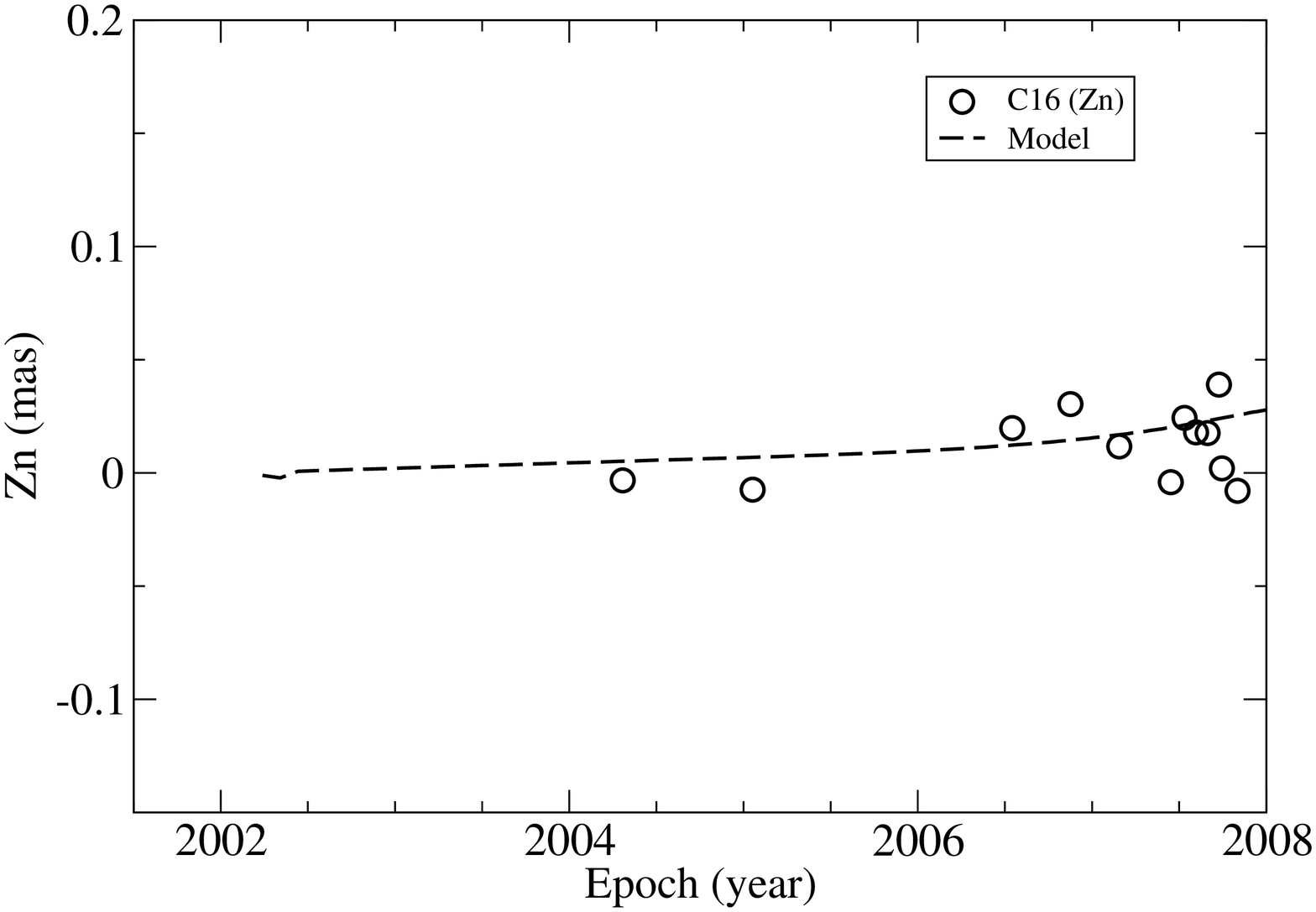}
   \includegraphics[width=6cm,angle=-90]{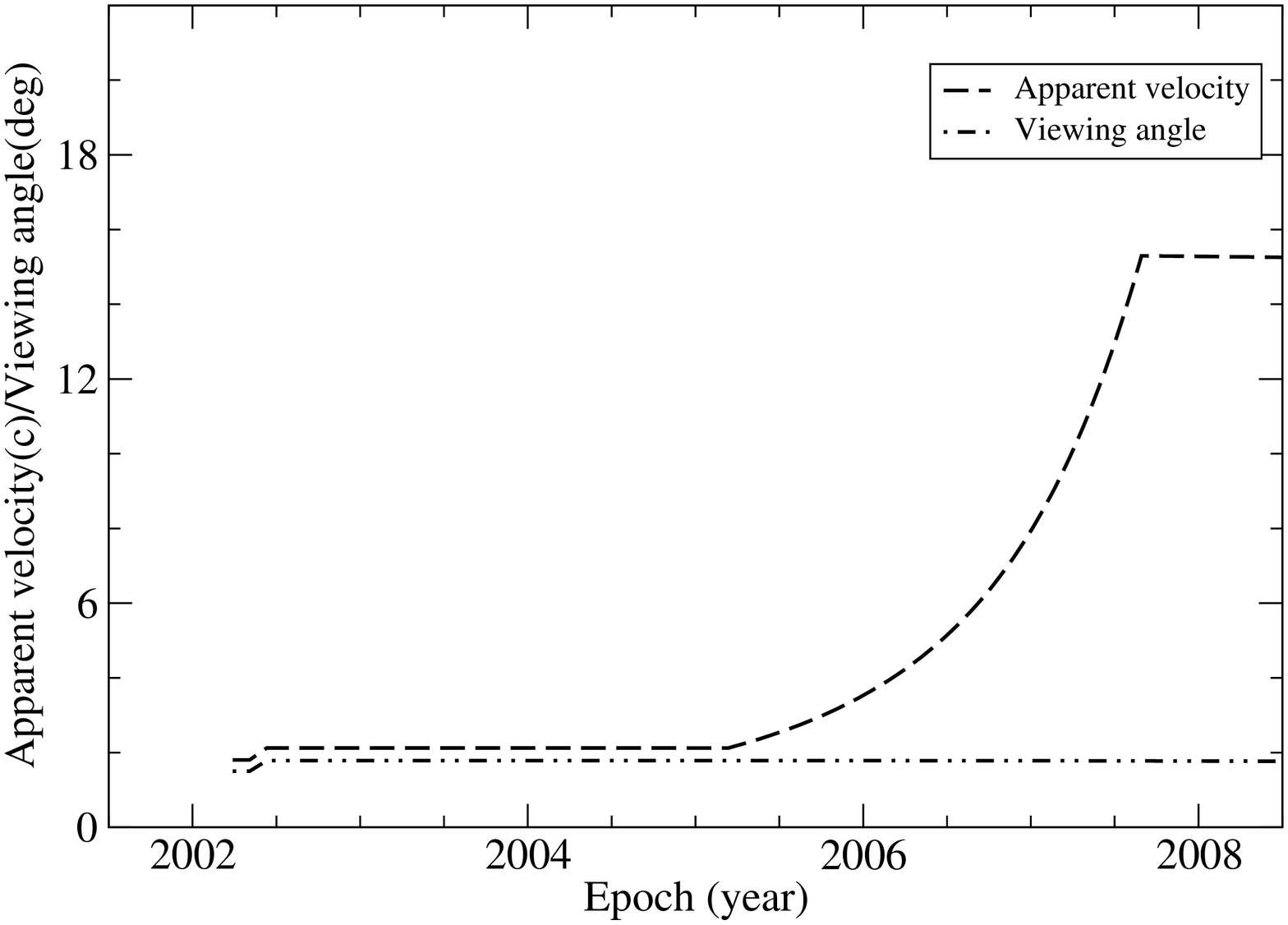}
   \includegraphics[width=6cm,angle=-90]{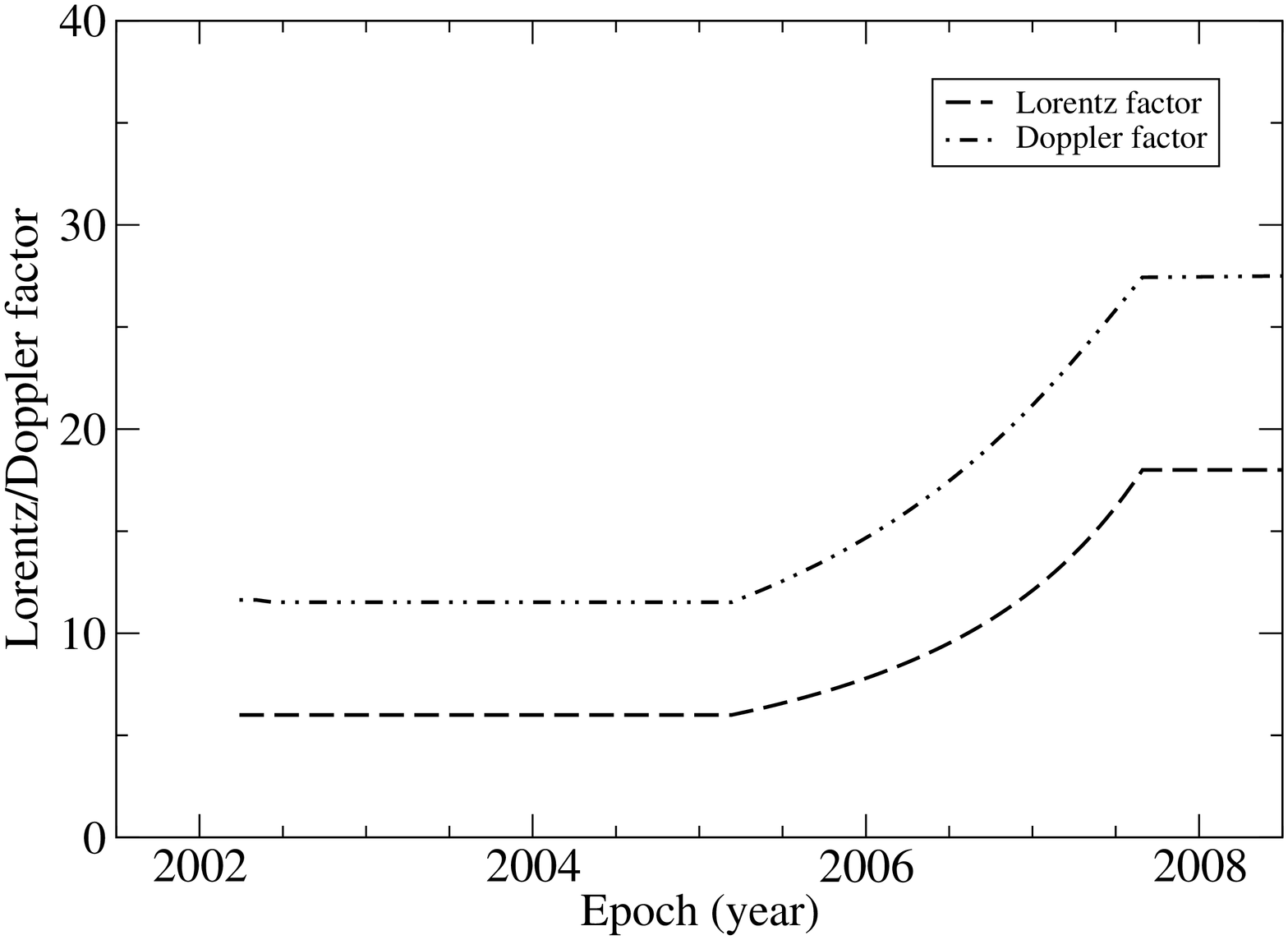}
   \caption{Knot C16: precession phase $\phi_0$=5.80\,rad and ejection time
   $t_0$=2002.24. Model-fitting results: trajectory $Z_n(X_n)$, coordinates
   $X_n(t)$ and $Z_n(t)$, core separation $r_n(t)$, modeled apparent velocity
   $\beta_a(t)$ and viewing angle $\theta(t)$, bulk Lorentz factor $\Gamma(t)$
   and Doppler factor $\delta(t)$. Its kinematics within $r_n{\sim}$0.8\,mas
   could be well model-simulated in terms of the precessing nozzle scenario.}
   \end{figure*}
   \begin{figure*}
   \centering
   \includegraphics[width=6cm,angle=-90]{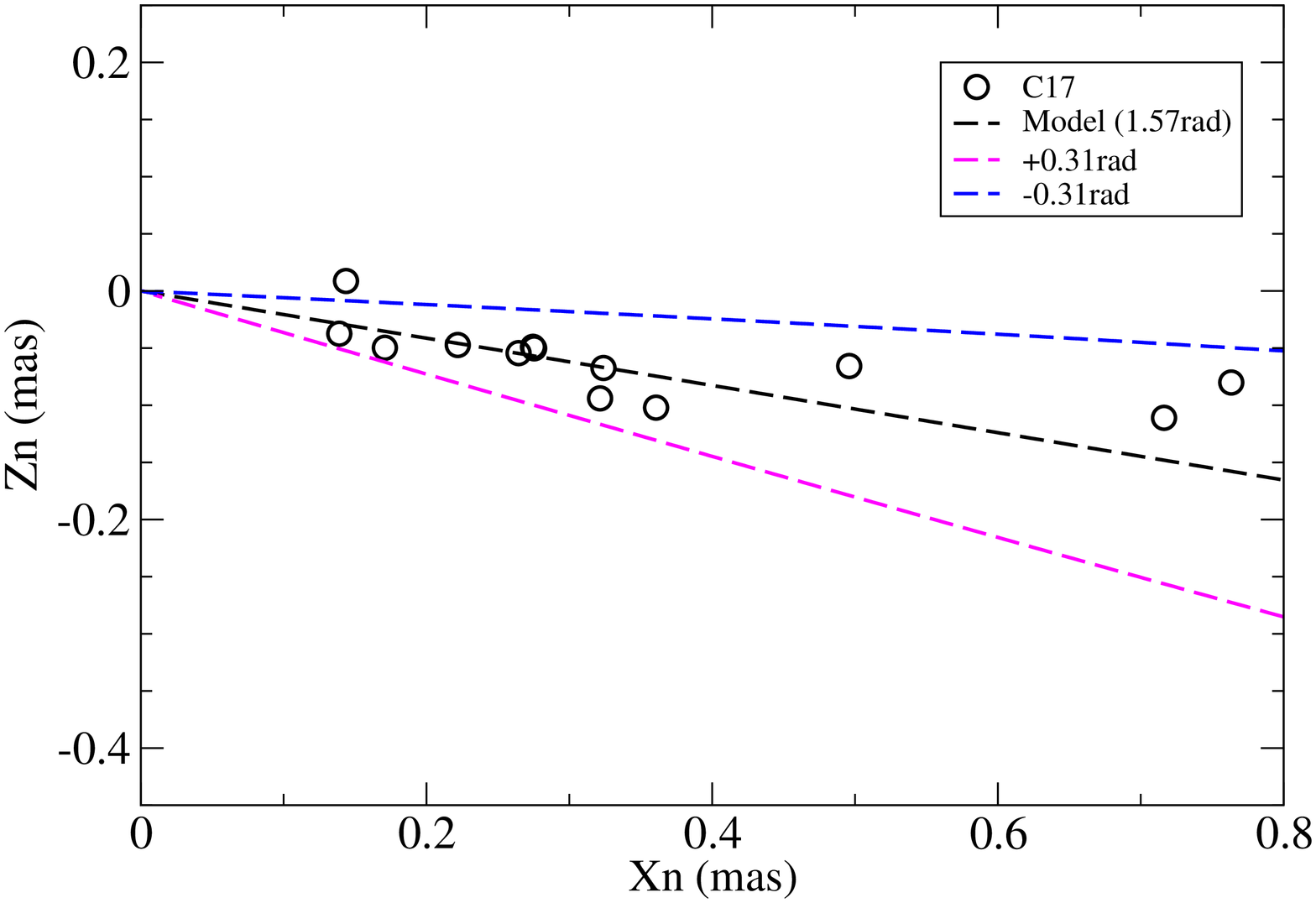}
   \includegraphics[width=6cm,angle=-90]{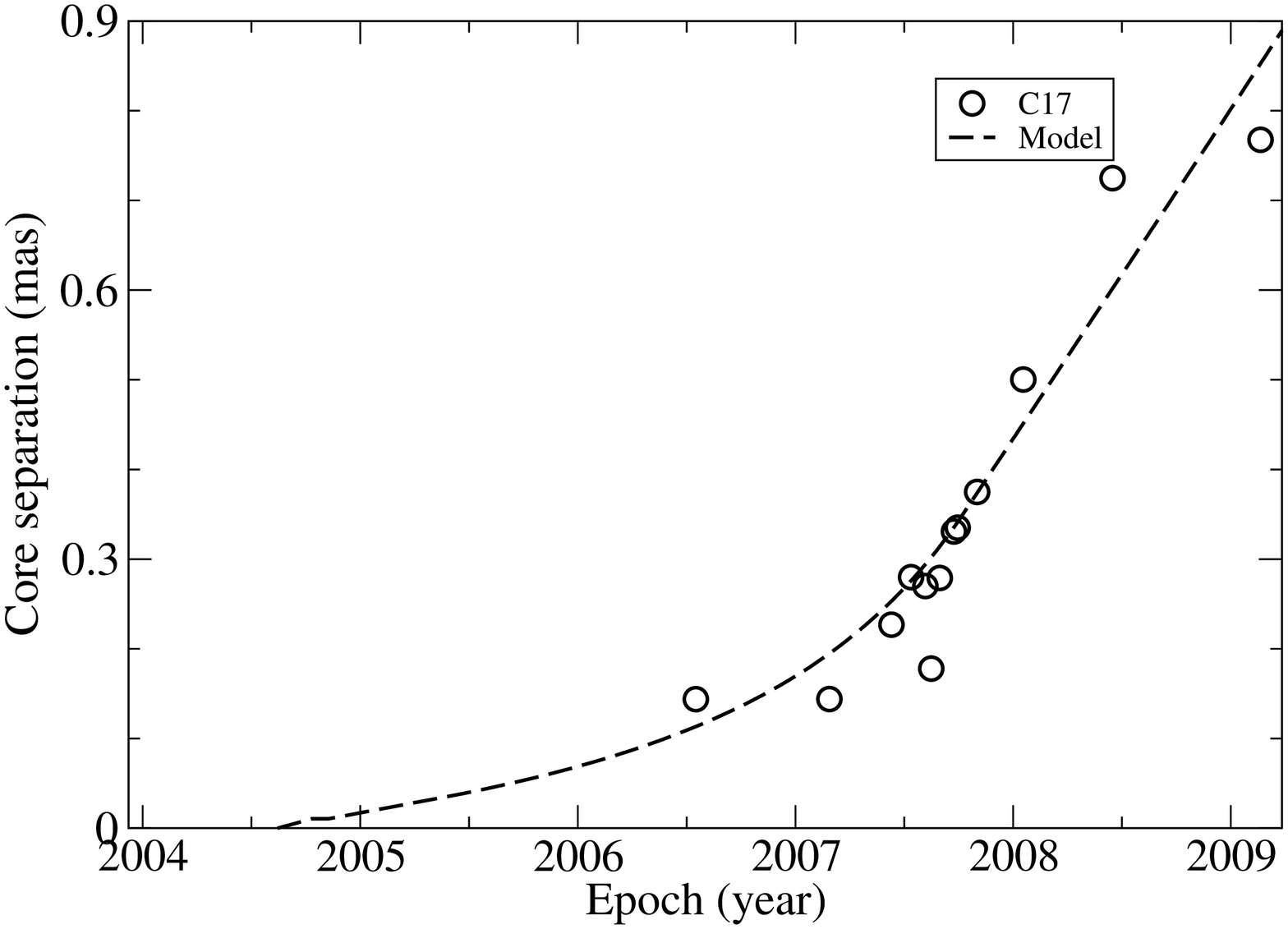}
   \includegraphics[width=6cm,angle=-90]{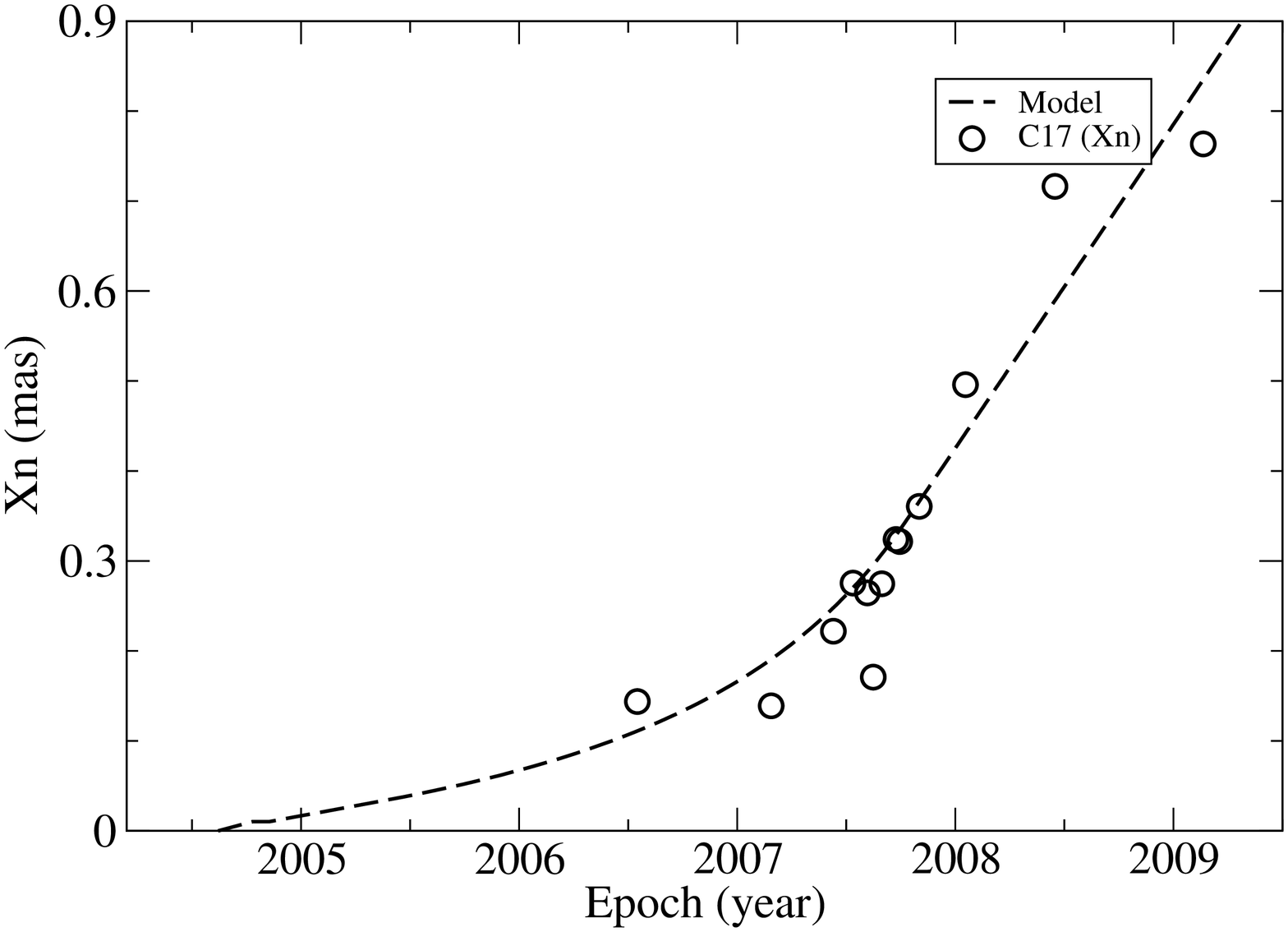}
   \includegraphics[width=6cm,angle=-90]{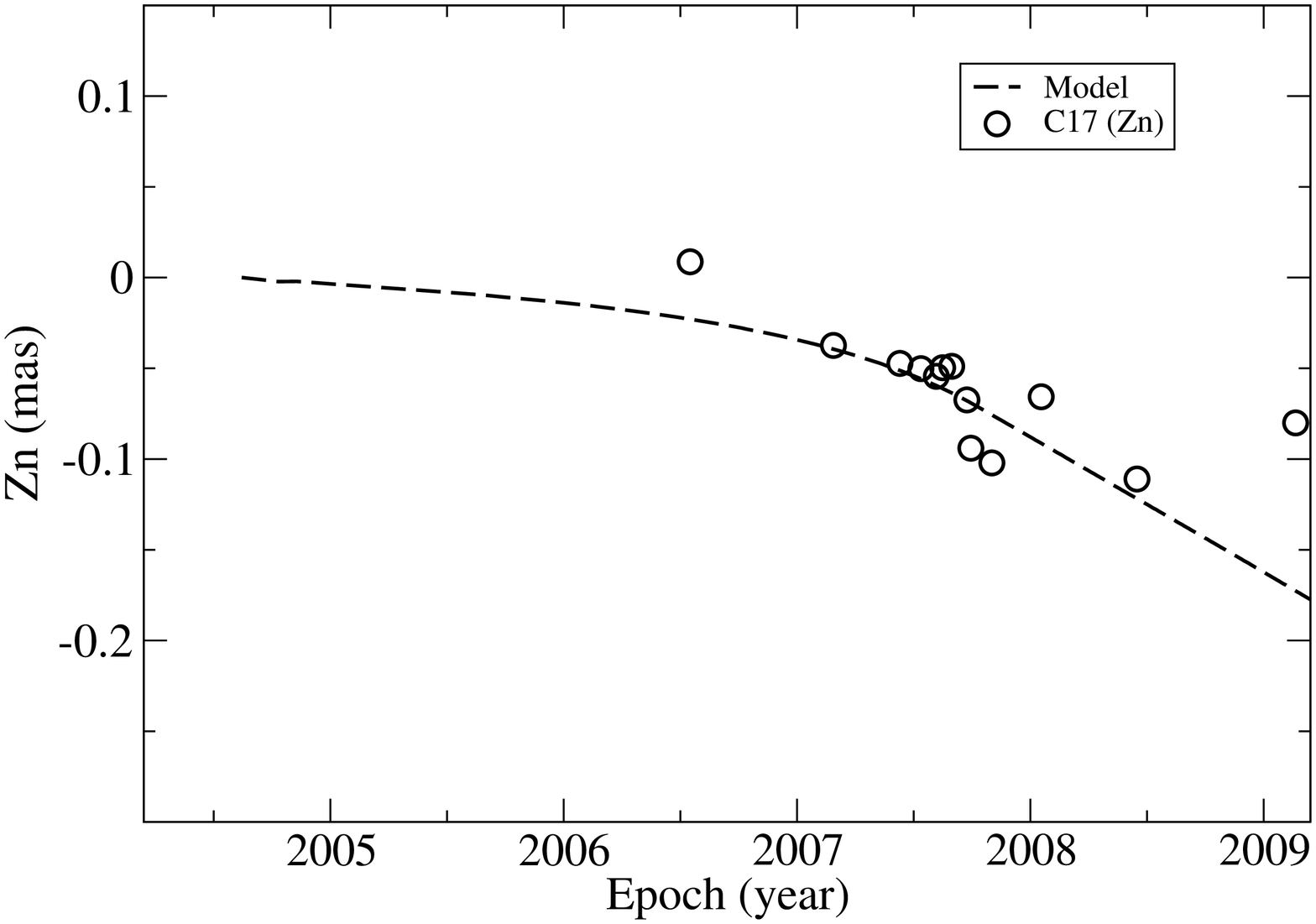}
   \includegraphics[width=6cm,angle=-90]{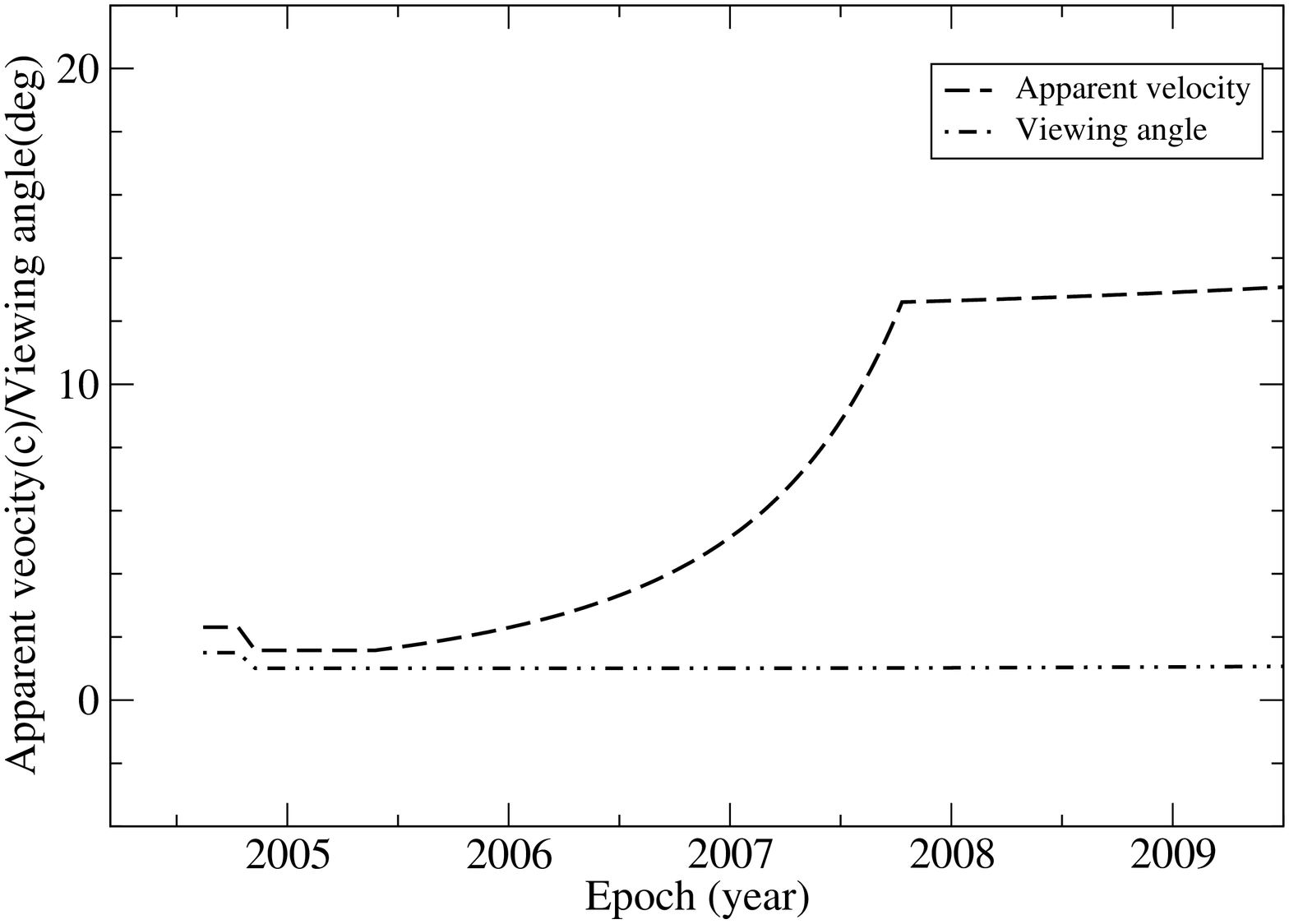}
   \includegraphics[width=6cm,angle=-90]{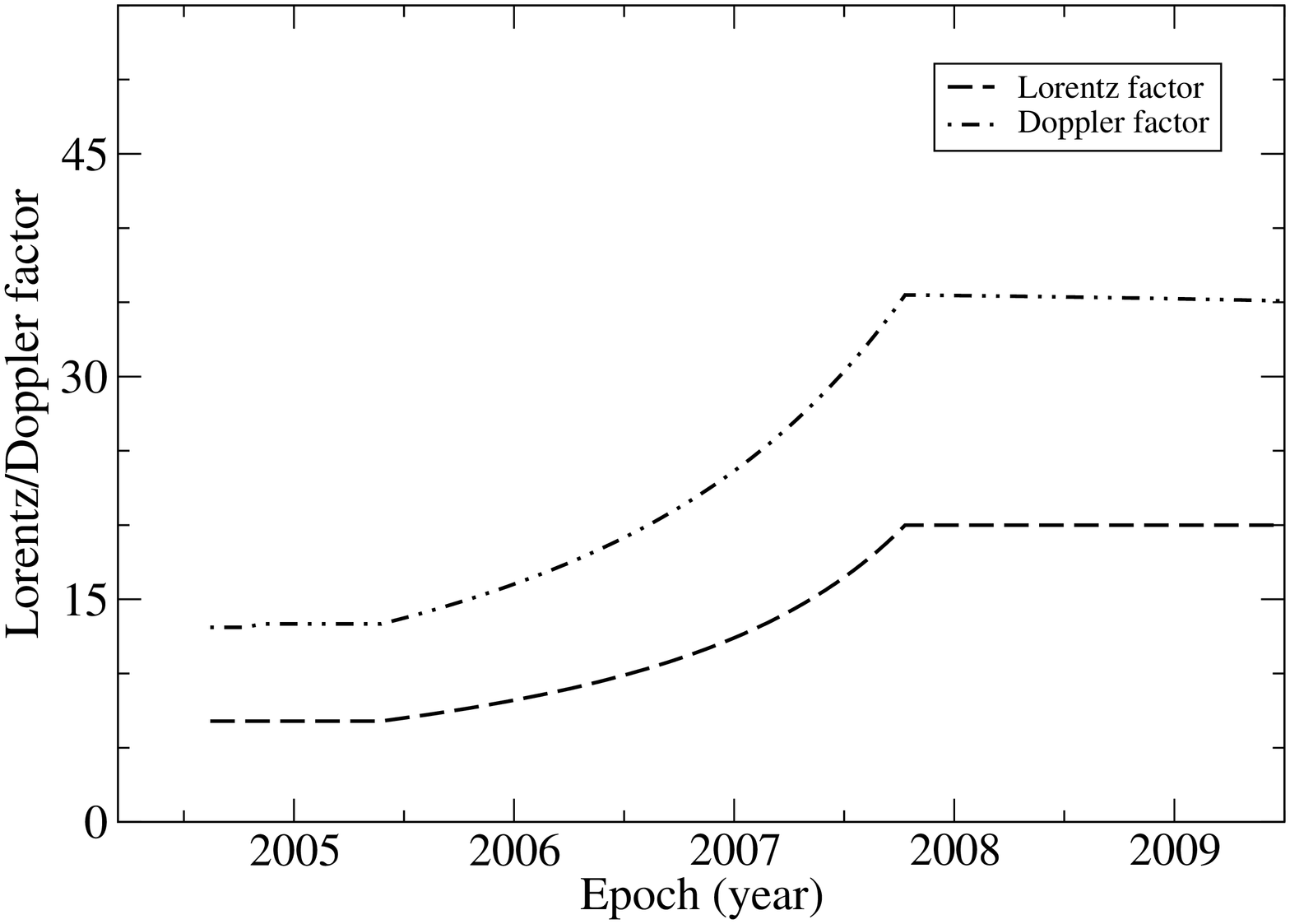}
   \caption{Knot C17: precession phase $\phi_0$(rad)=1.57+2$\pi$ and ejection 
  time $t_0$=2004.62. Model-fitting results: trajectory $Z_n(X_n)$, coordinates
  $X_n(t)$ and $Z_n(t)$, core separation $r_n(t)$, modeled apparent velocity
  $\beta_a(t)$ and viewing angle $\theta$(t), bulk Lorentz factor $\Gamma$(t)
  and Doppler factor $\delta(t)$. Its kinematics within $r_n{\sim}$0.8\,mas
  could be well model-simulated in terms of the precessing nozzle scenario.}
   \end{figure*}
   \begin{figure*}
   \centering
   \includegraphics[width=6cm,angle=-90]{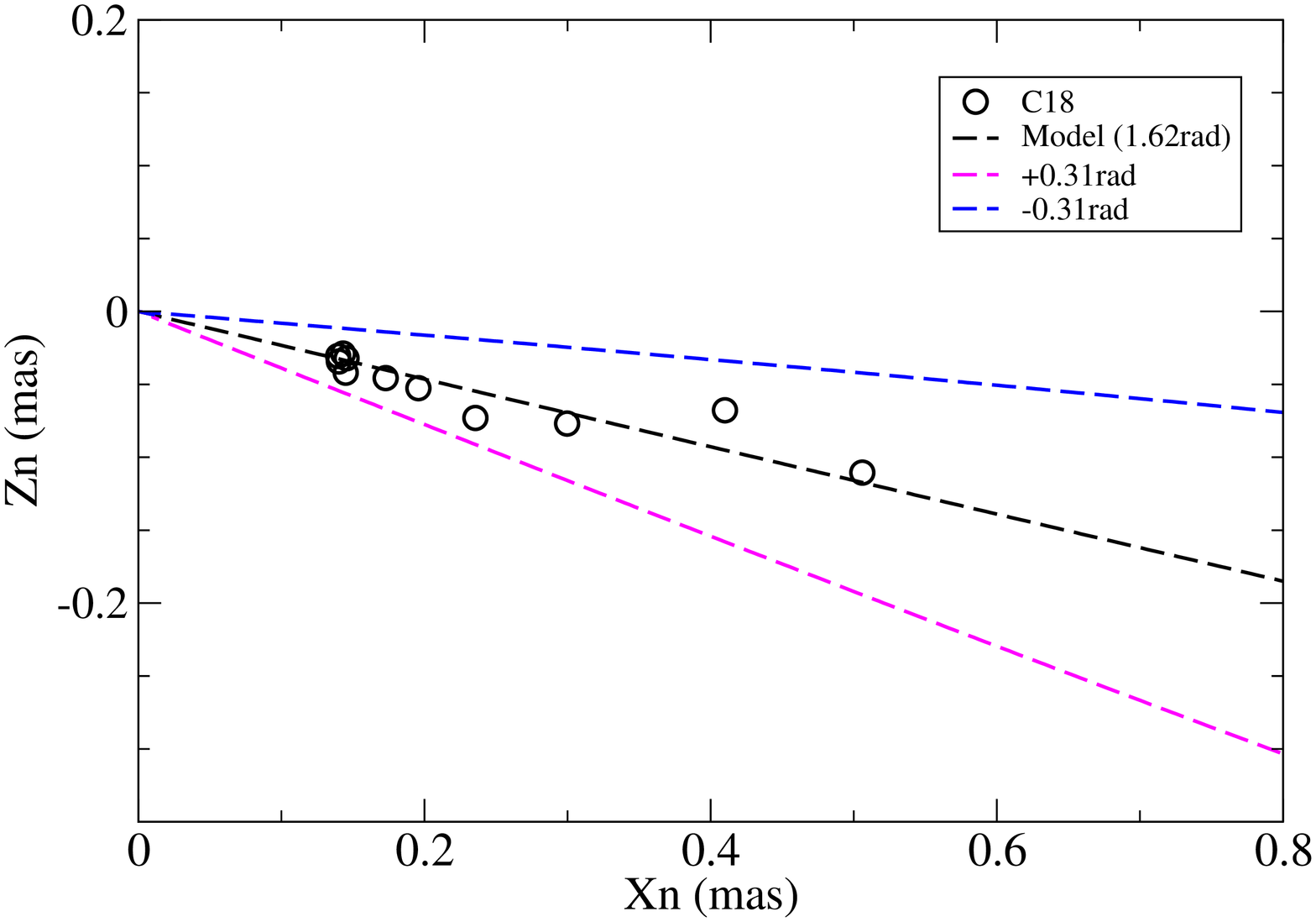}
   \includegraphics[width=6cm,angle=-90]{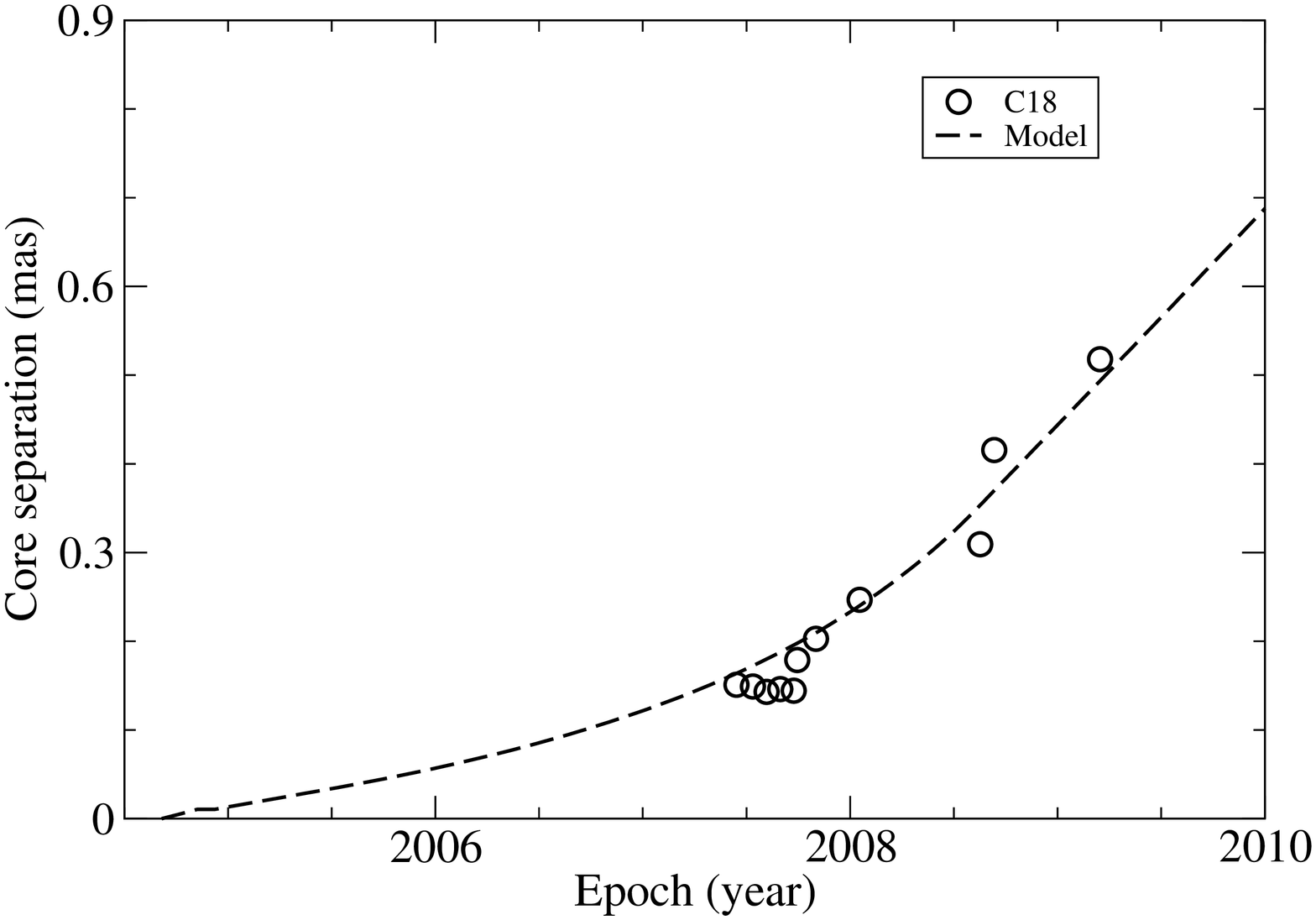}
   \includegraphics[width=6cm,angle=-90]{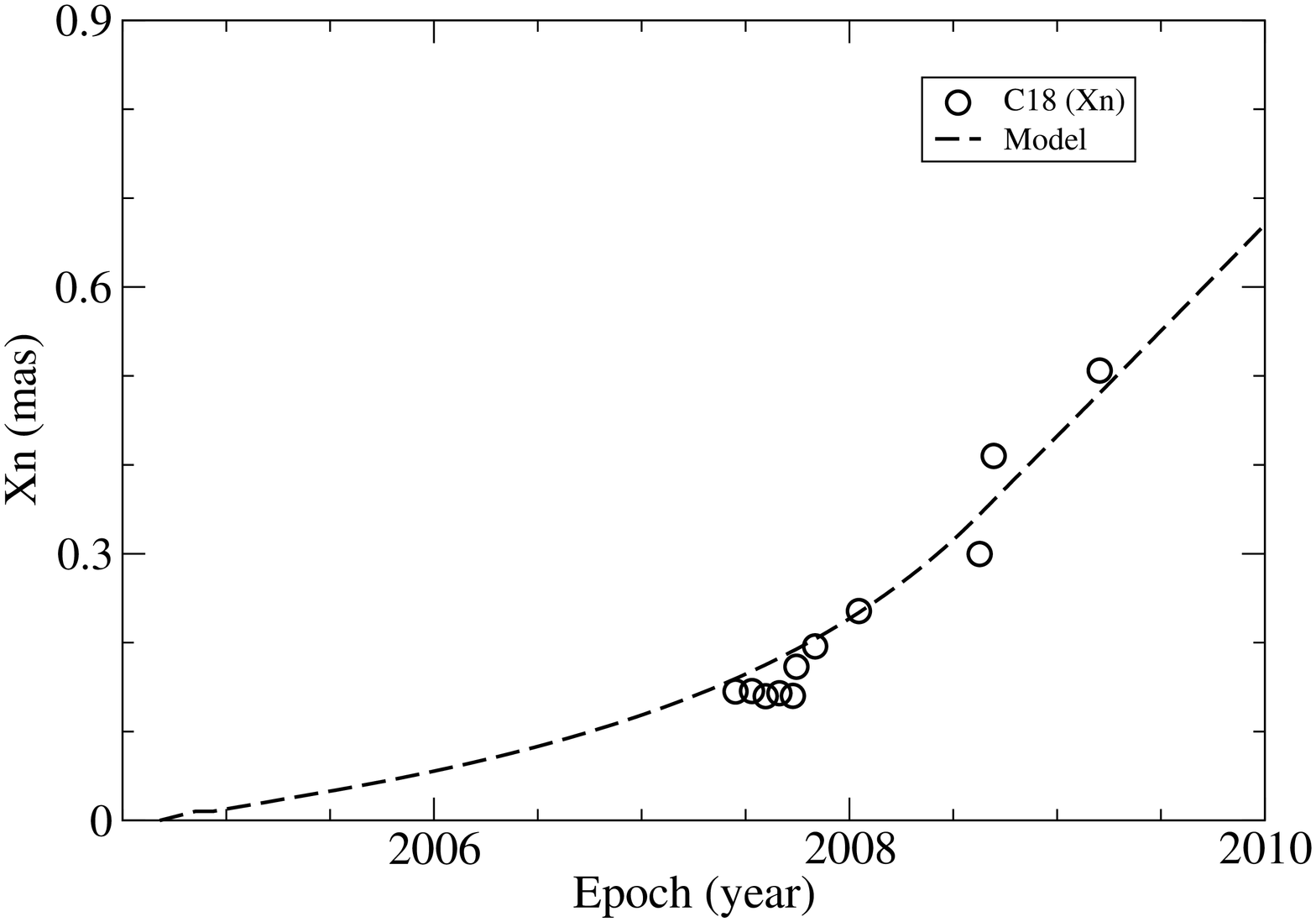}
   \includegraphics[width=6cm,angle=-90]{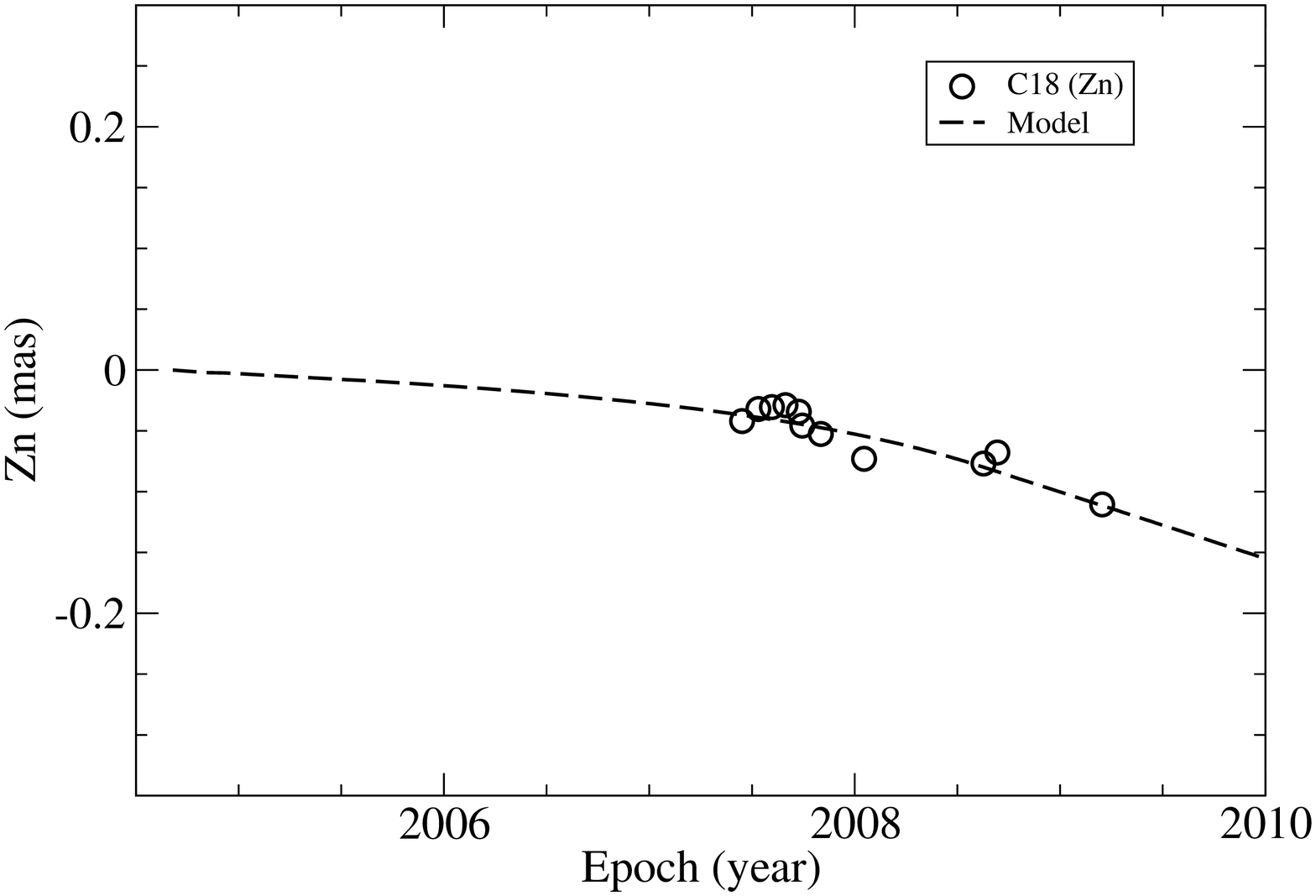}
   \includegraphics[width=6cm,angle=-90]{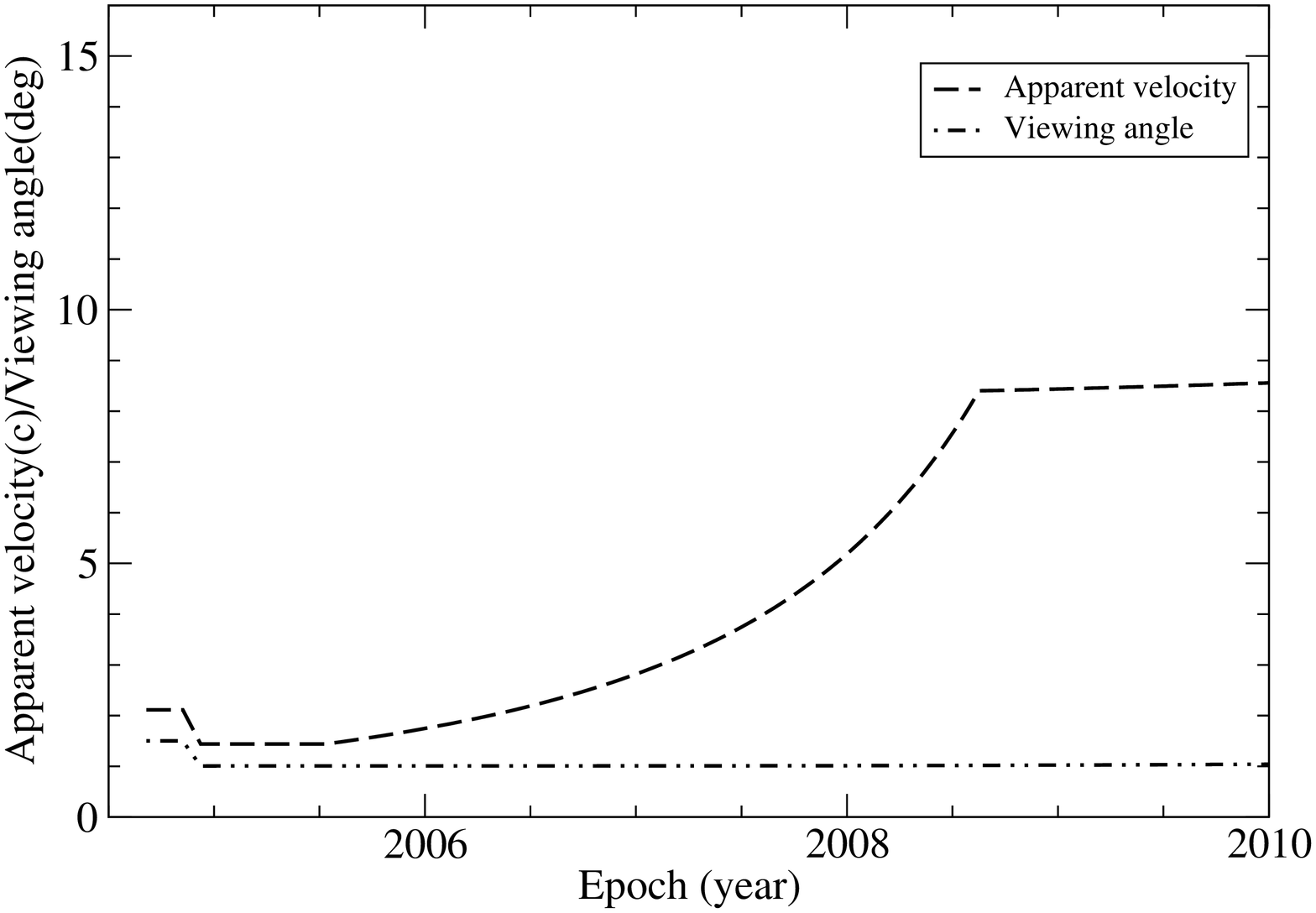}
   \includegraphics[width=6cm,angle=-90]{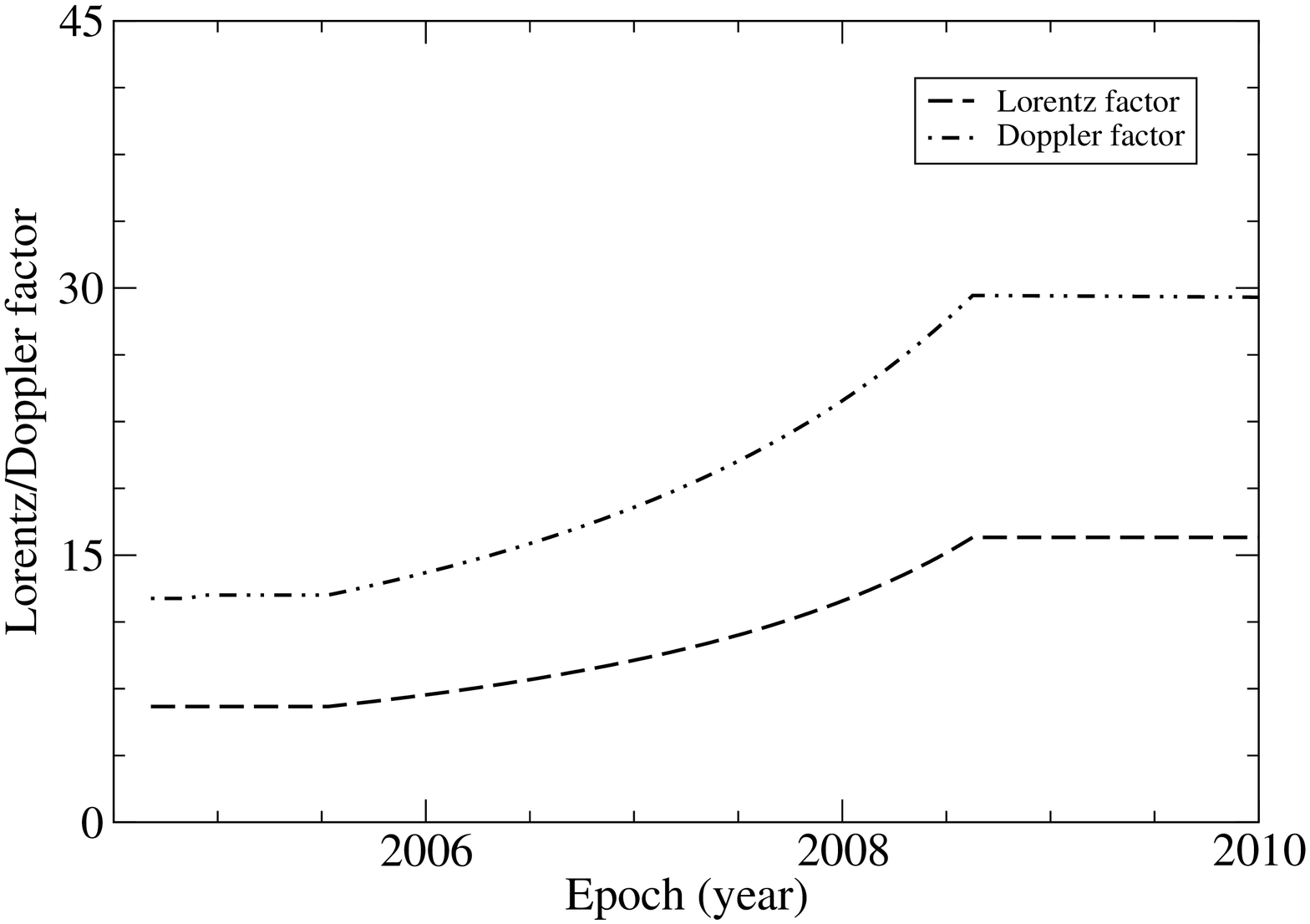}
   \caption{Knot C18: precession phase $\phi_0$(rad)=1.62+2$\pi$ and ejection 
  time $t_0$=2004.68. Model-fitting results: trajectory $Z_n(X_n)$, coordinates
  $X_n(t)$ and $Z_n(t)$, core separation $r_n(t)$, modeled apparent velocity
   $\beta_a(t)$ and viewing angle $\theta(t)$.Its kinematics within core 
   separation $r_n{\sim}$0.60\,mas could be well simulated by the precessing
    nozzle scenario.}
   \end{figure*}
   \begin{figure*}
   \centering
   \includegraphics[width=6cm,angle=-90]{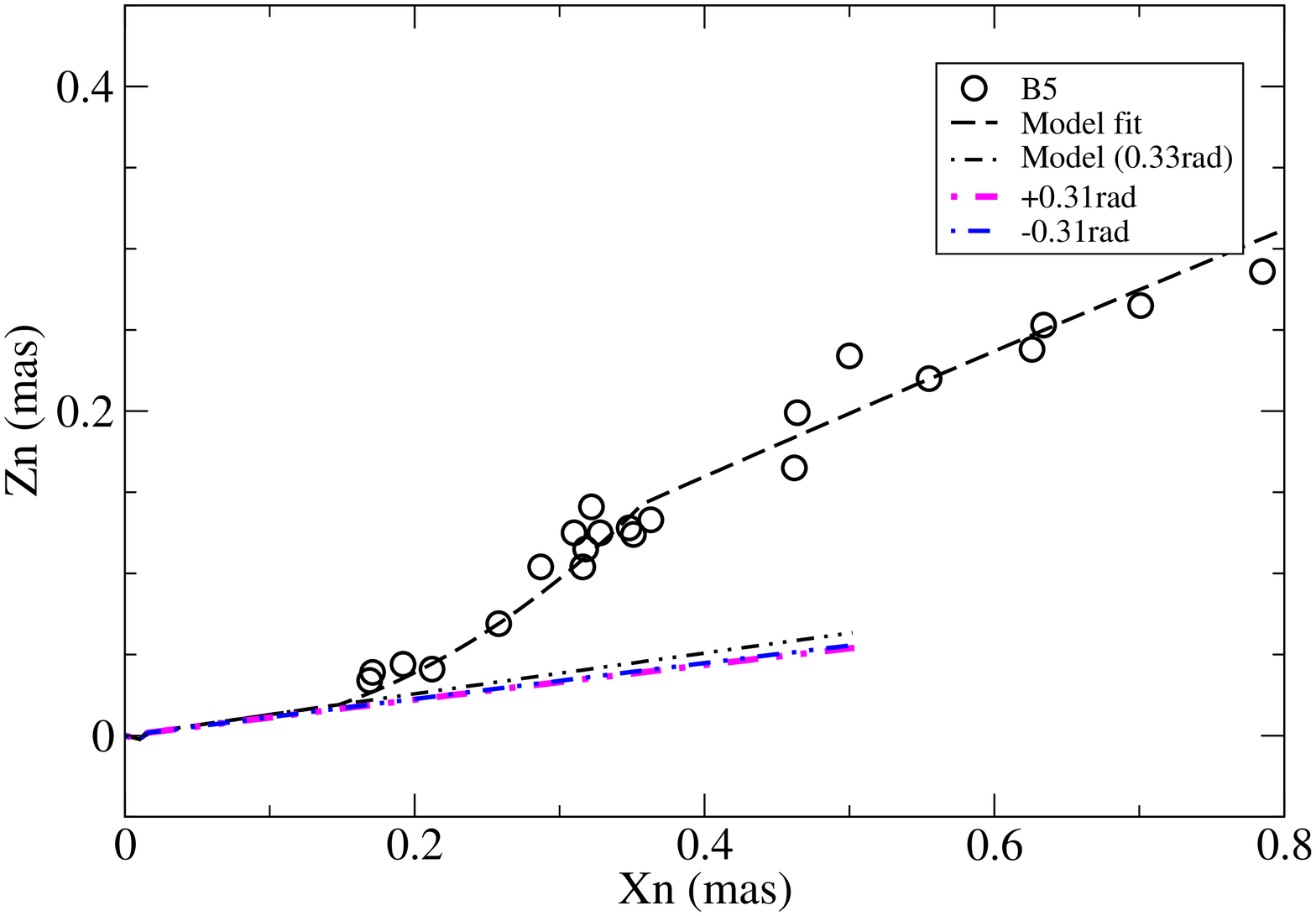}
   \includegraphics[width=6cm,angle=-90]{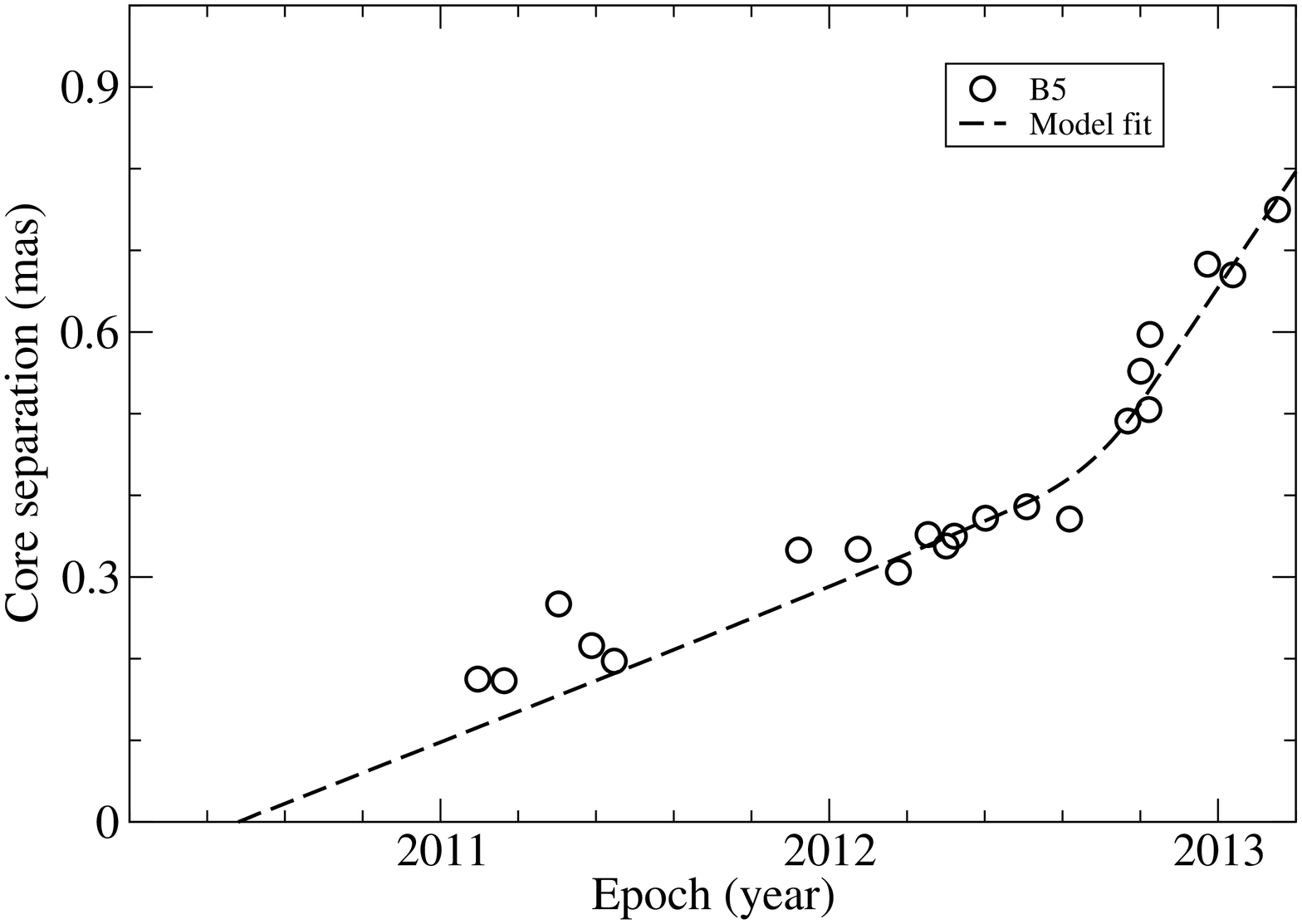}
   \includegraphics[width=6cm,angle=-90]{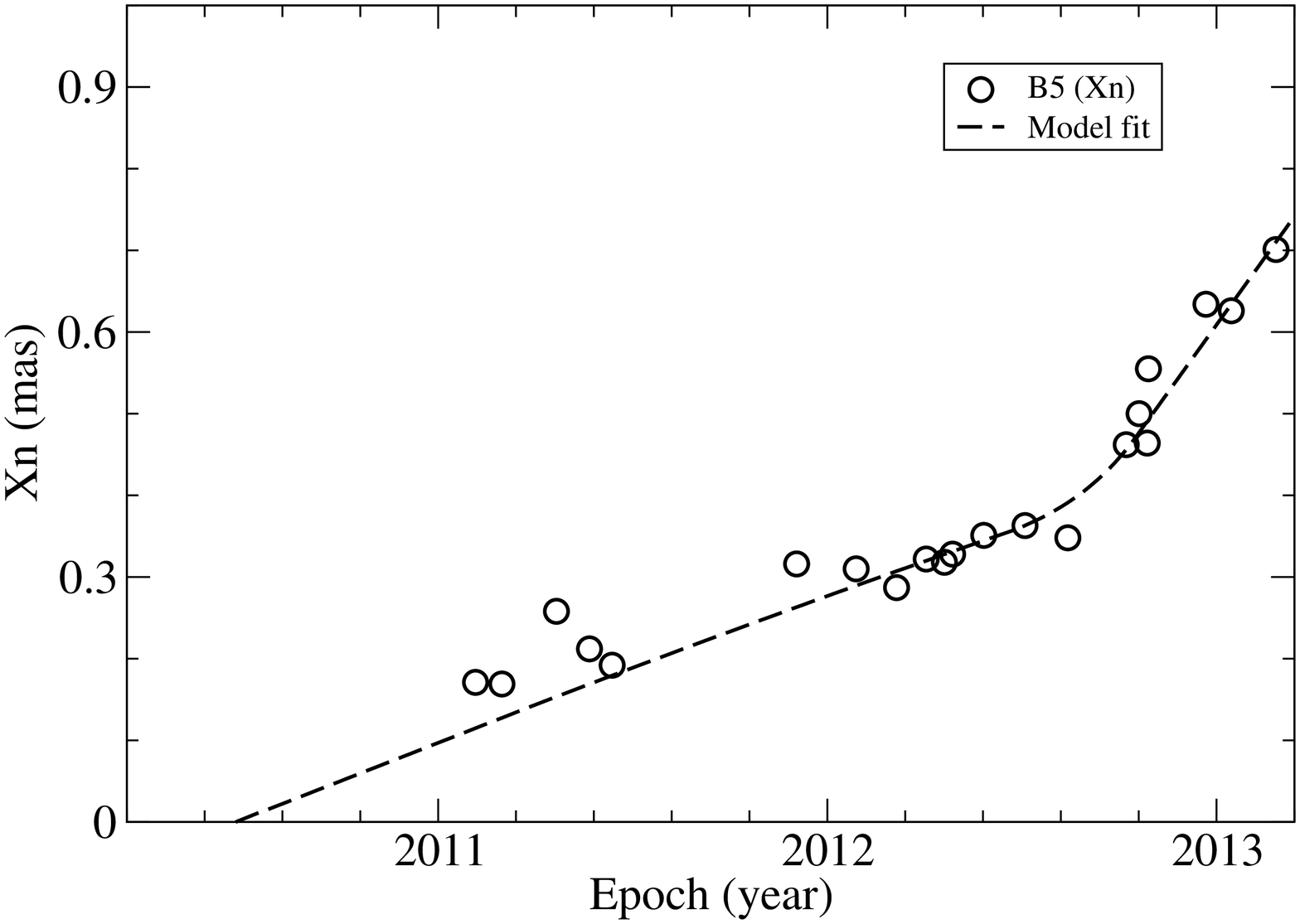}
   \includegraphics[width=6cm,angle=-90]{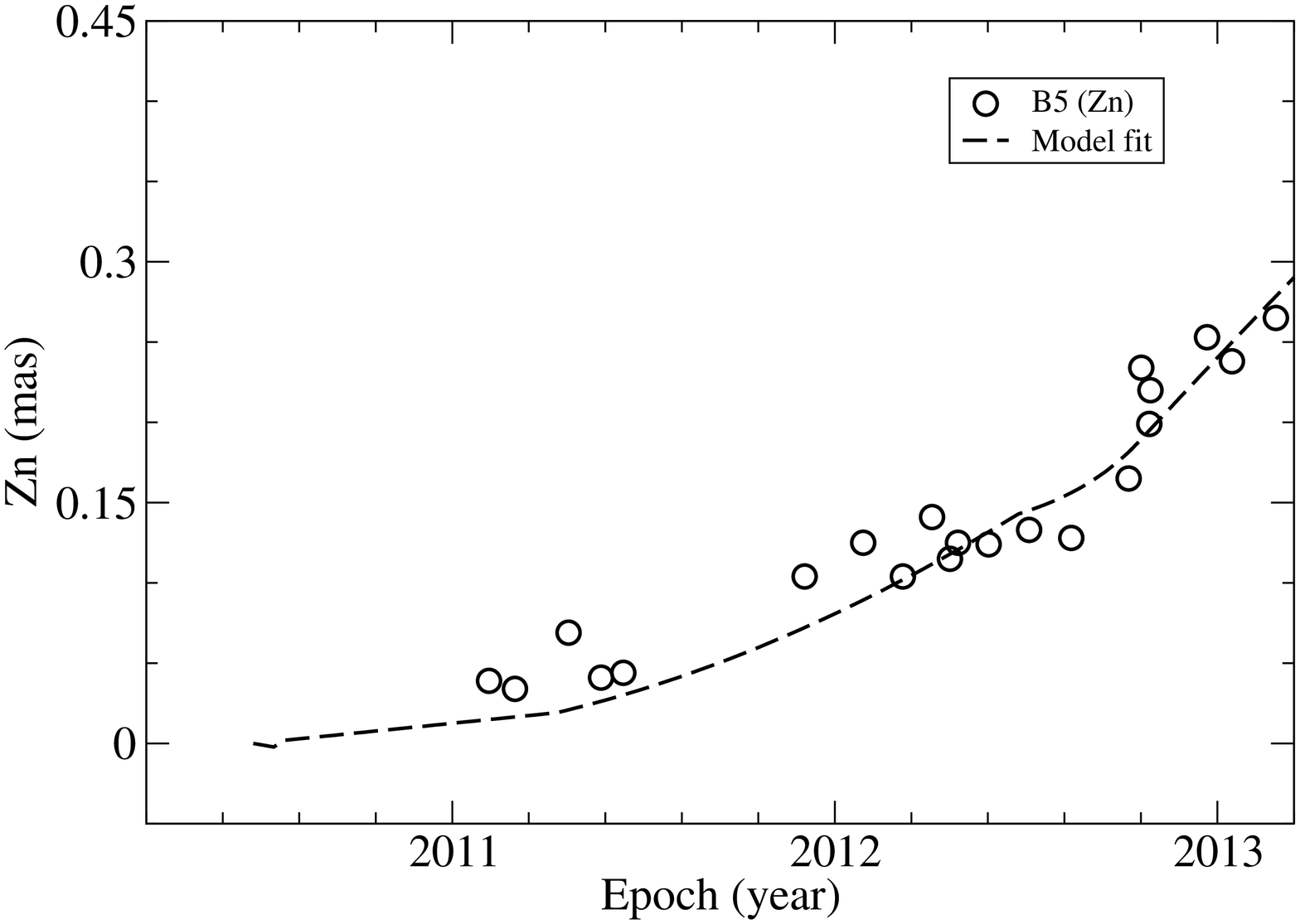}
   \includegraphics[width=6cm,angle=-90]{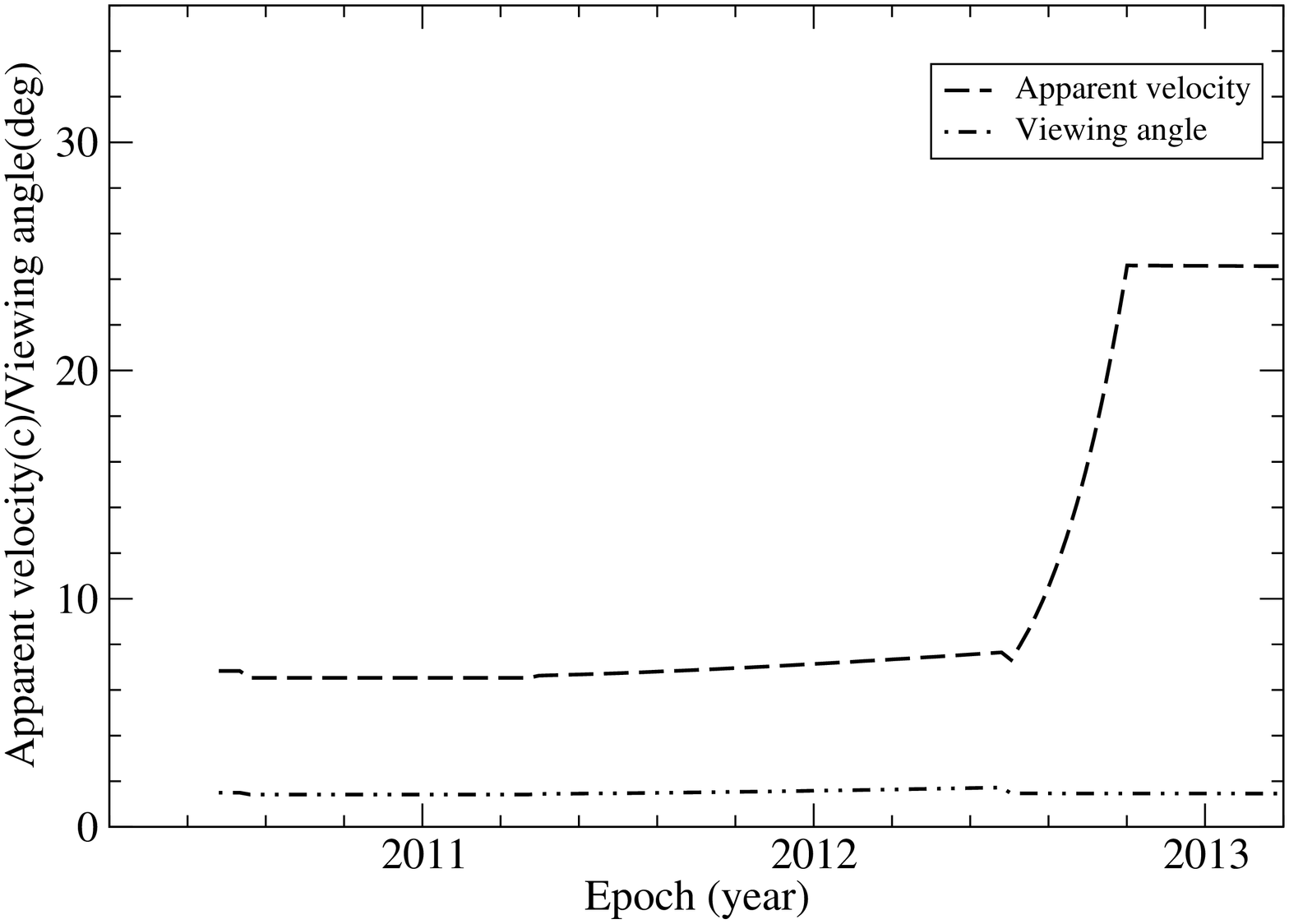}
   \includegraphics[width=6cm,angle=-90]{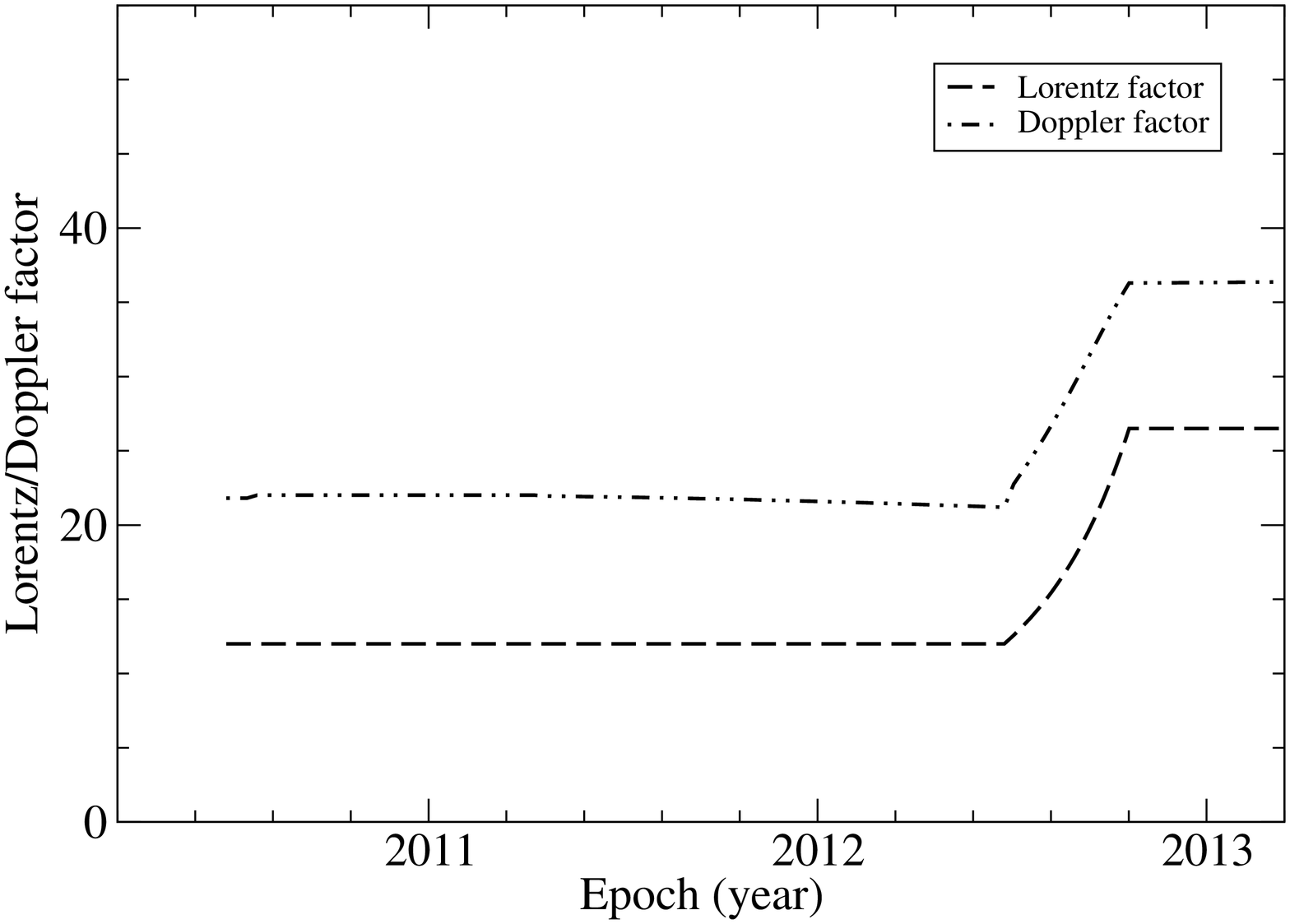}
   \caption{Knot B5: precession phase $\phi_0$(rad)=0.33+4$\pi$ and ejection 
   time $t_0$=2010.48. Its trajectory was observed near the edge of the 
   jet-cone and the data-points within $X_n{\sim}$0.2\,mas are near the
   modeled trajectories defined by phases 0.33$\pm$0.31\,rad, which are almost
   overlapped with each other in this case. Its precessing common 
   trajectory might be assumed to extend to $r_n{\sim}$0.15\,mas.}
   \end{figure*}
   \begin{figure*}
   \centering
   \includegraphics[width=6cm,angle=-90]{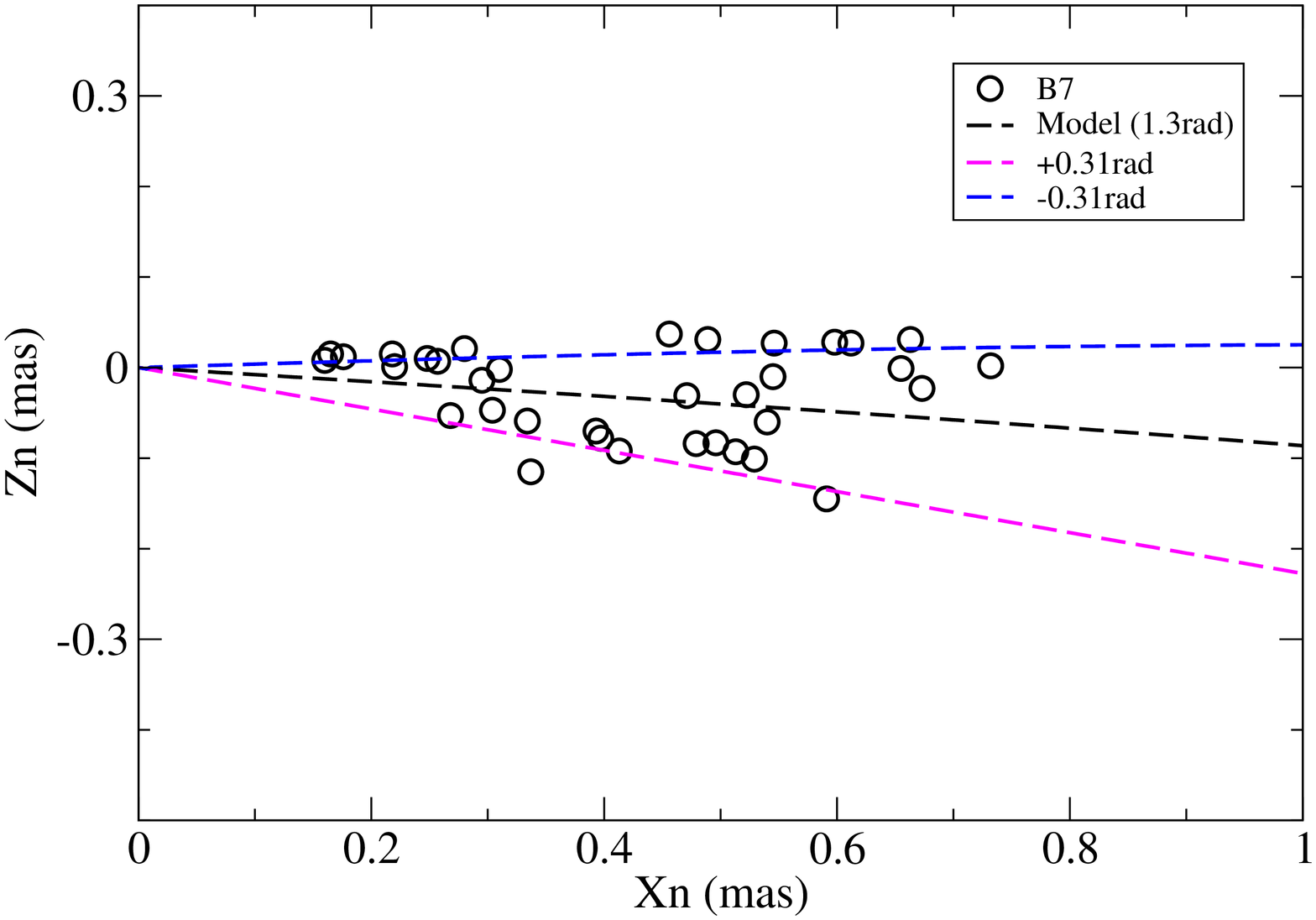}
   \includegraphics[width=6cm,angle=-90]{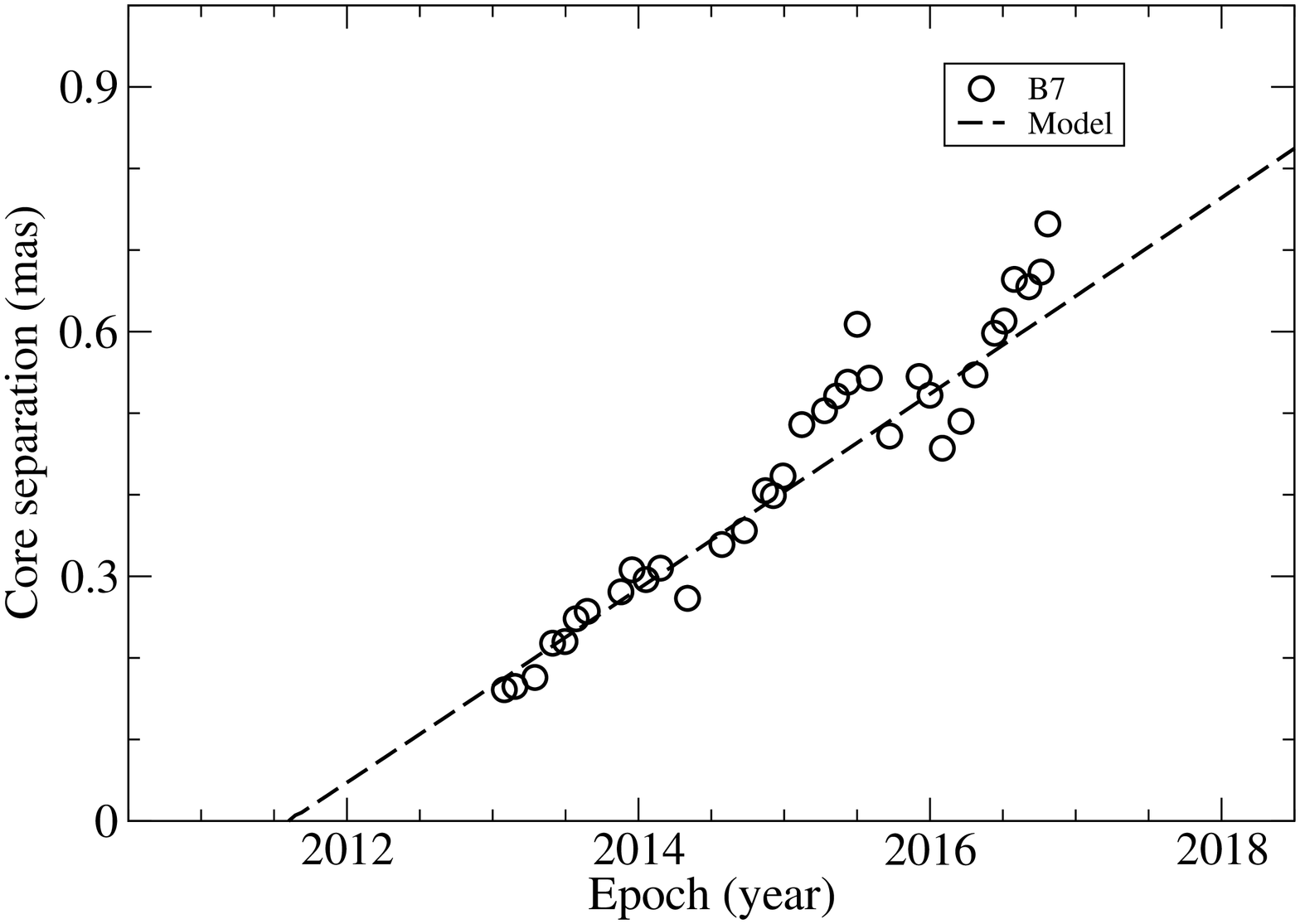}
   \includegraphics[width=6cm,angle=-90]{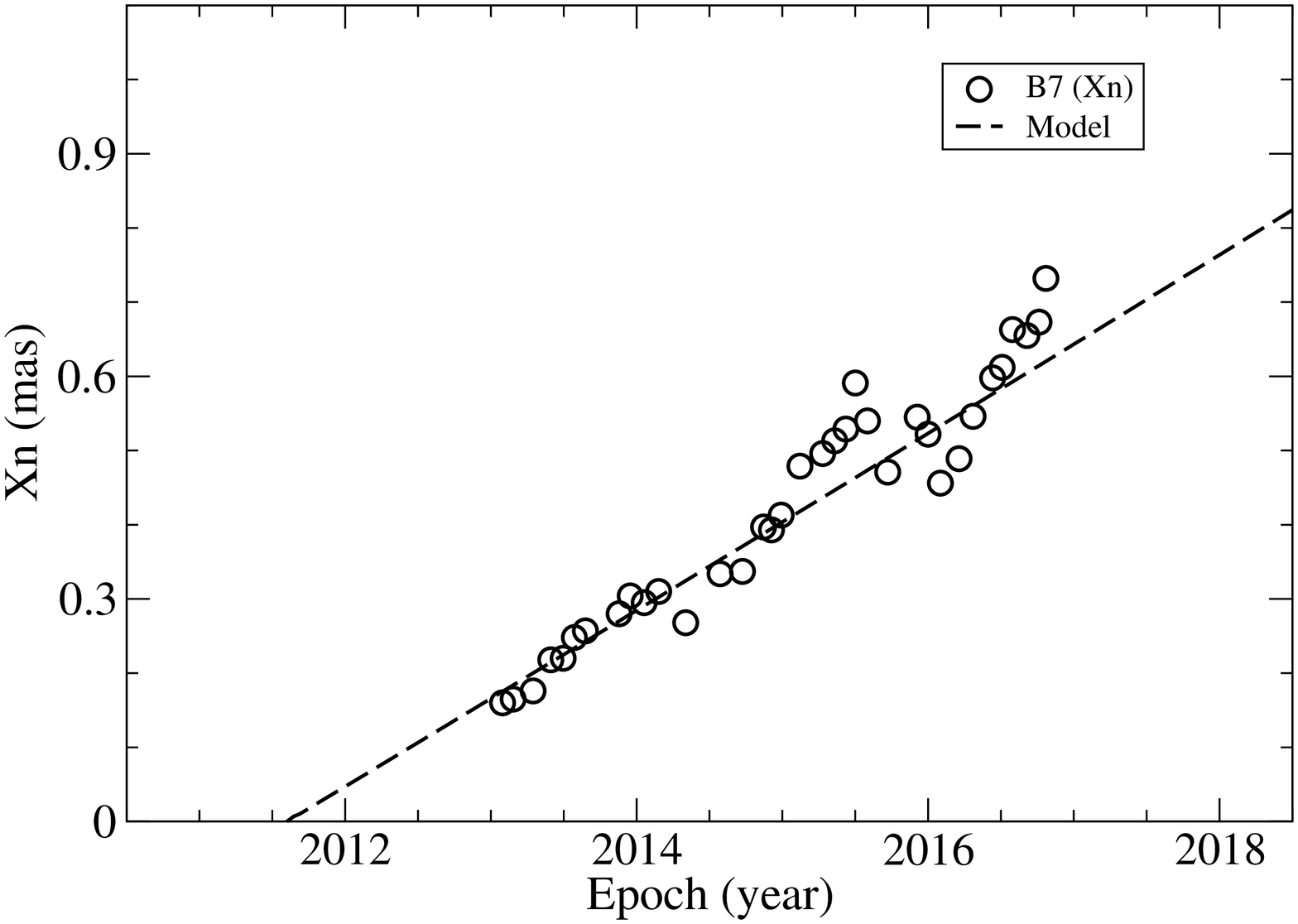}
   \includegraphics[width=6cm,angle=-90]{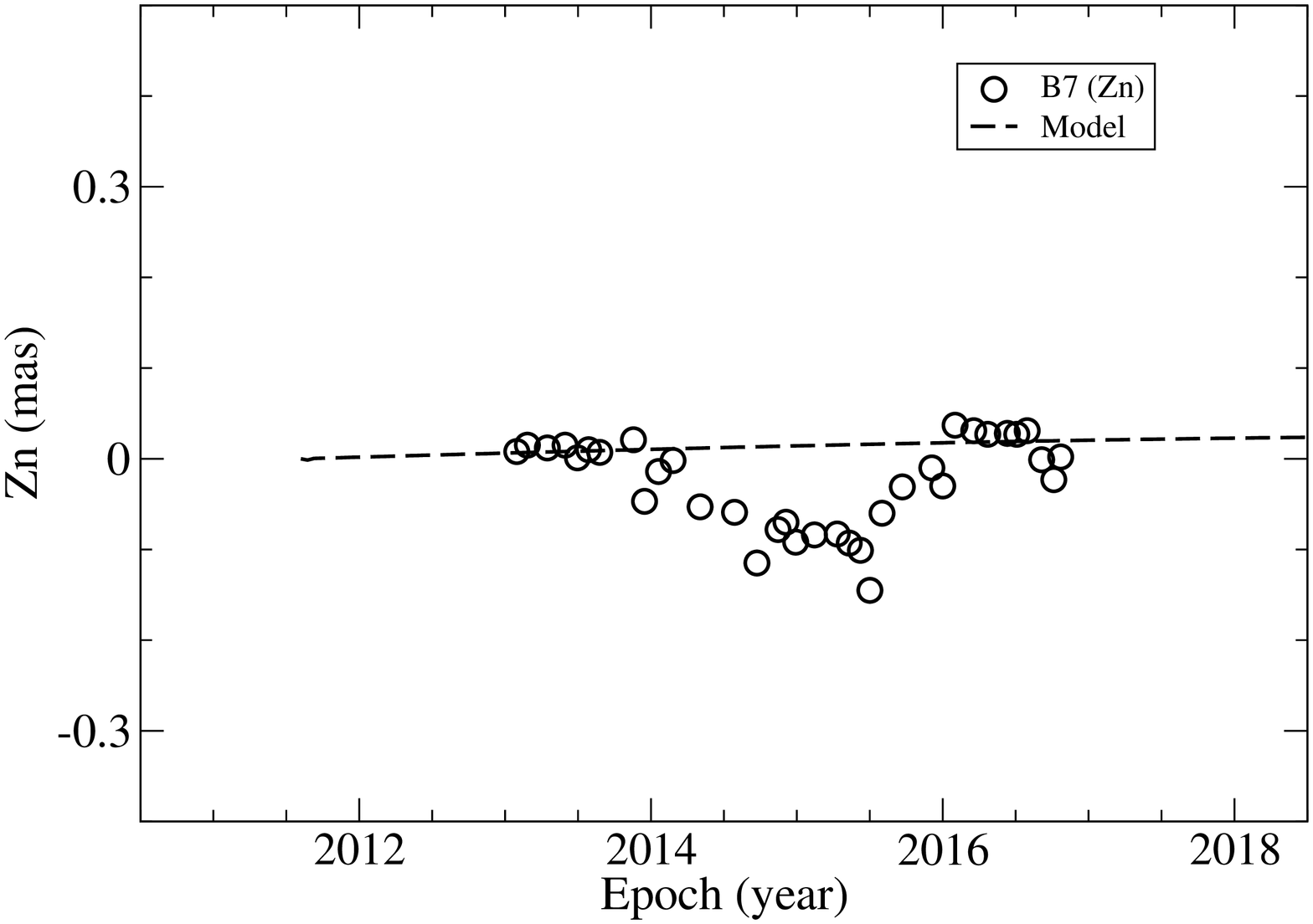}
   \includegraphics[width=6cm,angle=-90]{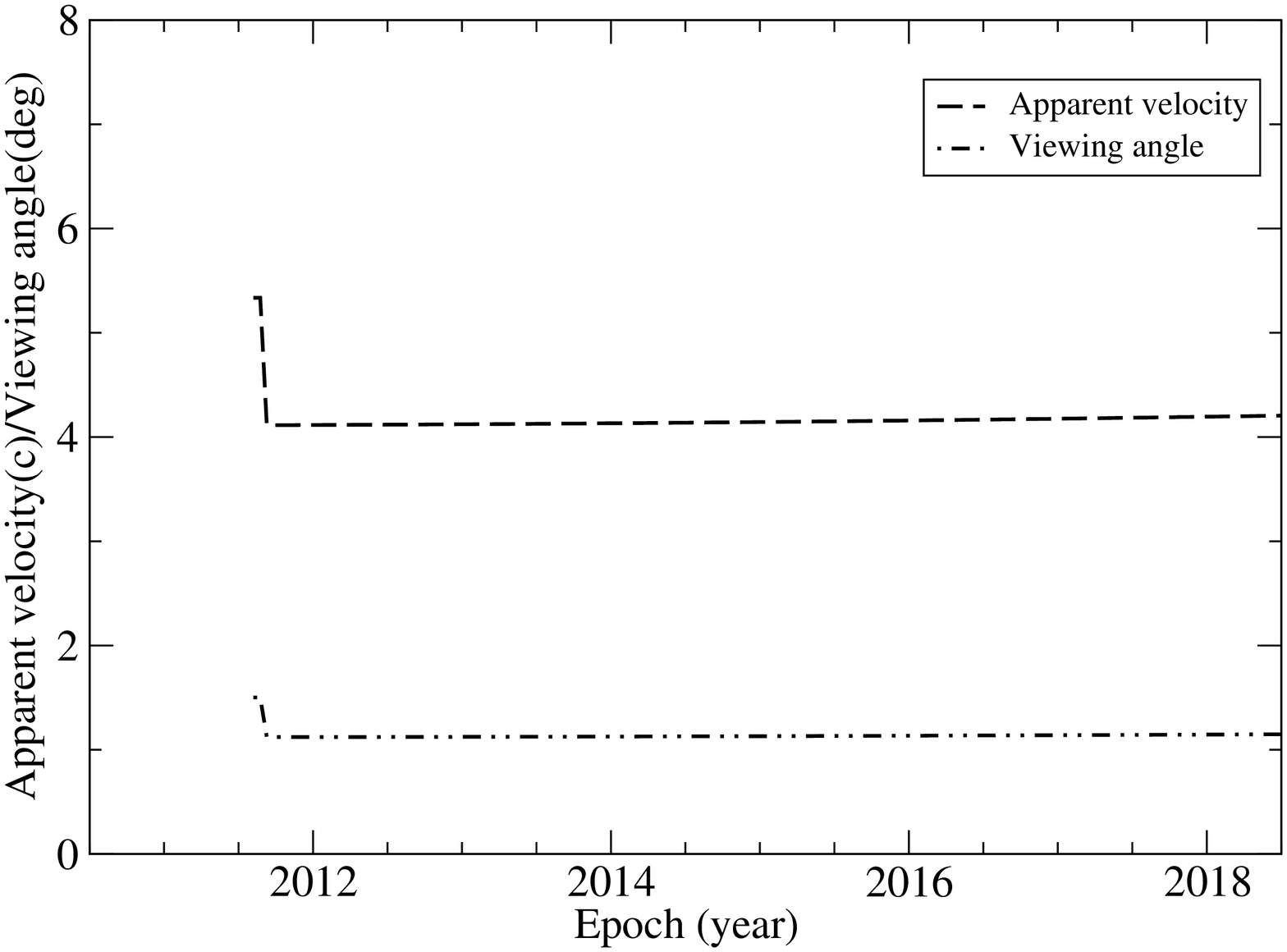}
   \includegraphics[width=6cm,angle=-90]{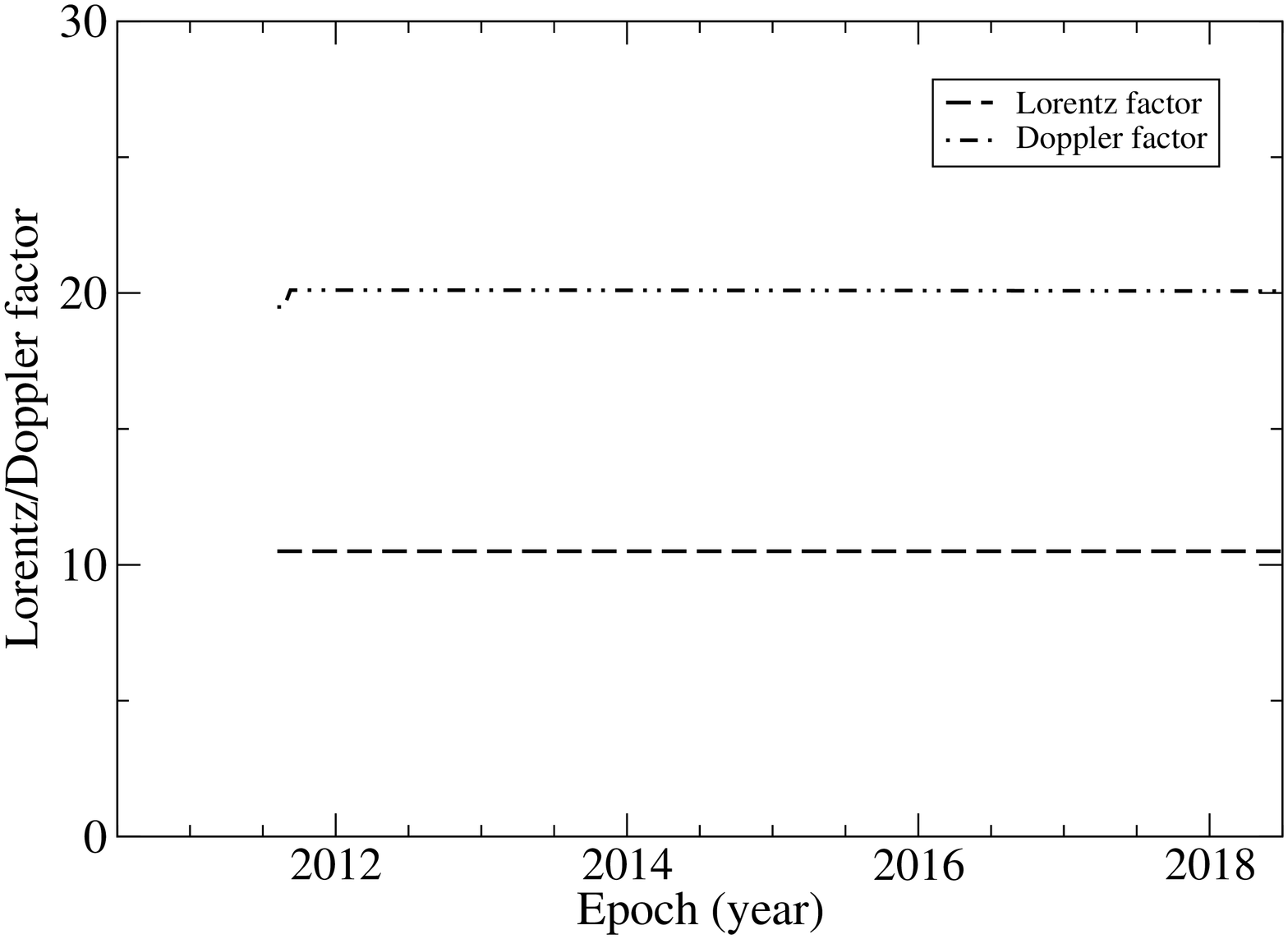}
   \caption{Knot B7: precession phase $\phi_0$(rad)=1.30+4$\pi$ and 
   ejection time $t_0$=2011.60. Model-fitting results: trajectory $Z_n(X_n)$,
   coordinates $X_n(t)$ and $Z_n(t)$, core separation $r_n(t)$, modeled
   apparent velocity $\beta_a(t)$ and viewing angle $\theta(t)$, bulk Lorentz 
   factor $\Gamma(t)$ and Doppler factor $\delta(t)$. Its kinematics within 
   core separation $r_n$$\sim$0.80\,mas could be well model-simulated in 
   terms of the precessing nozzle scenario. }
   \end{figure*}
   \begin{figure*}
   \centering
   \includegraphics[width=6cm,angle=-90]{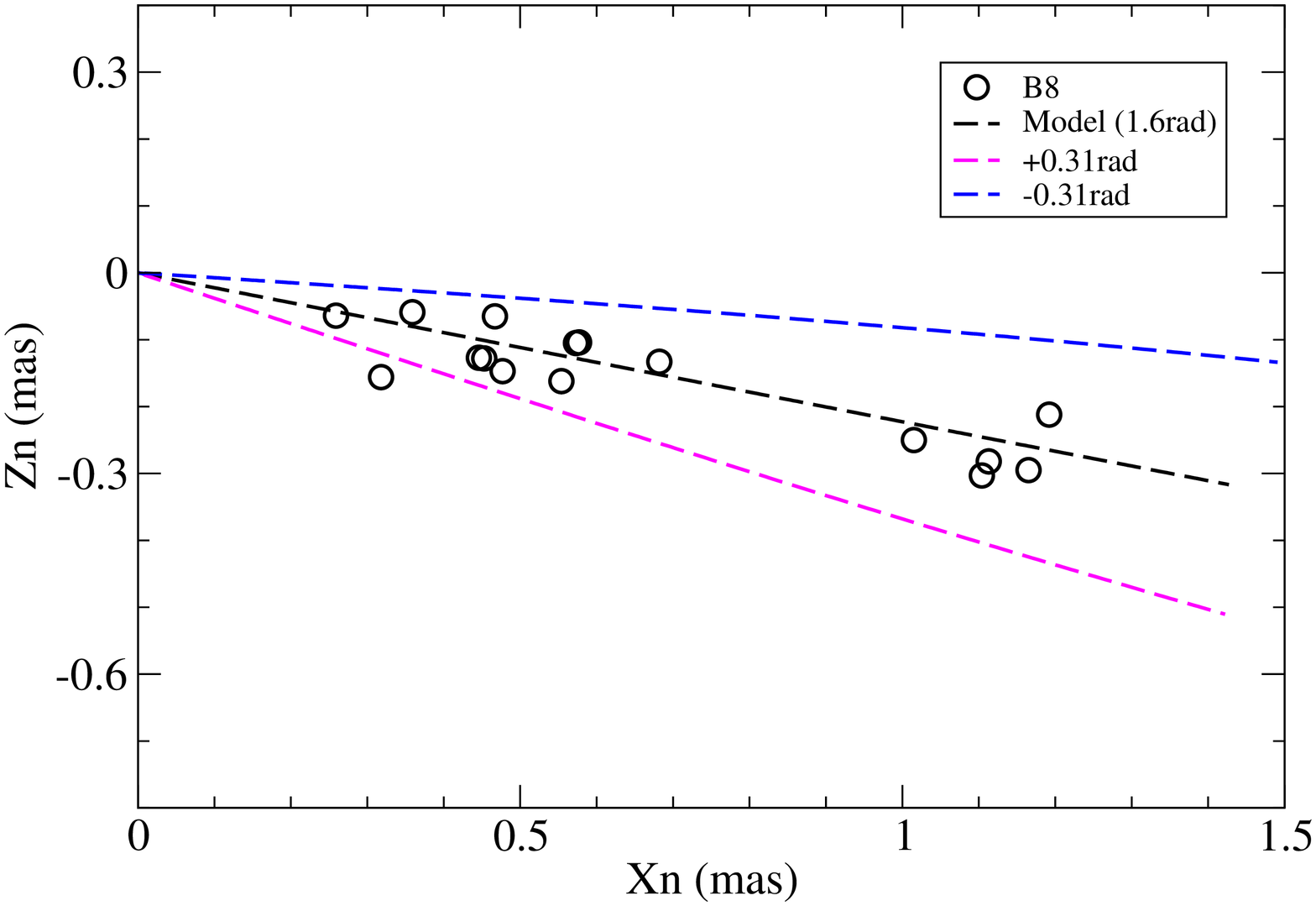}
   \includegraphics[width=6cm,angle=-90]{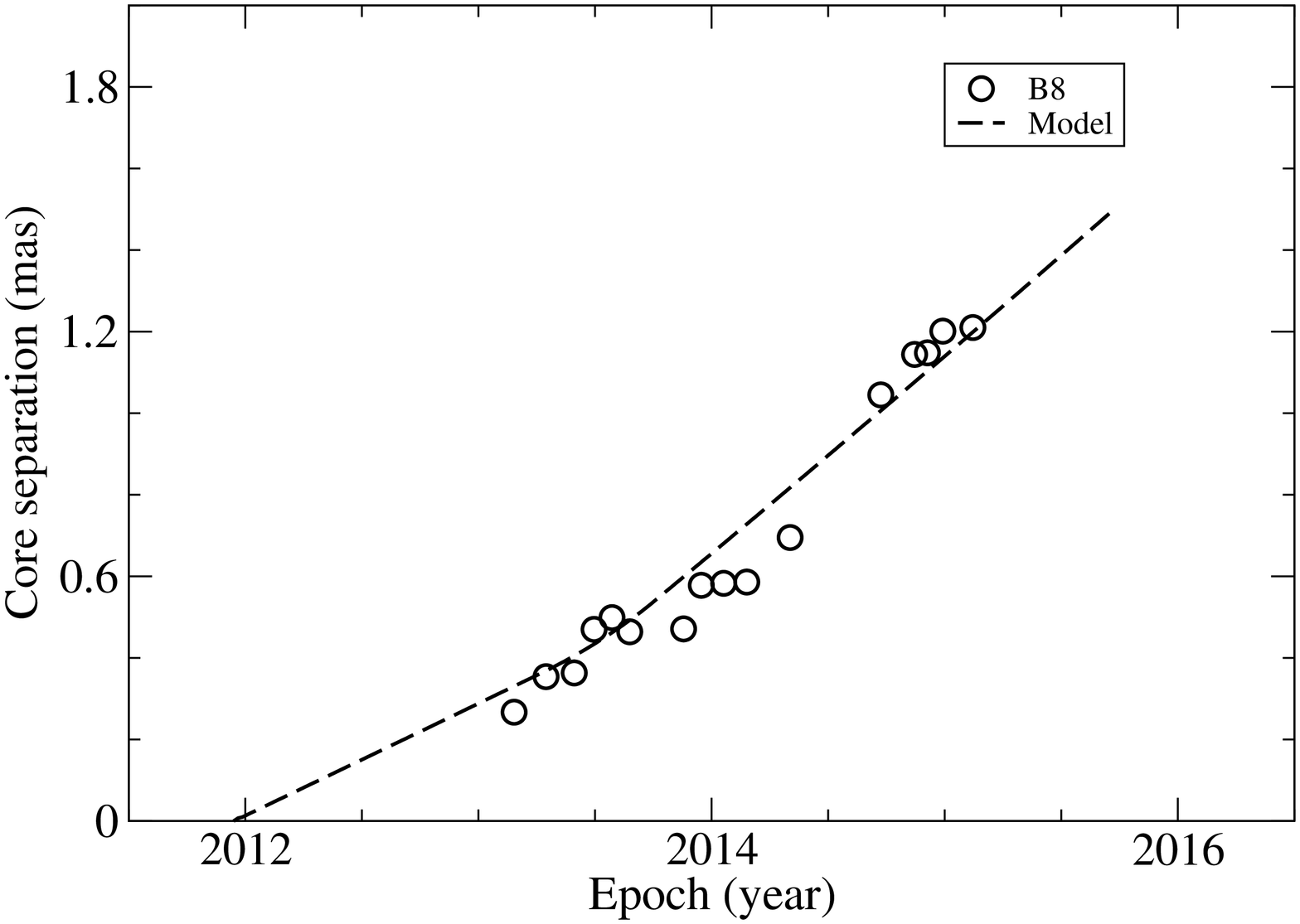}
   \includegraphics[width=6cm,angle=-90]{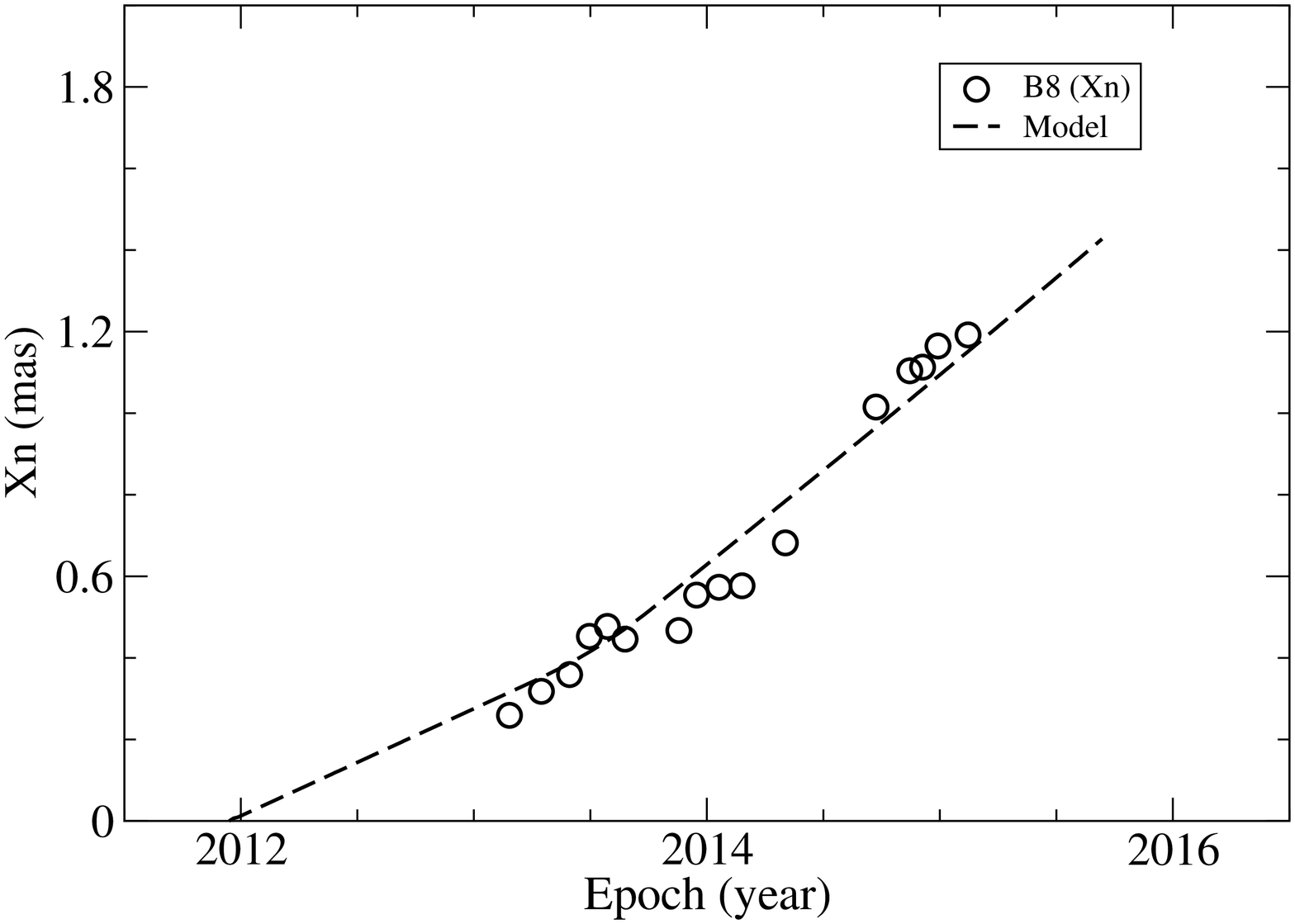}
   \includegraphics[width=6cm,angle=-90]{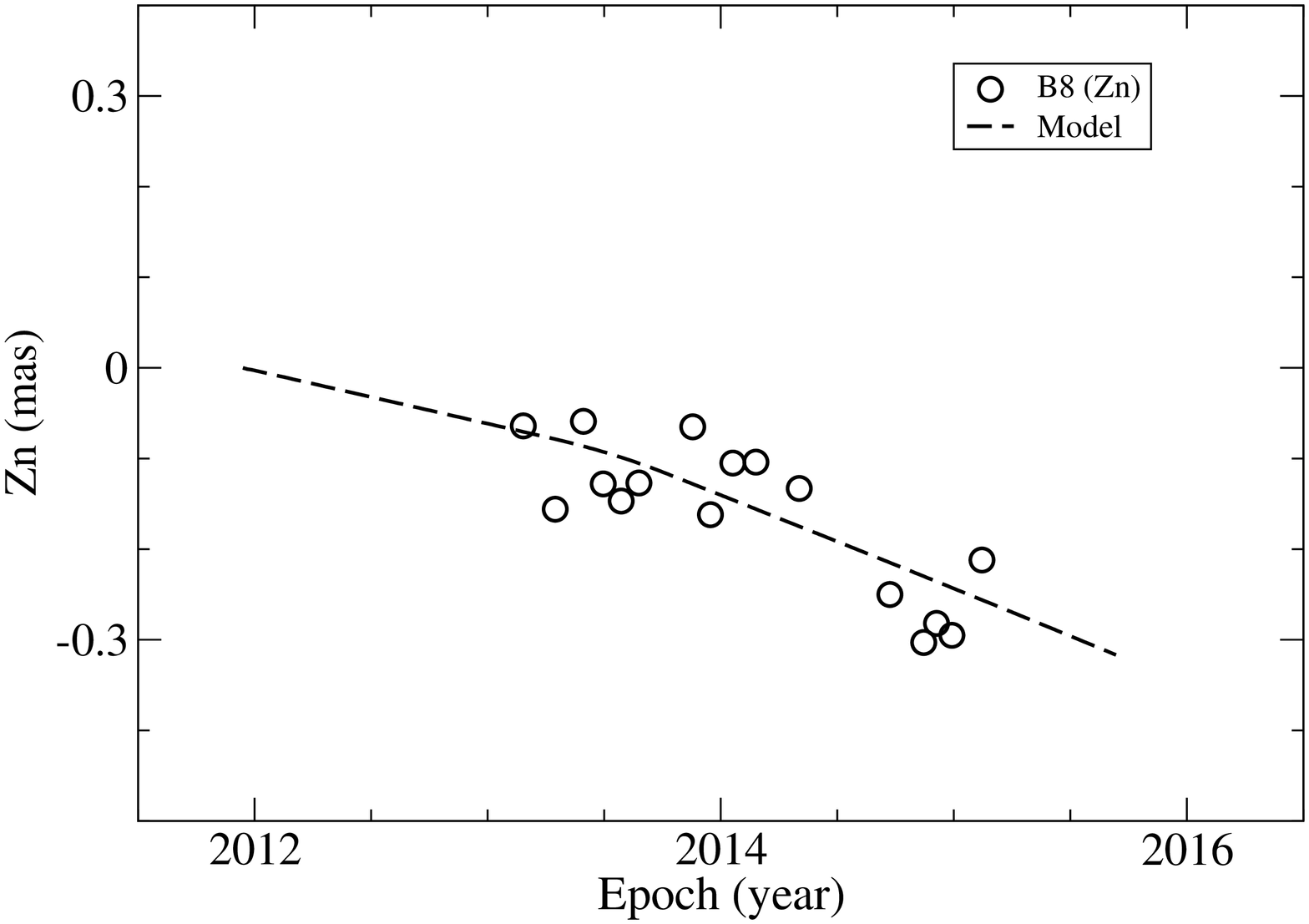}
   \includegraphics[width=6cm,angle=-90]{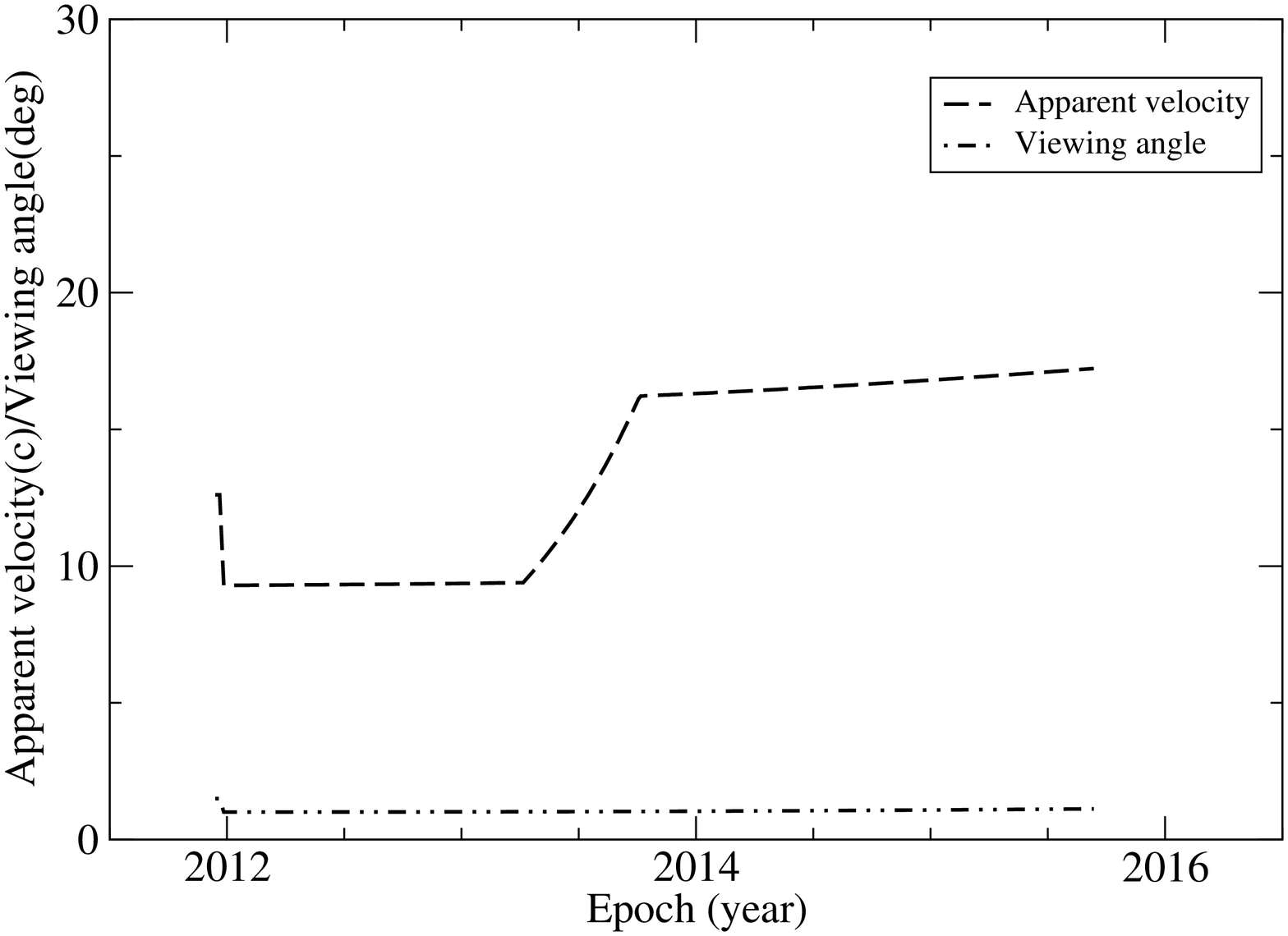}
   \includegraphics[width=6cm,angle=-90]{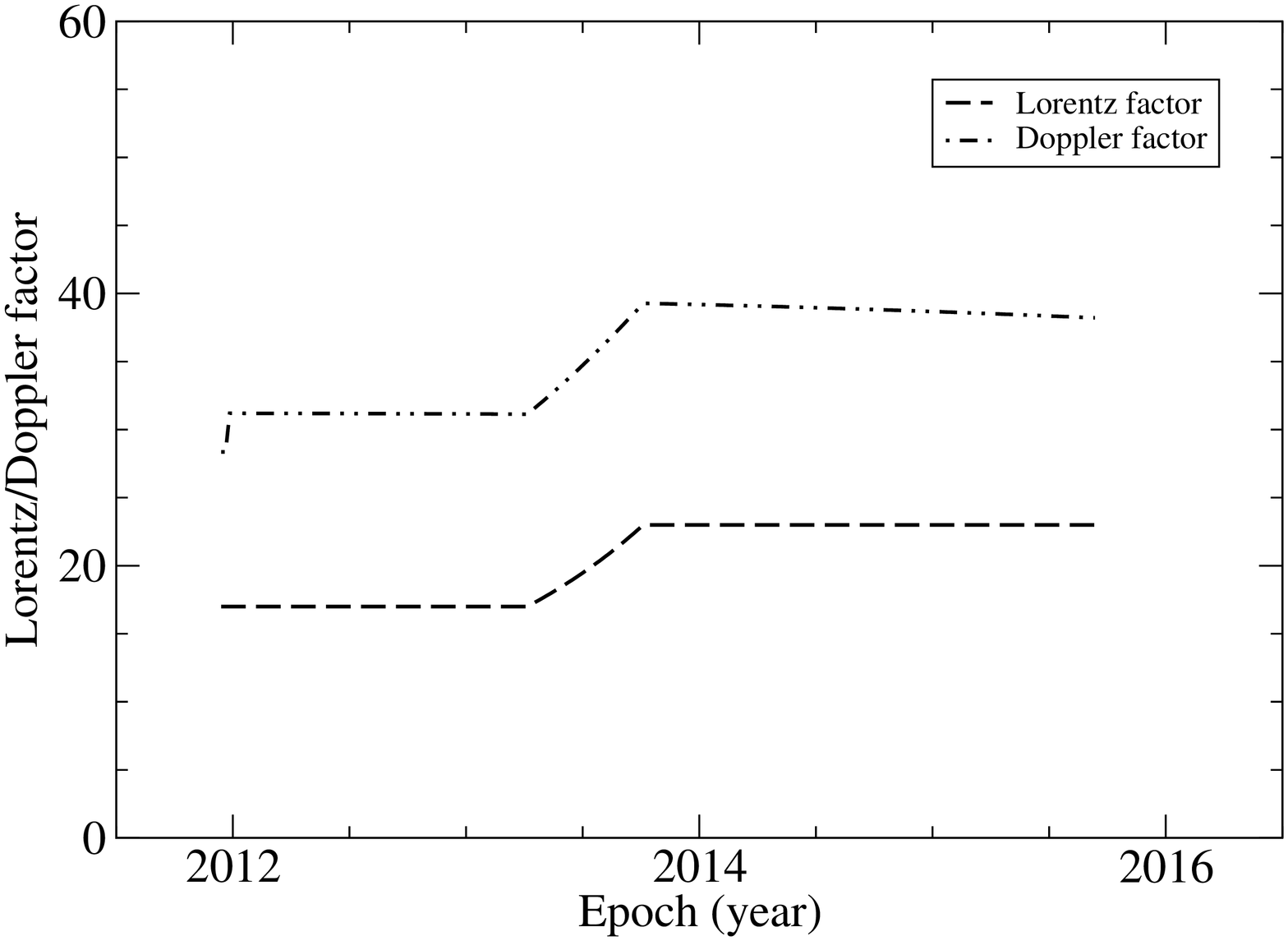}
   \caption{Knot B8: precession phase $\phi_0$(rad)=1.60+4$\pi$ and ejection 
   time $t_0$=2011.95. Model-fitting results: trajectory $Z_n(X_n)$, 
   coordinates $X_n(t)$ and $Z_n(t)$, core separation $r_n(t)$, modeled 
   apparent velocity $\beta_a$(t) and viewing angle $\theta$(t), bulk Lorentz
   factor$\Gamma(t)$ and Doppler factor $\delta(t)$. Its kinematics within core
    separation $r_n{\sim}$1.20\,mas could be well model-simulated in terms of
    the precessing nozzle scenario.}
   \end{figure*}
   \begin{figure*}
   \centering
   \includegraphics[width=6cm,angle=-90]{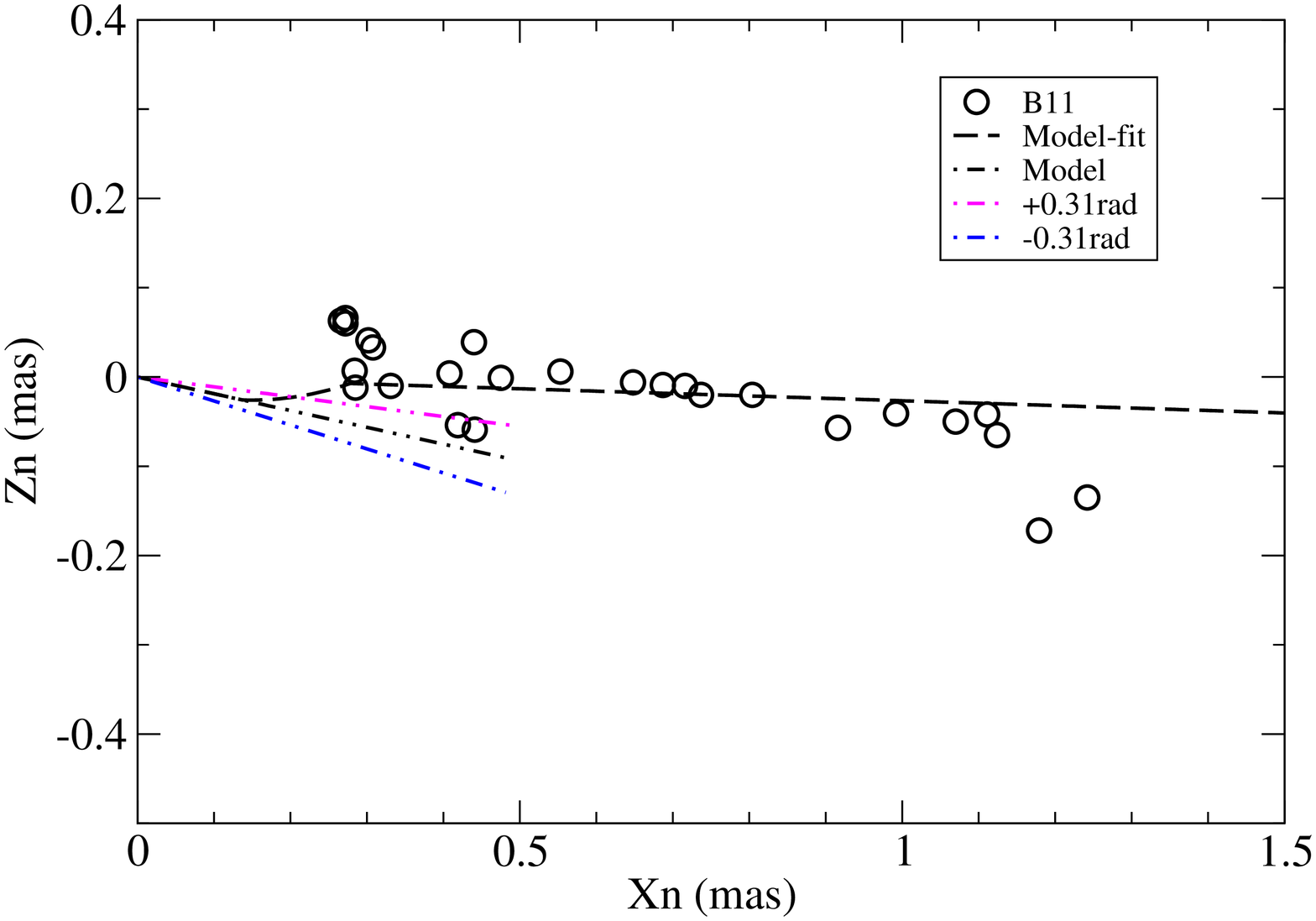}
   \includegraphics[width=6cm,angle=-90]{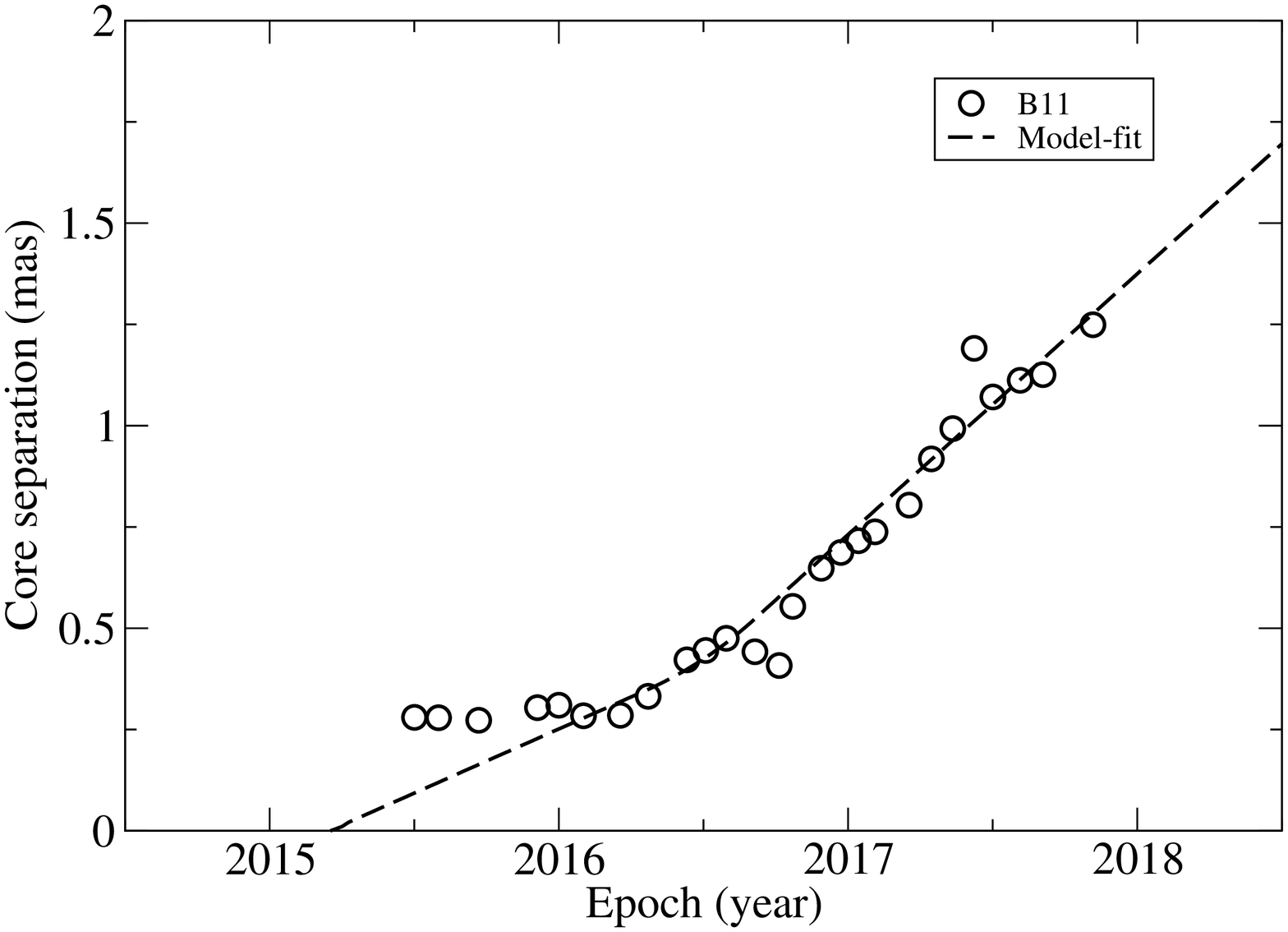}
   \includegraphics[width=6cm,angle=-90]{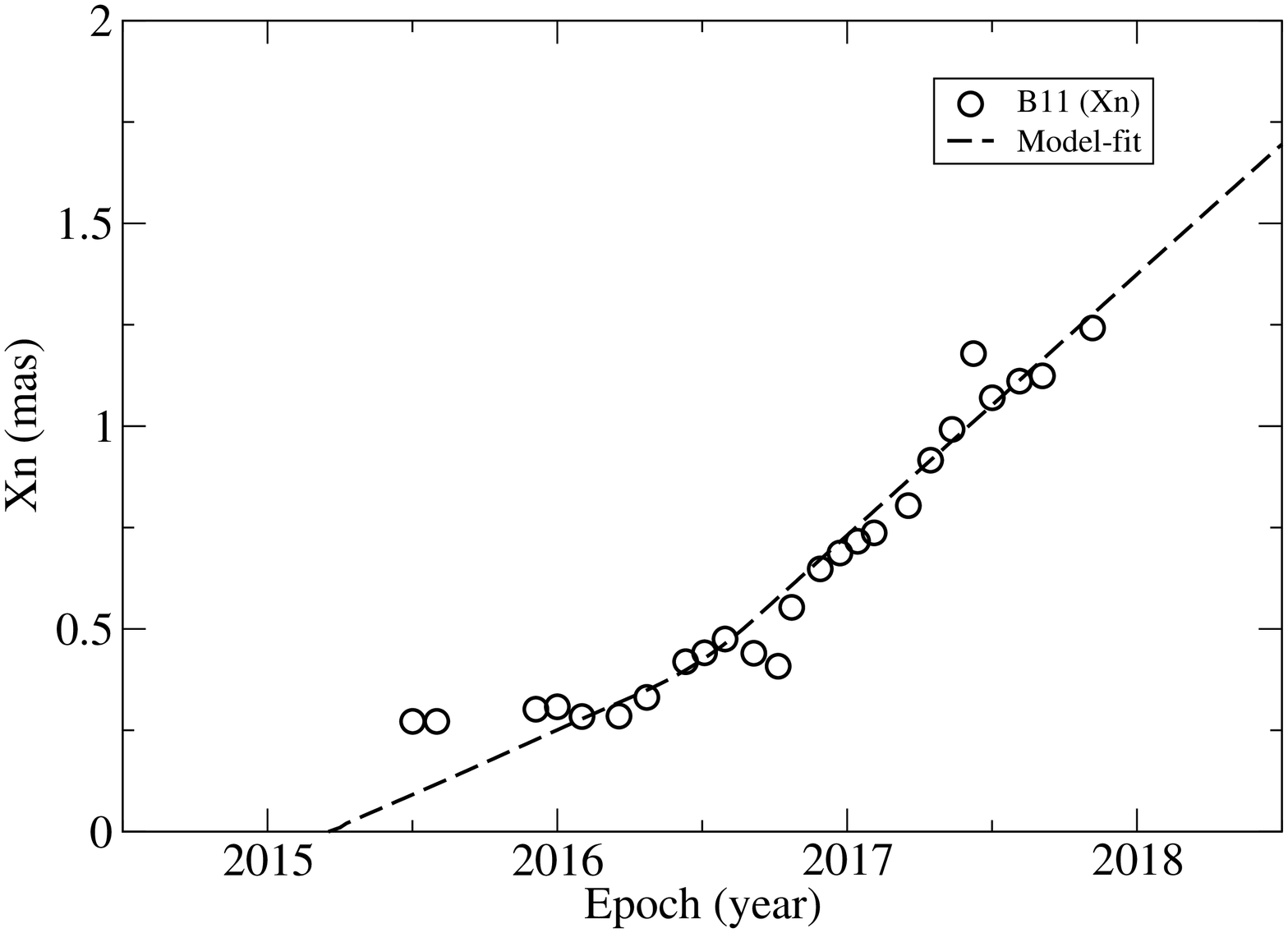}
   \includegraphics[width=6cm,angle=-90]{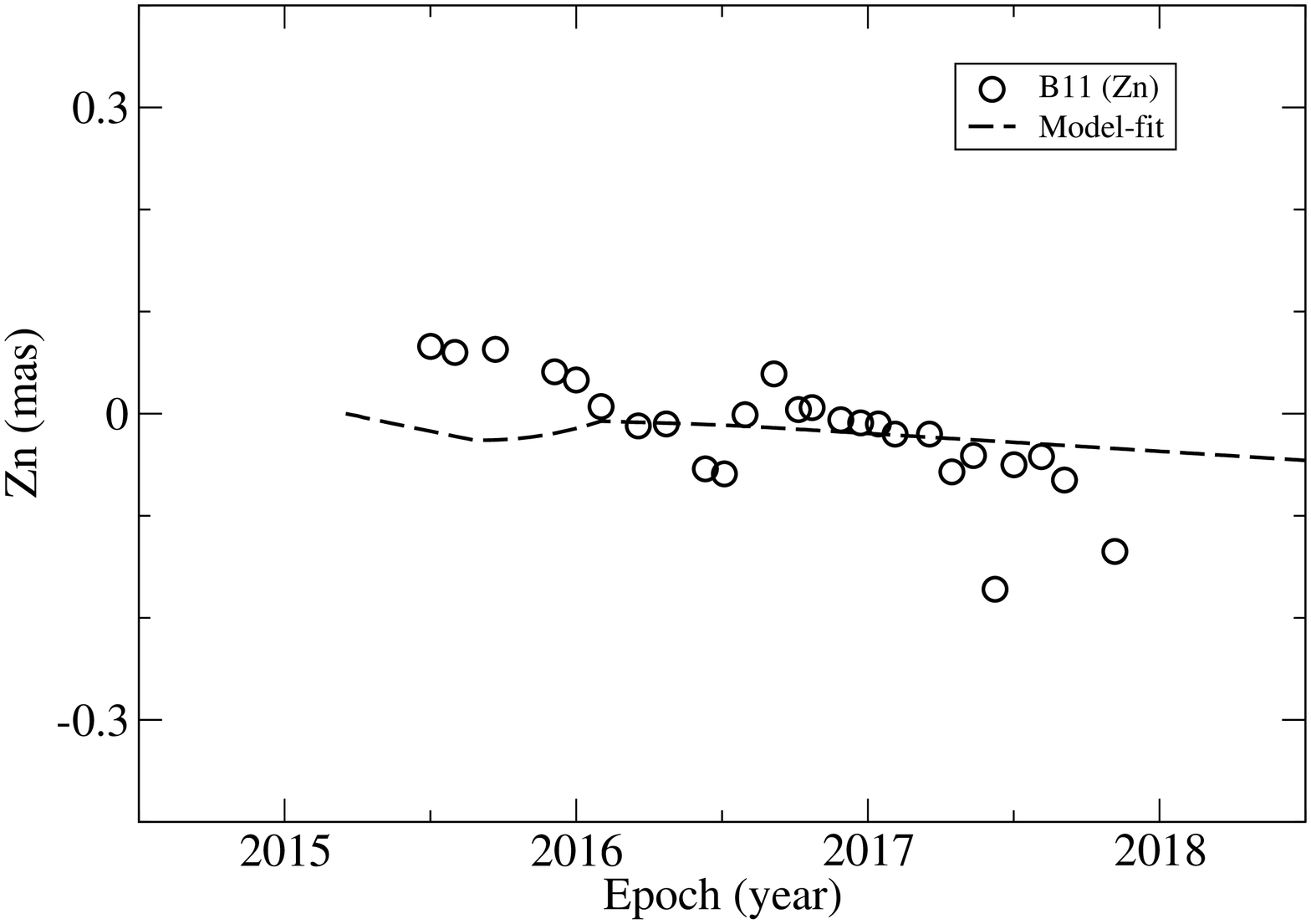}
   \includegraphics[width=6cm,angle=-90]{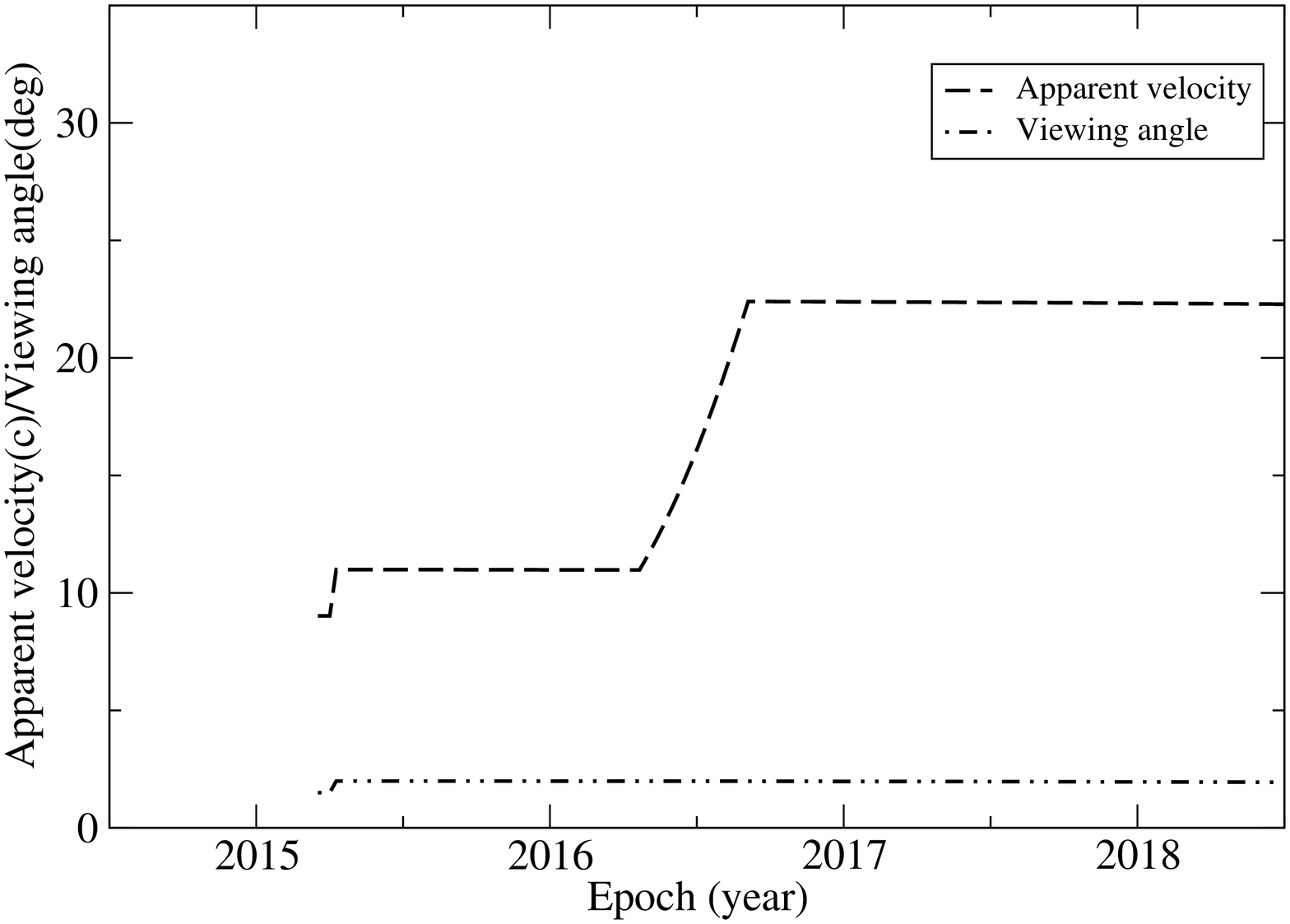}
   \includegraphics[width=6cm,angle=-90]{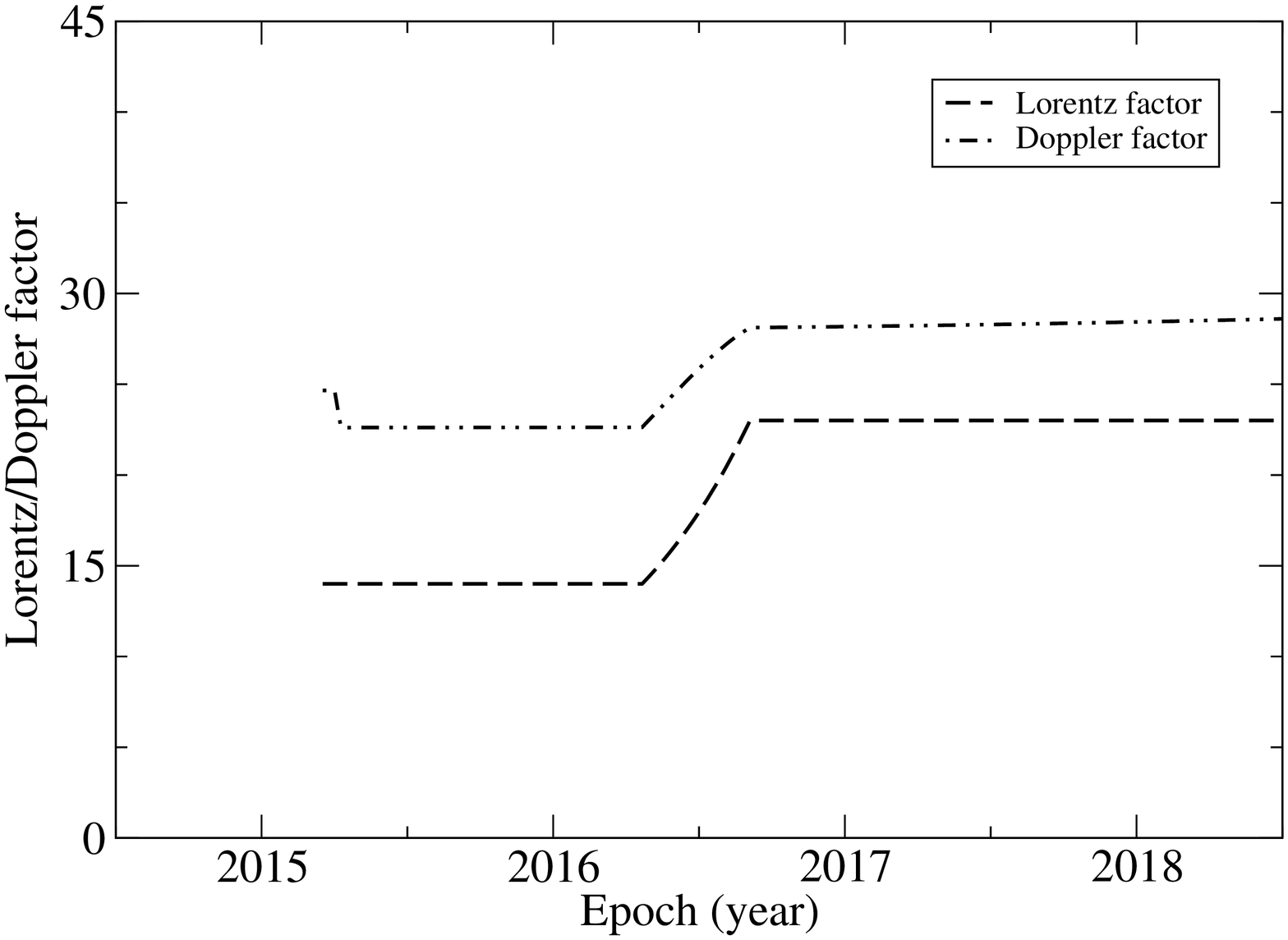}
   \caption{Knot B11: precession phase $\phi_0$(rad)=4.80+4$\pi$ and ejection 
   time $t_0$=2015.67. Model-fitting results: trajectory $Z_n(X_n)$, 
    coordinates $X_n(t)$ and $Z_n(t)$, core separation $r_n(t)$, modeled
    apparent velocity $\beta_a$(t) and viewing angle $\theta$(t), bulk Lorentz
   factor $\Gamma(t)$ and Doppler factor $\delta(t)$. Its innermost trajectory
   within core separation $r_n{\sim}$0.12\,mas was coincident with the 
   precessing common trajectory, but no observational data available there.
    Beyond $r_n{\sim}$0.12\,mas the trajectory was modeled by taking its 
   rotation into account.}
   \end{figure*}
  \end{appendix}
 \end{document}